\documentclass[apjl]{emulateapj}
\usepackage{comment}
\usepackage{ifthen}

\newcommand{\forloop}[5][1]%
{%
\setcounter{#2}{#3}%
\ifthenelse{#4}%
	{%
	#5%
	\addtocounter{#2}{#1}%
	\forloop[#1]{#2}{\value{#2}}{#4}{#5}%
	}%
	{%
	}%
}%

\newcommand{\ctbd}[1]{}

\newcommand{\lc}{light curve}
\newcommand{\lcs}{light curves}
\newcommand{\Lc}{Light curve}

\newcommand{\band}[1]{\ensuremath{#1}~band}

\newcommand{\masy}{\ensuremath{\rm mas\,yr^{-1}}}
\newcommand{\kms}{\ensuremath{\rm km\,s^{-1}}}
\newcommand{\ms}{\ensuremath{\rm m\,s^{-1}}}

\newcommand{\gcmc}{\ensuremath{\rm g\,cm^{-3}}}
\newcommand{\ergscmsq}{\ensuremath{\rm erg\,s^{-1}\,cm^{-2}}}

\newcommand{\teff}{\ensuremath{T_{\rm eff}}}

\newcommand{\vsini}{\ensuremath{v \sin{i}}}
\newcommand{\feh}{\ensuremath{\rm [Fe/H]}}

\newcommand{\rsun}{\ensuremath{R_\sun}}
\newcommand{\msun}{\ensuremath{M_\sun}}
\newcommand{\lsun}{\ensuremath{L_\sun}}

\newcommand{\rstar}{\ensuremath{R_\star}}
\newcommand{\mstar}{\ensuremath{M_\star}}
\newcommand{\lstar}{\ensuremath{L_\star}}

\newcommand{\teffstar}{\ensuremath{T_{\rm eff\star}}}
\newcommand{\rhostar}{\ensuremath{\rho_\star}}
\newcommand{\loggstar}{\ensuremath{\log{g_{\star}}}}

\newcommand{\rpl}{\ensuremath{R_{p}}}
\newcommand{\mpl}{\ensuremath{M_{p}}}

\newcommand{\rhopl}{\ensuremath{\rho_{p}}}

\newcommand{\arstar}{\ensuremath{a/\rstar}}
\newcommand{\zrstar}{\ensuremath{\zeta/\rstar}}

\newcommand{\rjup}{\ensuremath{R_{\rm J}}}
\newcommand{\mjup}{\ensuremath{M_{\rm J}}}

\newcommand{\refsecl}[1]{\mbox{Section \ref{sec:#1}}}

\newcommand{\hj}{hot Jupiter}

\newcommand{\hatcurhtrCandxxxxxA}{HATS578-002}
\newcommand{\hatcurCCraCandxxxxxA}{\ensuremath{19^{\mathrm h}12^{\mathrm m}51.48{\mathrm s}}}
\newcommand{\hatcurCCdecCandxxxxxA}{\ensuremath{-19^{\arcdeg}12^{\arcmin}47.3{\arcsec}}}
\newcommand{\hatcurCLASSCandxxxxxA}{EB}
\newcommand{\hatcurCCEPICCandxxxxxA}{EPIC~218210199}
\newcommand{\hatcurLCPCandxxxxxA}{\ensuremath{1.9858275\pm0.0000019}} 
\newcommand{\hatcurLCTCandxxxxxA}{\ensuremath{2457170.51503\pm0.00047}} 
\newcommand{\hatcurLCdurCandxxxxxA}{\ensuremath{0.09642\pm0.00726}} 
\newcommand{\hatcurLCrprstarCandxxxxxA}{\ensuremath{0.11261\pm0.00155}} 
\newcommand{\hatcurLCbsqCandxxxxxA}{\ensuremath{0.770\pm0.029}} 
\newcommand{\hatcurCCmagBCandxxxxxA}{\ensuremath{14.704}} 
\newcommand{\hatcurCCmagVCandxxxxxA}{\ensuremath{14.039}} 
\newcommand{\hatcurCCmaggCandxxxxxA}{\ensuremath{14.317}} 
\newcommand{\hatcurCCmagrCandxxxxxA}{\ensuremath{13.872}} 
\newcommand{\hatcurCCmagiCandxxxxxA}{\ensuremath{13.649}} 
\newcommand{\hatcurCCmagJCandxxxxxA}{\ensuremath{12.720\pm0.02}} 
\newcommand{\hatcurCCmagHCandxxxxxA}{\ensuremath{12.410\pm0.02}} 
\newcommand{\hatcurCCmagKCandxxxxxA}{\ensuremath{12.314\pm0.03}} 
\newcommand{\hatcurSPECWiFeSTeffCandxxxxxA}{\ensuremath{6014\pm300}} 
\newcommand{\hatcurSPECWiFeSloggCandxxxxxA}{\ensuremath{3.10\pm0.30}} 
\newcommand{\hatcurSPECWiFeSzfehCandxxxxxA}{\ensuremath{-0.5\pm0.5}} 
\newcommand{\hatcurSPECWiFeSNlowresCandxxxxxA}{\ensuremath{1}} 
\newcommand{\hatcurSPECWiFeSNmedresCandxxxxxA}{\ensuremath{3}} 
\newcommand{\hatcurSPECWiFeSsnrangelowresCandxxxxxA}{139.0} 
\newcommand{\hatcurSPECWiFeSsnrangemedresCandxxxxxA}{31.1--59.5} 
\newcommand{\hatcurSPECWiFeSdaterangelowresCandxxxxxA}{2012-09-08} 
\newcommand{\hatcurSPECWiFeSdaterangemedresCandxxxxxA}{2012-09-24--2012-09-25} 
\newcommand{\hatcurSPECWiFeSRVKCandxxxxxA}{\ensuremath{-16.64\pm0.63}} 
\newcommand{\hatcurSPECWiFeSRVGCandxxxxxA}{\ensuremath{-10.06\pm0.62}} 
\newcommand{\hatcurSPECarcesTeffCandxxxxxA}{\ensuremath{NULL}} 
\newcommand{\hatcurSPECarcesloggCandxxxxxA}{\ensuremath{NULL}} 
\newcommand{\hatcurSPECarceszfehCandxxxxxA}{\ensuremath{NULL}} 
\newcommand{\hatcurSPECarcesvsiniCandxxxxxA}{\ensuremath{NULL}} 
\newcommand{\hatcurSPECarcesNCandxxxxxA}{\ensuremath{0}} 
\newcommand{\hatcurSPECarcessnrangeCandxxxxxA}{NULL} 
\newcommand{\hatcurSPECarcesdaterangelowresCandxxxxxA}{NULL} 
\newcommand{\hatcurSPECarcesRVKCandxxxxxA}{\ensuremath{NULL}} 
\newcommand{\hatcurSPECarcesRVGCandxxxxxA}{\ensuremath{NULL}} 
\newcommand{\hatcurSPECdupontTeffCandxxxxxA}{\ensuremath{NULL}} 
\newcommand{\hatcurSPECdupontloggCandxxxxxA}{\ensuremath{NULL}} 
\newcommand{\hatcurSPECdupontzfehCandxxxxxA}{\ensuremath{NULL}} 
\newcommand{\hatcurSPECdupontvsiniCandxxxxxA}{\ensuremath{NULL}} 
\newcommand{\hatcurSPECdupontNCandxxxxxA}{\ensuremath{0}} 
\newcommand{\hatcurSPECdupontsnrangeCandxxxxxA}{NULL} 
\newcommand{\hatcurSPECdupontdaterangelowresCandxxxxxA}{NULL} 
\newcommand{\hatcurSPECdupontRVKCandxxxxxA}{\ensuremath{NULL}} 
\newcommand{\hatcurSPECdupontRVGCandxxxxxA}{\ensuremath{NULL}} 
\newcommand{\hatcurSPECharpsTeffCandxxxxxA}{\ensuremath{NULL}} 
\newcommand{\hatcurSPECharpsloggCandxxxxxA}{\ensuremath{NULL}} 
\newcommand{\hatcurSPECharpszfehCandxxxxxA}{\ensuremath{NULL}} 
\newcommand{\hatcurSPECharpsvsiniCandxxxxxA}{\ensuremath{NULL}} 
\newcommand{\hatcurSPECharpsNCandxxxxxA}{\ensuremath{0}} 
\newcommand{\hatcurSPECharpssnrangeCandxxxxxA}{NULL} 
\newcommand{\hatcurSPECharpsdaterangelowresCandxxxxxA}{NULL} 
\newcommand{\hatcurSPECharpsRVKCandxxxxxA}{\ensuremath{NULL}} 
\newcommand{\hatcurSPECharpsRVGCandxxxxxA}{\ensuremath{NULL}} 
\newcommand{\hatcurSPECpfsTeffCandxxxxxA}{\ensuremath{NULL}} 
\newcommand{\hatcurSPECpfsloggCandxxxxxA}{\ensuremath{NULL}} 
\newcommand{\hatcurSPECpfszfehCandxxxxxA}{\ensuremath{NULL}} 
\newcommand{\hatcurSPECpfsvsiniCandxxxxxA}{\ensuremath{NULL}} 
\newcommand{\hatcurSPECpfsNCandxxxxxA}{\ensuremath{0}} 
\newcommand{\hatcurSPECpfssnrangeCandxxxxxA}{NULL} 
\newcommand{\hatcurSPECpfsdaterangelowresCandxxxxxA}{NULL} 
\newcommand{\hatcurSPECpfsRVKCandxxxxxA}{\ensuremath{NULL}} 
\newcommand{\hatcurSPECpfsRVGCandxxxxxA}{\ensuremath{NULL}} 
\newcommand{\hatcurSPECferosTeffCandxxxxxA}{\ensuremath{NULL}} 
\newcommand{\hatcurSPECferosloggCandxxxxxA}{\ensuremath{NULL}} 
\newcommand{\hatcurSPECferoszfehCandxxxxxA}{\ensuremath{NULL}} 
\newcommand{\hatcurSPECferosvsiniCandxxxxxA}{\ensuremath{NULL}} 
\newcommand{\hatcurSPECferosNCandxxxxxA}{\ensuremath{0}} 
\newcommand{\hatcurSPECferossnrangeCandxxxxxA}{NULL} 
\newcommand{\hatcurSPECferosdaterangelowresCandxxxxxA}{NULL} 
\newcommand{\hatcurSPECferosRVKCandxxxxxA}{\ensuremath{NULL}} 
\newcommand{\hatcurSPECferosRVGCandxxxxxA}{\ensuremath{NULL}} 
\newcommand{\hatcurSPECfiesTeffCandxxxxxA}{\ensuremath{NULL}} 
\newcommand{\hatcurSPECfiesloggCandxxxxxA}{\ensuremath{NULL}} 
\newcommand{\hatcurSPECfieszfehCandxxxxxA}{\ensuremath{NULL}} 
\newcommand{\hatcurSPECfiesvsiniCandxxxxxA}{\ensuremath{NULL}} 
\newcommand{\hatcurSPECfiesNCandxxxxxA}{\ensuremath{0}} 
\newcommand{\hatcurSPECfiessnrangeCandxxxxxA}{NULL} 
\newcommand{\hatcurSPECfiesdaterangelowresCandxxxxxA}{NULL} 
\newcommand{\hatcurSPECfiesRVKCandxxxxxA}{\ensuremath{NULL}} 
\newcommand{\hatcurSPECfiesRVGCandxxxxxA}{\ensuremath{NULL}} 
\newcommand{\hatcurSPECcoralieTeffCandxxxxxA}{\ensuremath{NULL}} 
\newcommand{\hatcurSPECcoralieloggCandxxxxxA}{\ensuremath{NULL}} 
\newcommand{\hatcurSPECcoraliezfehCandxxxxxA}{\ensuremath{NULL}} 
\newcommand{\hatcurSPECcoralievsiniCandxxxxxA}{\ensuremath{NULL}} 
\newcommand{\hatcurSPECcoralieNCandxxxxxA}{\ensuremath{0}} 
\newcommand{\hatcurSPECcoraliesnrangeCandxxxxxA}{NULL} 
\newcommand{\hatcurSPECcoraliedaterangelowresCandxxxxxA}{NULL} 
\newcommand{\hatcurSPECcoralieRVKCandxxxxxA}{\ensuremath{NULL}} 
\newcommand{\hatcurSPECcoralieRVGCandxxxxxA}{\ensuremath{NULL}} 
\newcommand{\hatcurhtrCandxxxxxB}{HATS578-003}
\newcommand{\hatcurCCraCandxxxxxB}{\ensuremath{19^{\mathrm h}12^{\mathrm m}55.08{\mathrm s}}}
\newcommand{\hatcurCCdecCandxxxxxB}{\ensuremath{-20^{\arcdeg}56^{\arcmin}22.0{\arcsec}}}
\newcommand{\hatcurCLASSCandxxxxxB}{EB}
\newcommand{\hatcurCCEPICCandxxxxxB}{EPIC~217231249}
\newcommand{\hatcurLCPCandxxxxxB}{\ensuremath{4.8332040\pm0.0000091}} 
\newcommand{\hatcurLCTCandxxxxxB}{\ensuremath{2456589.65837\pm0.00149}} 
\newcommand{\hatcurLCdurCandxxxxxB}{\ensuremath{0.21162\pm0.01695}} 
\newcommand{\hatcurLCrprstarCandxxxxxB}{\ensuremath{0.13627\pm0.00202}} 
\newcommand{\hatcurLCbsqCandxxxxxB}{\ensuremath{0.448\pm0.144}} 
\newcommand{\hatcurCCmagBCandxxxxxB}{\ensuremath{14.842}} 
\newcommand{\hatcurCCmagVCandxxxxxB}{\ensuremath{14.079}} 
\newcommand{\hatcurCCmaggCandxxxxxB}{\ensuremath{14.425}} 
\newcommand{\hatcurCCmagrCandxxxxxB}{\ensuremath{13.881}} 
\newcommand{\hatcurCCmagiCandxxxxxB}{\ensuremath{13.583}} 
\newcommand{\hatcurCCmagJCandxxxxxB}{\ensuremath{12.677\pm0.02}} 
\newcommand{\hatcurCCmagHCandxxxxxB}{\ensuremath{12.355\pm0.02}} 
\newcommand{\hatcurCCmagKCandxxxxxB}{\ensuremath{12.208\pm0.02}} 
\newcommand{\hatcurSPECWiFeSTeffCandxxxxxB}{\ensuremath{5928\pm300}} 
\newcommand{\hatcurSPECWiFeSloggCandxxxxxB}{\ensuremath{3.50\pm0.30}} 
\newcommand{\hatcurSPECWiFeSzfehCandxxxxxB}{\ensuremath{0.0\pm0.5}} 
\newcommand{\hatcurSPECWiFeSNlowresCandxxxxxB}{\ensuremath{1}} 
\newcommand{\hatcurSPECWiFeSNmedresCandxxxxxB}{\ensuremath{2}} 
\newcommand{\hatcurSPECWiFeSsnrangelowresCandxxxxxB}{115.0} 
\newcommand{\hatcurSPECWiFeSsnrangemedresCandxxxxxB}{53.2--53.4} 
\newcommand{\hatcurSPECWiFeSdaterangelowresCandxxxxxB}{2012-09-08} 
\newcommand{\hatcurSPECWiFeSdaterangemedresCandxxxxxB}{2012-09-24--2012-09-25} 
\newcommand{\hatcurSPECWiFeSRVKCandxxxxxB}{\ensuremath{33.04\pm1.41}} 
\newcommand{\hatcurSPECWiFeSRVGCandxxxxxB}{\ensuremath{16.30\pm0.60}} 
\newcommand{\hatcurSPECarcesTeffCandxxxxxB}{\ensuremath{NULL}} 
\newcommand{\hatcurSPECarcesloggCandxxxxxB}{\ensuremath{NULL}} 
\newcommand{\hatcurSPECarceszfehCandxxxxxB}{\ensuremath{NULL}} 
\newcommand{\hatcurSPECarcesvsiniCandxxxxxB}{\ensuremath{NULL}} 
\newcommand{\hatcurSPECarcesNCandxxxxxB}{\ensuremath{0}} 
\newcommand{\hatcurSPECarcessnrangeCandxxxxxB}{NULL} 
\newcommand{\hatcurSPECarcesdaterangelowresCandxxxxxB}{NULL} 
\newcommand{\hatcurSPECarcesRVKCandxxxxxB}{\ensuremath{NULL}} 
\newcommand{\hatcurSPECarcesRVGCandxxxxxB}{\ensuremath{NULL}} 
\newcommand{\hatcurSPECdupontTeffCandxxxxxB}{\ensuremath{NULL}} 
\newcommand{\hatcurSPECdupontloggCandxxxxxB}{\ensuremath{NULL}} 
\newcommand{\hatcurSPECdupontzfehCandxxxxxB}{\ensuremath{NULL}} 
\newcommand{\hatcurSPECdupontvsiniCandxxxxxB}{\ensuremath{NULL}} 
\newcommand{\hatcurSPECdupontNCandxxxxxB}{\ensuremath{0}} 
\newcommand{\hatcurSPECdupontsnrangeCandxxxxxB}{NULL} 
\newcommand{\hatcurSPECdupontdaterangelowresCandxxxxxB}{NULL} 
\newcommand{\hatcurSPECdupontRVKCandxxxxxB}{\ensuremath{NULL}} 
\newcommand{\hatcurSPECdupontRVGCandxxxxxB}{\ensuremath{NULL}} 
\newcommand{\hatcurSPECharpsTeffCandxxxxxB}{\ensuremath{NULL}} 
\newcommand{\hatcurSPECharpsloggCandxxxxxB}{\ensuremath{NULL}} 
\newcommand{\hatcurSPECharpszfehCandxxxxxB}{\ensuremath{NULL}} 
\newcommand{\hatcurSPECharpsvsiniCandxxxxxB}{\ensuremath{NULL}} 
\newcommand{\hatcurSPECharpsNCandxxxxxB}{\ensuremath{0}} 
\newcommand{\hatcurSPECharpssnrangeCandxxxxxB}{NULL} 
\newcommand{\hatcurSPECharpsdaterangelowresCandxxxxxB}{NULL} 
\newcommand{\hatcurSPECharpsRVKCandxxxxxB}{\ensuremath{NULL}} 
\newcommand{\hatcurSPECharpsRVGCandxxxxxB}{\ensuremath{NULL}} 
\newcommand{\hatcurSPECpfsTeffCandxxxxxB}{\ensuremath{NULL}} 
\newcommand{\hatcurSPECpfsloggCandxxxxxB}{\ensuremath{NULL}} 
\newcommand{\hatcurSPECpfszfehCandxxxxxB}{\ensuremath{NULL}} 
\newcommand{\hatcurSPECpfsvsiniCandxxxxxB}{\ensuremath{NULL}} 
\newcommand{\hatcurSPECpfsNCandxxxxxB}{\ensuremath{0}} 
\newcommand{\hatcurSPECpfssnrangeCandxxxxxB}{NULL} 
\newcommand{\hatcurSPECpfsdaterangelowresCandxxxxxB}{NULL} 
\newcommand{\hatcurSPECpfsRVKCandxxxxxB}{\ensuremath{NULL}} 
\newcommand{\hatcurSPECpfsRVGCandxxxxxB}{\ensuremath{NULL}} 
\newcommand{\hatcurSPECferosTeffCandxxxxxB}{\ensuremath{NULL}} 
\newcommand{\hatcurSPECferosloggCandxxxxxB}{\ensuremath{NULL}} 
\newcommand{\hatcurSPECferoszfehCandxxxxxB}{\ensuremath{NULL}} 
\newcommand{\hatcurSPECferosvsiniCandxxxxxB}{\ensuremath{NULL}} 
\newcommand{\hatcurSPECferosNCandxxxxxB}{\ensuremath{0}} 
\newcommand{\hatcurSPECferossnrangeCandxxxxxB}{NULL} 
\newcommand{\hatcurSPECferosdaterangelowresCandxxxxxB}{NULL} 
\newcommand{\hatcurSPECferosRVKCandxxxxxB}{\ensuremath{NULL}} 
\newcommand{\hatcurSPECferosRVGCandxxxxxB}{\ensuremath{NULL}} 
\newcommand{\hatcurSPECfiesTeffCandxxxxxB}{\ensuremath{NULL}} 
\newcommand{\hatcurSPECfiesloggCandxxxxxB}{\ensuremath{NULL}} 
\newcommand{\hatcurSPECfieszfehCandxxxxxB}{\ensuremath{NULL}} 
\newcommand{\hatcurSPECfiesvsiniCandxxxxxB}{\ensuremath{NULL}} 
\newcommand{\hatcurSPECfiesNCandxxxxxB}{\ensuremath{0}} 
\newcommand{\hatcurSPECfiessnrangeCandxxxxxB}{NULL} 
\newcommand{\hatcurSPECfiesdaterangelowresCandxxxxxB}{NULL} 
\newcommand{\hatcurSPECfiesRVKCandxxxxxB}{\ensuremath{NULL}} 
\newcommand{\hatcurSPECfiesRVGCandxxxxxB}{\ensuremath{NULL}} 
\newcommand{\hatcurSPECcoralieTeffCandxxxxxB}{\ensuremath{NULL}} 
\newcommand{\hatcurSPECcoralieloggCandxxxxxB}{\ensuremath{NULL}} 
\newcommand{\hatcurSPECcoraliezfehCandxxxxxB}{\ensuremath{NULL}} 
\newcommand{\hatcurSPECcoralievsiniCandxxxxxB}{\ensuremath{NULL}} 
\newcommand{\hatcurSPECcoralieNCandxxxxxB}{\ensuremath{0}} 
\newcommand{\hatcurSPECcoraliesnrangeCandxxxxxB}{NULL} 
\newcommand{\hatcurSPECcoraliedaterangelowresCandxxxxxB}{NULL} 
\newcommand{\hatcurSPECcoralieRVKCandxxxxxB}{\ensuremath{NULL}} 
\newcommand{\hatcurSPECcoralieRVGCandxxxxxB}{\ensuremath{NULL}} 
\newcommand{\hatcurhtrCandxxxxxC}{HATS578-004}
\newcommand{\hatcurCCraCandxxxxxC}{\ensuremath{19^{\mathrm h}13^{\mathrm m}20.64{\mathrm s}}}
\newcommand{\hatcurCCdecCandxxxxxC}{\ensuremath{-22^{\arcdeg}05^{\arcmin}40.4{\arcsec}}}
\newcommand{\hatcurCLASSCandxxxxxC}{BEB}
\newcommand{\hatcurCCEPICCandxxxxxC}{EPIC~216579956}
\newcommand{\hatcurLCPCandxxxxxC}{\ensuremath{0.7057527\pm0.0000063}} 
\newcommand{\hatcurLCTCandxxxxxC}{\ensuremath{2457332.86183\pm0.00106}} 
\newcommand{\hatcurLCdurCandxxxxxC}{\ensuremath{0.06691\pm0.00957}} 
\newcommand{\hatcurLCrprstarCandxxxxxC}{\ensuremath{0.04817\pm0.00197}} 
\newcommand{\hatcurLCbsqCandxxxxxC}{\ensuremath{0.905\pm0.085}} 
\newcommand{\hatcurCCmagBCandxxxxxC}{\ensuremath{15.999}} 
\newcommand{\hatcurCCmagVCandxxxxxC}{\ensuremath{15.464}} 
\newcommand{\hatcurCCmaggCandxxxxxC}{\ensuremath{15.739}} 
\newcommand{\hatcurCCmagrCandxxxxxC}{\ensuremath{15.217}} 
\newcommand{\hatcurCCmagiCandxxxxxC}{\ensuremath{14.950}} 
\newcommand{\hatcurCCmagJCandxxxxxC}{\ensuremath{14.431\pm0.03}} 
\newcommand{\hatcurCCmagHCandxxxxxC}{\ensuremath{14.122\pm0.04}} 
\newcommand{\hatcurCCmagKCandxxxxxC}{\ensuremath{14.021\pm0.07}} 
\newcommand{\hatcurSPECWiFeSTeffCandxxxxxC}{\ensuremath{5728\pm300}} 
\newcommand{\hatcurSPECWiFeSloggCandxxxxxC}{\ensuremath{3.50\pm0.30}} 
\newcommand{\hatcurSPECWiFeSzfehCandxxxxxC}{\ensuremath{-1.0\pm0.5}} 
\newcommand{\hatcurSPECWiFeSNlowresCandxxxxxC}{\ensuremath{1}} 
\newcommand{\hatcurSPECWiFeSNmedresCandxxxxxC}{\ensuremath{2}} 
\newcommand{\hatcurSPECWiFeSsnrangelowresCandxxxxxC}{26.0} 
\newcommand{\hatcurSPECWiFeSsnrangemedresCandxxxxxC}{7.7--9.4} 
\newcommand{\hatcurSPECWiFeSdaterangelowresCandxxxxxC}{2016-03-14} 
\newcommand{\hatcurSPECWiFeSdaterangemedresCandxxxxxC}{2016-03-14--2016-03-25} 
\newcommand{\hatcurSPECWiFeSRVKCandxxxxxC}{\ensuremath{<2.0}} 
\newcommand{\hatcurSPECWiFeSRVGCandxxxxxC}{\ensuremath{-9.42\pm2.79}} 
\newcommand{\hatcurSPECarcesTeffCandxxxxxC}{\ensuremath{NULL}} 
\newcommand{\hatcurSPECarcesloggCandxxxxxC}{\ensuremath{NULL}} 
\newcommand{\hatcurSPECarceszfehCandxxxxxC}{\ensuremath{NULL}} 
\newcommand{\hatcurSPECarcesvsiniCandxxxxxC}{\ensuremath{NULL}} 
\newcommand{\hatcurSPECarcesNCandxxxxxC}{\ensuremath{0}} 
\newcommand{\hatcurSPECarcessnrangeCandxxxxxC}{NULL} 
\newcommand{\hatcurSPECarcesdaterangelowresCandxxxxxC}{NULL} 
\newcommand{\hatcurSPECarcesRVKCandxxxxxC}{\ensuremath{NULL}} 
\newcommand{\hatcurSPECarcesRVGCandxxxxxC}{\ensuremath{NULL}} 
\newcommand{\hatcurSPECdupontTeffCandxxxxxC}{\ensuremath{NULL}} 
\newcommand{\hatcurSPECdupontloggCandxxxxxC}{\ensuremath{NULL}} 
\newcommand{\hatcurSPECdupontzfehCandxxxxxC}{\ensuremath{NULL}} 
\newcommand{\hatcurSPECdupontvsiniCandxxxxxC}{\ensuremath{NULL}} 
\newcommand{\hatcurSPECdupontNCandxxxxxC}{\ensuremath{0}} 
\newcommand{\hatcurSPECdupontsnrangeCandxxxxxC}{NULL} 
\newcommand{\hatcurSPECdupontdaterangelowresCandxxxxxC}{NULL} 
\newcommand{\hatcurSPECdupontRVKCandxxxxxC}{\ensuremath{NULL}} 
\newcommand{\hatcurSPECdupontRVGCandxxxxxC}{\ensuremath{NULL}} 
\newcommand{\hatcurSPECharpsTeffCandxxxxxC}{\ensuremath{NULL}} 
\newcommand{\hatcurSPECharpsloggCandxxxxxC}{\ensuremath{NULL}} 
\newcommand{\hatcurSPECharpszfehCandxxxxxC}{\ensuremath{NULL}} 
\newcommand{\hatcurSPECharpsvsiniCandxxxxxC}{\ensuremath{NULL}} 
\newcommand{\hatcurSPECharpsNCandxxxxxC}{\ensuremath{0}} 
\newcommand{\hatcurSPECharpssnrangeCandxxxxxC}{NULL} 
\newcommand{\hatcurSPECharpsdaterangelowresCandxxxxxC}{NULL} 
\newcommand{\hatcurSPECharpsRVKCandxxxxxC}{\ensuremath{NULL}} 
\newcommand{\hatcurSPECharpsRVGCandxxxxxC}{\ensuremath{NULL}} 
\newcommand{\hatcurSPECpfsTeffCandxxxxxC}{\ensuremath{NULL}} 
\newcommand{\hatcurSPECpfsloggCandxxxxxC}{\ensuremath{NULL}} 
\newcommand{\hatcurSPECpfszfehCandxxxxxC}{\ensuremath{NULL}} 
\newcommand{\hatcurSPECpfsvsiniCandxxxxxC}{\ensuremath{NULL}} 
\newcommand{\hatcurSPECpfsNCandxxxxxC}{\ensuremath{0}} 
\newcommand{\hatcurSPECpfssnrangeCandxxxxxC}{NULL} 
\newcommand{\hatcurSPECpfsdaterangelowresCandxxxxxC}{NULL} 
\newcommand{\hatcurSPECpfsRVKCandxxxxxC}{\ensuremath{NULL}} 
\newcommand{\hatcurSPECpfsRVGCandxxxxxC}{\ensuremath{NULL}} 
\newcommand{\hatcurSPECferosTeffCandxxxxxC}{\ensuremath{NULL}} 
\newcommand{\hatcurSPECferosloggCandxxxxxC}{\ensuremath{NULL}} 
\newcommand{\hatcurSPECferoszfehCandxxxxxC}{\ensuremath{NULL}} 
\newcommand{\hatcurSPECferosvsiniCandxxxxxC}{\ensuremath{NULL}} 
\newcommand{\hatcurSPECferosNCandxxxxxC}{\ensuremath{0}} 
\newcommand{\hatcurSPECferossnrangeCandxxxxxC}{NULL} 
\newcommand{\hatcurSPECferosdaterangelowresCandxxxxxC}{NULL} 
\newcommand{\hatcurSPECferosRVKCandxxxxxC}{\ensuremath{NULL}} 
\newcommand{\hatcurSPECferosRVGCandxxxxxC}{\ensuremath{NULL}} 
\newcommand{\hatcurSPECfiesTeffCandxxxxxC}{\ensuremath{NULL}} 
\newcommand{\hatcurSPECfiesloggCandxxxxxC}{\ensuremath{NULL}} 
\newcommand{\hatcurSPECfieszfehCandxxxxxC}{\ensuremath{NULL}} 
\newcommand{\hatcurSPECfiesvsiniCandxxxxxC}{\ensuremath{NULL}} 
\newcommand{\hatcurSPECfiesNCandxxxxxC}{\ensuremath{0}} 
\newcommand{\hatcurSPECfiessnrangeCandxxxxxC}{NULL} 
\newcommand{\hatcurSPECfiesdaterangelowresCandxxxxxC}{NULL} 
\newcommand{\hatcurSPECfiesRVKCandxxxxxC}{\ensuremath{NULL}} 
\newcommand{\hatcurSPECfiesRVGCandxxxxxC}{\ensuremath{NULL}} 
\newcommand{\hatcurSPECcoralieTeffCandxxxxxC}{\ensuremath{NULL}} 
\newcommand{\hatcurSPECcoralieloggCandxxxxxC}{\ensuremath{NULL}} 
\newcommand{\hatcurSPECcoraliezfehCandxxxxxC}{\ensuremath{NULL}} 
\newcommand{\hatcurSPECcoralievsiniCandxxxxxC}{\ensuremath{NULL}} 
\newcommand{\hatcurSPECcoralieNCandxxxxxC}{\ensuremath{0}} 
\newcommand{\hatcurSPECcoraliesnrangeCandxxxxxC}{NULL} 
\newcommand{\hatcurSPECcoraliedaterangelowresCandxxxxxC}{NULL} 
\newcommand{\hatcurSPECcoralieRVKCandxxxxxC}{\ensuremath{NULL}} 
\newcommand{\hatcurSPECcoralieRVGCandxxxxxC}{\ensuremath{NULL}} 
\newcommand{\hatcurhtrCandxxxxxD}{HATS579-001}
\newcommand{\hatcurCCraCandxxxxxD}{\ensuremath{19^{\mathrm h}21^{\mathrm m}10.44{\mathrm s}}}
\newcommand{\hatcurCCdecCandxxxxxD}{\ensuremath{-25^{\arcdeg}57^{\arcmin}59.8{\arcsec}}}
\newcommand{\hatcurCLASSCandxxxxxD}{EB}
\newcommand{\hatcurCCEPICCandxxxxxD}{EPIC~214652580}
\newcommand{\hatcurLCPCandxxxxxD}{\ensuremath{8.9120143\pm0.0000230}} 
\newcommand{\hatcurLCTCandxxxxxD}{\ensuremath{2457320.69926\pm0.00047}} 
\newcommand{\hatcurLCdurCandxxxxxD}{\ensuremath{0.16508\pm0.00803}} 
\newcommand{\hatcurLCrprstarCandxxxxxD}{\ensuremath{0.14571\pm0.00100}} 
\newcommand{\hatcurLCbsqCandxxxxxD}{\ensuremath{0.692\pm0.017}} 
\newcommand{\hatcurCCmagBCandxxxxxD}{\ensuremath{15.304}} 
\newcommand{\hatcurCCmagVCandxxxxxD}{\ensuremath{14.474}} 
\newcommand{\hatcurCCmaggCandxxxxxD}{\ensuremath{14.847}} 
\newcommand{\hatcurCCmagrCandxxxxxD}{\ensuremath{14.312}} 
\newcommand{\hatcurCCmagiCandxxxxxD}{\ensuremath{14.074}} 
\newcommand{\hatcurCCmagJCandxxxxxD}{\ensuremath{13.078\pm0.03}} 
\newcommand{\hatcurCCmagHCandxxxxxD}{\ensuremath{12.718\pm0.02}} 
\newcommand{\hatcurCCmagKCandxxxxxD}{\ensuremath{12.603\pm0.03}} 
\newcommand{\hatcurSPECWiFeSTeffCandxxxxxD}{\ensuremath{5861\pm300}} 
\newcommand{\hatcurSPECWiFeSloggCandxxxxxD}{\ensuremath{3.90\pm0.30}} 
\newcommand{\hatcurSPECWiFeSzfehCandxxxxxD}{\ensuremath{0.0\pm0.5}} 
\newcommand{\hatcurSPECWiFeSNlowresCandxxxxxD}{\ensuremath{1}} 
\newcommand{\hatcurSPECWiFeSNmedresCandxxxxxD}{\ensuremath{4}} 
\newcommand{\hatcurSPECWiFeSsnrangelowresCandxxxxxD}{104.2} 
\newcommand{\hatcurSPECWiFeSsnrangemedresCandxxxxxD}{19.4--36.8} 
\newcommand{\hatcurSPECWiFeSdaterangelowresCandxxxxxD}{2012-09-09} 
\newcommand{\hatcurSPECWiFeSdaterangemedresCandxxxxxD}{2012-09-24--2012-10-25} 
\newcommand{\hatcurSPECWiFeSRVKCandxxxxxD}{\ensuremath{44.22\pm0.81}} 
\newcommand{\hatcurSPECWiFeSRVGCandxxxxxD}{\ensuremath{31.56\pm0.73}} 
\newcommand{\hatcurSPECarcesTeffCandxxxxxD}{\ensuremath{NULL}} 
\newcommand{\hatcurSPECarcesloggCandxxxxxD}{\ensuremath{NULL}} 
\newcommand{\hatcurSPECarceszfehCandxxxxxD}{\ensuremath{NULL}} 
\newcommand{\hatcurSPECarcesvsiniCandxxxxxD}{\ensuremath{NULL}} 
\newcommand{\hatcurSPECarcesNCandxxxxxD}{\ensuremath{0}} 
\newcommand{\hatcurSPECarcessnrangeCandxxxxxD}{NULL} 
\newcommand{\hatcurSPECarcesdaterangelowresCandxxxxxD}{NULL} 
\newcommand{\hatcurSPECarcesRVKCandxxxxxD}{\ensuremath{NULL}} 
\newcommand{\hatcurSPECarcesRVGCandxxxxxD}{\ensuremath{NULL}} 
\newcommand{\hatcurSPECdupontTeffCandxxxxxD}{\ensuremath{NULL}} 
\newcommand{\hatcurSPECdupontloggCandxxxxxD}{\ensuremath{NULL}} 
\newcommand{\hatcurSPECdupontzfehCandxxxxxD}{\ensuremath{NULL}} 
\newcommand{\hatcurSPECdupontvsiniCandxxxxxD}{\ensuremath{NULL}} 
\newcommand{\hatcurSPECdupontNCandxxxxxD}{\ensuremath{0}} 
\newcommand{\hatcurSPECdupontsnrangeCandxxxxxD}{NULL} 
\newcommand{\hatcurSPECdupontdaterangelowresCandxxxxxD}{NULL} 
\newcommand{\hatcurSPECdupontRVKCandxxxxxD}{\ensuremath{NULL}} 
\newcommand{\hatcurSPECdupontRVGCandxxxxxD}{\ensuremath{NULL}} 
\newcommand{\hatcurSPECharpsTeffCandxxxxxD}{\ensuremath{NULL}} 
\newcommand{\hatcurSPECharpsloggCandxxxxxD}{\ensuremath{NULL}} 
\newcommand{\hatcurSPECharpszfehCandxxxxxD}{\ensuremath{NULL}} 
\newcommand{\hatcurSPECharpsvsiniCandxxxxxD}{\ensuremath{NULL}} 
\newcommand{\hatcurSPECharpsNCandxxxxxD}{\ensuremath{0}} 
\newcommand{\hatcurSPECharpssnrangeCandxxxxxD}{NULL} 
\newcommand{\hatcurSPECharpsdaterangelowresCandxxxxxD}{NULL} 
\newcommand{\hatcurSPECharpsRVKCandxxxxxD}{\ensuremath{NULL}} 
\newcommand{\hatcurSPECharpsRVGCandxxxxxD}{\ensuremath{NULL}} 
\newcommand{\hatcurSPECpfsTeffCandxxxxxD}{\ensuremath{NULL}} 
\newcommand{\hatcurSPECpfsloggCandxxxxxD}{\ensuremath{NULL}} 
\newcommand{\hatcurSPECpfszfehCandxxxxxD}{\ensuremath{NULL}} 
\newcommand{\hatcurSPECpfsvsiniCandxxxxxD}{\ensuremath{NULL}} 
\newcommand{\hatcurSPECpfsNCandxxxxxD}{\ensuremath{0}} 
\newcommand{\hatcurSPECpfssnrangeCandxxxxxD}{NULL} 
\newcommand{\hatcurSPECpfsdaterangelowresCandxxxxxD}{NULL} 
\newcommand{\hatcurSPECpfsRVKCandxxxxxD}{\ensuremath{NULL}} 
\newcommand{\hatcurSPECpfsRVGCandxxxxxD}{\ensuremath{NULL}} 
\newcommand{\hatcurSPECferosTeffCandxxxxxD}{\ensuremath{NULL}} 
\newcommand{\hatcurSPECferosloggCandxxxxxD}{\ensuremath{NULL}} 
\newcommand{\hatcurSPECferoszfehCandxxxxxD}{\ensuremath{NULL}} 
\newcommand{\hatcurSPECferosvsiniCandxxxxxD}{\ensuremath{NULL}} 
\newcommand{\hatcurSPECferosNCandxxxxxD}{\ensuremath{0}} 
\newcommand{\hatcurSPECferossnrangeCandxxxxxD}{NULL} 
\newcommand{\hatcurSPECferosdaterangelowresCandxxxxxD}{NULL} 
\newcommand{\hatcurSPECferosRVKCandxxxxxD}{\ensuremath{NULL}} 
\newcommand{\hatcurSPECferosRVGCandxxxxxD}{\ensuremath{NULL}} 
\newcommand{\hatcurSPECfiesTeffCandxxxxxD}{\ensuremath{NULL}} 
\newcommand{\hatcurSPECfiesloggCandxxxxxD}{\ensuremath{NULL}} 
\newcommand{\hatcurSPECfieszfehCandxxxxxD}{\ensuremath{NULL}} 
\newcommand{\hatcurSPECfiesvsiniCandxxxxxD}{\ensuremath{NULL}} 
\newcommand{\hatcurSPECfiesNCandxxxxxD}{\ensuremath{0}} 
\newcommand{\hatcurSPECfiessnrangeCandxxxxxD}{NULL} 
\newcommand{\hatcurSPECfiesdaterangelowresCandxxxxxD}{NULL} 
\newcommand{\hatcurSPECfiesRVKCandxxxxxD}{\ensuremath{NULL}} 
\newcommand{\hatcurSPECfiesRVGCandxxxxxD}{\ensuremath{NULL}} 
\newcommand{\hatcurSPECcoralieTeffCandxxxxxD}{\ensuremath{NULL}} 
\newcommand{\hatcurSPECcoralieloggCandxxxxxD}{\ensuremath{NULL}} 
\newcommand{\hatcurSPECcoraliezfehCandxxxxxD}{\ensuremath{NULL}} 
\newcommand{\hatcurSPECcoralievsiniCandxxxxxD}{\ensuremath{NULL}} 
\newcommand{\hatcurSPECcoralieNCandxxxxxD}{\ensuremath{0}} 
\newcommand{\hatcurSPECcoraliesnrangeCandxxxxxD}{NULL} 
\newcommand{\hatcurSPECcoraliedaterangelowresCandxxxxxD}{NULL} 
\newcommand{\hatcurSPECcoralieRVKCandxxxxxD}{\ensuremath{NULL}} 
\newcommand{\hatcurSPECcoralieRVGCandxxxxxD}{\ensuremath{NULL}} 
\newcommand{\hatcurhtrCandxxxxxE}{HATS579-007}
\newcommand{\hatcurCCraCandxxxxxE}{\ensuremath{19^{\mathrm h}17^{\mathrm m}44.16{\mathrm s}}}
\newcommand{\hatcurCCdecCandxxxxxE}{\ensuremath{-23^{\arcdeg}50^{\arcmin}51.2{\arcsec}}}
\newcommand{\hatcurCLASSCandxxxxxE}{BEB}
\newcommand{\hatcurCCEPICCandxxxxxE}{EPIC~215626177}
\newcommand{\hatcurLCPCandxxxxxE}{\ensuremath{2.0772332\pm0.0000157}} 
\newcommand{\hatcurLCTCandxxxxxE}{\ensuremath{2455943.10546\pm0.00243}} 
\newcommand{\hatcurLCdurCandxxxxxE}{\ensuremath{0.09895\pm0.01384}} 
\newcommand{\hatcurLCrprstarCandxxxxxE}{\ensuremath{0.06156\pm0.00357}} 
\newcommand{\hatcurLCbsqCandxxxxxE}{\ensuremath{0.768\pm0.523}} 
\newcommand{\hatcurCCmagBCandxxxxxE}{\ensuremath{13.759}} 
\newcommand{\hatcurCCmagVCandxxxxxE}{\ensuremath{13.145}} 
\newcommand{\hatcurCCmaggCandxxxxxE}{\ensuremath{13.421}} 
\newcommand{\hatcurCCmagrCandxxxxxE}{\ensuremath{12.971}} 
\newcommand{\hatcurCCmagiCandxxxxxE}{\ensuremath{12.771}} 
\newcommand{\hatcurCCmagJCandxxxxxE}{\ensuremath{11.935\pm0.02}} 
\newcommand{\hatcurCCmagHCandxxxxxE}{\ensuremath{11.625\pm0.03}} 
\newcommand{\hatcurCCmagKCandxxxxxE}{\ensuremath{11.569\pm0.03}} 
\newcommand{\hatcurSPECWiFeSTeffCandxxxxxE}{\ensuremath{6245\pm300}} 
\newcommand{\hatcurSPECWiFeSloggCandxxxxxE}{\ensuremath{3.90\pm0.30}} 
\newcommand{\hatcurSPECWiFeSzfehCandxxxxxE}{\ensuremath{-0.5\pm0.5}} 
\newcommand{\hatcurSPECWiFeSNlowresCandxxxxxE}{\ensuremath{1}} 
\newcommand{\hatcurSPECWiFeSNmedresCandxxxxxE}{\ensuremath{5}} 
\newcommand{\hatcurSPECWiFeSsnrangelowresCandxxxxxE}{179.9} 
\newcommand{\hatcurSPECWiFeSsnrangemedresCandxxxxxE}{34.8--103.7} 
\newcommand{\hatcurSPECWiFeSdaterangelowresCandxxxxxE}{2012-09-09} 
\newcommand{\hatcurSPECWiFeSdaterangemedresCandxxxxxE}{2012-09-24--2013-05-17} 
\newcommand{\hatcurSPECWiFeSRVKCandxxxxxE}{\ensuremath{<2.0}} 
\newcommand{\hatcurSPECWiFeSRVGCandxxxxxE}{\ensuremath{7.48\pm0.38}} 
\newcommand{\hatcurSPECarcesTeffCandxxxxxE}{\ensuremath{6056\pm50}} 
\newcommand{\hatcurSPECarcesloggCandxxxxxE}{\ensuremath{3.48\pm0.10}} 
\newcommand{\hatcurSPECarceszfehCandxxxxxE}{\ensuremath{-0.2\pm0.1}} 
\newcommand{\hatcurSPECarcesvsiniCandxxxxxE}{\ensuremath{6.3\pm0.5}} 
\newcommand{\hatcurSPECarcesNCandxxxxxE}{\ensuremath{1}} 
\newcommand{\hatcurSPECarcessnrangeCandxxxxxE}{27.0} 
\newcommand{\hatcurSPECarcesdaterangelowresCandxxxxxE}{2012-08-25} 
\newcommand{\hatcurSPECarcesRVKCandxxxxxE}{\ensuremath{NULL}} 
\newcommand{\hatcurSPECarcesRVGCandxxxxxE}{\ensuremath{10.97}} 
\newcommand{\hatcurSPECdupontTeffCandxxxxxE}{\ensuremath{6600\pm100}} 
\newcommand{\hatcurSPECdupontloggCandxxxxxE}{\ensuremath{4.80\pm0.50}} 
\newcommand{\hatcurSPECdupontzfehCandxxxxxE}{\ensuremath{0.0\pm0.5}} 
\newcommand{\hatcurSPECdupontvsiniCandxxxxxE}{\ensuremath{5.0\pm2.0}} 
\newcommand{\hatcurSPECdupontNCandxxxxxE}{\ensuremath{1}} 
\newcommand{\hatcurSPECdupontsnrangeCandxxxxxE}{48.0} 
\newcommand{\hatcurSPECdupontdaterangelowresCandxxxxxE}{2013-08-21} 
\newcommand{\hatcurSPECdupontRVKCandxxxxxE}{\ensuremath{NULL}} 
\newcommand{\hatcurSPECdupontRVGCandxxxxxE}{\ensuremath{10.57}} 
\newcommand{\hatcurSPECharpsTeffCandxxxxxE}{\ensuremath{0\pm0}} 
\newcommand{\hatcurSPECharpsloggCandxxxxxE}{\ensuremath{0.00\pm0.00}} 
\newcommand{\hatcurSPECharpszfehCandxxxxxE}{\ensuremath{0.0\pm0.0}} 
\newcommand{\hatcurSPECharpsvsiniCandxxxxxE}{\ensuremath{0.0\pm0.0}} 
\newcommand{\hatcurSPECharpsNCandxxxxxE}{\ensuremath{5}} 
\newcommand{\hatcurSPECharpssnrangeCandxxxxxE}{23.0--36.0} 
\newcommand{\hatcurSPECharpsdaterangelowresCandxxxxxE}{2014-08-16--2014-08-19} 
\newcommand{\hatcurSPECharpsRVKCandxxxxxE}{\ensuremath{-0.00\pm0.00}} 
\newcommand{\hatcurSPECharpsRVGCandxxxxxE}{\ensuremath{10.84\pm0.00}} 
\newcommand{\hatcurSPECpfsTeffCandxxxxxE}{\ensuremath{NULL}} 
\newcommand{\hatcurSPECpfsloggCandxxxxxE}{\ensuremath{NULL}} 
\newcommand{\hatcurSPECpfszfehCandxxxxxE}{\ensuremath{NULL}} 
\newcommand{\hatcurSPECpfsvsiniCandxxxxxE}{\ensuremath{NULL}} 
\newcommand{\hatcurSPECpfsNCandxxxxxE}{\ensuremath{5}} 
\newcommand{\hatcurSPECpfssnrangeCandxxxxxE}{0.0--0.0} 
\newcommand{\hatcurSPECpfsdaterangelowresCandxxxxxE}{2014-08-03--2014-08-13} 
\newcommand{\hatcurSPECpfsRVKCandxxxxxE}{\ensuremath{0.00\pm0.00}} 
\newcommand{\hatcurSPECpfsRVGCandxxxxxE}{\ensuremath{-0.00\pm0.00}} 
\newcommand{\hatcurSPECferosTeffCandxxxxxE}{\ensuremath{6295\pm30}} 
\newcommand{\hatcurSPECferosloggCandxxxxxE}{\ensuremath{4.20\pm0.15}} 
\newcommand{\hatcurSPECferoszfehCandxxxxxE}{\ensuremath{-0.5\pm0.2}} 
\newcommand{\hatcurSPECferosvsiniCandxxxxxE}{\ensuremath{6.4\pm0.6}} 
\newcommand{\hatcurSPECferosNCandxxxxxE}{\ensuremath{22}} 
\newcommand{\hatcurSPECferossnrangeCandxxxxxE}{0.6--80.0} 
\newcommand{\hatcurSPECferosdaterangelowresCandxxxxxE}{2012-08-23--2013-09-17} 
\newcommand{\hatcurSPECferosRVKCandxxxxxE}{\ensuremath{0.02\pm0.01}} 
\newcommand{\hatcurSPECferosRVGCandxxxxxE}{\ensuremath{10.92\pm0.01}} 
\newcommand{\hatcurSPECfiesTeffCandxxxxxE}{\ensuremath{6104\pm50}} 
\newcommand{\hatcurSPECfiesloggCandxxxxxE}{\ensuremath{3.66\pm0.10}} 
\newcommand{\hatcurSPECfieszfehCandxxxxxE}{\ensuremath{-0.1\pm0.1}} 
\newcommand{\hatcurSPECfiesvsiniCandxxxxxE}{\ensuremath{6.2\pm0.5}} 
\newcommand{\hatcurSPECfiesNCandxxxxxE}{\ensuremath{1}} 
\newcommand{\hatcurSPECfiessnrangeCandxxxxxE}{34.8} 
\newcommand{\hatcurSPECfiesdaterangelowresCandxxxxxE}{2013-08-24} 
\newcommand{\hatcurSPECfiesRVKCandxxxxxE}{\ensuremath{NULL}} 
\newcommand{\hatcurSPECfiesRVGCandxxxxxE}{\ensuremath{11.53}} 
\newcommand{\hatcurSPECcoralieTeffCandxxxxxE}{\ensuremath{5877\pm30}} 
\newcommand{\hatcurSPECcoralieloggCandxxxxxE}{\ensuremath{3.46\pm0.15}} 
\newcommand{\hatcurSPECcoraliezfehCandxxxxxE}{\ensuremath{-0.5\pm0.2}} 
\newcommand{\hatcurSPECcoralievsiniCandxxxxxE}{\ensuremath{7.5\pm0.6}} 
\newcommand{\hatcurSPECcoralieNCandxxxxxE}{\ensuremath{11}} 
\newcommand{\hatcurSPECcoraliesnrangeCandxxxxxE}{11.0--22.0} 
\newcommand{\hatcurSPECcoraliedaterangelowresCandxxxxxE}{2013-06-20--2013-08-21} 
\newcommand{\hatcurSPECcoralieRVKCandxxxxxE}{\ensuremath{-0.04\pm0.03}} 
\newcommand{\hatcurSPECcoralieRVGCandxxxxxE}{\ensuremath{10.75\pm0.02}} 
\newcommand{\hatcurhtrCandxxxxxF}{HATS579-008}
\newcommand{\hatcurCCraCandxxxxxF}{\ensuremath{19^{\mathrm h}20^{\mathrm m}29.04{\mathrm s}}}
\newcommand{\hatcurCCdecCandxxxxxF}{\ensuremath{-23^{\arcdeg}40^{\arcmin}28.6{\arcsec}}}
\newcommand{\hatcurCLASSCandxxxxxF}{EB}
\newcommand{\hatcurCCEPICCandxxxxxF}{EPIC~215716837}
\newcommand{\hatcurLCPCandxxxxxF}{\ensuremath{8.6829946\pm0.0000362}} 
\newcommand{\hatcurLCTCandxxxxxF}{\ensuremath{2457285.32239\pm0.00121}} 
\newcommand{\hatcurLCdurCandxxxxxF}{\ensuremath{0.28869\pm0.01814}} 
\newcommand{\hatcurLCrprstarCandxxxxxF}{\ensuremath{0.14714\pm0.00160}} 
\newcommand{\hatcurLCbsqCandxxxxxF}{\ensuremath{0.711\pm0.022}} 
\newcommand{\hatcurCCmagBCandxxxxxF}{\ensuremath{15.953}} 
\newcommand{\hatcurCCmagVCandxxxxxF}{\ensuremath{15.258}} 
\newcommand{\hatcurCCmaggCandxxxxxF}{\ensuremath{15.552}} 
\newcommand{\hatcurCCmagrCandxxxxxF}{\ensuremath{15.027}} 
\newcommand{\hatcurCCmagiCandxxxxxF}{\ensuremath{14.874}} 
\newcommand{\hatcurCCmagJCandxxxxxF}{\ensuremath{13.853\pm0.03}} 
\newcommand{\hatcurCCmagHCandxxxxxF}{\ensuremath{13.510\pm0.03}} 
\newcommand{\hatcurCCmagKCandxxxxxF}{\ensuremath{13.447\pm0.04}} 
\newcommand{\hatcurSPECWiFeSTeffCandxxxxxF}{\ensuremath{5334\pm300}} 
\newcommand{\hatcurSPECWiFeSloggCandxxxxxF}{\ensuremath{4.40\pm0.30}} 
\newcommand{\hatcurSPECWiFeSzfehCandxxxxxF}{\ensuremath{-0.5\pm0.5}} 
\newcommand{\hatcurSPECWiFeSNlowresCandxxxxxF}{\ensuremath{1}} 
\newcommand{\hatcurSPECWiFeSNmedresCandxxxxxF}{\ensuremath{2}} 
\newcommand{\hatcurSPECWiFeSsnrangelowresCandxxxxxF}{32.2} 
\newcommand{\hatcurSPECWiFeSsnrangemedresCandxxxxxF}{12.7--27.5} 
\newcommand{\hatcurSPECWiFeSdaterangelowresCandxxxxxF}{2016-03-24} 
\newcommand{\hatcurSPECWiFeSdaterangemedresCandxxxxxF}{2016-03-24--2016-03-30} 
\newcommand{\hatcurSPECWiFeSRVKCandxxxxxF}{\ensuremath{<2.0}} 
\newcommand{\hatcurSPECWiFeSRVGCandxxxxxF}{\ensuremath{-26.90\pm1.96}} 
\newcommand{\hatcurSPECarcesTeffCandxxxxxF}{\ensuremath{NULL}} 
\newcommand{\hatcurSPECarcesloggCandxxxxxF}{\ensuremath{NULL}} 
\newcommand{\hatcurSPECarceszfehCandxxxxxF}{\ensuremath{NULL}} 
\newcommand{\hatcurSPECarcesvsiniCandxxxxxF}{\ensuremath{NULL}} 
\newcommand{\hatcurSPECarcesNCandxxxxxF}{\ensuremath{0}} 
\newcommand{\hatcurSPECarcessnrangeCandxxxxxF}{NULL} 
\newcommand{\hatcurSPECarcesdaterangelowresCandxxxxxF}{NULL} 
\newcommand{\hatcurSPECarcesRVKCandxxxxxF}{\ensuremath{NULL}} 
\newcommand{\hatcurSPECarcesRVGCandxxxxxF}{\ensuremath{NULL}} 
\newcommand{\hatcurSPECdupontTeffCandxxxxxF}{\ensuremath{NULL}} 
\newcommand{\hatcurSPECdupontloggCandxxxxxF}{\ensuremath{NULL}} 
\newcommand{\hatcurSPECdupontzfehCandxxxxxF}{\ensuremath{NULL}} 
\newcommand{\hatcurSPECdupontvsiniCandxxxxxF}{\ensuremath{NULL}} 
\newcommand{\hatcurSPECdupontNCandxxxxxF}{\ensuremath{0}} 
\newcommand{\hatcurSPECdupontsnrangeCandxxxxxF}{NULL} 
\newcommand{\hatcurSPECdupontdaterangelowresCandxxxxxF}{NULL} 
\newcommand{\hatcurSPECdupontRVKCandxxxxxF}{\ensuremath{NULL}} 
\newcommand{\hatcurSPECdupontRVGCandxxxxxF}{\ensuremath{NULL}} 
\newcommand{\hatcurSPECharpsTeffCandxxxxxF}{\ensuremath{NULL}} 
\newcommand{\hatcurSPECharpsloggCandxxxxxF}{\ensuremath{NULL}} 
\newcommand{\hatcurSPECharpszfehCandxxxxxF}{\ensuremath{NULL}} 
\newcommand{\hatcurSPECharpsvsiniCandxxxxxF}{\ensuremath{NULL}} 
\newcommand{\hatcurSPECharpsNCandxxxxxF}{\ensuremath{0}} 
\newcommand{\hatcurSPECharpssnrangeCandxxxxxF}{NULL} 
\newcommand{\hatcurSPECharpsdaterangelowresCandxxxxxF}{NULL} 
\newcommand{\hatcurSPECharpsRVKCandxxxxxF}{\ensuremath{NULL}} 
\newcommand{\hatcurSPECharpsRVGCandxxxxxF}{\ensuremath{NULL}} 
\newcommand{\hatcurSPECpfsTeffCandxxxxxF}{\ensuremath{NULL}} 
\newcommand{\hatcurSPECpfsloggCandxxxxxF}{\ensuremath{NULL}} 
\newcommand{\hatcurSPECpfszfehCandxxxxxF}{\ensuremath{NULL}} 
\newcommand{\hatcurSPECpfsvsiniCandxxxxxF}{\ensuremath{NULL}} 
\newcommand{\hatcurSPECpfsNCandxxxxxF}{\ensuremath{0}} 
\newcommand{\hatcurSPECpfssnrangeCandxxxxxF}{NULL} 
\newcommand{\hatcurSPECpfsdaterangelowresCandxxxxxF}{NULL} 
\newcommand{\hatcurSPECpfsRVKCandxxxxxF}{\ensuremath{NULL}} 
\newcommand{\hatcurSPECpfsRVGCandxxxxxF}{\ensuremath{NULL}} 
\newcommand{\hatcurSPECferosTeffCandxxxxxF}{\ensuremath{NULL}} 
\newcommand{\hatcurSPECferosloggCandxxxxxF}{\ensuremath{NULL}} 
\newcommand{\hatcurSPECferoszfehCandxxxxxF}{\ensuremath{NULL}} 
\newcommand{\hatcurSPECferosvsiniCandxxxxxF}{\ensuremath{NULL}} 
\newcommand{\hatcurSPECferosNCandxxxxxF}{\ensuremath{0}} 
\newcommand{\hatcurSPECferossnrangeCandxxxxxF}{NULL} 
\newcommand{\hatcurSPECferosdaterangelowresCandxxxxxF}{NULL} 
\newcommand{\hatcurSPECferosRVKCandxxxxxF}{\ensuremath{NULL}} 
\newcommand{\hatcurSPECferosRVGCandxxxxxF}{\ensuremath{NULL}} 
\newcommand{\hatcurSPECfiesTeffCandxxxxxF}{\ensuremath{NULL}} 
\newcommand{\hatcurSPECfiesloggCandxxxxxF}{\ensuremath{NULL}} 
\newcommand{\hatcurSPECfieszfehCandxxxxxF}{\ensuremath{NULL}} 
\newcommand{\hatcurSPECfiesvsiniCandxxxxxF}{\ensuremath{NULL}} 
\newcommand{\hatcurSPECfiesNCandxxxxxF}{\ensuremath{0}} 
\newcommand{\hatcurSPECfiessnrangeCandxxxxxF}{NULL} 
\newcommand{\hatcurSPECfiesdaterangelowresCandxxxxxF}{NULL} 
\newcommand{\hatcurSPECfiesRVKCandxxxxxF}{\ensuremath{NULL}} 
\newcommand{\hatcurSPECfiesRVGCandxxxxxF}{\ensuremath{NULL}} 
\newcommand{\hatcurSPECcoralieTeffCandxxxxxF}{\ensuremath{NULL}} 
\newcommand{\hatcurSPECcoralieloggCandxxxxxF}{\ensuremath{NULL}} 
\newcommand{\hatcurSPECcoraliezfehCandxxxxxF}{\ensuremath{NULL}} 
\newcommand{\hatcurSPECcoralievsiniCandxxxxxF}{\ensuremath{NULL}} 
\newcommand{\hatcurSPECcoralieNCandxxxxxF}{\ensuremath{0}} 
\newcommand{\hatcurSPECcoraliesnrangeCandxxxxxF}{NULL} 
\newcommand{\hatcurSPECcoraliedaterangelowresCandxxxxxF}{NULL} 
\newcommand{\hatcurSPECcoralieRVKCandxxxxxF}{\ensuremath{NULL}} 
\newcommand{\hatcurSPECcoralieRVGCandxxxxxF}{\ensuremath{NULL}} 
\newcommand{\hatcurhtrCandxxxxxG}{HATS579-009}
\newcommand{\hatcurCCraCandxxxxxG}{\ensuremath{19^{\mathrm h}17^{\mathrm m}38.76{\mathrm s}}}
\newcommand{\hatcurCCdecCandxxxxxG}{\ensuremath{-24^{\arcdeg}56^{\arcmin}09.0{\arcsec}}}
\newcommand{\hatcurCLASSCandxxxxxG}{EB}
\newcommand{\hatcurCCEPICCandxxxxxG}{EPIC~215101303}
\newcommand{\hatcurLCPCandxxxxxG}{\ensuremath{15.2073500\pm0.0000307}} 
\newcommand{\hatcurLCTCandxxxxxG}{\ensuremath{2457047.10940\pm0.00117}} 
\newcommand{\hatcurLCdurCandxxxxxG}{\ensuremath{0.13596\pm0.00703}} 
\newcommand{\hatcurLCrprstarCandxxxxxG}{\ensuremath{0.16459\pm0.00179}} 
\newcommand{\hatcurLCbsqCandxxxxxG}{\ensuremath{0.238\pm0.148}} 
\newcommand{\hatcurCCmagBCandxxxxxG}{\ensuremath{15.792}} 
\newcommand{\hatcurCCmagVCandxxxxxG}{\ensuremath{15.079}} 
\newcommand{\hatcurCCmaggCandxxxxxG}{\ensuremath{15.486}} 
\newcommand{\hatcurCCmagrCandxxxxxG}{\ensuremath{14.945}} 
\newcommand{\hatcurCCmagiCandxxxxxG}{\ensuremath{14.631}} 
\newcommand{\hatcurCCmagJCandxxxxxG}{\ensuremath{13.756\pm0.03}} 
\newcommand{\hatcurCCmagHCandxxxxxG}{\ensuremath{13.418\pm0.03}} 
\newcommand{\hatcurCCmagKCandxxxxxG}{\ensuremath{13.306\pm0.04}} 
\newcommand{\hatcurSPECWiFeSTeffCandxxxxxG}{\ensuremath{6260\pm300}} 
\newcommand{\hatcurSPECWiFeSloggCandxxxxxG}{\ensuremath{3.60\pm0.30}} 
\newcommand{\hatcurSPECWiFeSzfehCandxxxxxG}{\ensuremath{-0.5\pm0.5}} 
\newcommand{\hatcurSPECWiFeSNlowresCandxxxxxG}{\ensuremath{1}} 
\newcommand{\hatcurSPECWiFeSNmedresCandxxxxxG}{\ensuremath{2}} 
\newcommand{\hatcurSPECWiFeSsnrangelowresCandxxxxxG}{13.0} 
\newcommand{\hatcurSPECWiFeSsnrangemedresCandxxxxxG}{9.2--14.7} 
\newcommand{\hatcurSPECWiFeSdaterangelowresCandxxxxxG}{2016-03-14} 
\newcommand{\hatcurSPECWiFeSdaterangemedresCandxxxxxG}{2016-03-14--2016-03-24} 
\newcommand{\hatcurSPECWiFeSRVKCandxxxxxG}{\ensuremath{20.38\pm2.71}} 
\newcommand{\hatcurSPECWiFeSRVGCandxxxxxG}{\ensuremath{-62.11\pm1.78}} 
\newcommand{\hatcurSPECarcesTeffCandxxxxxG}{\ensuremath{NULL}} 
\newcommand{\hatcurSPECarcesloggCandxxxxxG}{\ensuremath{NULL}} 
\newcommand{\hatcurSPECarceszfehCandxxxxxG}{\ensuremath{NULL}} 
\newcommand{\hatcurSPECarcesvsiniCandxxxxxG}{\ensuremath{NULL}} 
\newcommand{\hatcurSPECarcesNCandxxxxxG}{\ensuremath{0}} 
\newcommand{\hatcurSPECarcessnrangeCandxxxxxG}{NULL} 
\newcommand{\hatcurSPECarcesdaterangelowresCandxxxxxG}{NULL} 
\newcommand{\hatcurSPECarcesRVKCandxxxxxG}{\ensuremath{NULL}} 
\newcommand{\hatcurSPECarcesRVGCandxxxxxG}{\ensuremath{NULL}} 
\newcommand{\hatcurSPECdupontTeffCandxxxxxG}{\ensuremath{NULL}} 
\newcommand{\hatcurSPECdupontloggCandxxxxxG}{\ensuremath{NULL}} 
\newcommand{\hatcurSPECdupontzfehCandxxxxxG}{\ensuremath{NULL}} 
\newcommand{\hatcurSPECdupontvsiniCandxxxxxG}{\ensuremath{NULL}} 
\newcommand{\hatcurSPECdupontNCandxxxxxG}{\ensuremath{0}} 
\newcommand{\hatcurSPECdupontsnrangeCandxxxxxG}{NULL} 
\newcommand{\hatcurSPECdupontdaterangelowresCandxxxxxG}{NULL} 
\newcommand{\hatcurSPECdupontRVKCandxxxxxG}{\ensuremath{NULL}} 
\newcommand{\hatcurSPECdupontRVGCandxxxxxG}{\ensuremath{NULL}} 
\newcommand{\hatcurSPECharpsTeffCandxxxxxG}{\ensuremath{NULL}} 
\newcommand{\hatcurSPECharpsloggCandxxxxxG}{\ensuremath{NULL}} 
\newcommand{\hatcurSPECharpszfehCandxxxxxG}{\ensuremath{NULL}} 
\newcommand{\hatcurSPECharpsvsiniCandxxxxxG}{\ensuremath{NULL}} 
\newcommand{\hatcurSPECharpsNCandxxxxxG}{\ensuremath{0}} 
\newcommand{\hatcurSPECharpssnrangeCandxxxxxG}{NULL} 
\newcommand{\hatcurSPECharpsdaterangelowresCandxxxxxG}{NULL} 
\newcommand{\hatcurSPECharpsRVKCandxxxxxG}{\ensuremath{NULL}} 
\newcommand{\hatcurSPECharpsRVGCandxxxxxG}{\ensuremath{NULL}} 
\newcommand{\hatcurSPECpfsTeffCandxxxxxG}{\ensuremath{NULL}} 
\newcommand{\hatcurSPECpfsloggCandxxxxxG}{\ensuremath{NULL}} 
\newcommand{\hatcurSPECpfszfehCandxxxxxG}{\ensuremath{NULL}} 
\newcommand{\hatcurSPECpfsvsiniCandxxxxxG}{\ensuremath{NULL}} 
\newcommand{\hatcurSPECpfsNCandxxxxxG}{\ensuremath{0}} 
\newcommand{\hatcurSPECpfssnrangeCandxxxxxG}{NULL} 
\newcommand{\hatcurSPECpfsdaterangelowresCandxxxxxG}{NULL} 
\newcommand{\hatcurSPECpfsRVKCandxxxxxG}{\ensuremath{NULL}} 
\newcommand{\hatcurSPECpfsRVGCandxxxxxG}{\ensuremath{NULL}} 
\newcommand{\hatcurSPECferosTeffCandxxxxxG}{\ensuremath{NULL}} 
\newcommand{\hatcurSPECferosloggCandxxxxxG}{\ensuremath{NULL}} 
\newcommand{\hatcurSPECferoszfehCandxxxxxG}{\ensuremath{NULL}} 
\newcommand{\hatcurSPECferosvsiniCandxxxxxG}{\ensuremath{NULL}} 
\newcommand{\hatcurSPECferosNCandxxxxxG}{\ensuremath{0}} 
\newcommand{\hatcurSPECferossnrangeCandxxxxxG}{NULL} 
\newcommand{\hatcurSPECferosdaterangelowresCandxxxxxG}{NULL} 
\newcommand{\hatcurSPECferosRVKCandxxxxxG}{\ensuremath{NULL}} 
\newcommand{\hatcurSPECferosRVGCandxxxxxG}{\ensuremath{NULL}} 
\newcommand{\hatcurSPECfiesTeffCandxxxxxG}{\ensuremath{NULL}} 
\newcommand{\hatcurSPECfiesloggCandxxxxxG}{\ensuremath{NULL}} 
\newcommand{\hatcurSPECfieszfehCandxxxxxG}{\ensuremath{NULL}} 
\newcommand{\hatcurSPECfiesvsiniCandxxxxxG}{\ensuremath{NULL}} 
\newcommand{\hatcurSPECfiesNCandxxxxxG}{\ensuremath{0}} 
\newcommand{\hatcurSPECfiessnrangeCandxxxxxG}{NULL} 
\newcommand{\hatcurSPECfiesdaterangelowresCandxxxxxG}{NULL} 
\newcommand{\hatcurSPECfiesRVKCandxxxxxG}{\ensuremath{NULL}} 
\newcommand{\hatcurSPECfiesRVGCandxxxxxG}{\ensuremath{NULL}} 
\newcommand{\hatcurSPECcoralieTeffCandxxxxxG}{\ensuremath{NULL}} 
\newcommand{\hatcurSPECcoralieloggCandxxxxxG}{\ensuremath{NULL}} 
\newcommand{\hatcurSPECcoraliezfehCandxxxxxG}{\ensuremath{NULL}} 
\newcommand{\hatcurSPECcoralievsiniCandxxxxxG}{\ensuremath{NULL}} 
\newcommand{\hatcurSPECcoralieNCandxxxxxG}{\ensuremath{0}} 
\newcommand{\hatcurSPECcoraliesnrangeCandxxxxxG}{NULL} 
\newcommand{\hatcurSPECcoraliedaterangelowresCandxxxxxG}{NULL} 
\newcommand{\hatcurSPECcoralieRVKCandxxxxxG}{\ensuremath{NULL}} 
\newcommand{\hatcurSPECcoralieRVGCandxxxxxG}{\ensuremath{NULL}} 
\newcommand{\hatcurhtrCandxxxxxU}{HATS579-010}
\newcommand{\hatcurCCraCandxxxxxU}{\ensuremath{19^{\mathrm h}19^{\mathrm m}18.12{\mathrm s}}}
\newcommand{\hatcurCCdecCandxxxxxU}{\ensuremath{-25^{\arcdeg}21^{\arcmin}21.2{\arcsec}}}
\newcommand{\hatcurCLASSCandxxxxxU}{BEB}
\newcommand{\hatcurCCEPICCandxxxxxU}{EPIC~214912104}
\newcommand{\hatcurLCPCandxxxxxU}{\ensuremath{13.3370855\pm0.0000020}} 
\newcommand{\hatcurLCTCandxxxxxU}{\ensuremath{2400010.45400\pm0.00000}} 
\newcommand{\hatcurLCdurCandxxxxxU}{\ensuremath{0.00000\pm16.12333}} 
\newcommand{\hatcurLCrprstarCandxxxxxU}{\ensuremath{0.1900\pm0.0010}} 
\newcommand{\hatcurLCbsqCandxxxxxU}{\ensuremath{\pm}} 
\newcommand{\hatcurCCmagBCandxxxxxU}{\ensuremath{15.011}} 
\newcommand{\hatcurCCmagVCandxxxxxU}{\ensuremath{14.316}} 
\newcommand{\hatcurCCmaggCandxxxxxU}{\ensuremath{14.735}} 
\newcommand{\hatcurCCmagrCandxxxxxU}{\ensuremath{14.210}} 
\newcommand{\hatcurCCmagiCandxxxxxU}{\ensuremath{14.042}} 
\newcommand{\hatcurCCmagJCandxxxxxU}{\ensuremath{13.087\pm0.02}} 
\newcommand{\hatcurCCmagHCandxxxxxU}{\ensuremath{12.730\pm0.02}} 
\newcommand{\hatcurCCmagKCandxxxxxU}{\ensuremath{12.689\pm0.03}} 
\newcommand{\hatcurSPECWiFeSTeffCandxxxxxU}{\ensuremath{5945\pm300}} 
\newcommand{\hatcurSPECWiFeSloggCandxxxxxU}{\ensuremath{4.00\pm0.30}} 
\newcommand{\hatcurSPECWiFeSzfehCandxxxxxU}{\ensuremath{-0.5\pm0.5}} 
\newcommand{\hatcurSPECWiFeSNlowresCandxxxxxU}{\ensuremath{1}} 
\newcommand{\hatcurSPECWiFeSNmedresCandxxxxxU}{\ensuremath{4}} 
\newcommand{\hatcurSPECWiFeSsnrangelowresCandxxxxxU}{162.8} 
\newcommand{\hatcurSPECWiFeSsnrangemedresCandxxxxxU}{11.0--36.2} 
\newcommand{\hatcurSPECWiFeSdaterangelowresCandxxxxxU}{2013-04-01} 
\newcommand{\hatcurSPECWiFeSdaterangemedresCandxxxxxU}{2013-03-24--2013-09-21} 
\newcommand{\hatcurSPECWiFeSRVKCandxxxxxU}{\ensuremath{<2.0}} 
\newcommand{\hatcurSPECWiFeSRVGCandxxxxxU}{\ensuremath{40.64\pm0.92}} 
\newcommand{\hatcurSPECarcesTeffCandxxxxxU}{\ensuremath{NULL}} 
\newcommand{\hatcurSPECarcesloggCandxxxxxU}{\ensuremath{NULL}} 
\newcommand{\hatcurSPECarceszfehCandxxxxxU}{\ensuremath{NULL}} 
\newcommand{\hatcurSPECarcesvsiniCandxxxxxU}{\ensuremath{NULL}} 
\newcommand{\hatcurSPECarcesNCandxxxxxU}{\ensuremath{0}} 
\newcommand{\hatcurSPECarcessnrangeCandxxxxxU}{NULL} 
\newcommand{\hatcurSPECarcesdaterangelowresCandxxxxxU}{NULL} 
\newcommand{\hatcurSPECarcesRVKCandxxxxxU}{\ensuremath{NULL}} 
\newcommand{\hatcurSPECarcesRVGCandxxxxxU}{\ensuremath{NULL}} 
\newcommand{\hatcurSPECdupontTeffCandxxxxxU}{\ensuremath{NULL}} 
\newcommand{\hatcurSPECdupontloggCandxxxxxU}{\ensuremath{NULL}} 
\newcommand{\hatcurSPECdupontzfehCandxxxxxU}{\ensuremath{NULL}} 
\newcommand{\hatcurSPECdupontvsiniCandxxxxxU}{\ensuremath{NULL}} 
\newcommand{\hatcurSPECdupontNCandxxxxxU}{\ensuremath{0}} 
\newcommand{\hatcurSPECdupontsnrangeCandxxxxxU}{NULL} 
\newcommand{\hatcurSPECdupontdaterangelowresCandxxxxxU}{NULL} 
\newcommand{\hatcurSPECdupontRVKCandxxxxxU}{\ensuremath{NULL}} 
\newcommand{\hatcurSPECdupontRVGCandxxxxxU}{\ensuremath{NULL}} 
\newcommand{\hatcurSPECharpsTeffCandxxxxxU}{\ensuremath{NULL}} 
\newcommand{\hatcurSPECharpsloggCandxxxxxU}{\ensuremath{NULL}} 
\newcommand{\hatcurSPECharpszfehCandxxxxxU}{\ensuremath{NULL}} 
\newcommand{\hatcurSPECharpsvsiniCandxxxxxU}{\ensuremath{NULL}} 
\newcommand{\hatcurSPECharpsNCandxxxxxU}{\ensuremath{0}} 
\newcommand{\hatcurSPECharpssnrangeCandxxxxxU}{NULL} 
\newcommand{\hatcurSPECharpsdaterangelowresCandxxxxxU}{NULL} 
\newcommand{\hatcurSPECharpsRVKCandxxxxxU}{\ensuremath{NULL}} 
\newcommand{\hatcurSPECharpsRVGCandxxxxxU}{\ensuremath{NULL}} 
\newcommand{\hatcurSPECpfsTeffCandxxxxxU}{\ensuremath{NULL}} 
\newcommand{\hatcurSPECpfsloggCandxxxxxU}{\ensuremath{NULL}} 
\newcommand{\hatcurSPECpfszfehCandxxxxxU}{\ensuremath{NULL}} 
\newcommand{\hatcurSPECpfsvsiniCandxxxxxU}{\ensuremath{NULL}} 
\newcommand{\hatcurSPECpfsNCandxxxxxU}{\ensuremath{0}} 
\newcommand{\hatcurSPECpfssnrangeCandxxxxxU}{NULL} 
\newcommand{\hatcurSPECpfsdaterangelowresCandxxxxxU}{NULL} 
\newcommand{\hatcurSPECpfsRVKCandxxxxxU}{\ensuremath{NULL}} 
\newcommand{\hatcurSPECpfsRVGCandxxxxxU}{\ensuremath{NULL}} 
\newcommand{\hatcurSPECferosTeffCandxxxxxU}{\ensuremath{NULL}} 
\newcommand{\hatcurSPECferosloggCandxxxxxU}{\ensuremath{NULL}} 
\newcommand{\hatcurSPECferoszfehCandxxxxxU}{\ensuremath{NULL}} 
\newcommand{\hatcurSPECferosvsiniCandxxxxxU}{\ensuremath{NULL}} 
\newcommand{\hatcurSPECferosNCandxxxxxU}{\ensuremath{0}} 
\newcommand{\hatcurSPECferossnrangeCandxxxxxU}{NULL} 
\newcommand{\hatcurSPECferosdaterangelowresCandxxxxxU}{NULL} 
\newcommand{\hatcurSPECferosRVKCandxxxxxU}{\ensuremath{NULL}} 
\newcommand{\hatcurSPECferosRVGCandxxxxxU}{\ensuremath{NULL}} 
\newcommand{\hatcurSPECfiesTeffCandxxxxxU}{\ensuremath{NULL}} 
\newcommand{\hatcurSPECfiesloggCandxxxxxU}{\ensuremath{NULL}} 
\newcommand{\hatcurSPECfieszfehCandxxxxxU}{\ensuremath{NULL}} 
\newcommand{\hatcurSPECfiesvsiniCandxxxxxU}{\ensuremath{NULL}} 
\newcommand{\hatcurSPECfiesNCandxxxxxU}{\ensuremath{0}} 
\newcommand{\hatcurSPECfiessnrangeCandxxxxxU}{NULL} 
\newcommand{\hatcurSPECfiesdaterangelowresCandxxxxxU}{NULL} 
\newcommand{\hatcurSPECfiesRVKCandxxxxxU}{\ensuremath{NULL}} 
\newcommand{\hatcurSPECfiesRVGCandxxxxxU}{\ensuremath{NULL}} 
\newcommand{\hatcurSPECcoralieTeffCandxxxxxU}{\ensuremath{NULL}} 
\newcommand{\hatcurSPECcoralieloggCandxxxxxU}{\ensuremath{NULL}} 
\newcommand{\hatcurSPECcoraliezfehCandxxxxxU}{\ensuremath{NULL}} 
\newcommand{\hatcurSPECcoralievsiniCandxxxxxU}{\ensuremath{NULL}} 
\newcommand{\hatcurSPECcoralieNCandxxxxxU}{\ensuremath{0}} 
\newcommand{\hatcurSPECcoraliesnrangeCandxxxxxU}{NULL} 
\newcommand{\hatcurSPECcoraliedaterangelowresCandxxxxxU}{NULL} 
\newcommand{\hatcurSPECcoralieRVKCandxxxxxU}{\ensuremath{NULL}} 
\newcommand{\hatcurSPECcoralieRVGCandxxxxxU}{\ensuremath{NULL}} 
\newcommand{\hatcurhtrCandxxxxxH}{HATS579-014}
\newcommand{\hatcurCCraCandxxxxxH}{\ensuremath{19^{\mathrm h}24^{\mathrm m}10.08{\mathrm s}}}
\newcommand{\hatcurCCdecCandxxxxxH}{\ensuremath{-22^{\arcdeg}20^{\arcmin}27.2{\arcsec}}}
\newcommand{\hatcurCLASSCandxxxxxH}{EB}
\newcommand{\hatcurCCEPICCandxxxxxH}{EPIC~216442060}
\newcommand{\hatcurLCPCandxxxxxH}{\ensuremath{5.2027430\pm0.0000044}} 
\newcommand{\hatcurLCTCandxxxxxH}{\ensuremath{2457265.46459\pm0.00028}} 
\newcommand{\hatcurLCdurCandxxxxxH}{\ensuremath{0.13160\pm0.00613}} 
\newcommand{\hatcurLCrprstarCandxxxxxH}{\ensuremath{0.14502\pm0.00086}} 
\newcommand{\hatcurLCbsqCandxxxxxH}{\ensuremath{0.721\pm0.011}} 
\newcommand{\hatcurCCmagBCandxxxxxH}{\ensuremath{13.264}} 
\newcommand{\hatcurCCmagVCandxxxxxH}{\ensuremath{12.834}} 
\newcommand{\hatcurCCmaggCandxxxxxH}{\ensuremath{12.959}} 
\newcommand{\hatcurCCmagrCandxxxxxH}{\ensuremath{12.747}} 
\newcommand{\hatcurCCmagiCandxxxxxH}{\ensuremath{12.526}} 
\newcommand{\hatcurCCmagJCandxxxxxH}{\ensuremath{11.815\pm0.02}} 
\newcommand{\hatcurCCmagHCandxxxxxH}{\ensuremath{11.603\pm0.02}} 
\newcommand{\hatcurCCmagKCandxxxxxH}{\ensuremath{11.532\pm0.02}} 
\newcommand{\hatcurSPECWiFeSTeffCandxxxxxH}{\ensuremath{SB2}} 
\newcommand{\hatcurSPECWiFeSloggCandxxxxxH}{\ensuremath{5.00\pm0.50}} 
\newcommand{\hatcurSPECWiFeSzfehCandxxxxxH}{\ensuremath{NULL}} 
\newcommand{\hatcurSPECWiFeSNlowresCandxxxxxH}{\ensuremath{0}} 
\newcommand{\hatcurSPECWiFeSNmedresCandxxxxxH}{\ensuremath{0}} 
\newcommand{\hatcurSPECWiFeSsnrangelowresCandxxxxxH}{NULL} 
\newcommand{\hatcurSPECWiFeSsnrangemedresCandxxxxxH}{NULL} 
\newcommand{\hatcurSPECWiFeSdaterangelowresCandxxxxxH}{NULL} 
\newcommand{\hatcurSPECWiFeSdaterangemedresCandxxxxxH}{NULL} 
\newcommand{\hatcurSPECWiFeSRVKCandxxxxxH}{\ensuremath{...}} 
\newcommand{\hatcurSPECWiFeSRVGCandxxxxxH}{\ensuremath{NULL}} 
\newcommand{\hatcurSPECarcesTeffCandxxxxxH}{\ensuremath{NULL}} 
\newcommand{\hatcurSPECarcesloggCandxxxxxH}{\ensuremath{NULL}} 
\newcommand{\hatcurSPECarceszfehCandxxxxxH}{\ensuremath{NULL}} 
\newcommand{\hatcurSPECarcesvsiniCandxxxxxH}{\ensuremath{NULL}} 
\newcommand{\hatcurSPECarcesNCandxxxxxH}{\ensuremath{0}} 
\newcommand{\hatcurSPECarcessnrangeCandxxxxxH}{NULL} 
\newcommand{\hatcurSPECarcesdaterangelowresCandxxxxxH}{NULL} 
\newcommand{\hatcurSPECarcesRVKCandxxxxxH}{\ensuremath{NULL}} 
\newcommand{\hatcurSPECarcesRVGCandxxxxxH}{\ensuremath{NULL}} 
\newcommand{\hatcurSPECdupontTeffCandxxxxxH}{\ensuremath{NULL}} 
\newcommand{\hatcurSPECdupontloggCandxxxxxH}{\ensuremath{NULL}} 
\newcommand{\hatcurSPECdupontzfehCandxxxxxH}{\ensuremath{NULL}} 
\newcommand{\hatcurSPECdupontvsiniCandxxxxxH}{\ensuremath{NULL}} 
\newcommand{\hatcurSPECdupontNCandxxxxxH}{\ensuremath{0}} 
\newcommand{\hatcurSPECdupontsnrangeCandxxxxxH}{NULL} 
\newcommand{\hatcurSPECdupontdaterangelowresCandxxxxxH}{NULL} 
\newcommand{\hatcurSPECdupontRVKCandxxxxxH}{\ensuremath{NULL}} 
\newcommand{\hatcurSPECdupontRVGCandxxxxxH}{\ensuremath{NULL}} 
\newcommand{\hatcurSPECharpsTeffCandxxxxxH}{\ensuremath{NULL}} 
\newcommand{\hatcurSPECharpsloggCandxxxxxH}{\ensuremath{NULL}} 
\newcommand{\hatcurSPECharpszfehCandxxxxxH}{\ensuremath{NULL}} 
\newcommand{\hatcurSPECharpsvsiniCandxxxxxH}{\ensuremath{NULL}} 
\newcommand{\hatcurSPECharpsNCandxxxxxH}{\ensuremath{0}} 
\newcommand{\hatcurSPECharpssnrangeCandxxxxxH}{NULL} 
\newcommand{\hatcurSPECharpsdaterangelowresCandxxxxxH}{NULL} 
\newcommand{\hatcurSPECharpsRVKCandxxxxxH}{\ensuremath{NULL}} 
\newcommand{\hatcurSPECharpsRVGCandxxxxxH}{\ensuremath{NULL}} 
\newcommand{\hatcurSPECpfsTeffCandxxxxxH}{\ensuremath{NULL}} 
\newcommand{\hatcurSPECpfsloggCandxxxxxH}{\ensuremath{NULL}} 
\newcommand{\hatcurSPECpfszfehCandxxxxxH}{\ensuremath{NULL}} 
\newcommand{\hatcurSPECpfsvsiniCandxxxxxH}{\ensuremath{NULL}} 
\newcommand{\hatcurSPECpfsNCandxxxxxH}{\ensuremath{0}} 
\newcommand{\hatcurSPECpfssnrangeCandxxxxxH}{NULL} 
\newcommand{\hatcurSPECpfsdaterangelowresCandxxxxxH}{NULL} 
\newcommand{\hatcurSPECpfsRVKCandxxxxxH}{\ensuremath{NULL}} 
\newcommand{\hatcurSPECpfsRVGCandxxxxxH}{\ensuremath{NULL}} 
\newcommand{\hatcurSPECferosTeffCandxxxxxH}{\ensuremath{SB2}} 
\newcommand{\hatcurSPECferosloggCandxxxxxH}{\ensuremath{5.00\pm0.50}} 
\newcommand{\hatcurSPECferoszfehCandxxxxxH}{\ensuremath{-0.5\pm0.5}} 
\newcommand{\hatcurSPECferosvsiniCandxxxxxH}{\ensuremath{30.0\pm2.0}} 
\newcommand{\hatcurSPECferosNCandxxxxxH}{\ensuremath{1}} 
\newcommand{\hatcurSPECferossnrangeCandxxxxxH}{60.0} 
\newcommand{\hatcurSPECferosdaterangelowresCandxxxxxH}{2012-08-22} 
\newcommand{\hatcurSPECferosRVKCandxxxxxH}{\ensuremath{NULL}} 
\newcommand{\hatcurSPECferosRVGCandxxxxxH}{\ensuremath{26.71}} 
\newcommand{\hatcurSPECfiesTeffCandxxxxxH}{\ensuremath{NULL}} 
\newcommand{\hatcurSPECfiesloggCandxxxxxH}{\ensuremath{NULL}} 
\newcommand{\hatcurSPECfieszfehCandxxxxxH}{\ensuremath{NULL}} 
\newcommand{\hatcurSPECfiesvsiniCandxxxxxH}{\ensuremath{NULL}} 
\newcommand{\hatcurSPECfiesNCandxxxxxH}{\ensuremath{0}} 
\newcommand{\hatcurSPECfiessnrangeCandxxxxxH}{NULL} 
\newcommand{\hatcurSPECfiesdaterangelowresCandxxxxxH}{NULL} 
\newcommand{\hatcurSPECfiesRVKCandxxxxxH}{\ensuremath{NULL}} 
\newcommand{\hatcurSPECfiesRVGCandxxxxxH}{\ensuremath{NULL}} 
\newcommand{\hatcurSPECcoralieTeffCandxxxxxH}{\ensuremath{NULL}} 
\newcommand{\hatcurSPECcoralieloggCandxxxxxH}{\ensuremath{NULL}} 
\newcommand{\hatcurSPECcoraliezfehCandxxxxxH}{\ensuremath{NULL}} 
\newcommand{\hatcurSPECcoralievsiniCandxxxxxH}{\ensuremath{NULL}} 
\newcommand{\hatcurSPECcoralieNCandxxxxxH}{\ensuremath{0}} 
\newcommand{\hatcurSPECcoraliesnrangeCandxxxxxH}{NULL} 
\newcommand{\hatcurSPECcoraliedaterangelowresCandxxxxxH}{NULL} 
\newcommand{\hatcurSPECcoralieRVKCandxxxxxH}{\ensuremath{NULL}} 
\newcommand{\hatcurSPECcoralieRVGCandxxxxxH}{\ensuremath{NULL}} 
\newcommand{\hatcurhtrCandxxxxxI}{HATS579-015}
\newcommand{\hatcurCCraCandxxxxxI}{\ensuremath{19^{\mathrm h}22^{\mathrm m}38.28{\mathrm s}}}
\newcommand{\hatcurCCdecCandxxxxxI}{\ensuremath{-21^{\arcdeg}05^{\arcmin}01.9{\arcsec}}}
\newcommand{\hatcurCLASSCandxxxxxI}{EB}
\newcommand{\hatcurCCEPICCandxxxxxI}{EPIC~217149884}
\newcommand{\hatcurLCPCandxxxxxI}{\ensuremath{16.6924091\pm0.0000246}} 
\newcommand{\hatcurLCTCandxxxxxI}{\ensuremath{2457248.29441\pm0.00053}} 
\newcommand{\hatcurLCdurCandxxxxxI}{\ensuremath{0.26737\pm0.01102}} 
\newcommand{\hatcurLCrprstarCandxxxxxI}{\ensuremath{0.19491\pm0.00131}} 
\newcommand{\hatcurLCbsqCandxxxxxI}{\ensuremath{0.614\pm0.015}} 
\newcommand{\hatcurCCmagBCandxxxxxI}{\ensuremath{15.154}} 
\newcommand{\hatcurCCmagVCandxxxxxI}{\ensuremath{14.421}} 
\newcommand{\hatcurCCmaggCandxxxxxI}{\ensuremath{14.718}} 
\newcommand{\hatcurCCmagrCandxxxxxI}{\ensuremath{14.204}} 
\newcommand{\hatcurCCmagiCandxxxxxI}{\ensuremath{14.001}} 
\newcommand{\hatcurCCmagJCandxxxxxI}{\ensuremath{13.055\pm0.02}} 
\newcommand{\hatcurCCmagHCandxxxxxI}{\ensuremath{12.703\pm0.03}} 
\newcommand{\hatcurCCmagKCandxxxxxI}{\ensuremath{12.660\pm0.03}} 
\newcommand{\hatcurSPECWiFeSTeffCandxxxxxI}{\ensuremath{5751\pm300}} 
\newcommand{\hatcurSPECWiFeSloggCandxxxxxI}{\ensuremath{4.00\pm0.30}} 
\newcommand{\hatcurSPECWiFeSzfehCandxxxxxI}{\ensuremath{0.0\pm0.5}} 
\newcommand{\hatcurSPECWiFeSNlowresCandxxxxxI}{\ensuremath{1}} 
\newcommand{\hatcurSPECWiFeSNmedresCandxxxxxI}{\ensuremath{5}} 
\newcommand{\hatcurSPECWiFeSsnrangelowresCandxxxxxI}{67.1} 
\newcommand{\hatcurSPECWiFeSsnrangemedresCandxxxxxI}{23.5--54.4} 
\newcommand{\hatcurSPECWiFeSdaterangelowresCandxxxxxI}{2012-09-09} 
\newcommand{\hatcurSPECWiFeSdaterangemedresCandxxxxxI}{2012-09-24--2012-10-25} 
\newcommand{\hatcurSPECWiFeSRVKCandxxxxxI}{\ensuremath{-5.47\pm1.05}} 
\newcommand{\hatcurSPECWiFeSRVGCandxxxxxI}{\ensuremath{-8.34\pm0.56}} 
\newcommand{\hatcurSPECarcesTeffCandxxxxxI}{\ensuremath{NULL}} 
\newcommand{\hatcurSPECarcesloggCandxxxxxI}{\ensuremath{NULL}} 
\newcommand{\hatcurSPECarceszfehCandxxxxxI}{\ensuremath{NULL}} 
\newcommand{\hatcurSPECarcesvsiniCandxxxxxI}{\ensuremath{NULL}} 
\newcommand{\hatcurSPECarcesNCandxxxxxI}{\ensuremath{0}} 
\newcommand{\hatcurSPECarcessnrangeCandxxxxxI}{NULL} 
\newcommand{\hatcurSPECarcesdaterangelowresCandxxxxxI}{NULL} 
\newcommand{\hatcurSPECarcesRVKCandxxxxxI}{\ensuremath{NULL}} 
\newcommand{\hatcurSPECarcesRVGCandxxxxxI}{\ensuremath{NULL}} 
\newcommand{\hatcurSPECdupontTeffCandxxxxxI}{\ensuremath{NULL}} 
\newcommand{\hatcurSPECdupontloggCandxxxxxI}{\ensuremath{NULL}} 
\newcommand{\hatcurSPECdupontzfehCandxxxxxI}{\ensuremath{NULL}} 
\newcommand{\hatcurSPECdupontvsiniCandxxxxxI}{\ensuremath{NULL}} 
\newcommand{\hatcurSPECdupontNCandxxxxxI}{\ensuremath{0}} 
\newcommand{\hatcurSPECdupontsnrangeCandxxxxxI}{NULL} 
\newcommand{\hatcurSPECdupontdaterangelowresCandxxxxxI}{NULL} 
\newcommand{\hatcurSPECdupontRVKCandxxxxxI}{\ensuremath{NULL}} 
\newcommand{\hatcurSPECdupontRVGCandxxxxxI}{\ensuremath{NULL}} 
\newcommand{\hatcurSPECharpsTeffCandxxxxxI}{\ensuremath{NULL}} 
\newcommand{\hatcurSPECharpsloggCandxxxxxI}{\ensuremath{NULL}} 
\newcommand{\hatcurSPECharpszfehCandxxxxxI}{\ensuremath{NULL}} 
\newcommand{\hatcurSPECharpsvsiniCandxxxxxI}{\ensuremath{NULL}} 
\newcommand{\hatcurSPECharpsNCandxxxxxI}{\ensuremath{0}} 
\newcommand{\hatcurSPECharpssnrangeCandxxxxxI}{NULL} 
\newcommand{\hatcurSPECharpsdaterangelowresCandxxxxxI}{NULL} 
\newcommand{\hatcurSPECharpsRVKCandxxxxxI}{\ensuremath{NULL}} 
\newcommand{\hatcurSPECharpsRVGCandxxxxxI}{\ensuremath{NULL}} 
\newcommand{\hatcurSPECpfsTeffCandxxxxxI}{\ensuremath{NULL}} 
\newcommand{\hatcurSPECpfsloggCandxxxxxI}{\ensuremath{NULL}} 
\newcommand{\hatcurSPECpfszfehCandxxxxxI}{\ensuremath{NULL}} 
\newcommand{\hatcurSPECpfsvsiniCandxxxxxI}{\ensuremath{NULL}} 
\newcommand{\hatcurSPECpfsNCandxxxxxI}{\ensuremath{0}} 
\newcommand{\hatcurSPECpfssnrangeCandxxxxxI}{NULL} 
\newcommand{\hatcurSPECpfsdaterangelowresCandxxxxxI}{NULL} 
\newcommand{\hatcurSPECpfsRVKCandxxxxxI}{\ensuremath{NULL}} 
\newcommand{\hatcurSPECpfsRVGCandxxxxxI}{\ensuremath{NULL}} 
\newcommand{\hatcurSPECferosTeffCandxxxxxI}{\ensuremath{5817\pm58}} 
\newcommand{\hatcurSPECferosloggCandxxxxxI}{\ensuremath{4.93\pm0.29}} 
\newcommand{\hatcurSPECferoszfehCandxxxxxI}{\ensuremath{0.2\pm0.3}} 
\newcommand{\hatcurSPECferosvsiniCandxxxxxI}{\ensuremath{2.5\pm1.2}} 
\newcommand{\hatcurSPECferosNCandxxxxxI}{\ensuremath{3}} 
\newcommand{\hatcurSPECferossnrangeCandxxxxxI}{18.0--32.0} 
\newcommand{\hatcurSPECferosdaterangelowresCandxxxxxI}{2016-05-18--2016-06-19} 
\newcommand{\hatcurSPECferosRVKCandxxxxxI}{\ensuremath{17.69\pm0.05}} 
\newcommand{\hatcurSPECferosRVGCandxxxxxI}{\ensuremath{-1.40\pm0.04}} 
\newcommand{\hatcurSPECfiesTeffCandxxxxxI}{\ensuremath{NULL}} 
\newcommand{\hatcurSPECfiesloggCandxxxxxI}{\ensuremath{NULL}} 
\newcommand{\hatcurSPECfieszfehCandxxxxxI}{\ensuremath{NULL}} 
\newcommand{\hatcurSPECfiesvsiniCandxxxxxI}{\ensuremath{NULL}} 
\newcommand{\hatcurSPECfiesNCandxxxxxI}{\ensuremath{0}} 
\newcommand{\hatcurSPECfiessnrangeCandxxxxxI}{NULL} 
\newcommand{\hatcurSPECfiesdaterangelowresCandxxxxxI}{NULL} 
\newcommand{\hatcurSPECfiesRVKCandxxxxxI}{\ensuremath{NULL}} 
\newcommand{\hatcurSPECfiesRVGCandxxxxxI}{\ensuremath{NULL}} 
\newcommand{\hatcurSPECcoralieTeffCandxxxxxI}{\ensuremath{NULL}} 
\newcommand{\hatcurSPECcoralieloggCandxxxxxI}{\ensuremath{NULL}} 
\newcommand{\hatcurSPECcoraliezfehCandxxxxxI}{\ensuremath{NULL}} 
\newcommand{\hatcurSPECcoralievsiniCandxxxxxI}{\ensuremath{NULL}} 
\newcommand{\hatcurSPECcoralieNCandxxxxxI}{\ensuremath{0}} 
\newcommand{\hatcurSPECcoraliesnrangeCandxxxxxI}{NULL} 
\newcommand{\hatcurSPECcoraliedaterangelowresCandxxxxxI}{NULL} 
\newcommand{\hatcurSPECcoralieRVKCandxxxxxI}{\ensuremath{NULL}} 
\newcommand{\hatcurSPECcoralieRVGCandxxxxxI}{\ensuremath{NULL}} 
\newcommand{\hatcurhtrCandxxxxxJ}{HATS579-036}
\newcommand{\hatcurCCraCandxxxxxJ}{\ensuremath{19^{\mathrm h}37^{\mathrm m}55.20{\mathrm s}}}
\newcommand{\hatcurCCdecCandxxxxxJ}{\ensuremath{-24^{\arcdeg}23^{\arcmin}05.9{\arcsec}}}
\newcommand{\hatcurCLASSCandxxxxxJ}{EB}
\newcommand{\hatcurCCEPICCandxxxxxJ}{EPIC~215358983}
\newcommand{\hatcurLCPCandxxxxxJ}{\ensuremath{6.4218981\pm0.0000077}} 
\newcommand{\hatcurLCTCandxxxxxJ}{\ensuremath{2457001.60205\pm0.00079}} 
\newcommand{\hatcurLCdurCandxxxxxJ}{\ensuremath{0.26826\pm0.00627}} 
\newcommand{\hatcurLCrprstarCandxxxxxJ}{\ensuremath{0.15515\pm0.00084}} 
\newcommand{\hatcurLCbsqCandxxxxxJ}{\ensuremath{0.434\pm0.047}} 
\newcommand{\hatcurCCmagBCandxxxxxJ}{\ensuremath{14.585}} 
\newcommand{\hatcurCCmagVCandxxxxxJ}{\ensuremath{13.987}} 
\newcommand{\hatcurCCmaggCandxxxxxJ}{\ensuremath{14.202}} 
\newcommand{\hatcurCCmagrCandxxxxxJ}{\ensuremath{13.788}} 
\newcommand{\hatcurCCmagiCandxxxxxJ}{\ensuremath{13.612}} 
\newcommand{\hatcurCCmagJCandxxxxxJ}{\ensuremath{12.689\pm0.03}} 
\newcommand{\hatcurCCmagHCandxxxxxJ}{\ensuremath{12.366\pm0.03}} 
\newcommand{\hatcurCCmagKCandxxxxxJ}{\ensuremath{12.328\pm0.03}} 
\newcommand{\hatcurSPECWiFeSTeffCandxxxxxJ}{\ensuremath{6700\pm100}} 
\newcommand{\hatcurSPECWiFeSloggCandxxxxxJ}{\ensuremath{5.00\pm0.50}} 
\newcommand{\hatcurSPECWiFeSzfehCandxxxxxJ}{\ensuremath{NULL}} 
\newcommand{\hatcurSPECWiFeSNlowresCandxxxxxJ}{\ensuremath{0}} 
\newcommand{\hatcurSPECWiFeSNmedresCandxxxxxJ}{\ensuremath{0}} 
\newcommand{\hatcurSPECWiFeSsnrangelowresCandxxxxxJ}{NULL} 
\newcommand{\hatcurSPECWiFeSsnrangemedresCandxxxxxJ}{NULL} 
\newcommand{\hatcurSPECWiFeSdaterangelowresCandxxxxxJ}{NULL} 
\newcommand{\hatcurSPECWiFeSdaterangemedresCandxxxxxJ}{NULL} 
\newcommand{\hatcurSPECWiFeSRVKCandxxxxxJ}{\ensuremath{22.49\pm0.01}} 
\newcommand{\hatcurSPECWiFeSRVGCandxxxxxJ}{\ensuremath{NULL}} 
\newcommand{\hatcurSPECarcesTeffCandxxxxxJ}{\ensuremath{NULL}} 
\newcommand{\hatcurSPECarcesloggCandxxxxxJ}{\ensuremath{NULL}} 
\newcommand{\hatcurSPECarceszfehCandxxxxxJ}{\ensuremath{NULL}} 
\newcommand{\hatcurSPECarcesvsiniCandxxxxxJ}{\ensuremath{NULL}} 
\newcommand{\hatcurSPECarcesNCandxxxxxJ}{\ensuremath{0}} 
\newcommand{\hatcurSPECarcessnrangeCandxxxxxJ}{NULL} 
\newcommand{\hatcurSPECarcesdaterangelowresCandxxxxxJ}{NULL} 
\newcommand{\hatcurSPECarcesRVKCandxxxxxJ}{\ensuremath{NULL}} 
\newcommand{\hatcurSPECarcesRVGCandxxxxxJ}{\ensuremath{NULL}} 
\newcommand{\hatcurSPECdupontTeffCandxxxxxJ}{\ensuremath{NULL}} 
\newcommand{\hatcurSPECdupontloggCandxxxxxJ}{\ensuremath{NULL}} 
\newcommand{\hatcurSPECdupontzfehCandxxxxxJ}{\ensuremath{NULL}} 
\newcommand{\hatcurSPECdupontvsiniCandxxxxxJ}{\ensuremath{NULL}} 
\newcommand{\hatcurSPECdupontNCandxxxxxJ}{\ensuremath{0}} 
\newcommand{\hatcurSPECdupontsnrangeCandxxxxxJ}{NULL} 
\newcommand{\hatcurSPECdupontdaterangelowresCandxxxxxJ}{NULL} 
\newcommand{\hatcurSPECdupontRVKCandxxxxxJ}{\ensuremath{NULL}} 
\newcommand{\hatcurSPECdupontRVGCandxxxxxJ}{\ensuremath{NULL}} 
\newcommand{\hatcurSPECharpsTeffCandxxxxxJ}{\ensuremath{NULL}} 
\newcommand{\hatcurSPECharpsloggCandxxxxxJ}{\ensuremath{NULL}} 
\newcommand{\hatcurSPECharpszfehCandxxxxxJ}{\ensuremath{NULL}} 
\newcommand{\hatcurSPECharpsvsiniCandxxxxxJ}{\ensuremath{NULL}} 
\newcommand{\hatcurSPECharpsNCandxxxxxJ}{\ensuremath{0}} 
\newcommand{\hatcurSPECharpssnrangeCandxxxxxJ}{NULL} 
\newcommand{\hatcurSPECharpsdaterangelowresCandxxxxxJ}{NULL} 
\newcommand{\hatcurSPECharpsRVKCandxxxxxJ}{\ensuremath{NULL}} 
\newcommand{\hatcurSPECharpsRVGCandxxxxxJ}{\ensuremath{NULL}} 
\newcommand{\hatcurSPECpfsTeffCandxxxxxJ}{\ensuremath{NULL}} 
\newcommand{\hatcurSPECpfsloggCandxxxxxJ}{\ensuremath{NULL}} 
\newcommand{\hatcurSPECpfszfehCandxxxxxJ}{\ensuremath{NULL}} 
\newcommand{\hatcurSPECpfsvsiniCandxxxxxJ}{\ensuremath{NULL}} 
\newcommand{\hatcurSPECpfsNCandxxxxxJ}{\ensuremath{0}} 
\newcommand{\hatcurSPECpfssnrangeCandxxxxxJ}{NULL} 
\newcommand{\hatcurSPECpfsdaterangelowresCandxxxxxJ}{NULL} 
\newcommand{\hatcurSPECpfsRVKCandxxxxxJ}{\ensuremath{NULL}} 
\newcommand{\hatcurSPECpfsRVGCandxxxxxJ}{\ensuremath{NULL}} 
\newcommand{\hatcurSPECferosTeffCandxxxxxJ}{\ensuremath{6550\pm71}} 
\newcommand{\hatcurSPECferosloggCandxxxxxJ}{\ensuremath{4.75\pm0.35}} 
\newcommand{\hatcurSPECferoszfehCandxxxxxJ}{\ensuremath{-0.7\pm0.4}} 
\newcommand{\hatcurSPECferosvsiniCandxxxxxJ}{\ensuremath{12.5\pm1.4}} 
\newcommand{\hatcurSPECferosNCandxxxxxJ}{\ensuremath{4}} 
\newcommand{\hatcurSPECferossnrangeCandxxxxxJ}{11.8--48.0} 
\newcommand{\hatcurSPECferosdaterangelowresCandxxxxxJ}{2012-08-26--2012-08-29} 
\newcommand{\hatcurSPECferosRVKCandxxxxxJ}{\ensuremath{22.53\pm0.03}} 
\newcommand{\hatcurSPECferosRVGCandxxxxxJ}{\ensuremath{-11.97\pm0.03}} 
\newcommand{\hatcurSPECfiesTeffCandxxxxxJ}{\ensuremath{NULL}} 
\newcommand{\hatcurSPECfiesloggCandxxxxxJ}{\ensuremath{NULL}} 
\newcommand{\hatcurSPECfieszfehCandxxxxxJ}{\ensuremath{NULL}} 
\newcommand{\hatcurSPECfiesvsiniCandxxxxxJ}{\ensuremath{NULL}} 
\newcommand{\hatcurSPECfiesNCandxxxxxJ}{\ensuremath{0}} 
\newcommand{\hatcurSPECfiessnrangeCandxxxxxJ}{NULL} 
\newcommand{\hatcurSPECfiesdaterangelowresCandxxxxxJ}{NULL} 
\newcommand{\hatcurSPECfiesRVKCandxxxxxJ}{\ensuremath{NULL}} 
\newcommand{\hatcurSPECfiesRVGCandxxxxxJ}{\ensuremath{NULL}} 
\newcommand{\hatcurSPECcoralieTeffCandxxxxxJ}{\ensuremath{NULL}} 
\newcommand{\hatcurSPECcoralieloggCandxxxxxJ}{\ensuremath{NULL}} 
\newcommand{\hatcurSPECcoraliezfehCandxxxxxJ}{\ensuremath{NULL}} 
\newcommand{\hatcurSPECcoralievsiniCandxxxxxJ}{\ensuremath{NULL}} 
\newcommand{\hatcurSPECcoralieNCandxxxxxJ}{\ensuremath{0}} 
\newcommand{\hatcurSPECcoraliesnrangeCandxxxxxJ}{NULL} 
\newcommand{\hatcurSPECcoraliedaterangelowresCandxxxxxJ}{NULL} 
\newcommand{\hatcurSPECcoralieRVKCandxxxxxJ}{\ensuremath{NULL}} 
\newcommand{\hatcurSPECcoralieRVGCandxxxxxJ}{\ensuremath{NULL}} 
\newcommand{\hatcurhtrCandxxxxxK}{HATS579-037}
\newcommand{\hatcurCCraCandxxxxxK}{\ensuremath{19^{\mathrm h}44^{\mathrm m}53.16{\mathrm s}}}
\newcommand{\hatcurCCdecCandxxxxxK}{\ensuremath{-24^{\arcdeg}38^{\arcmin}57.4{\arcsec}}}
\newcommand{\hatcurCLASSCandxxxxxK}{BEB}
\newcommand{\hatcurCCEPICCandxxxxxK}{EPIC~215234145}
\newcommand{\hatcurLCPCandxxxxxK}{\ensuremath{1.2539910\pm0.0000020}} 
\newcommand{\hatcurLCTCandxxxxxK}{\ensuremath{2455879.30405\pm0.00069}} 
\newcommand{\hatcurLCdurCandxxxxxK}{\ensuremath{0.07682\pm0.00735}} 
\newcommand{\hatcurLCrprstarCandxxxxxK}{\ensuremath{0.09970\pm0.00226}} 
\newcommand{\hatcurLCbsqCandxxxxxK}{\ensuremath{0.805\pm0.045}} 
\newcommand{\hatcurCCmagBCandxxxxxK}{\ensuremath{14.238}} 
\newcommand{\hatcurCCmagVCandxxxxxK}{\ensuremath{13.536}} 
\newcommand{\hatcurCCmaggCandxxxxxK}{\ensuremath{13.775}} 
\newcommand{\hatcurCCmagrCandxxxxxK}{\ensuremath{13.140}} 
\newcommand{\hatcurCCmagiCandxxxxxK}{\ensuremath{12.823}} 
\newcommand{\hatcurCCmagJCandxxxxxK}{\ensuremath{11.718\pm0.02}} 
\newcommand{\hatcurCCmagHCandxxxxxK}{\ensuremath{11.278\pm0.03}} 
\newcommand{\hatcurCCmagKCandxxxxxK}{\ensuremath{11.170\pm0.02}} 
\newcommand{\hatcurSPECWiFeSTeffCandxxxxxK}{\ensuremath{5524\pm300}} 
\newcommand{\hatcurSPECWiFeSloggCandxxxxxK}{\ensuremath{4.50\pm0.30}} 
\newcommand{\hatcurSPECWiFeSzfehCandxxxxxK}{\ensuremath{0.0\pm0.5}} 
\newcommand{\hatcurSPECWiFeSNlowresCandxxxxxK}{\ensuremath{1}} 
\newcommand{\hatcurSPECWiFeSNmedresCandxxxxxK}{\ensuremath{3}} 
\newcommand{\hatcurSPECWiFeSsnrangelowresCandxxxxxK}{101.3} 
\newcommand{\hatcurSPECWiFeSsnrangemedresCandxxxxxK}{52.6--85.9} 
\newcommand{\hatcurSPECWiFeSdaterangelowresCandxxxxxK}{2012-09-08} 
\newcommand{\hatcurSPECWiFeSdaterangemedresCandxxxxxK}{2012-09-24--2012-09-26} 
\newcommand{\hatcurSPECWiFeSRVKCandxxxxxK}{\ensuremath{<2.0}} 
\newcommand{\hatcurSPECWiFeSRVGCandxxxxxK}{\ensuremath{-28.39\pm0.29}} 
\newcommand{\hatcurSPECarcesTeffCandxxxxxK}{\ensuremath{NULL}} 
\newcommand{\hatcurSPECarcesloggCandxxxxxK}{\ensuremath{NULL}} 
\newcommand{\hatcurSPECarceszfehCandxxxxxK}{\ensuremath{NULL}} 
\newcommand{\hatcurSPECarcesvsiniCandxxxxxK}{\ensuremath{NULL}} 
\newcommand{\hatcurSPECarcesNCandxxxxxK}{\ensuremath{0}} 
\newcommand{\hatcurSPECarcessnrangeCandxxxxxK}{NULL} 
\newcommand{\hatcurSPECarcesdaterangelowresCandxxxxxK}{NULL} 
\newcommand{\hatcurSPECarcesRVKCandxxxxxK}{\ensuremath{NULL}} 
\newcommand{\hatcurSPECarcesRVGCandxxxxxK}{\ensuremath{NULL}} 
\newcommand{\hatcurSPECdupontTeffCandxxxxxK}{\ensuremath{NULL}} 
\newcommand{\hatcurSPECdupontloggCandxxxxxK}{\ensuremath{NULL}} 
\newcommand{\hatcurSPECdupontzfehCandxxxxxK}{\ensuremath{NULL}} 
\newcommand{\hatcurSPECdupontvsiniCandxxxxxK}{\ensuremath{NULL}} 
\newcommand{\hatcurSPECdupontNCandxxxxxK}{\ensuremath{0}} 
\newcommand{\hatcurSPECdupontsnrangeCandxxxxxK}{NULL} 
\newcommand{\hatcurSPECdupontdaterangelowresCandxxxxxK}{NULL} 
\newcommand{\hatcurSPECdupontRVKCandxxxxxK}{\ensuremath{NULL}} 
\newcommand{\hatcurSPECdupontRVGCandxxxxxK}{\ensuremath{NULL}} 
\newcommand{\hatcurSPECharpsTeffCandxxxxxK}{\ensuremath{NULL}} 
\newcommand{\hatcurSPECharpsloggCandxxxxxK}{\ensuremath{NULL}} 
\newcommand{\hatcurSPECharpszfehCandxxxxxK}{\ensuremath{NULL}} 
\newcommand{\hatcurSPECharpsvsiniCandxxxxxK}{\ensuremath{NULL}} 
\newcommand{\hatcurSPECharpsNCandxxxxxK}{\ensuremath{0}} 
\newcommand{\hatcurSPECharpssnrangeCandxxxxxK}{NULL} 
\newcommand{\hatcurSPECharpsdaterangelowresCandxxxxxK}{NULL} 
\newcommand{\hatcurSPECharpsRVKCandxxxxxK}{\ensuremath{NULL}} 
\newcommand{\hatcurSPECharpsRVGCandxxxxxK}{\ensuremath{NULL}} 
\newcommand{\hatcurSPECpfsTeffCandxxxxxK}{\ensuremath{NULL}} 
\newcommand{\hatcurSPECpfsloggCandxxxxxK}{\ensuremath{NULL}} 
\newcommand{\hatcurSPECpfszfehCandxxxxxK}{\ensuremath{NULL}} 
\newcommand{\hatcurSPECpfsvsiniCandxxxxxK}{\ensuremath{NULL}} 
\newcommand{\hatcurSPECpfsNCandxxxxxK}{\ensuremath{0}} 
\newcommand{\hatcurSPECpfssnrangeCandxxxxxK}{NULL} 
\newcommand{\hatcurSPECpfsdaterangelowresCandxxxxxK}{NULL} 
\newcommand{\hatcurSPECpfsRVKCandxxxxxK}{\ensuremath{NULL}} 
\newcommand{\hatcurSPECpfsRVGCandxxxxxK}{\ensuremath{NULL}} 
\newcommand{\hatcurSPECferosTeffCandxxxxxK}{\ensuremath{5558\pm41}} 
\newcommand{\hatcurSPECferosloggCandxxxxxK}{\ensuremath{4.60\pm0.20}} 
\newcommand{\hatcurSPECferoszfehCandxxxxxK}{\ensuremath{-0.4\pm0.2}} 
\newcommand{\hatcurSPECferosvsiniCandxxxxxK}{\ensuremath{3.3\pm0.8}} 
\newcommand{\hatcurSPECferosNCandxxxxxK}{\ensuremath{10}} 
\newcommand{\hatcurSPECferossnrangeCandxxxxxK}{12.6--59.0} 
\newcommand{\hatcurSPECferosdaterangelowresCandxxxxxK}{2013-04-16--2014-07-14} 
\newcommand{\hatcurSPECferosRVKCandxxxxxK}{\ensuremath{0.18\pm0.01}} 
\newcommand{\hatcurSPECferosRVGCandxxxxxK}{\ensuremath{-28.83\pm0.01}} 
\newcommand{\hatcurSPECfiesTeffCandxxxxxK}{\ensuremath{NULL}} 
\newcommand{\hatcurSPECfiesloggCandxxxxxK}{\ensuremath{NULL}} 
\newcommand{\hatcurSPECfieszfehCandxxxxxK}{\ensuremath{NULL}} 
\newcommand{\hatcurSPECfiesvsiniCandxxxxxK}{\ensuremath{NULL}} 
\newcommand{\hatcurSPECfiesNCandxxxxxK}{\ensuremath{0}} 
\newcommand{\hatcurSPECfiessnrangeCandxxxxxK}{NULL} 
\newcommand{\hatcurSPECfiesdaterangelowresCandxxxxxK}{NULL} 
\newcommand{\hatcurSPECfiesRVKCandxxxxxK}{\ensuremath{NULL}} 
\newcommand{\hatcurSPECfiesRVGCandxxxxxK}{\ensuremath{NULL}} 
\newcommand{\hatcurSPECcoralieTeffCandxxxxxK}{\ensuremath{5175\pm50}} 
\newcommand{\hatcurSPECcoralieloggCandxxxxxK}{\ensuremath{3.85\pm0.25}} 
\newcommand{\hatcurSPECcoraliezfehCandxxxxxK}{\ensuremath{-0.8\pm0.2}} 
\newcommand{\hatcurSPECcoralievsiniCandxxxxxK}{\ensuremath{2.5\pm1.0}} 
\newcommand{\hatcurSPECcoralieNCandxxxxxK}{\ensuremath{4}} 
\newcommand{\hatcurSPECcoraliesnrangeCandxxxxxK}{8.0--18.0} 
\newcommand{\hatcurSPECcoraliedaterangelowresCandxxxxxK}{2012-08-21--2012-08-25} 
\newcommand{\hatcurSPECcoralieRVKCandxxxxxK}{\ensuremath{59.47\pm0.04}} 
\newcommand{\hatcurSPECcoralieRVGCandxxxxxK}{\ensuremath{-27.02\pm0.02}} 
\newcommand{\hatcurhtrCandxxxxxL}{HATS579-039}
\newcommand{\hatcurCCraCandxxxxxL}{\ensuremath{19^{\mathrm h}36^{\mathrm m}18.72{\mathrm s}}}
\newcommand{\hatcurCCdecCandxxxxxL}{\ensuremath{-24^{\arcdeg}23^{\arcmin}47.0{\arcsec}}}
\newcommand{\hatcurCLASSCandxxxxxL}{EB}
\newcommand{\hatcurCCEPICCandxxxxxL}{EPIC~215353525}
\newcommand{\hatcurLCPCandxxxxxL}{\ensuremath{0.9085948\pm0.0000042}} 
\newcommand{\hatcurLCTCandxxxxxL}{\ensuremath{2455844.28763\pm0.00128}} 
\newcommand{\hatcurLCdurCandxxxxxL}{\ensuremath{0.06313\pm0.00847}} 
\newcommand{\hatcurLCrprstarCandxxxxxL}{\ensuremath{0.10881\pm0.00469}} 
\newcommand{\hatcurLCbsqCandxxxxxL}{\ensuremath{0.755\pm0.157}} 
\newcommand{\hatcurCCmagBCandxxxxxL}{\ensuremath{15.981}} 
\newcommand{\hatcurCCmagVCandxxxxxL}{\ensuremath{14.946}} 
\newcommand{\hatcurCCmaggCandxxxxxL}{\ensuremath{15.104}} 
\newcommand{\hatcurCCmagrCandxxxxxL}{\ensuremath{14.508}} 
\newcommand{\hatcurCCmagiCandxxxxxL}{\ensuremath{14.556}} 
\newcommand{\hatcurCCmagJCandxxxxxL}{\ensuremath{13.265\pm0.02}} 
\newcommand{\hatcurCCmagHCandxxxxxL}{\ensuremath{12.872\pm0.03}} 
\newcommand{\hatcurCCmagKCandxxxxxL}{\ensuremath{12.784\pm0.03}} 
\newcommand{\hatcurSPECWiFeSTeffCandxxxxxL}{\ensuremath{5229\pm212}} 
\newcommand{\hatcurSPECWiFeSloggCandxxxxxL}{\ensuremath{4.50\pm0.21}} 
\newcommand{\hatcurSPECWiFeSzfehCandxxxxxL}{\ensuremath{-1.0\pm0.4}} 
\newcommand{\hatcurSPECWiFeSNlowresCandxxxxxL}{\ensuremath{2}} 
\newcommand{\hatcurSPECWiFeSNmedresCandxxxxxL}{\ensuremath{9}} 
\newcommand{\hatcurSPECWiFeSsnrangelowresCandxxxxxL}{33.7--44.4} 
\newcommand{\hatcurSPECWiFeSsnrangemedresCandxxxxxL}{1.6--57.8} 
\newcommand{\hatcurSPECWiFeSdaterangelowresCandxxxxxL}{2013-05-25--2013-09-26} 
\newcommand{\hatcurSPECWiFeSdaterangemedresCandxxxxxL}{2013-05-17--2013-09-27} 
\newcommand{\hatcurSPECWiFeSRVKCandxxxxxL}{\ensuremath{<2.0}} 
\newcommand{\hatcurSPECWiFeSRVGCandxxxxxL}{\ensuremath{18.18\pm0.72}} 
\newcommand{\hatcurSPECarcesTeffCandxxxxxL}{\ensuremath{NULL}} 
\newcommand{\hatcurSPECarcesloggCandxxxxxL}{\ensuremath{NULL}} 
\newcommand{\hatcurSPECarceszfehCandxxxxxL}{\ensuremath{NULL}} 
\newcommand{\hatcurSPECarcesvsiniCandxxxxxL}{\ensuremath{NULL}} 
\newcommand{\hatcurSPECarcesNCandxxxxxL}{\ensuremath{0}} 
\newcommand{\hatcurSPECarcessnrangeCandxxxxxL}{NULL} 
\newcommand{\hatcurSPECarcesdaterangelowresCandxxxxxL}{NULL} 
\newcommand{\hatcurSPECarcesRVKCandxxxxxL}{\ensuremath{NULL}} 
\newcommand{\hatcurSPECarcesRVGCandxxxxxL}{\ensuremath{NULL}} 
\newcommand{\hatcurSPECdupontTeffCandxxxxxL}{\ensuremath{NULL}} 
\newcommand{\hatcurSPECdupontloggCandxxxxxL}{\ensuremath{NULL}} 
\newcommand{\hatcurSPECdupontzfehCandxxxxxL}{\ensuremath{NULL}} 
\newcommand{\hatcurSPECdupontvsiniCandxxxxxL}{\ensuremath{NULL}} 
\newcommand{\hatcurSPECdupontNCandxxxxxL}{\ensuremath{0}} 
\newcommand{\hatcurSPECdupontsnrangeCandxxxxxL}{NULL} 
\newcommand{\hatcurSPECdupontdaterangelowresCandxxxxxL}{NULL} 
\newcommand{\hatcurSPECdupontRVKCandxxxxxL}{\ensuremath{NULL}} 
\newcommand{\hatcurSPECdupontRVGCandxxxxxL}{\ensuremath{NULL}} 
\newcommand{\hatcurSPECharpsTeffCandxxxxxL}{\ensuremath{NULL}} 
\newcommand{\hatcurSPECharpsloggCandxxxxxL}{\ensuremath{NULL}} 
\newcommand{\hatcurSPECharpszfehCandxxxxxL}{\ensuremath{NULL}} 
\newcommand{\hatcurSPECharpsvsiniCandxxxxxL}{\ensuremath{NULL}} 
\newcommand{\hatcurSPECharpsNCandxxxxxL}{\ensuremath{0}} 
\newcommand{\hatcurSPECharpssnrangeCandxxxxxL}{NULL} 
\newcommand{\hatcurSPECharpsdaterangelowresCandxxxxxL}{NULL} 
\newcommand{\hatcurSPECharpsRVKCandxxxxxL}{\ensuremath{NULL}} 
\newcommand{\hatcurSPECharpsRVGCandxxxxxL}{\ensuremath{NULL}} 
\newcommand{\hatcurSPECpfsTeffCandxxxxxL}{\ensuremath{NULL}} 
\newcommand{\hatcurSPECpfsloggCandxxxxxL}{\ensuremath{NULL}} 
\newcommand{\hatcurSPECpfszfehCandxxxxxL}{\ensuremath{NULL}} 
\newcommand{\hatcurSPECpfsvsiniCandxxxxxL}{\ensuremath{NULL}} 
\newcommand{\hatcurSPECpfsNCandxxxxxL}{\ensuremath{0}} 
\newcommand{\hatcurSPECpfssnrangeCandxxxxxL}{NULL} 
\newcommand{\hatcurSPECpfsdaterangelowresCandxxxxxL}{NULL} 
\newcommand{\hatcurSPECpfsRVKCandxxxxxL}{\ensuremath{NULL}} 
\newcommand{\hatcurSPECpfsRVGCandxxxxxL}{\ensuremath{NULL}} 
\newcommand{\hatcurSPECferosTeffCandxxxxxL}{\ensuremath{NULL}} 
\newcommand{\hatcurSPECferosloggCandxxxxxL}{\ensuremath{NULL}} 
\newcommand{\hatcurSPECferoszfehCandxxxxxL}{\ensuremath{NULL}} 
\newcommand{\hatcurSPECferosvsiniCandxxxxxL}{\ensuremath{NULL}} 
\newcommand{\hatcurSPECferosNCandxxxxxL}{\ensuremath{0}} 
\newcommand{\hatcurSPECferossnrangeCandxxxxxL}{NULL} 
\newcommand{\hatcurSPECferosdaterangelowresCandxxxxxL}{NULL} 
\newcommand{\hatcurSPECferosRVKCandxxxxxL}{\ensuremath{NULL}} 
\newcommand{\hatcurSPECferosRVGCandxxxxxL}{\ensuremath{NULL}} 
\newcommand{\hatcurSPECfiesTeffCandxxxxxL}{\ensuremath{NULL}} 
\newcommand{\hatcurSPECfiesloggCandxxxxxL}{\ensuremath{NULL}} 
\newcommand{\hatcurSPECfieszfehCandxxxxxL}{\ensuremath{NULL}} 
\newcommand{\hatcurSPECfiesvsiniCandxxxxxL}{\ensuremath{NULL}} 
\newcommand{\hatcurSPECfiesNCandxxxxxL}{\ensuremath{0}} 
\newcommand{\hatcurSPECfiessnrangeCandxxxxxL}{NULL} 
\newcommand{\hatcurSPECfiesdaterangelowresCandxxxxxL}{NULL} 
\newcommand{\hatcurSPECfiesRVKCandxxxxxL}{\ensuremath{NULL}} 
\newcommand{\hatcurSPECfiesRVGCandxxxxxL}{\ensuremath{NULL}} 
\newcommand{\hatcurSPECcoralieTeffCandxxxxxL}{\ensuremath{NULL}} 
\newcommand{\hatcurSPECcoralieloggCandxxxxxL}{\ensuremath{NULL}} 
\newcommand{\hatcurSPECcoraliezfehCandxxxxxL}{\ensuremath{NULL}} 
\newcommand{\hatcurSPECcoralievsiniCandxxxxxL}{\ensuremath{NULL}} 
\newcommand{\hatcurSPECcoralieNCandxxxxxL}{\ensuremath{0}} 
\newcommand{\hatcurSPECcoraliesnrangeCandxxxxxL}{NULL} 
\newcommand{\hatcurSPECcoraliedaterangelowresCandxxxxxL}{NULL} 
\newcommand{\hatcurSPECcoralieRVKCandxxxxxL}{\ensuremath{NULL}} 
\newcommand{\hatcurSPECcoralieRVGCandxxxxxL}{\ensuremath{NULL}} 
\newcommand{\hatcurhtrCandxxxxxM}{HATS579-040}
\newcommand{\hatcurCCraCandxxxxxM}{\ensuremath{19^{\mathrm h}40^{\mathrm m}37.56{\mathrm s}}}
\newcommand{\hatcurCCdecCandxxxxxM}{\ensuremath{-22^{\arcdeg}43^{\arcmin}18.0{\arcsec}}}
\newcommand{\hatcurCLASSCandxxxxxM}{CAND}
\newcommand{\hatcurCCEPICCandxxxxxM}{EPIC~216231580}
\newcommand{\hatcurLCPCandxxxxxM}{\ensuremath{3.9052839\pm0.0000085}} 
\newcommand{\hatcurLCTCandxxxxxM}{\ensuremath{2457247.84650\pm0.00081}} 
\newcommand{\hatcurLCdurCandxxxxxM}{\ensuremath{0.17444\pm0.01112}} 
\newcommand{\hatcurLCrprstarCandxxxxxM}{\ensuremath{0.13867\pm0.00141}} 
\newcommand{\hatcurLCbsqCandxxxxxM}{\ensuremath{0.688\pm0.030}} 
\newcommand{\hatcurCCmagBCandxxxxxM}{\ensuremath{15.785}} 
\newcommand{\hatcurCCmagVCandxxxxxM}{\ensuremath{14.963}} 
\newcommand{\hatcurCCmaggCandxxxxxM}{\ensuremath{15.388}} 
\newcommand{\hatcurCCmagrCandxxxxxM}{\ensuremath{14.825}} 
\newcommand{\hatcurCCmagiCandxxxxxM}{\ensuremath{14.404}} 
\newcommand{\hatcurCCmagJCandxxxxxM}{\ensuremath{13.493\pm0.03}} 
\newcommand{\hatcurCCmagHCandxxxxxM}{\ensuremath{13.185\pm0.03}} 
\newcommand{\hatcurCCmagKCandxxxxxM}{\ensuremath{13.091\pm0.03}} 
\newcommand{\hatcurSPECWiFeSTeffCandxxxxxM}{\ensuremath{5304\pm300}} 
\newcommand{\hatcurSPECWiFeSloggCandxxxxxM}{\ensuremath{4.00\pm0.30}} 
\newcommand{\hatcurSPECWiFeSzfehCandxxxxxM}{\ensuremath{0.0\pm0.5}} 
\newcommand{\hatcurSPECWiFeSNlowresCandxxxxxM}{\ensuremath{1}} 
\newcommand{\hatcurSPECWiFeSNmedresCandxxxxxM}{\ensuremath{4}} 
\newcommand{\hatcurSPECWiFeSsnrangelowresCandxxxxxM}{113.2} 
\newcommand{\hatcurSPECWiFeSsnrangemedresCandxxxxxM}{10.0--37.1} 
\newcommand{\hatcurSPECWiFeSdaterangelowresCandxxxxxM}{2012-10-30} 
\newcommand{\hatcurSPECWiFeSdaterangemedresCandxxxxxM}{2013-05-17--2013-09-24} 
\newcommand{\hatcurSPECWiFeSRVKCandxxxxxM}{\ensuremath{<2.0}} 
\newcommand{\hatcurSPECWiFeSRVGCandxxxxxM}{\ensuremath{-48.77\pm0.72}} 
\newcommand{\hatcurSPECarcesTeffCandxxxxxM}{\ensuremath{NULL}} 
\newcommand{\hatcurSPECarcesloggCandxxxxxM}{\ensuremath{NULL}} 
\newcommand{\hatcurSPECarceszfehCandxxxxxM}{\ensuremath{NULL}} 
\newcommand{\hatcurSPECarcesvsiniCandxxxxxM}{\ensuremath{NULL}} 
\newcommand{\hatcurSPECarcesNCandxxxxxM}{\ensuremath{0}} 
\newcommand{\hatcurSPECarcessnrangeCandxxxxxM}{NULL} 
\newcommand{\hatcurSPECarcesdaterangelowresCandxxxxxM}{NULL} 
\newcommand{\hatcurSPECarcesRVKCandxxxxxM}{\ensuremath{NULL}} 
\newcommand{\hatcurSPECarcesRVGCandxxxxxM}{\ensuremath{NULL}} 
\newcommand{\hatcurSPECdupontTeffCandxxxxxM}{\ensuremath{NULL}} 
\newcommand{\hatcurSPECdupontloggCandxxxxxM}{\ensuremath{NULL}} 
\newcommand{\hatcurSPECdupontzfehCandxxxxxM}{\ensuremath{NULL}} 
\newcommand{\hatcurSPECdupontvsiniCandxxxxxM}{\ensuremath{NULL}} 
\newcommand{\hatcurSPECdupontNCandxxxxxM}{\ensuremath{0}} 
\newcommand{\hatcurSPECdupontsnrangeCandxxxxxM}{NULL} 
\newcommand{\hatcurSPECdupontdaterangelowresCandxxxxxM}{NULL} 
\newcommand{\hatcurSPECdupontRVKCandxxxxxM}{\ensuremath{NULL}} 
\newcommand{\hatcurSPECdupontRVGCandxxxxxM}{\ensuremath{NULL}} 
\newcommand{\hatcurSPECharpsTeffCandxxxxxM}{\ensuremath{NULL}} 
\newcommand{\hatcurSPECharpsloggCandxxxxxM}{\ensuremath{NULL}} 
\newcommand{\hatcurSPECharpszfehCandxxxxxM}{\ensuremath{NULL}} 
\newcommand{\hatcurSPECharpsvsiniCandxxxxxM}{\ensuremath{NULL}} 
\newcommand{\hatcurSPECharpsNCandxxxxxM}{\ensuremath{0}} 
\newcommand{\hatcurSPECharpssnrangeCandxxxxxM}{NULL} 
\newcommand{\hatcurSPECharpsdaterangelowresCandxxxxxM}{NULL} 
\newcommand{\hatcurSPECharpsRVKCandxxxxxM}{\ensuremath{NULL}} 
\newcommand{\hatcurSPECharpsRVGCandxxxxxM}{\ensuremath{NULL}} 
\newcommand{\hatcurSPECpfsTeffCandxxxxxM}{\ensuremath{NULL}} 
\newcommand{\hatcurSPECpfsloggCandxxxxxM}{\ensuremath{NULL}} 
\newcommand{\hatcurSPECpfszfehCandxxxxxM}{\ensuremath{NULL}} 
\newcommand{\hatcurSPECpfsvsiniCandxxxxxM}{\ensuremath{NULL}} 
\newcommand{\hatcurSPECpfsNCandxxxxxM}{\ensuremath{0}} 
\newcommand{\hatcurSPECpfssnrangeCandxxxxxM}{NULL} 
\newcommand{\hatcurSPECpfsdaterangelowresCandxxxxxM}{NULL} 
\newcommand{\hatcurSPECpfsRVKCandxxxxxM}{\ensuremath{NULL}} 
\newcommand{\hatcurSPECpfsRVGCandxxxxxM}{\ensuremath{NULL}} 
\newcommand{\hatcurSPECferosTeffCandxxxxxM}{\ensuremath{NULL}} 
\newcommand{\hatcurSPECferosloggCandxxxxxM}{\ensuremath{NULL}} 
\newcommand{\hatcurSPECferoszfehCandxxxxxM}{\ensuremath{NULL}} 
\newcommand{\hatcurSPECferosvsiniCandxxxxxM}{\ensuremath{NULL}} 
\newcommand{\hatcurSPECferosNCandxxxxxM}{\ensuremath{0}} 
\newcommand{\hatcurSPECferossnrangeCandxxxxxM}{NULL} 
\newcommand{\hatcurSPECferosdaterangelowresCandxxxxxM}{NULL} 
\newcommand{\hatcurSPECferosRVKCandxxxxxM}{\ensuremath{NULL}} 
\newcommand{\hatcurSPECferosRVGCandxxxxxM}{\ensuremath{NULL}} 
\newcommand{\hatcurSPECfiesTeffCandxxxxxM}{\ensuremath{NULL}} 
\newcommand{\hatcurSPECfiesloggCandxxxxxM}{\ensuremath{NULL}} 
\newcommand{\hatcurSPECfieszfehCandxxxxxM}{\ensuremath{NULL}} 
\newcommand{\hatcurSPECfiesvsiniCandxxxxxM}{\ensuremath{NULL}} 
\newcommand{\hatcurSPECfiesNCandxxxxxM}{\ensuremath{0}} 
\newcommand{\hatcurSPECfiessnrangeCandxxxxxM}{NULL} 
\newcommand{\hatcurSPECfiesdaterangelowresCandxxxxxM}{NULL} 
\newcommand{\hatcurSPECfiesRVKCandxxxxxM}{\ensuremath{NULL}} 
\newcommand{\hatcurSPECfiesRVGCandxxxxxM}{\ensuremath{NULL}} 
\newcommand{\hatcurSPECcoralieTeffCandxxxxxM}{\ensuremath{NULL}} 
\newcommand{\hatcurSPECcoralieloggCandxxxxxM}{\ensuremath{NULL}} 
\newcommand{\hatcurSPECcoraliezfehCandxxxxxM}{\ensuremath{NULL}} 
\newcommand{\hatcurSPECcoralievsiniCandxxxxxM}{\ensuremath{NULL}} 
\newcommand{\hatcurSPECcoralieNCandxxxxxM}{\ensuremath{0}} 
\newcommand{\hatcurSPECcoraliesnrangeCandxxxxxM}{NULL} 
\newcommand{\hatcurSPECcoraliedaterangelowresCandxxxxxM}{NULL} 
\newcommand{\hatcurSPECcoralieRVKCandxxxxxM}{\ensuremath{NULL}} 
\newcommand{\hatcurSPECcoralieRVGCandxxxxxM}{\ensuremath{NULL}} 
\newcommand{\hatcurhtrCandxxxxxN}{HATS579-041}
\newcommand{\hatcurCCraCandxxxxxN}{\ensuremath{19^{\mathrm h}35^{\mathrm m}41.64{\mathrm s}}}
\newcommand{\hatcurCCdecCandxxxxxN}{\ensuremath{-23^{\arcdeg}40^{\arcmin}42.5{\arcsec}}}
\newcommand{\hatcurCLASSCandxxxxxN}{EB}
\newcommand{\hatcurCCEPICCandxxxxxN}{EPIC~215714765}
\newcommand{\hatcurLCPCandxxxxxN}{\ensuremath{6.6890947\pm0.0000092}} 
\newcommand{\hatcurLCTCandxxxxxN}{\ensuremath{2457086.78510\pm0.00081}} 
\newcommand{\hatcurLCdurCandxxxxxN}{\ensuremath{0.09195\pm0.00854}} 
\newcommand{\hatcurLCrprstarCandxxxxxN}{\ensuremath{0.16474\pm0.00313}} 
\newcommand{\hatcurLCbsqCandxxxxxN}{\ensuremath{0.558\pm0.106}} 
\newcommand{\hatcurCCmagBCandxxxxxN}{\ensuremath{15.345}} 
\newcommand{\hatcurCCmagVCandxxxxxN}{\ensuremath{14.660}} 
\newcommand{\hatcurCCmaggCandxxxxxN}{\ensuremath{14.950}} 
\newcommand{\hatcurCCmagrCandxxxxxN}{\ensuremath{14.349}} 
\newcommand{\hatcurCCmagiCandxxxxxN}{\ensuremath{14.143}} 
\newcommand{\hatcurCCmagJCandxxxxxN}{\ensuremath{13.371\pm0.02}} 
\newcommand{\hatcurCCmagHCandxxxxxN}{\ensuremath{12.929\pm0.03}} 
\newcommand{\hatcurCCmagKCandxxxxxN}{\ensuremath{12.871\pm0.03}} 
\newcommand{\hatcurSPECWiFeSTeffCandxxxxxN}{\ensuremath{5645\pm300}} 
\newcommand{\hatcurSPECWiFeSloggCandxxxxxN}{\ensuremath{4.50\pm0.30}} 
\newcommand{\hatcurSPECWiFeSzfehCandxxxxxN}{\ensuremath{0.0\pm0.5}} 
\newcommand{\hatcurSPECWiFeSNlowresCandxxxxxN}{\ensuremath{1}} 
\newcommand{\hatcurSPECWiFeSNmedresCandxxxxxN}{\ensuremath{4}} 
\newcommand{\hatcurSPECWiFeSsnrangelowresCandxxxxxN}{19.8} 
\newcommand{\hatcurSPECWiFeSsnrangemedresCandxxxxxN}{4.8--48.2} 
\newcommand{\hatcurSPECWiFeSdaterangelowresCandxxxxxN}{2013-09-24} 
\newcommand{\hatcurSPECWiFeSdaterangemedresCandxxxxxN}{2013-05-17--2013-09-24} 
\newcommand{\hatcurSPECWiFeSRVKCandxxxxxN}{\ensuremath{54.92\pm1.14}} 
\newcommand{\hatcurSPECWiFeSRVGCandxxxxxN}{\ensuremath{28.51\pm1.09}} 
\newcommand{\hatcurSPECarcesTeffCandxxxxxN}{\ensuremath{NULL}} 
\newcommand{\hatcurSPECarcesloggCandxxxxxN}{\ensuremath{NULL}} 
\newcommand{\hatcurSPECarceszfehCandxxxxxN}{\ensuremath{NULL}} 
\newcommand{\hatcurSPECarcesvsiniCandxxxxxN}{\ensuremath{NULL}} 
\newcommand{\hatcurSPECarcesNCandxxxxxN}{\ensuremath{0}} 
\newcommand{\hatcurSPECarcessnrangeCandxxxxxN}{NULL} 
\newcommand{\hatcurSPECarcesdaterangelowresCandxxxxxN}{NULL} 
\newcommand{\hatcurSPECarcesRVKCandxxxxxN}{\ensuremath{NULL}} 
\newcommand{\hatcurSPECarcesRVGCandxxxxxN}{\ensuremath{NULL}} 
\newcommand{\hatcurSPECdupontTeffCandxxxxxN}{\ensuremath{NULL}} 
\newcommand{\hatcurSPECdupontloggCandxxxxxN}{\ensuremath{NULL}} 
\newcommand{\hatcurSPECdupontzfehCandxxxxxN}{\ensuremath{NULL}} 
\newcommand{\hatcurSPECdupontvsiniCandxxxxxN}{\ensuremath{NULL}} 
\newcommand{\hatcurSPECdupontNCandxxxxxN}{\ensuremath{0}} 
\newcommand{\hatcurSPECdupontsnrangeCandxxxxxN}{NULL} 
\newcommand{\hatcurSPECdupontdaterangelowresCandxxxxxN}{NULL} 
\newcommand{\hatcurSPECdupontRVKCandxxxxxN}{\ensuremath{NULL}} 
\newcommand{\hatcurSPECdupontRVGCandxxxxxN}{\ensuremath{NULL}} 
\newcommand{\hatcurSPECharpsTeffCandxxxxxN}{\ensuremath{NULL}} 
\newcommand{\hatcurSPECharpsloggCandxxxxxN}{\ensuremath{NULL}} 
\newcommand{\hatcurSPECharpszfehCandxxxxxN}{\ensuremath{NULL}} 
\newcommand{\hatcurSPECharpsvsiniCandxxxxxN}{\ensuremath{NULL}} 
\newcommand{\hatcurSPECharpsNCandxxxxxN}{\ensuremath{0}} 
\newcommand{\hatcurSPECharpssnrangeCandxxxxxN}{NULL} 
\newcommand{\hatcurSPECharpsdaterangelowresCandxxxxxN}{NULL} 
\newcommand{\hatcurSPECharpsRVKCandxxxxxN}{\ensuremath{NULL}} 
\newcommand{\hatcurSPECharpsRVGCandxxxxxN}{\ensuremath{NULL}} 
\newcommand{\hatcurSPECpfsTeffCandxxxxxN}{\ensuremath{NULL}} 
\newcommand{\hatcurSPECpfsloggCandxxxxxN}{\ensuremath{NULL}} 
\newcommand{\hatcurSPECpfszfehCandxxxxxN}{\ensuremath{NULL}} 
\newcommand{\hatcurSPECpfsvsiniCandxxxxxN}{\ensuremath{NULL}} 
\newcommand{\hatcurSPECpfsNCandxxxxxN}{\ensuremath{0}} 
\newcommand{\hatcurSPECpfssnrangeCandxxxxxN}{NULL} 
\newcommand{\hatcurSPECpfsdaterangelowresCandxxxxxN}{NULL} 
\newcommand{\hatcurSPECpfsRVKCandxxxxxN}{\ensuremath{NULL}} 
\newcommand{\hatcurSPECpfsRVGCandxxxxxN}{\ensuremath{NULL}} 
\newcommand{\hatcurSPECferosTeffCandxxxxxN}{\ensuremath{NULL}} 
\newcommand{\hatcurSPECferosloggCandxxxxxN}{\ensuremath{NULL}} 
\newcommand{\hatcurSPECferoszfehCandxxxxxN}{\ensuremath{NULL}} 
\newcommand{\hatcurSPECferosvsiniCandxxxxxN}{\ensuremath{NULL}} 
\newcommand{\hatcurSPECferosNCandxxxxxN}{\ensuremath{0}} 
\newcommand{\hatcurSPECferossnrangeCandxxxxxN}{NULL} 
\newcommand{\hatcurSPECferosdaterangelowresCandxxxxxN}{NULL} 
\newcommand{\hatcurSPECferosRVKCandxxxxxN}{\ensuremath{NULL}} 
\newcommand{\hatcurSPECferosRVGCandxxxxxN}{\ensuremath{NULL}} 
\newcommand{\hatcurSPECfiesTeffCandxxxxxN}{\ensuremath{NULL}} 
\newcommand{\hatcurSPECfiesloggCandxxxxxN}{\ensuremath{NULL}} 
\newcommand{\hatcurSPECfieszfehCandxxxxxN}{\ensuremath{NULL}} 
\newcommand{\hatcurSPECfiesvsiniCandxxxxxN}{\ensuremath{NULL}} 
\newcommand{\hatcurSPECfiesNCandxxxxxN}{\ensuremath{0}} 
\newcommand{\hatcurSPECfiessnrangeCandxxxxxN}{NULL} 
\newcommand{\hatcurSPECfiesdaterangelowresCandxxxxxN}{NULL} 
\newcommand{\hatcurSPECfiesRVKCandxxxxxN}{\ensuremath{NULL}} 
\newcommand{\hatcurSPECfiesRVGCandxxxxxN}{\ensuremath{NULL}} 
\newcommand{\hatcurSPECcoralieTeffCandxxxxxN}{\ensuremath{NULL}} 
\newcommand{\hatcurSPECcoralieloggCandxxxxxN}{\ensuremath{NULL}} 
\newcommand{\hatcurSPECcoraliezfehCandxxxxxN}{\ensuremath{NULL}} 
\newcommand{\hatcurSPECcoralievsiniCandxxxxxN}{\ensuremath{NULL}} 
\newcommand{\hatcurSPECcoralieNCandxxxxxN}{\ensuremath{0}} 
\newcommand{\hatcurSPECcoraliesnrangeCandxxxxxN}{NULL} 
\newcommand{\hatcurSPECcoraliedaterangelowresCandxxxxxN}{NULL} 
\newcommand{\hatcurSPECcoralieRVKCandxxxxxN}{\ensuremath{NULL}} 
\newcommand{\hatcurSPECcoralieRVGCandxxxxxN}{\ensuremath{NULL}} 
\newcommand{\hatcurhtrCandxxxxxO}{HATS579-043}
\newcommand{\hatcurCCraCandxxxxxO}{\ensuremath{19^{\mathrm h}25^{\mathrm m}51.60{\mathrm s}}}
\newcommand{\hatcurCCdecCandxxxxxO}{\ensuremath{-22^{\arcdeg}07^{\arcmin}30.3{\arcsec}}}
\newcommand{\hatcurCLASSCandxxxxxO}{EB}
\newcommand{\hatcurCCEPICCandxxxxxO}{EPIC~216562832}
\newcommand{\hatcurLCPCandxxxxxO}{\ensuremath{0.6633600\pm0.0000010}} 
\newcommand{\hatcurLCTCandxxxxxO}{\ensuremath{2457303.61090\pm0.00036}} 
\newcommand{\hatcurLCdurCandxxxxxO}{\ensuremath{0.09496\pm0.00638}} 
\newcommand{\hatcurLCrprstarCandxxxxxO}{\ensuremath{0.10784\pm0.00111}} 
\newcommand{\hatcurLCbsqCandxxxxxO}{\ensuremath{0.783\pm0.021}} 
\newcommand{\hatcurCCmagBCandxxxxxO}{\ensuremath{16.717}} 
\newcommand{\hatcurCCmagVCandxxxxxO}{\ensuremath{16.007}} 
\newcommand{\hatcurCCmaggCandxxxxxO}{\ensuremath{16.241}} 
\newcommand{\hatcurCCmagrCandxxxxxO}{\ensuremath{15.714}} 
\newcommand{\hatcurCCmagiCandxxxxxO}{\ensuremath{15.510}} 
\newcommand{\hatcurCCmagJCandxxxxxO}{\ensuremath{14.566\pm0.04}} 
\newcommand{\hatcurCCmagHCandxxxxxO}{\ensuremath{14.203\pm0.03}} 
\newcommand{\hatcurCCmagKCandxxxxxO}{\ensuremath{14.116\pm0.05}} 
\newcommand{\hatcurSPECWiFeSTeffCandxxxxxO}{\ensuremath{6300\pm300}} 
\newcommand{\hatcurSPECWiFeSloggCandxxxxxO}{\ensuremath{4.90\pm0.30}} 
\newcommand{\hatcurSPECWiFeSzfehCandxxxxxO}{\ensuremath{-0.7\pm0.4}} 
\newcommand{\hatcurSPECWiFeSNlowresCandxxxxxO}{\ensuremath{2}} 
\newcommand{\hatcurSPECWiFeSNmedresCandxxxxxO}{\ensuremath{2}} 
\newcommand{\hatcurSPECWiFeSsnrangelowresCandxxxxxO}{10.4--22.8} 
\newcommand{\hatcurSPECWiFeSsnrangemedresCandxxxxxO}{4.9--14.7} 
\newcommand{\hatcurSPECWiFeSdaterangelowresCandxxxxxO}{2016-03-26--2016-04-12} 
\newcommand{\hatcurSPECWiFeSdaterangemedresCandxxxxxO}{2016-03-26--2016-04-12} 
\newcommand{\hatcurSPECWiFeSRVKCandxxxxxO}{\ensuremath{<2.0}} 
\newcommand{\hatcurSPECWiFeSRVGCandxxxxxO}{\ensuremath{-103.58\pm2.18}} 
\newcommand{\hatcurSPECarcesTeffCandxxxxxO}{\ensuremath{NULL}} 
\newcommand{\hatcurSPECarcesloggCandxxxxxO}{\ensuremath{NULL}} 
\newcommand{\hatcurSPECarceszfehCandxxxxxO}{\ensuremath{NULL}} 
\newcommand{\hatcurSPECarcesvsiniCandxxxxxO}{\ensuremath{NULL}} 
\newcommand{\hatcurSPECarcesNCandxxxxxO}{\ensuremath{0}} 
\newcommand{\hatcurSPECarcessnrangeCandxxxxxO}{NULL} 
\newcommand{\hatcurSPECarcesdaterangelowresCandxxxxxO}{NULL} 
\newcommand{\hatcurSPECarcesRVKCandxxxxxO}{\ensuremath{NULL}} 
\newcommand{\hatcurSPECarcesRVGCandxxxxxO}{\ensuremath{NULL}} 
\newcommand{\hatcurSPECdupontTeffCandxxxxxO}{\ensuremath{NULL}} 
\newcommand{\hatcurSPECdupontloggCandxxxxxO}{\ensuremath{NULL}} 
\newcommand{\hatcurSPECdupontzfehCandxxxxxO}{\ensuremath{NULL}} 
\newcommand{\hatcurSPECdupontvsiniCandxxxxxO}{\ensuremath{NULL}} 
\newcommand{\hatcurSPECdupontNCandxxxxxO}{\ensuremath{0}} 
\newcommand{\hatcurSPECdupontsnrangeCandxxxxxO}{NULL} 
\newcommand{\hatcurSPECdupontdaterangelowresCandxxxxxO}{NULL} 
\newcommand{\hatcurSPECdupontRVKCandxxxxxO}{\ensuremath{NULL}} 
\newcommand{\hatcurSPECdupontRVGCandxxxxxO}{\ensuremath{NULL}} 
\newcommand{\hatcurSPECharpsTeffCandxxxxxO}{\ensuremath{NULL}} 
\newcommand{\hatcurSPECharpsloggCandxxxxxO}{\ensuremath{NULL}} 
\newcommand{\hatcurSPECharpszfehCandxxxxxO}{\ensuremath{NULL}} 
\newcommand{\hatcurSPECharpsvsiniCandxxxxxO}{\ensuremath{NULL}} 
\newcommand{\hatcurSPECharpsNCandxxxxxO}{\ensuremath{0}} 
\newcommand{\hatcurSPECharpssnrangeCandxxxxxO}{NULL} 
\newcommand{\hatcurSPECharpsdaterangelowresCandxxxxxO}{NULL} 
\newcommand{\hatcurSPECharpsRVKCandxxxxxO}{\ensuremath{NULL}} 
\newcommand{\hatcurSPECharpsRVGCandxxxxxO}{\ensuremath{NULL}} 
\newcommand{\hatcurSPECpfsTeffCandxxxxxO}{\ensuremath{NULL}} 
\newcommand{\hatcurSPECpfsloggCandxxxxxO}{\ensuremath{NULL}} 
\newcommand{\hatcurSPECpfszfehCandxxxxxO}{\ensuremath{NULL}} 
\newcommand{\hatcurSPECpfsvsiniCandxxxxxO}{\ensuremath{NULL}} 
\newcommand{\hatcurSPECpfsNCandxxxxxO}{\ensuremath{0}} 
\newcommand{\hatcurSPECpfssnrangeCandxxxxxO}{NULL} 
\newcommand{\hatcurSPECpfsdaterangelowresCandxxxxxO}{NULL} 
\newcommand{\hatcurSPECpfsRVKCandxxxxxO}{\ensuremath{NULL}} 
\newcommand{\hatcurSPECpfsRVGCandxxxxxO}{\ensuremath{NULL}} 
\newcommand{\hatcurSPECferosTeffCandxxxxxO}{\ensuremath{NULL}} 
\newcommand{\hatcurSPECferosloggCandxxxxxO}{\ensuremath{NULL}} 
\newcommand{\hatcurSPECferoszfehCandxxxxxO}{\ensuremath{NULL}} 
\newcommand{\hatcurSPECferosvsiniCandxxxxxO}{\ensuremath{NULL}} 
\newcommand{\hatcurSPECferosNCandxxxxxO}{\ensuremath{0}} 
\newcommand{\hatcurSPECferossnrangeCandxxxxxO}{NULL} 
\newcommand{\hatcurSPECferosdaterangelowresCandxxxxxO}{NULL} 
\newcommand{\hatcurSPECferosRVKCandxxxxxO}{\ensuremath{NULL}} 
\newcommand{\hatcurSPECferosRVGCandxxxxxO}{\ensuremath{NULL}} 
\newcommand{\hatcurSPECfiesTeffCandxxxxxO}{\ensuremath{NULL}} 
\newcommand{\hatcurSPECfiesloggCandxxxxxO}{\ensuremath{NULL}} 
\newcommand{\hatcurSPECfieszfehCandxxxxxO}{\ensuremath{NULL}} 
\newcommand{\hatcurSPECfiesvsiniCandxxxxxO}{\ensuremath{NULL}} 
\newcommand{\hatcurSPECfiesNCandxxxxxO}{\ensuremath{0}} 
\newcommand{\hatcurSPECfiessnrangeCandxxxxxO}{NULL} 
\newcommand{\hatcurSPECfiesdaterangelowresCandxxxxxO}{NULL} 
\newcommand{\hatcurSPECfiesRVKCandxxxxxO}{\ensuremath{NULL}} 
\newcommand{\hatcurSPECfiesRVGCandxxxxxO}{\ensuremath{NULL}} 
\newcommand{\hatcurSPECcoralieTeffCandxxxxxO}{\ensuremath{NULL}} 
\newcommand{\hatcurSPECcoralieloggCandxxxxxO}{\ensuremath{NULL}} 
\newcommand{\hatcurSPECcoraliezfehCandxxxxxO}{\ensuremath{NULL}} 
\newcommand{\hatcurSPECcoralievsiniCandxxxxxO}{\ensuremath{NULL}} 
\newcommand{\hatcurSPECcoralieNCandxxxxxO}{\ensuremath{0}} 
\newcommand{\hatcurSPECcoraliesnrangeCandxxxxxO}{NULL} 
\newcommand{\hatcurSPECcoraliedaterangelowresCandxxxxxO}{NULL} 
\newcommand{\hatcurSPECcoralieRVKCandxxxxxO}{\ensuremath{NULL}} 
\newcommand{\hatcurSPECcoralieRVGCandxxxxxO}{\ensuremath{NULL}} 
\newcommand{\hatcurhtrCandxxxxxP}{HATS579-044}
\newcommand{\hatcurCCraCandxxxxxP}{\ensuremath{19^{\mathrm h}17^{\mathrm m}45.24{\mathrm s}}}
\newcommand{\hatcurCCdecCandxxxxxP}{\ensuremath{-20^{\arcdeg}39^{\arcmin}15.6{\arcsec}}}
\newcommand{\hatcurCLASSCandxxxxxP}{CAND}
\newcommand{\hatcurCCEPICCandxxxxxP}{EPIC~217393088}
\newcommand{\hatcurLCPCandxxxxxP}{\ensuremath{1.3194747\pm0.0000021}} 
\newcommand{\hatcurLCTCandxxxxxP}{\ensuremath{2457235.14088\pm0.00065}} 
\newcommand{\hatcurLCdurCandxxxxxP}{\ensuremath{0.12818\pm0.00401}} 
\newcommand{\hatcurLCrprstarCandxxxxxP}{\ensuremath{0.10742\pm0.00135}} 
\newcommand{\hatcurLCbsqCandxxxxxP}{\ensuremath{0.240\pm0.139}} 
\newcommand{\hatcurCCmagBCandxxxxxP}{\ensuremath{16.398}} 
\newcommand{\hatcurCCmagVCandxxxxxP}{\ensuremath{15.575}} 
\newcommand{\hatcurCCmaggCandxxxxxP}{\ensuremath{15.909}} 
\newcommand{\hatcurCCmagrCandxxxxxP}{\ensuremath{15.363}} 
\newcommand{\hatcurCCmagiCandxxxxxP}{\ensuremath{15.024}} 
\newcommand{\hatcurCCmagJCandxxxxxP}{\ensuremath{14.279\pm0.03}} 
\newcommand{\hatcurCCmagHCandxxxxxP}{\ensuremath{13.965\pm0.03}} 
\newcommand{\hatcurCCmagKCandxxxxxP}{\ensuremath{13.971\pm0.05}} 
\newcommand{\hatcurSPECWiFeSTeffCandxxxxxP}{\ensuremath{5945\pm300}} 
\newcommand{\hatcurSPECWiFeSloggCandxxxxxP}{\ensuremath{4.50\pm0.30}} 
\newcommand{\hatcurSPECWiFeSzfehCandxxxxxP}{\ensuremath{0.0\pm0.5}} 
\newcommand{\hatcurSPECWiFeSNlowresCandxxxxxP}{\ensuremath{1}} 
\newcommand{\hatcurSPECWiFeSNmedresCandxxxxxP}{\ensuremath{3}} 
\newcommand{\hatcurSPECWiFeSsnrangelowresCandxxxxxP}{24.3} 
\newcommand{\hatcurSPECWiFeSsnrangemedresCandxxxxxP}{6.1--22.7} 
\newcommand{\hatcurSPECWiFeSdaterangelowresCandxxxxxP}{2016-03-25} 
\newcommand{\hatcurSPECWiFeSdaterangemedresCandxxxxxP}{2016-03-25--2016-03-27} 
\newcommand{\hatcurSPECWiFeSRVKCandxxxxxP}{\ensuremath{<2.0}} 
\newcommand{\hatcurSPECWiFeSRVGCandxxxxxP}{\ensuremath{-61.97\pm1.26}} 
\newcommand{\hatcurSPECarcesTeffCandxxxxxP}{\ensuremath{NULL}} 
\newcommand{\hatcurSPECarcesloggCandxxxxxP}{\ensuremath{NULL}} 
\newcommand{\hatcurSPECarceszfehCandxxxxxP}{\ensuremath{NULL}} 
\newcommand{\hatcurSPECarcesvsiniCandxxxxxP}{\ensuremath{NULL}} 
\newcommand{\hatcurSPECarcesNCandxxxxxP}{\ensuremath{0}} 
\newcommand{\hatcurSPECarcessnrangeCandxxxxxP}{NULL} 
\newcommand{\hatcurSPECarcesdaterangelowresCandxxxxxP}{NULL} 
\newcommand{\hatcurSPECarcesRVKCandxxxxxP}{\ensuremath{NULL}} 
\newcommand{\hatcurSPECarcesRVGCandxxxxxP}{\ensuremath{NULL}} 
\newcommand{\hatcurSPECdupontTeffCandxxxxxP}{\ensuremath{NULL}} 
\newcommand{\hatcurSPECdupontloggCandxxxxxP}{\ensuremath{NULL}} 
\newcommand{\hatcurSPECdupontzfehCandxxxxxP}{\ensuremath{NULL}} 
\newcommand{\hatcurSPECdupontvsiniCandxxxxxP}{\ensuremath{NULL}} 
\newcommand{\hatcurSPECdupontNCandxxxxxP}{\ensuremath{0}} 
\newcommand{\hatcurSPECdupontsnrangeCandxxxxxP}{NULL} 
\newcommand{\hatcurSPECdupontdaterangelowresCandxxxxxP}{NULL} 
\newcommand{\hatcurSPECdupontRVKCandxxxxxP}{\ensuremath{NULL}} 
\newcommand{\hatcurSPECdupontRVGCandxxxxxP}{\ensuremath{NULL}} 
\newcommand{\hatcurSPECharpsTeffCandxxxxxP}{\ensuremath{NULL}} 
\newcommand{\hatcurSPECharpsloggCandxxxxxP}{\ensuremath{NULL}} 
\newcommand{\hatcurSPECharpszfehCandxxxxxP}{\ensuremath{NULL}} 
\newcommand{\hatcurSPECharpsvsiniCandxxxxxP}{\ensuremath{NULL}} 
\newcommand{\hatcurSPECharpsNCandxxxxxP}{\ensuremath{0}} 
\newcommand{\hatcurSPECharpssnrangeCandxxxxxP}{NULL} 
\newcommand{\hatcurSPECharpsdaterangelowresCandxxxxxP}{NULL} 
\newcommand{\hatcurSPECharpsRVKCandxxxxxP}{\ensuremath{NULL}} 
\newcommand{\hatcurSPECharpsRVGCandxxxxxP}{\ensuremath{NULL}} 
\newcommand{\hatcurSPECpfsTeffCandxxxxxP}{\ensuremath{NULL}} 
\newcommand{\hatcurSPECpfsloggCandxxxxxP}{\ensuremath{NULL}} 
\newcommand{\hatcurSPECpfszfehCandxxxxxP}{\ensuremath{NULL}} 
\newcommand{\hatcurSPECpfsvsiniCandxxxxxP}{\ensuremath{NULL}} 
\newcommand{\hatcurSPECpfsNCandxxxxxP}{\ensuremath{0}} 
\newcommand{\hatcurSPECpfssnrangeCandxxxxxP}{NULL} 
\newcommand{\hatcurSPECpfsdaterangelowresCandxxxxxP}{NULL} 
\newcommand{\hatcurSPECpfsRVKCandxxxxxP}{\ensuremath{NULL}} 
\newcommand{\hatcurSPECpfsRVGCandxxxxxP}{\ensuremath{NULL}} 
\newcommand{\hatcurSPECferosTeffCandxxxxxP}{\ensuremath{NULL}} 
\newcommand{\hatcurSPECferosloggCandxxxxxP}{\ensuremath{NULL}} 
\newcommand{\hatcurSPECferoszfehCandxxxxxP}{\ensuremath{NULL}} 
\newcommand{\hatcurSPECferosvsiniCandxxxxxP}{\ensuremath{NULL}} 
\newcommand{\hatcurSPECferosNCandxxxxxP}{\ensuremath{0}} 
\newcommand{\hatcurSPECferossnrangeCandxxxxxP}{NULL} 
\newcommand{\hatcurSPECferosdaterangelowresCandxxxxxP}{NULL} 
\newcommand{\hatcurSPECferosRVKCandxxxxxP}{\ensuremath{NULL}} 
\newcommand{\hatcurSPECferosRVGCandxxxxxP}{\ensuremath{NULL}} 
\newcommand{\hatcurSPECfiesTeffCandxxxxxP}{\ensuremath{NULL}} 
\newcommand{\hatcurSPECfiesloggCandxxxxxP}{\ensuremath{NULL}} 
\newcommand{\hatcurSPECfieszfehCandxxxxxP}{\ensuremath{NULL}} 
\newcommand{\hatcurSPECfiesvsiniCandxxxxxP}{\ensuremath{NULL}} 
\newcommand{\hatcurSPECfiesNCandxxxxxP}{\ensuremath{0}} 
\newcommand{\hatcurSPECfiessnrangeCandxxxxxP}{NULL} 
\newcommand{\hatcurSPECfiesdaterangelowresCandxxxxxP}{NULL} 
\newcommand{\hatcurSPECfiesRVKCandxxxxxP}{\ensuremath{NULL}} 
\newcommand{\hatcurSPECfiesRVGCandxxxxxP}{\ensuremath{NULL}} 
\newcommand{\hatcurSPECcoralieTeffCandxxxxxP}{\ensuremath{NULL}} 
\newcommand{\hatcurSPECcoralieloggCandxxxxxP}{\ensuremath{NULL}} 
\newcommand{\hatcurSPECcoraliezfehCandxxxxxP}{\ensuremath{NULL}} 
\newcommand{\hatcurSPECcoralievsiniCandxxxxxP}{\ensuremath{NULL}} 
\newcommand{\hatcurSPECcoralieNCandxxxxxP}{\ensuremath{0}} 
\newcommand{\hatcurSPECcoraliesnrangeCandxxxxxP}{NULL} 
\newcommand{\hatcurSPECcoraliedaterangelowresCandxxxxxP}{NULL} 
\newcommand{\hatcurSPECcoralieRVKCandxxxxxP}{\ensuremath{NULL}} 
\newcommand{\hatcurSPECcoralieRVGCandxxxxxP}{\ensuremath{NULL}} 
\newcommand{\hatcurhtrCandxxxxxQ}{HATS579-048}
\newcommand{\hatcurCCraCandxxxxxQ}{\ensuremath{19^{\mathrm h}43^{\mathrm m}39.36{\mathrm s}}}
\newcommand{\hatcurCCdecCandxxxxxQ}{\ensuremath{-23^{\arcdeg}29^{\arcmin}13.2{\arcsec}}}
\newcommand{\hatcurCLASSCandxxxxxQ}{CAND}
\newcommand{\hatcurCCEPICCandxxxxxQ}{EPIC~215816368}
\newcommand{\hatcurLCPCandxxxxxQ}{\ensuremath{10.1460176\pm0.0000464}} 
\newcommand{\hatcurLCTCandxxxxxQ}{\ensuremath{2457263.96758\pm0.00162}} 
\newcommand{\hatcurLCdurCandxxxxxQ}{\ensuremath{0.26501\pm0.01877}} 
\newcommand{\hatcurLCrprstarCandxxxxxQ}{\ensuremath{0.16817\pm0.00220}} 
\newcommand{\hatcurLCbsqCandxxxxxQ}{\ensuremath{0.649\pm0.036}} 
\newcommand{\hatcurCCmagBCandxxxxxQ}{\ensuremath{16.692}} 
\newcommand{\hatcurCCmagVCandxxxxxQ}{\ensuremath{15.835}} 
\newcommand{\hatcurCCmaggCandxxxxxQ}{\ensuremath{16.287}} 
\newcommand{\hatcurCCmagrCandxxxxxQ}{\ensuremath{15.642}} 
\newcommand{\hatcurCCmagiCandxxxxxQ}{\ensuremath{15.339}} 
\newcommand{\hatcurCCmagJCandxxxxxQ}{\ensuremath{14.462\pm0.03}} 
\newcommand{\hatcurCCmagHCandxxxxxQ}{\ensuremath{14.044\pm0.03}} 
\newcommand{\hatcurCCmagKCandxxxxxQ}{\ensuremath{13.878\pm0.05}} 
\newcommand{\hatcurSPECWiFeSTeffCandxxxxxQ}{\ensuremath{5320\pm300}} 
\newcommand{\hatcurSPECWiFeSloggCandxxxxxQ}{\ensuremath{4.90\pm0.30}} 
\newcommand{\hatcurSPECWiFeSzfehCandxxxxxQ}{\ensuremath{-0.5\pm0.5}} 
\newcommand{\hatcurSPECWiFeSNlowresCandxxxxxQ}{\ensuremath{1}} 
\newcommand{\hatcurSPECWiFeSNmedresCandxxxxxQ}{\ensuremath{3}} 
\newcommand{\hatcurSPECWiFeSsnrangelowresCandxxxxxQ}{25.4} 
\newcommand{\hatcurSPECWiFeSsnrangemedresCandxxxxxQ}{5.2--9.7} 
\newcommand{\hatcurSPECWiFeSdaterangelowresCandxxxxxQ}{2016-03-26} 
\newcommand{\hatcurSPECWiFeSdaterangemedresCandxxxxxQ}{2016-03-26--2016-04-12} 
\newcommand{\hatcurSPECWiFeSRVKCandxxxxxQ}{\ensuremath{7.24\pm2.21}} 
\newcommand{\hatcurSPECWiFeSRVGCandxxxxxQ}{\ensuremath{-29.13\pm1.82}} 
\newcommand{\hatcurSPECarcesTeffCandxxxxxQ}{\ensuremath{NULL}} 
\newcommand{\hatcurSPECarcesloggCandxxxxxQ}{\ensuremath{NULL}} 
\newcommand{\hatcurSPECarceszfehCandxxxxxQ}{\ensuremath{NULL}} 
\newcommand{\hatcurSPECarcesvsiniCandxxxxxQ}{\ensuremath{NULL}} 
\newcommand{\hatcurSPECarcesNCandxxxxxQ}{\ensuremath{0}} 
\newcommand{\hatcurSPECarcessnrangeCandxxxxxQ}{NULL} 
\newcommand{\hatcurSPECarcesdaterangelowresCandxxxxxQ}{NULL} 
\newcommand{\hatcurSPECarcesRVKCandxxxxxQ}{\ensuremath{NULL}} 
\newcommand{\hatcurSPECarcesRVGCandxxxxxQ}{\ensuremath{NULL}} 
\newcommand{\hatcurSPECdupontTeffCandxxxxxQ}{\ensuremath{NULL}} 
\newcommand{\hatcurSPECdupontloggCandxxxxxQ}{\ensuremath{NULL}} 
\newcommand{\hatcurSPECdupontzfehCandxxxxxQ}{\ensuremath{NULL}} 
\newcommand{\hatcurSPECdupontvsiniCandxxxxxQ}{\ensuremath{NULL}} 
\newcommand{\hatcurSPECdupontNCandxxxxxQ}{\ensuremath{0}} 
\newcommand{\hatcurSPECdupontsnrangeCandxxxxxQ}{NULL} 
\newcommand{\hatcurSPECdupontdaterangelowresCandxxxxxQ}{NULL} 
\newcommand{\hatcurSPECdupontRVKCandxxxxxQ}{\ensuremath{NULL}} 
\newcommand{\hatcurSPECdupontRVGCandxxxxxQ}{\ensuremath{NULL}} 
\newcommand{\hatcurSPECharpsTeffCandxxxxxQ}{\ensuremath{NULL}} 
\newcommand{\hatcurSPECharpsloggCandxxxxxQ}{\ensuremath{NULL}} 
\newcommand{\hatcurSPECharpszfehCandxxxxxQ}{\ensuremath{NULL}} 
\newcommand{\hatcurSPECharpsvsiniCandxxxxxQ}{\ensuremath{NULL}} 
\newcommand{\hatcurSPECharpsNCandxxxxxQ}{\ensuremath{0}} 
\newcommand{\hatcurSPECharpssnrangeCandxxxxxQ}{NULL} 
\newcommand{\hatcurSPECharpsdaterangelowresCandxxxxxQ}{NULL} 
\newcommand{\hatcurSPECharpsRVKCandxxxxxQ}{\ensuremath{NULL}} 
\newcommand{\hatcurSPECharpsRVGCandxxxxxQ}{\ensuremath{NULL}} 
\newcommand{\hatcurSPECpfsTeffCandxxxxxQ}{\ensuremath{NULL}} 
\newcommand{\hatcurSPECpfsloggCandxxxxxQ}{\ensuremath{NULL}} 
\newcommand{\hatcurSPECpfszfehCandxxxxxQ}{\ensuremath{NULL}} 
\newcommand{\hatcurSPECpfsvsiniCandxxxxxQ}{\ensuremath{NULL}} 
\newcommand{\hatcurSPECpfsNCandxxxxxQ}{\ensuremath{0}} 
\newcommand{\hatcurSPECpfssnrangeCandxxxxxQ}{NULL} 
\newcommand{\hatcurSPECpfsdaterangelowresCandxxxxxQ}{NULL} 
\newcommand{\hatcurSPECpfsRVKCandxxxxxQ}{\ensuremath{NULL}} 
\newcommand{\hatcurSPECpfsRVGCandxxxxxQ}{\ensuremath{NULL}} 
\newcommand{\hatcurSPECferosTeffCandxxxxxQ}{\ensuremath{NULL}} 
\newcommand{\hatcurSPECferosloggCandxxxxxQ}{\ensuremath{NULL}} 
\newcommand{\hatcurSPECferoszfehCandxxxxxQ}{\ensuremath{NULL}} 
\newcommand{\hatcurSPECferosvsiniCandxxxxxQ}{\ensuremath{NULL}} 
\newcommand{\hatcurSPECferosNCandxxxxxQ}{\ensuremath{0}} 
\newcommand{\hatcurSPECferossnrangeCandxxxxxQ}{NULL} 
\newcommand{\hatcurSPECferosdaterangelowresCandxxxxxQ}{NULL} 
\newcommand{\hatcurSPECferosRVKCandxxxxxQ}{\ensuremath{NULL}} 
\newcommand{\hatcurSPECferosRVGCandxxxxxQ}{\ensuremath{NULL}} 
\newcommand{\hatcurSPECfiesTeffCandxxxxxQ}{\ensuremath{NULL}} 
\newcommand{\hatcurSPECfiesloggCandxxxxxQ}{\ensuremath{NULL}} 
\newcommand{\hatcurSPECfieszfehCandxxxxxQ}{\ensuremath{NULL}} 
\newcommand{\hatcurSPECfiesvsiniCandxxxxxQ}{\ensuremath{NULL}} 
\newcommand{\hatcurSPECfiesNCandxxxxxQ}{\ensuremath{0}} 
\newcommand{\hatcurSPECfiessnrangeCandxxxxxQ}{NULL} 
\newcommand{\hatcurSPECfiesdaterangelowresCandxxxxxQ}{NULL} 
\newcommand{\hatcurSPECfiesRVKCandxxxxxQ}{\ensuremath{NULL}} 
\newcommand{\hatcurSPECfiesRVGCandxxxxxQ}{\ensuremath{NULL}} 
\newcommand{\hatcurSPECcoralieTeffCandxxxxxQ}{\ensuremath{NULL}} 
\newcommand{\hatcurSPECcoralieloggCandxxxxxQ}{\ensuremath{NULL}} 
\newcommand{\hatcurSPECcoraliezfehCandxxxxxQ}{\ensuremath{NULL}} 
\newcommand{\hatcurSPECcoralievsiniCandxxxxxQ}{\ensuremath{NULL}} 
\newcommand{\hatcurSPECcoralieNCandxxxxxQ}{\ensuremath{0}} 
\newcommand{\hatcurSPECcoraliesnrangeCandxxxxxQ}{NULL} 
\newcommand{\hatcurSPECcoraliedaterangelowresCandxxxxxQ}{NULL} 
\newcommand{\hatcurSPECcoralieRVKCandxxxxxQ}{\ensuremath{NULL}} 
\newcommand{\hatcurSPECcoralieRVGCandxxxxxQ}{\ensuremath{NULL}} 
\newcommand{\hatcurhtrCandxxxxxR}{HATS579-050}
\newcommand{\hatcurCCraCandxxxxxR}{\ensuremath{19^{\mathrm h}15^{\mathrm m}34.92{\mathrm s}}}
\newcommand{\hatcurCCdecCandxxxxxR}{\ensuremath{-24^{\arcdeg}08^{\arcmin}34.9{\arcsec}}}
\newcommand{\hatcurCLASSCandxxxxxR}{EB}
\newcommand{\hatcurCCEPICCandxxxxxR}{EPIC~215474548}
\newcommand{\hatcurLCPCandxxxxxR}{\ensuremath{1.2085467\pm0.0000024}} 
\newcommand{\hatcurLCTCandxxxxxR}{\ensuremath{2457322.57557\pm0.00035}} 
\newcommand{\hatcurLCdurCandxxxxxR}{\ensuremath{0.08691\pm0.00455}} 
\newcommand{\hatcurLCrprstarCandxxxxxR}{\ensuremath{0.12716\pm0.00114}} 
\newcommand{\hatcurLCbsqCandxxxxxR}{\ensuremath{0.674\pm0.032}} 
\newcommand{\hatcurCCmagBCandxxxxxR}{\ensuremath{17.458}} 
\newcommand{\hatcurCCmagVCandxxxxxR}{\ensuremath{16.221}} 
\newcommand{\hatcurCCmaggCandxxxxxR}{\ensuremath{16.819}} 
\newcommand{\hatcurCCmagrCandxxxxxR}{\ensuremath{15.881}} 
\newcommand{\hatcurCCmagiCandxxxxxR}{\ensuremath{15.496}} 
\newcommand{\hatcurCCmagJCandxxxxxR}{\ensuremath{14.142\pm0.04}} 
\newcommand{\hatcurCCmagHCandxxxxxR}{\ensuremath{13.470\pm0.03}} 
\newcommand{\hatcurCCmagKCandxxxxxR}{\ensuremath{13.418\pm0.04}} 
\newcommand{\hatcurSPECWiFeSTeffCandxxxxxR}{\ensuremath{...}} 
\newcommand{\hatcurSPECWiFeSloggCandxxxxxR}{\ensuremath{4.705\pm0.084}} 
\newcommand{\hatcurSPECWiFeSzfehCandxxxxxR}{\ensuremath{NULL}} 
\newcommand{\hatcurSPECWiFeSNlowresCandxxxxxR}{\ensuremath{0}} 
\newcommand{\hatcurSPECWiFeSNmedresCandxxxxxR}{\ensuremath{0}} 
\newcommand{\hatcurSPECWiFeSsnrangelowresCandxxxxxR}{NULL} 
\newcommand{\hatcurSPECWiFeSsnrangemedresCandxxxxxR}{NULL} 
\newcommand{\hatcurSPECWiFeSdaterangelowresCandxxxxxR}{NULL} 
\newcommand{\hatcurSPECWiFeSdaterangemedresCandxxxxxR}{NULL} 
\newcommand{\hatcurSPECWiFeSRVKCandxxxxxR}{\ensuremath{...}} 
\newcommand{\hatcurSPECWiFeSRVGCandxxxxxR}{\ensuremath{NULL}} 
\newcommand{\hatcurSPECarcesTeffCandxxxxxR}{\ensuremath{NULL}} 
\newcommand{\hatcurSPECarcesloggCandxxxxxR}{\ensuremath{NULL}} 
\newcommand{\hatcurSPECarceszfehCandxxxxxR}{\ensuremath{NULL}} 
\newcommand{\hatcurSPECarcesvsiniCandxxxxxR}{\ensuremath{NULL}} 
\newcommand{\hatcurSPECarcesNCandxxxxxR}{\ensuremath{0}} 
\newcommand{\hatcurSPECarcessnrangeCandxxxxxR}{NULL} 
\newcommand{\hatcurSPECarcesdaterangelowresCandxxxxxR}{NULL} 
\newcommand{\hatcurSPECarcesRVKCandxxxxxR}{\ensuremath{NULL}} 
\newcommand{\hatcurSPECarcesRVGCandxxxxxR}{\ensuremath{NULL}} 
\newcommand{\hatcurSPECdupontTeffCandxxxxxR}{\ensuremath{NULL}} 
\newcommand{\hatcurSPECdupontloggCandxxxxxR}{\ensuremath{NULL}} 
\newcommand{\hatcurSPECdupontzfehCandxxxxxR}{\ensuremath{NULL}} 
\newcommand{\hatcurSPECdupontvsiniCandxxxxxR}{\ensuremath{NULL}} 
\newcommand{\hatcurSPECdupontNCandxxxxxR}{\ensuremath{0}} 
\newcommand{\hatcurSPECdupontsnrangeCandxxxxxR}{NULL} 
\newcommand{\hatcurSPECdupontdaterangelowresCandxxxxxR}{NULL} 
\newcommand{\hatcurSPECdupontRVKCandxxxxxR}{\ensuremath{NULL}} 
\newcommand{\hatcurSPECdupontRVGCandxxxxxR}{\ensuremath{NULL}} 
\newcommand{\hatcurSPECharpsTeffCandxxxxxR}{\ensuremath{NULL}} 
\newcommand{\hatcurSPECharpsloggCandxxxxxR}{\ensuremath{NULL}} 
\newcommand{\hatcurSPECharpszfehCandxxxxxR}{\ensuremath{NULL}} 
\newcommand{\hatcurSPECharpsvsiniCandxxxxxR}{\ensuremath{NULL}} 
\newcommand{\hatcurSPECharpsNCandxxxxxR}{\ensuremath{0}} 
\newcommand{\hatcurSPECharpssnrangeCandxxxxxR}{NULL} 
\newcommand{\hatcurSPECharpsdaterangelowresCandxxxxxR}{NULL} 
\newcommand{\hatcurSPECharpsRVKCandxxxxxR}{\ensuremath{NULL}} 
\newcommand{\hatcurSPECharpsRVGCandxxxxxR}{\ensuremath{NULL}} 
\newcommand{\hatcurSPECpfsTeffCandxxxxxR}{\ensuremath{NULL}} 
\newcommand{\hatcurSPECpfsloggCandxxxxxR}{\ensuremath{NULL}} 
\newcommand{\hatcurSPECpfszfehCandxxxxxR}{\ensuremath{NULL}} 
\newcommand{\hatcurSPECpfsvsiniCandxxxxxR}{\ensuremath{NULL}} 
\newcommand{\hatcurSPECpfsNCandxxxxxR}{\ensuremath{0}} 
\newcommand{\hatcurSPECpfssnrangeCandxxxxxR}{NULL} 
\newcommand{\hatcurSPECpfsdaterangelowresCandxxxxxR}{NULL} 
\newcommand{\hatcurSPECpfsRVKCandxxxxxR}{\ensuremath{NULL}} 
\newcommand{\hatcurSPECpfsRVGCandxxxxxR}{\ensuremath{NULL}} 
\newcommand{\hatcurSPECferosTeffCandxxxxxR}{\ensuremath{NULL}} 
\newcommand{\hatcurSPECferosloggCandxxxxxR}{\ensuremath{NULL}} 
\newcommand{\hatcurSPECferoszfehCandxxxxxR}{\ensuremath{NULL}} 
\newcommand{\hatcurSPECferosvsiniCandxxxxxR}{\ensuremath{NULL}} 
\newcommand{\hatcurSPECferosNCandxxxxxR}{\ensuremath{0}} 
\newcommand{\hatcurSPECferossnrangeCandxxxxxR}{NULL} 
\newcommand{\hatcurSPECferosdaterangelowresCandxxxxxR}{NULL} 
\newcommand{\hatcurSPECferosRVKCandxxxxxR}{\ensuremath{NULL}} 
\newcommand{\hatcurSPECferosRVGCandxxxxxR}{\ensuremath{NULL}} 
\newcommand{\hatcurSPECfiesTeffCandxxxxxR}{\ensuremath{NULL}} 
\newcommand{\hatcurSPECfiesloggCandxxxxxR}{\ensuremath{NULL}} 
\newcommand{\hatcurSPECfieszfehCandxxxxxR}{\ensuremath{NULL}} 
\newcommand{\hatcurSPECfiesvsiniCandxxxxxR}{\ensuremath{NULL}} 
\newcommand{\hatcurSPECfiesNCandxxxxxR}{\ensuremath{0}} 
\newcommand{\hatcurSPECfiessnrangeCandxxxxxR}{NULL} 
\newcommand{\hatcurSPECfiesdaterangelowresCandxxxxxR}{NULL} 
\newcommand{\hatcurSPECfiesRVKCandxxxxxR}{\ensuremath{NULL}} 
\newcommand{\hatcurSPECfiesRVGCandxxxxxR}{\ensuremath{NULL}} 
\newcommand{\hatcurSPECcoralieTeffCandxxxxxR}{\ensuremath{NULL}} 
\newcommand{\hatcurSPECcoralieloggCandxxxxxR}{\ensuremath{NULL}} 
\newcommand{\hatcurSPECcoraliezfehCandxxxxxR}{\ensuremath{NULL}} 
\newcommand{\hatcurSPECcoralievsiniCandxxxxxR}{\ensuremath{NULL}} 
\newcommand{\hatcurSPECcoralieNCandxxxxxR}{\ensuremath{0}} 
\newcommand{\hatcurSPECcoraliesnrangeCandxxxxxR}{NULL} 
\newcommand{\hatcurSPECcoraliedaterangelowresCandxxxxxR}{NULL} 
\newcommand{\hatcurSPECcoralieRVKCandxxxxxR}{\ensuremath{NULL}} 
\newcommand{\hatcurSPECcoralieRVGCandxxxxxR}{\ensuremath{NULL}} 
\newcommand{\hatcurhtrCandxxxxxS}{HATS624-002}
\newcommand{\hatcurCCraCandxxxxxS}{\ensuremath{19^{\mathrm h}14^{\mathrm m}33.72{\mathrm s}}}
\newcommand{\hatcurCCdecCandxxxxxS}{\ensuremath{-26^{\arcdeg}18^{\arcmin}32.6{\arcsec}}}
\newcommand{\hatcurCLASSCandxxxxxS}{EB}
\newcommand{\hatcurCCEPICCandxxxxxS}{EPIC~214512594}
\newcommand{\hatcurLCPCandxxxxxS}{\ensuremath{1.8769829\pm0.0000021}} 
\newcommand{\hatcurLCTCandxxxxxS}{\ensuremath{2456676.27256\pm0.00087}} 
\newcommand{\hatcurLCdurCandxxxxxS}{\ensuremath{0.18007\pm0.00993}} 
\newcommand{\hatcurLCrprstarCandxxxxxS}{\ensuremath{0.25501\pm0.00292}} 
\newcommand{\hatcurLCbsqCandxxxxxS}{\ensuremath{0.413\pm0.057}} 
\newcommand{\hatcurCCmagBCandxxxxxS}{\ensuremath{17.329}} 
\newcommand{\hatcurCCmagVCandxxxxxS}{\ensuremath{16.374}} 
\newcommand{\hatcurCCmaggCandxxxxxS}{\ensuremath{16.515}} 
\newcommand{\hatcurCCmagrCandxxxxxS}{\ensuremath{15.774}} 
\newcommand{\hatcurCCmagiCandxxxxxS}{\ensuremath{15.455}} 
\newcommand{\hatcurCCmagJCandxxxxxS}{\ensuremath{14.282\pm0.06}} 
\newcommand{\hatcurCCmagHCandxxxxxS}{\ensuremath{13.805\pm0.05}} 
\newcommand{\hatcurCCmagKCandxxxxxS}{\ensuremath{13.622\pm0.06}} 
\newcommand{\hatcurSPECWiFeSTeffCandxxxxxS}{\ensuremath{4765\pm300}} 
\newcommand{\hatcurSPECWiFeSloggCandxxxxxS}{\ensuremath{4.40\pm0.30}} 
\newcommand{\hatcurSPECWiFeSzfehCandxxxxxS}{\ensuremath{-0.5\pm0.5}} 
\newcommand{\hatcurSPECWiFeSNlowresCandxxxxxS}{\ensuremath{1}} 
\newcommand{\hatcurSPECWiFeSNmedresCandxxxxxS}{\ensuremath{2}} 
\newcommand{\hatcurSPECWiFeSsnrangelowresCandxxxxxS}{17.3} 
\newcommand{\hatcurSPECWiFeSsnrangemedresCandxxxxxS}{5.5--13.7} 
\newcommand{\hatcurSPECWiFeSdaterangelowresCandxxxxxS}{2016-03-31} 
\newcommand{\hatcurSPECWiFeSdaterangemedresCandxxxxxS}{2016-03-30--2016-03-31} 
\newcommand{\hatcurSPECWiFeSRVKCandxxxxxS}{\ensuremath{<2.0}} 
\newcommand{\hatcurSPECWiFeSRVGCandxxxxxS}{\ensuremath{-12.86\pm1.44}} 
\newcommand{\hatcurSPECarcesTeffCandxxxxxS}{\ensuremath{NULL}} 
\newcommand{\hatcurSPECarcesloggCandxxxxxS}{\ensuremath{NULL}} 
\newcommand{\hatcurSPECarceszfehCandxxxxxS}{\ensuremath{NULL}} 
\newcommand{\hatcurSPECarcesvsiniCandxxxxxS}{\ensuremath{NULL}} 
\newcommand{\hatcurSPECarcesNCandxxxxxS}{\ensuremath{0}} 
\newcommand{\hatcurSPECarcessnrangeCandxxxxxS}{NULL} 
\newcommand{\hatcurSPECarcesdaterangelowresCandxxxxxS}{NULL} 
\newcommand{\hatcurSPECarcesRVKCandxxxxxS}{\ensuremath{NULL}} 
\newcommand{\hatcurSPECarcesRVGCandxxxxxS}{\ensuremath{NULL}} 
\newcommand{\hatcurSPECdupontTeffCandxxxxxS}{\ensuremath{NULL}} 
\newcommand{\hatcurSPECdupontloggCandxxxxxS}{\ensuremath{NULL}} 
\newcommand{\hatcurSPECdupontzfehCandxxxxxS}{\ensuremath{NULL}} 
\newcommand{\hatcurSPECdupontvsiniCandxxxxxS}{\ensuremath{NULL}} 
\newcommand{\hatcurSPECdupontNCandxxxxxS}{\ensuremath{0}} 
\newcommand{\hatcurSPECdupontsnrangeCandxxxxxS}{NULL} 
\newcommand{\hatcurSPECdupontdaterangelowresCandxxxxxS}{NULL} 
\newcommand{\hatcurSPECdupontRVKCandxxxxxS}{\ensuremath{NULL}} 
\newcommand{\hatcurSPECdupontRVGCandxxxxxS}{\ensuremath{NULL}} 
\newcommand{\hatcurSPECharpsTeffCandxxxxxS}{\ensuremath{NULL}} 
\newcommand{\hatcurSPECharpsloggCandxxxxxS}{\ensuremath{NULL}} 
\newcommand{\hatcurSPECharpszfehCandxxxxxS}{\ensuremath{NULL}} 
\newcommand{\hatcurSPECharpsvsiniCandxxxxxS}{\ensuremath{NULL}} 
\newcommand{\hatcurSPECharpsNCandxxxxxS}{\ensuremath{0}} 
\newcommand{\hatcurSPECharpssnrangeCandxxxxxS}{NULL} 
\newcommand{\hatcurSPECharpsdaterangelowresCandxxxxxS}{NULL} 
\newcommand{\hatcurSPECharpsRVKCandxxxxxS}{\ensuremath{NULL}} 
\newcommand{\hatcurSPECharpsRVGCandxxxxxS}{\ensuremath{NULL}} 
\newcommand{\hatcurSPECpfsTeffCandxxxxxS}{\ensuremath{NULL}} 
\newcommand{\hatcurSPECpfsloggCandxxxxxS}{\ensuremath{NULL}} 
\newcommand{\hatcurSPECpfszfehCandxxxxxS}{\ensuremath{NULL}} 
\newcommand{\hatcurSPECpfsvsiniCandxxxxxS}{\ensuremath{NULL}} 
\newcommand{\hatcurSPECpfsNCandxxxxxS}{\ensuremath{0}} 
\newcommand{\hatcurSPECpfssnrangeCandxxxxxS}{NULL} 
\newcommand{\hatcurSPECpfsdaterangelowresCandxxxxxS}{NULL} 
\newcommand{\hatcurSPECpfsRVKCandxxxxxS}{\ensuremath{NULL}} 
\newcommand{\hatcurSPECpfsRVGCandxxxxxS}{\ensuremath{NULL}} 
\newcommand{\hatcurSPECferosTeffCandxxxxxS}{\ensuremath{NULL}} 
\newcommand{\hatcurSPECferosloggCandxxxxxS}{\ensuremath{NULL}} 
\newcommand{\hatcurSPECferoszfehCandxxxxxS}{\ensuremath{NULL}} 
\newcommand{\hatcurSPECferosvsiniCandxxxxxS}{\ensuremath{NULL}} 
\newcommand{\hatcurSPECferosNCandxxxxxS}{\ensuremath{0}} 
\newcommand{\hatcurSPECferossnrangeCandxxxxxS}{NULL} 
\newcommand{\hatcurSPECferosdaterangelowresCandxxxxxS}{NULL} 
\newcommand{\hatcurSPECferosRVKCandxxxxxS}{\ensuremath{NULL}} 
\newcommand{\hatcurSPECferosRVGCandxxxxxS}{\ensuremath{NULL}} 
\newcommand{\hatcurSPECfiesTeffCandxxxxxS}{\ensuremath{NULL}} 
\newcommand{\hatcurSPECfiesloggCandxxxxxS}{\ensuremath{NULL}} 
\newcommand{\hatcurSPECfieszfehCandxxxxxS}{\ensuremath{NULL}} 
\newcommand{\hatcurSPECfiesvsiniCandxxxxxS}{\ensuremath{NULL}} 
\newcommand{\hatcurSPECfiesNCandxxxxxS}{\ensuremath{0}} 
\newcommand{\hatcurSPECfiessnrangeCandxxxxxS}{NULL} 
\newcommand{\hatcurSPECfiesdaterangelowresCandxxxxxS}{NULL} 
\newcommand{\hatcurSPECfiesRVKCandxxxxxS}{\ensuremath{NULL}} 
\newcommand{\hatcurSPECfiesRVGCandxxxxxS}{\ensuremath{NULL}} 
\newcommand{\hatcurSPECcoralieTeffCandxxxxxS}{\ensuremath{NULL}} 
\newcommand{\hatcurSPECcoralieloggCandxxxxxS}{\ensuremath{NULL}} 
\newcommand{\hatcurSPECcoraliezfehCandxxxxxS}{\ensuremath{NULL}} 
\newcommand{\hatcurSPECcoralievsiniCandxxxxxS}{\ensuremath{NULL}} 
\newcommand{\hatcurSPECcoralieNCandxxxxxS}{\ensuremath{0}} 
\newcommand{\hatcurSPECcoraliesnrangeCandxxxxxS}{NULL} 
\newcommand{\hatcurSPECcoraliedaterangelowresCandxxxxxS}{NULL} 
\newcommand{\hatcurSPECcoralieRVKCandxxxxxS}{\ensuremath{NULL}} 
\newcommand{\hatcurSPECcoralieRVGCandxxxxxS}{\ensuremath{NULL}} 
\newcommand{\hatcurhtrCandxxxxxT}{HATS624-003}
\newcommand{\hatcurCCraCandxxxxxT}{\ensuremath{19^{\mathrm h}17^{\mathrm m}11.76{\mathrm s}}}
\newcommand{\hatcurCCdecCandxxxxxT}{\ensuremath{-26^{\arcdeg}29^{\arcmin}21.4{\arcsec}}}
\newcommand{\hatcurCLASSCandxxxxxT}{BEB}
\newcommand{\hatcurCCEPICCandxxxxxT}{EPIC~214439239}
\newcommand{\hatcurLCPCandxxxxxT}{\ensuremath{0.6425813\pm0.0000014}} 
\newcommand{\hatcurLCTCandxxxxxT}{\ensuremath{2457316.03141\pm0.00044}} 
\newcommand{\hatcurLCdurCandxxxxxT}{\ensuremath{0.13731\pm0.00853}} 
\newcommand{\hatcurLCrprstarCandxxxxxT}{\ensuremath{0.11122\pm0.00094}} 
\newcommand{\hatcurLCbsqCandxxxxxT}{\ensuremath{0.773\pm0.018}} 
\newcommand{\hatcurCCmagBCandxxxxxT}{\ensuremath{16.378}} 
\newcommand{\hatcurCCmagVCandxxxxxT}{\ensuremath{15.537}} 
\newcommand{\hatcurCCmaggCandxxxxxT}{\ensuremath{15.967}} 
\newcommand{\hatcurCCmagrCandxxxxxT}{\ensuremath{15.453}} 
\newcommand{\hatcurCCmagiCandxxxxxT}{\ensuremath{15.253}} 
\newcommand{\hatcurCCmagJCandxxxxxT}{\ensuremath{14.321\pm0.05}} 
\newcommand{\hatcurCCmagHCandxxxxxT}{\ensuremath{13.986\pm0.05}} 
\newcommand{\hatcurCCmagKCandxxxxxT}{\ensuremath{13.909\pm0.06}} 
\newcommand{\hatcurSPECWiFeSTeffCandxxxxxT}{\ensuremath{...}} 
\newcommand{\hatcurSPECWiFeSloggCandxxxxxT}{\ensuremath{4.393\pm0.12}} 
\newcommand{\hatcurSPECWiFeSzfehCandxxxxxT}{\ensuremath{NULL}} 
\newcommand{\hatcurSPECWiFeSNlowresCandxxxxxT}{\ensuremath{0}} 
\newcommand{\hatcurSPECWiFeSNmedresCandxxxxxT}{\ensuremath{0}} 
\newcommand{\hatcurSPECWiFeSsnrangelowresCandxxxxxT}{NULL} 
\newcommand{\hatcurSPECWiFeSsnrangemedresCandxxxxxT}{NULL} 
\newcommand{\hatcurSPECWiFeSdaterangelowresCandxxxxxT}{NULL} 
\newcommand{\hatcurSPECWiFeSdaterangemedresCandxxxxxT}{NULL} 
\newcommand{\hatcurSPECWiFeSRVKCandxxxxxT}{\ensuremath{...}} 
\newcommand{\hatcurSPECWiFeSRVGCandxxxxxT}{\ensuremath{NULL}} 
\newcommand{\hatcurSPECarcesTeffCandxxxxxT}{\ensuremath{NULL}} 
\newcommand{\hatcurSPECarcesloggCandxxxxxT}{\ensuremath{NULL}} 
\newcommand{\hatcurSPECarceszfehCandxxxxxT}{\ensuremath{NULL}} 
\newcommand{\hatcurSPECarcesvsiniCandxxxxxT}{\ensuremath{NULL}} 
\newcommand{\hatcurSPECarcesNCandxxxxxT}{\ensuremath{0}} 
\newcommand{\hatcurSPECarcessnrangeCandxxxxxT}{NULL} 
\newcommand{\hatcurSPECarcesdaterangelowresCandxxxxxT}{NULL} 
\newcommand{\hatcurSPECarcesRVKCandxxxxxT}{\ensuremath{NULL}} 
\newcommand{\hatcurSPECarcesRVGCandxxxxxT}{\ensuremath{NULL}} 
\newcommand{\hatcurSPECdupontTeffCandxxxxxT}{\ensuremath{NULL}} 
\newcommand{\hatcurSPECdupontloggCandxxxxxT}{\ensuremath{NULL}} 
\newcommand{\hatcurSPECdupontzfehCandxxxxxT}{\ensuremath{NULL}} 
\newcommand{\hatcurSPECdupontvsiniCandxxxxxT}{\ensuremath{NULL}} 
\newcommand{\hatcurSPECdupontNCandxxxxxT}{\ensuremath{0}} 
\newcommand{\hatcurSPECdupontsnrangeCandxxxxxT}{NULL} 
\newcommand{\hatcurSPECdupontdaterangelowresCandxxxxxT}{NULL} 
\newcommand{\hatcurSPECdupontRVKCandxxxxxT}{\ensuremath{NULL}} 
\newcommand{\hatcurSPECdupontRVGCandxxxxxT}{\ensuremath{NULL}} 
\newcommand{\hatcurSPECharpsTeffCandxxxxxT}{\ensuremath{NULL}} 
\newcommand{\hatcurSPECharpsloggCandxxxxxT}{\ensuremath{NULL}} 
\newcommand{\hatcurSPECharpszfehCandxxxxxT}{\ensuremath{NULL}} 
\newcommand{\hatcurSPECharpsvsiniCandxxxxxT}{\ensuremath{NULL}} 
\newcommand{\hatcurSPECharpsNCandxxxxxT}{\ensuremath{0}} 
\newcommand{\hatcurSPECharpssnrangeCandxxxxxT}{NULL} 
\newcommand{\hatcurSPECharpsdaterangelowresCandxxxxxT}{NULL} 
\newcommand{\hatcurSPECharpsRVKCandxxxxxT}{\ensuremath{NULL}} 
\newcommand{\hatcurSPECharpsRVGCandxxxxxT}{\ensuremath{NULL}} 
\newcommand{\hatcurSPECpfsTeffCandxxxxxT}{\ensuremath{NULL}} 
\newcommand{\hatcurSPECpfsloggCandxxxxxT}{\ensuremath{NULL}} 
\newcommand{\hatcurSPECpfszfehCandxxxxxT}{\ensuremath{NULL}} 
\newcommand{\hatcurSPECpfsvsiniCandxxxxxT}{\ensuremath{NULL}} 
\newcommand{\hatcurSPECpfsNCandxxxxxT}{\ensuremath{0}} 
\newcommand{\hatcurSPECpfssnrangeCandxxxxxT}{NULL} 
\newcommand{\hatcurSPECpfsdaterangelowresCandxxxxxT}{NULL} 
\newcommand{\hatcurSPECpfsRVKCandxxxxxT}{\ensuremath{NULL}} 
\newcommand{\hatcurSPECpfsRVGCandxxxxxT}{\ensuremath{NULL}} 
\newcommand{\hatcurSPECferosTeffCandxxxxxT}{\ensuremath{NULL}} 
\newcommand{\hatcurSPECferosloggCandxxxxxT}{\ensuremath{NULL}} 
\newcommand{\hatcurSPECferoszfehCandxxxxxT}{\ensuremath{NULL}} 
\newcommand{\hatcurSPECferosvsiniCandxxxxxT}{\ensuremath{NULL}} 
\newcommand{\hatcurSPECferosNCandxxxxxT}{\ensuremath{0}} 
\newcommand{\hatcurSPECferossnrangeCandxxxxxT}{NULL} 
\newcommand{\hatcurSPECferosdaterangelowresCandxxxxxT}{NULL} 
\newcommand{\hatcurSPECferosRVKCandxxxxxT}{\ensuremath{NULL}} 
\newcommand{\hatcurSPECferosRVGCandxxxxxT}{\ensuremath{NULL}} 
\newcommand{\hatcurSPECfiesTeffCandxxxxxT}{\ensuremath{NULL}} 
\newcommand{\hatcurSPECfiesloggCandxxxxxT}{\ensuremath{NULL}} 
\newcommand{\hatcurSPECfieszfehCandxxxxxT}{\ensuremath{NULL}} 
\newcommand{\hatcurSPECfiesvsiniCandxxxxxT}{\ensuremath{NULL}} 
\newcommand{\hatcurSPECfiesNCandxxxxxT}{\ensuremath{0}} 
\newcommand{\hatcurSPECfiessnrangeCandxxxxxT}{NULL} 
\newcommand{\hatcurSPECfiesdaterangelowresCandxxxxxT}{NULL} 
\newcommand{\hatcurSPECfiesRVKCandxxxxxT}{\ensuremath{NULL}} 
\newcommand{\hatcurSPECfiesRVGCandxxxxxT}{\ensuremath{NULL}} 
\newcommand{\hatcurSPECcoralieTeffCandxxxxxT}{\ensuremath{NULL}} 
\newcommand{\hatcurSPECcoralieloggCandxxxxxT}{\ensuremath{NULL}} 
\newcommand{\hatcurSPECcoraliezfehCandxxxxxT}{\ensuremath{NULL}} 
\newcommand{\hatcurSPECcoralievsiniCandxxxxxT}{\ensuremath{NULL}} 
\newcommand{\hatcurSPECcoralieNCandxxxxxT}{\ensuremath{0}} 
\newcommand{\hatcurSPECcoraliesnrangeCandxxxxxT}{NULL} 
\newcommand{\hatcurSPECcoraliedaterangelowresCandxxxxxT}{NULL} 
\newcommand{\hatcurSPECcoralieRVKCandxxxxxT}{\ensuremath{NULL}} 
\newcommand{\hatcurSPECcoralieRVGCandxxxxxT}{\ensuremath{NULL}} 

\newcommand{\hatcurnine}{HATS-9}                             
\newcommand{\hatcurnineb}{HATS-9b}                             
\newcommand{\hatcurEPICnine}{EPIC 217671466}                             
\newcommand{\hatcurCCranine}{\ensuremath{19^{\mathrm h}23^{\mathrm m}14.28{\mathrm s}}}                            
\newcommand{\hatcurCCdecnine}{\ensuremath{-20{\arcdeg}09{\arcmin}58.7{\arcsec}}}                           
\newcommand{\hatcurCCtassmrnine}{\ensuremath{13.072}}        
\newcommand{\hatcurLCrprstarnine}{\ensuremath{0.0725\pm0.0041}}      
\newcommand{\hatcurLCdurnine}{\ensuremath{0.1457\pm0.0024}}          
\newcommand{\hatcurLCPnine}{\ensuremath{1.9153073\pm0.0000052}}      
\newcommand{\hatcurLCTnine}{\ensuremath{2456124.25896\pm0.00086}}    
\newcommand{\hatcurPPinine}{\ensuremath{86.5_{-2.5}^{+1.6}}}         
\newcommand{\hatcurPPrhonine}{\ensuremath{0.85\pm0.19}}              
\newcommand{\hatcurPPmlongnine}{\ensuremath{0.837\pm0.029}}          
\newcommand{\hatcurPPrlongnine}{\ensuremath{1.065\pm0.098}}          
\newcommand{\hatcureleven}{HATS-11}                             
\newcommand{\hatcurelevenb}{HATS-11b}                             
\newcommand{\hatcurEPICeleven}{EPIC 216414930}                             
\newcommand{\hatcurCCraeleven}{\ensuremath{19^{\mathrm h}17^{\mathrm m}36.24{\mathrm s}}}                            
\newcommand{\hatcurCCdeceleven}{\ensuremath{-22{\arcdeg}23{\arcmin}23.7{\arcsec}}}                           
\newcommand{\hatcurCCtassmreleven}{\ensuremath{13.865}}        
\newcommand{\hatcurLCrprstareleven}{\ensuremath{0.1076\pm0.0028}}      
\newcommand{\hatcurLCdureleven}{\ensuremath{0.1819\pm0.0039}}          
\newcommand{\hatcurLCPeleven}{\ensuremath{3.6191613\pm0.0000099}}      
\newcommand{\hatcurLCTeleven}{\ensuremath{2456574.9657\pm0.0013}}      
\newcommand{\hatcurPPieleven}{\ensuremath{88.31\pm0.86}}               
\newcommand{\hatcurPPrhoeleven}{\ensuremath{0.299_{-0.050}^{+0.071}}}  
\newcommand{\hatcurPPmlongeleven}{\ensuremath{0.85\pm0.12}}            
\newcommand{\hatcurPPrlongeleven}{\ensuremath{1.510\pm0.078}}          
\newcommand{\hatcurtwelve}{HATS-12}                             
\newcommand{\hatcurtwelveb}{HATS-12b}                             
\newcommand{\hatcurEPICtwelve}{EPIC 218131080}                             
\newcommand{\hatcurCCratwelve}{\ensuremath{19^{\mathrm h}16^{\mathrm m}48.72{\mathrm s}}}                            
\newcommand{\hatcurCCdectwelve}{\ensuremath{-19{\arcdeg}21{\arcmin}21.2{\arcsec}}}                           
\newcommand{\hatcurCCtassmrtwelve}{\ensuremath{12.654}}        
\newcommand{\hatcurLCrprstartwelve}{\ensuremath{0.0630\pm0.0022}}      
\newcommand{\hatcurLCdurtwelve}{\ensuremath{0.1899\pm0.0031}}          
\newcommand{\hatcurLCPtwelve}{\ensuremath{3.1428330\pm0.000011}}        
\newcommand{\hatcurLCTtwelve}{\ensuremath{2456798.9556\pm0.0012}}      
\newcommand{\hatcurPPitwelve}{\ensuremath{82.7\pm1.9}}                 
\newcommand{\hatcurPPrhotwelve}{\ensuremath{1.19_{-0.32}^{+0.54}}}     
\newcommand{\hatcurPPmlongtwelve}{\ensuremath{2.38\pm0.11}}            
\newcommand{\hatcurPPrlongtwelve}{\ensuremath{1.35\pm0.17}}            

\newcommand{\hatcurCCdecCand}[1]{\ifnum#1=578002 %
\hatcurCCdecCandxxxxxA
\else
\ifnum#1=578003 %
\hatcurCCdecCandxxxxxB
\else
\ifnum#1=578004 %
\hatcurCCdecCandxxxxxC
\else
\ifnum#1=579001 %
\hatcurCCdecCandxxxxxD
\else
\ifnum#1=579007 %
\hatcurCCdecCandxxxxxE
\else
\ifnum#1=579008 %
\hatcurCCdecCandxxxxxF
\else
\ifnum#1=579009 %
\hatcurCCdecCandxxxxxG
\else
\ifnum#1=579010 %
\hatcurCCdecCandxxxxxU
\else
\ifnum#1=579014 %
\hatcurCCdecCandxxxxxH
\else
\ifnum#1=579015 %
\hatcurCCdecCandxxxxxI
\else
\ifnum#1=579036 %
\hatcurCCdecCandxxxxxJ
\else
\ifnum#1=579037 %
\hatcurCCdecCandxxxxxK
\else
\ifnum#1=579039 %
\hatcurCCdecCandxxxxxL
\else
\ifnum#1=579040 %
\hatcurCCdecCandxxxxxM
\else
\ifnum#1=579041 %
\hatcurCCdecCandxxxxxN
\else
\ifnum#1=579043 %
\hatcurCCdecCandxxxxxO
\else
\ifnum#1=579044 %
\hatcurCCdecCandxxxxxP
\else
\ifnum#1=579048 %
\hatcurCCdecCandxxxxxQ
\else
\ifnum#1=579050 %
\hatcurCCdecCandxxxxxR
\else
\ifnum#1=624002 %
\hatcurCCdecCandxxxxxS
\else
\ifnum#1=624003 %
\hatcurCCdecCandxxxxxT
\else
??????\fi
\fi
\fi
\fi
\fi
\fi
\fi
\fi
\fi
\fi
\fi
\fi
\fi
\fi
\fi
\fi
\fi
\fi
\fi
\fi
\fi
}
\newcommand{\hatcurCCEPICCand}[1]{\ifnum#1=578002 %
\hatcurCCEPICCandxxxxxA
\else
\ifnum#1=578003 %
\hatcurCCEPICCandxxxxxB
\else
\ifnum#1=578004 %
\hatcurCCEPICCandxxxxxC
\else
\ifnum#1=579001 %
\hatcurCCEPICCandxxxxxD
\else
\ifnum#1=579007 %
\hatcurCCEPICCandxxxxxE
\else
\ifnum#1=579008 %
\hatcurCCEPICCandxxxxxF
\else
\ifnum#1=579009 %
\hatcurCCEPICCandxxxxxG
\else
\ifnum#1=579010 %
\hatcurCCEPICCandxxxxxU
\else
\ifnum#1=579014 %
\hatcurCCEPICCandxxxxxH
\else
\ifnum#1=579015 %
\hatcurCCEPICCandxxxxxI
\else
\ifnum#1=579036 %
\hatcurCCEPICCandxxxxxJ
\else
\ifnum#1=579037 %
\hatcurCCEPICCandxxxxxK
\else
\ifnum#1=579039 %
\hatcurCCEPICCandxxxxxL
\else
\ifnum#1=579040 %
\hatcurCCEPICCandxxxxxM
\else
\ifnum#1=579041 %
\hatcurCCEPICCandxxxxxN
\else
\ifnum#1=579043 %
\hatcurCCEPICCandxxxxxO
\else
\ifnum#1=579044 %
\hatcurCCEPICCandxxxxxP
\else
\ifnum#1=579048 %
\hatcurCCEPICCandxxxxxQ
\else
\ifnum#1=579050 %
\hatcurCCEPICCandxxxxxR
\else
\ifnum#1=624002 %
\hatcurCCEPICCandxxxxxS
\else
\ifnum#1=624003 %
\hatcurCCEPICCandxxxxxT
\else
??????\fi
\fi
\fi
\fi
\fi
\fi
\fi
\fi
\fi
\fi
\fi
\fi
\fi
\fi
\fi
\fi
\fi
\fi
\fi
\fi
\fi
}
\newcommand{\hatcurCCmagBCand}[1]{\ifnum#1=578002 %
\hatcurCCmagBCandxxxxxA
\else
\ifnum#1=578003 %
\hatcurCCmagBCandxxxxxB
\else
\ifnum#1=578004 %
\hatcurCCmagBCandxxxxxC
\else
\ifnum#1=579001 %
\hatcurCCmagBCandxxxxxD
\else
\ifnum#1=579007 %
\hatcurCCmagBCandxxxxxE
\else
\ifnum#1=579008 %
\hatcurCCmagBCandxxxxxF
\else
\ifnum#1=579009 %
\hatcurCCmagBCandxxxxxG
\else
\ifnum#1=579010 %
\hatcurCCmagBCandxxxxxU
\else
\ifnum#1=579014 %
\hatcurCCmagBCandxxxxxH
\else
\ifnum#1=579015 %
\hatcurCCmagBCandxxxxxI
\else
\ifnum#1=579036 %
\hatcurCCmagBCandxxxxxJ
\else
\ifnum#1=579037 %
\hatcurCCmagBCandxxxxxK
\else
\ifnum#1=579039 %
\hatcurCCmagBCandxxxxxL
\else
\ifnum#1=579040 %
\hatcurCCmagBCandxxxxxM
\else
\ifnum#1=579041 %
\hatcurCCmagBCandxxxxxN
\else
\ifnum#1=579043 %
\hatcurCCmagBCandxxxxxO
\else
\ifnum#1=579044 %
\hatcurCCmagBCandxxxxxP
\else
\ifnum#1=579048 %
\hatcurCCmagBCandxxxxxQ
\else
\ifnum#1=579050 %
\hatcurCCmagBCandxxxxxR
\else
\ifnum#1=624002 %
\hatcurCCmagBCandxxxxxS
\else
\ifnum#1=624003 %
\hatcurCCmagBCandxxxxxT
\else
??????\fi
\fi
\fi
\fi
\fi
\fi
\fi
\fi
\fi
\fi
\fi
\fi
\fi
\fi
\fi
\fi
\fi
\fi
\fi
\fi
\fi
}
\newcommand{\hatcurCCmaggCand}[1]{\ifnum#1=578002 %
\hatcurCCmaggCandxxxxxA
\else
\ifnum#1=578003 %
\hatcurCCmaggCandxxxxxB
\else
\ifnum#1=578004 %
\hatcurCCmaggCandxxxxxC
\else
\ifnum#1=579001 %
\hatcurCCmaggCandxxxxxD
\else
\ifnum#1=579007 %
\hatcurCCmaggCandxxxxxE
\else
\ifnum#1=579008 %
\hatcurCCmaggCandxxxxxF
\else
\ifnum#1=579009 %
\hatcurCCmaggCandxxxxxG
\else
\ifnum#1=579010 %
\hatcurCCmaggCandxxxxxU
\else
\ifnum#1=579014 %
\hatcurCCmaggCandxxxxxH
\else
\ifnum#1=579015 %
\hatcurCCmaggCandxxxxxI
\else
\ifnum#1=579036 %
\hatcurCCmaggCandxxxxxJ
\else
\ifnum#1=579037 %
\hatcurCCmaggCandxxxxxK
\else
\ifnum#1=579039 %
\hatcurCCmaggCandxxxxxL
\else
\ifnum#1=579040 %
\hatcurCCmaggCandxxxxxM
\else
\ifnum#1=579041 %
\hatcurCCmaggCandxxxxxN
\else
\ifnum#1=579043 %
\hatcurCCmaggCandxxxxxO
\else
\ifnum#1=579044 %
\hatcurCCmaggCandxxxxxP
\else
\ifnum#1=579048 %
\hatcurCCmaggCandxxxxxQ
\else
\ifnum#1=579050 %
\hatcurCCmaggCandxxxxxR
\else
\ifnum#1=624002 %
\hatcurCCmaggCandxxxxxS
\else
\ifnum#1=624003 %
\hatcurCCmaggCandxxxxxT
\else
??????\fi
\fi
\fi
\fi
\fi
\fi
\fi
\fi
\fi
\fi
\fi
\fi
\fi
\fi
\fi
\fi
\fi
\fi
\fi
\fi
\fi
}
\newcommand{\hatcurCCmagHCand}[1]{\ifnum#1=578002 %
\hatcurCCmagHCandxxxxxA
\else
\ifnum#1=578003 %
\hatcurCCmagHCandxxxxxB
\else
\ifnum#1=578004 %
\hatcurCCmagHCandxxxxxC
\else
\ifnum#1=579001 %
\hatcurCCmagHCandxxxxxD
\else
\ifnum#1=579007 %
\hatcurCCmagHCandxxxxxE
\else
\ifnum#1=579008 %
\hatcurCCmagHCandxxxxxF
\else
\ifnum#1=579009 %
\hatcurCCmagHCandxxxxxG
\else
\ifnum#1=579010 %
\hatcurCCmagHCandxxxxxU
\else
\ifnum#1=579014 %
\hatcurCCmagHCandxxxxxH
\else
\ifnum#1=579015 %
\hatcurCCmagHCandxxxxxI
\else
\ifnum#1=579036 %
\hatcurCCmagHCandxxxxxJ
\else
\ifnum#1=579037 %
\hatcurCCmagHCandxxxxxK
\else
\ifnum#1=579039 %
\hatcurCCmagHCandxxxxxL
\else
\ifnum#1=579040 %
\hatcurCCmagHCandxxxxxM
\else
\ifnum#1=579041 %
\hatcurCCmagHCandxxxxxN
\else
\ifnum#1=579043 %
\hatcurCCmagHCandxxxxxO
\else
\ifnum#1=579044 %
\hatcurCCmagHCandxxxxxP
\else
\ifnum#1=579048 %
\hatcurCCmagHCandxxxxxQ
\else
\ifnum#1=579050 %
\hatcurCCmagHCandxxxxxR
\else
\ifnum#1=624002 %
\hatcurCCmagHCandxxxxxS
\else
\ifnum#1=624003 %
\hatcurCCmagHCandxxxxxT
\else
??????\fi
\fi
\fi
\fi
\fi
\fi
\fi
\fi
\fi
\fi
\fi
\fi
\fi
\fi
\fi
\fi
\fi
\fi
\fi
\fi
\fi
}
\newcommand{\hatcurCCmagiCand}[1]{\ifnum#1=578002 %
\hatcurCCmagiCandxxxxxA
\else
\ifnum#1=578003 %
\hatcurCCmagiCandxxxxxB
\else
\ifnum#1=578004 %
\hatcurCCmagiCandxxxxxC
\else
\ifnum#1=579001 %
\hatcurCCmagiCandxxxxxD
\else
\ifnum#1=579007 %
\hatcurCCmagiCandxxxxxE
\else
\ifnum#1=579008 %
\hatcurCCmagiCandxxxxxF
\else
\ifnum#1=579009 %
\hatcurCCmagiCandxxxxxG
\else
\ifnum#1=579010 %
\hatcurCCmagiCandxxxxxU
\else
\ifnum#1=579014 %
\hatcurCCmagiCandxxxxxH
\else
\ifnum#1=579015 %
\hatcurCCmagiCandxxxxxI
\else
\ifnum#1=579036 %
\hatcurCCmagiCandxxxxxJ
\else
\ifnum#1=579037 %
\hatcurCCmagiCandxxxxxK
\else
\ifnum#1=579039 %
\hatcurCCmagiCandxxxxxL
\else
\ifnum#1=579040 %
\hatcurCCmagiCandxxxxxM
\else
\ifnum#1=579041 %
\hatcurCCmagiCandxxxxxN
\else
\ifnum#1=579043 %
\hatcurCCmagiCandxxxxxO
\else
\ifnum#1=579044 %
\hatcurCCmagiCandxxxxxP
\else
\ifnum#1=579048 %
\hatcurCCmagiCandxxxxxQ
\else
\ifnum#1=579050 %
\hatcurCCmagiCandxxxxxR
\else
\ifnum#1=624002 %
\hatcurCCmagiCandxxxxxS
\else
\ifnum#1=624003 %
\hatcurCCmagiCandxxxxxT
\else
??????\fi
\fi
\fi
\fi
\fi
\fi
\fi
\fi
\fi
\fi
\fi
\fi
\fi
\fi
\fi
\fi
\fi
\fi
\fi
\fi
\fi
}
\newcommand{\hatcurCCmagJCand}[1]{\ifnum#1=578002 %
\hatcurCCmagJCandxxxxxA
\else
\ifnum#1=578003 %
\hatcurCCmagJCandxxxxxB
\else
\ifnum#1=578004 %
\hatcurCCmagJCandxxxxxC
\else
\ifnum#1=579001 %
\hatcurCCmagJCandxxxxxD
\else
\ifnum#1=579007 %
\hatcurCCmagJCandxxxxxE
\else
\ifnum#1=579008 %
\hatcurCCmagJCandxxxxxF
\else
\ifnum#1=579009 %
\hatcurCCmagJCandxxxxxG
\else
\ifnum#1=579010 %
\hatcurCCmagJCandxxxxxU
\else
\ifnum#1=579014 %
\hatcurCCmagJCandxxxxxH
\else
\ifnum#1=579015 %
\hatcurCCmagJCandxxxxxI
\else
\ifnum#1=579036 %
\hatcurCCmagJCandxxxxxJ
\else
\ifnum#1=579037 %
\hatcurCCmagJCandxxxxxK
\else
\ifnum#1=579039 %
\hatcurCCmagJCandxxxxxL
\else
\ifnum#1=579040 %
\hatcurCCmagJCandxxxxxM
\else
\ifnum#1=579041 %
\hatcurCCmagJCandxxxxxN
\else
\ifnum#1=579043 %
\hatcurCCmagJCandxxxxxO
\else
\ifnum#1=579044 %
\hatcurCCmagJCandxxxxxP
\else
\ifnum#1=579048 %
\hatcurCCmagJCandxxxxxQ
\else
\ifnum#1=579050 %
\hatcurCCmagJCandxxxxxR
\else
\ifnum#1=624002 %
\hatcurCCmagJCandxxxxxS
\else
\ifnum#1=624003 %
\hatcurCCmagJCandxxxxxT
\else
??????\fi
\fi
\fi
\fi
\fi
\fi
\fi
\fi
\fi
\fi
\fi
\fi
\fi
\fi
\fi
\fi
\fi
\fi
\fi
\fi
\fi
}
\newcommand{\hatcurCCmagKCand}[1]{\ifnum#1=578002 %
\hatcurCCmagKCandxxxxxA
\else
\ifnum#1=578003 %
\hatcurCCmagKCandxxxxxB
\else
\ifnum#1=578004 %
\hatcurCCmagKCandxxxxxC
\else
\ifnum#1=579001 %
\hatcurCCmagKCandxxxxxD
\else
\ifnum#1=579007 %
\hatcurCCmagKCandxxxxxE
\else
\ifnum#1=579008 %
\hatcurCCmagKCandxxxxxF
\else
\ifnum#1=579009 %
\hatcurCCmagKCandxxxxxG
\else
\ifnum#1=579010 %
\hatcurCCmagKCandxxxxxU
\else
\ifnum#1=579014 %
\hatcurCCmagKCandxxxxxH
\else
\ifnum#1=579015 %
\hatcurCCmagKCandxxxxxI
\else
\ifnum#1=579036 %
\hatcurCCmagKCandxxxxxJ
\else
\ifnum#1=579037 %
\hatcurCCmagKCandxxxxxK
\else
\ifnum#1=579039 %
\hatcurCCmagKCandxxxxxL
\else
\ifnum#1=579040 %
\hatcurCCmagKCandxxxxxM
\else
\ifnum#1=579041 %
\hatcurCCmagKCandxxxxxN
\else
\ifnum#1=579043 %
\hatcurCCmagKCandxxxxxO
\else
\ifnum#1=579044 %
\hatcurCCmagKCandxxxxxP
\else
\ifnum#1=579048 %
\hatcurCCmagKCandxxxxxQ
\else
\ifnum#1=579050 %
\hatcurCCmagKCandxxxxxR
\else
\ifnum#1=624002 %
\hatcurCCmagKCandxxxxxS
\else
\ifnum#1=624003 %
\hatcurCCmagKCandxxxxxT
\else
??????\fi
\fi
\fi
\fi
\fi
\fi
\fi
\fi
\fi
\fi
\fi
\fi
\fi
\fi
\fi
\fi
\fi
\fi
\fi
\fi
\fi
}
\newcommand{\hatcurCCmagrCand}[1]{\ifnum#1=578002 %
\hatcurCCmagrCandxxxxxA
\else
\ifnum#1=578003 %
\hatcurCCmagrCandxxxxxB
\else
\ifnum#1=578004 %
\hatcurCCmagrCandxxxxxC
\else
\ifnum#1=579001 %
\hatcurCCmagrCandxxxxxD
\else
\ifnum#1=579007 %
\hatcurCCmagrCandxxxxxE
\else
\ifnum#1=579008 %
\hatcurCCmagrCandxxxxxF
\else
\ifnum#1=579009 %
\hatcurCCmagrCandxxxxxG
\else
\ifnum#1=579010 %
\hatcurCCmagrCandxxxxxU
\else
\ifnum#1=579014 %
\hatcurCCmagrCandxxxxxH
\else
\ifnum#1=579015 %
\hatcurCCmagrCandxxxxxI
\else
\ifnum#1=579036 %
\hatcurCCmagrCandxxxxxJ
\else
\ifnum#1=579037 %
\hatcurCCmagrCandxxxxxK
\else
\ifnum#1=579039 %
\hatcurCCmagrCandxxxxxL
\else
\ifnum#1=579040 %
\hatcurCCmagrCandxxxxxM
\else
\ifnum#1=579041 %
\hatcurCCmagrCandxxxxxN
\else
\ifnum#1=579043 %
\hatcurCCmagrCandxxxxxO
\else
\ifnum#1=579044 %
\hatcurCCmagrCandxxxxxP
\else
\ifnum#1=579048 %
\hatcurCCmagrCandxxxxxQ
\else
\ifnum#1=579050 %
\hatcurCCmagrCandxxxxxR
\else
\ifnum#1=624002 %
\hatcurCCmagrCandxxxxxS
\else
\ifnum#1=624003 %
\hatcurCCmagrCandxxxxxT
\else
??????\fi
\fi
\fi
\fi
\fi
\fi
\fi
\fi
\fi
\fi
\fi
\fi
\fi
\fi
\fi
\fi
\fi
\fi
\fi
\fi
\fi
}
\newcommand{\hatcurCCmagVCand}[1]{\ifnum#1=578002 %
\hatcurCCmagVCandxxxxxA
\else
\ifnum#1=578003 %
\hatcurCCmagVCandxxxxxB
\else
\ifnum#1=578004 %
\hatcurCCmagVCandxxxxxC
\else
\ifnum#1=579001 %
\hatcurCCmagVCandxxxxxD
\else
\ifnum#1=579007 %
\hatcurCCmagVCandxxxxxE
\else
\ifnum#1=579008 %
\hatcurCCmagVCandxxxxxF
\else
\ifnum#1=579009 %
\hatcurCCmagVCandxxxxxG
\else
\ifnum#1=579010 %
\hatcurCCmagVCandxxxxxU
\else
\ifnum#1=579014 %
\hatcurCCmagVCandxxxxxH
\else
\ifnum#1=579015 %
\hatcurCCmagVCandxxxxxI
\else
\ifnum#1=579036 %
\hatcurCCmagVCandxxxxxJ
\else
\ifnum#1=579037 %
\hatcurCCmagVCandxxxxxK
\else
\ifnum#1=579039 %
\hatcurCCmagVCandxxxxxL
\else
\ifnum#1=579040 %
\hatcurCCmagVCandxxxxxM
\else
\ifnum#1=579041 %
\hatcurCCmagVCandxxxxxN
\else
\ifnum#1=579043 %
\hatcurCCmagVCandxxxxxO
\else
\ifnum#1=579044 %
\hatcurCCmagVCandxxxxxP
\else
\ifnum#1=579048 %
\hatcurCCmagVCandxxxxxQ
\else
\ifnum#1=579050 %
\hatcurCCmagVCandxxxxxR
\else
\ifnum#1=624002 %
\hatcurCCmagVCandxxxxxS
\else
\ifnum#1=624003 %
\hatcurCCmagVCandxxxxxT
\else
??????\fi
\fi
\fi
\fi
\fi
\fi
\fi
\fi
\fi
\fi
\fi
\fi
\fi
\fi
\fi
\fi
\fi
\fi
\fi
\fi
\fi
}
\newcommand{\hatcurCCraCand}[1]{\ifnum#1=578002 %
\hatcurCCraCandxxxxxA
\else
\ifnum#1=578003 %
\hatcurCCraCandxxxxxB
\else
\ifnum#1=578004 %
\hatcurCCraCandxxxxxC
\else
\ifnum#1=579001 %
\hatcurCCraCandxxxxxD
\else
\ifnum#1=579007 %
\hatcurCCraCandxxxxxE
\else
\ifnum#1=579008 %
\hatcurCCraCandxxxxxF
\else
\ifnum#1=579009 %
\hatcurCCraCandxxxxxG
\else
\ifnum#1=579010 %
\hatcurCCraCandxxxxxU
\else
\ifnum#1=579014 %
\hatcurCCraCandxxxxxH
\else
\ifnum#1=579015 %
\hatcurCCraCandxxxxxI
\else
\ifnum#1=579036 %
\hatcurCCraCandxxxxxJ
\else
\ifnum#1=579037 %
\hatcurCCraCandxxxxxK
\else
\ifnum#1=579039 %
\hatcurCCraCandxxxxxL
\else
\ifnum#1=579040 %
\hatcurCCraCandxxxxxM
\else
\ifnum#1=579041 %
\hatcurCCraCandxxxxxN
\else
\ifnum#1=579043 %
\hatcurCCraCandxxxxxO
\else
\ifnum#1=579044 %
\hatcurCCraCandxxxxxP
\else
\ifnum#1=579048 %
\hatcurCCraCandxxxxxQ
\else
\ifnum#1=579050 %
\hatcurCCraCandxxxxxR
\else
\ifnum#1=624002 %
\hatcurCCraCandxxxxxS
\else
\ifnum#1=624003 %
\hatcurCCraCandxxxxxT
\else
??????\fi
\fi
\fi
\fi
\fi
\fi
\fi
\fi
\fi
\fi
\fi
\fi
\fi
\fi
\fi
\fi
\fi
\fi
\fi
\fi
\fi
}
\newcommand{\hatcurCLASSCand}[1]{\ifnum#1=578002 %
\hatcurCLASSCandxxxxxA
\else
\ifnum#1=578003 %
\hatcurCLASSCandxxxxxB
\else
\ifnum#1=578004 %
\hatcurCLASSCandxxxxxC
\else
\ifnum#1=579001 %
\hatcurCLASSCandxxxxxD
\else
\ifnum#1=579007 %
\hatcurCLASSCandxxxxxE
\else
\ifnum#1=579008 %
\hatcurCLASSCandxxxxxF
\else
\ifnum#1=579009 %
\hatcurCLASSCandxxxxxG
\else
\ifnum#1=579010 %
\hatcurCLASSCandxxxxxU
\else
\ifnum#1=579014 %
\hatcurCLASSCandxxxxxH
\else
\ifnum#1=579015 %
\hatcurCLASSCandxxxxxI
\else
\ifnum#1=579036 %
\hatcurCLASSCandxxxxxJ
\else
\ifnum#1=579037 %
\hatcurCLASSCandxxxxxK
\else
\ifnum#1=579039 %
\hatcurCLASSCandxxxxxL
\else
\ifnum#1=579040 %
\hatcurCLASSCandxxxxxM
\else
\ifnum#1=579041 %
\hatcurCLASSCandxxxxxN
\else
\ifnum#1=579043 %
\hatcurCLASSCandxxxxxO
\else
\ifnum#1=579044 %
\hatcurCLASSCandxxxxxP
\else
\ifnum#1=579048 %
\hatcurCLASSCandxxxxxQ
\else
\ifnum#1=579050 %
\hatcurCLASSCandxxxxxR
\else
\ifnum#1=624002 %
\hatcurCLASSCandxxxxxS
\else
\ifnum#1=624003 %
\hatcurCLASSCandxxxxxT
\else
??????\fi
\fi
\fi
\fi
\fi
\fi
\fi
\fi
\fi
\fi
\fi
\fi
\fi
\fi
\fi
\fi
\fi
\fi
\fi
\fi
\fi
}
\newcommand{\hatcurhtrCand}[1]{\ifnum#1=578002 %
\hatcurhtrCandxxxxxA
\else
\ifnum#1=578003 %
\hatcurhtrCandxxxxxB
\else
\ifnum#1=578004 %
\hatcurhtrCandxxxxxC
\else
\ifnum#1=579001 %
\hatcurhtrCandxxxxxD
\else
\ifnum#1=579007 %
\hatcurhtrCandxxxxxE
\else
\ifnum#1=579008 %
\hatcurhtrCandxxxxxF
\else
\ifnum#1=579009 %
\hatcurhtrCandxxxxxG
\else
\ifnum#1=579010 %
\hatcurhtrCandxxxxxU
\else
\ifnum#1=579014 %
\hatcurhtrCandxxxxxH
\else
\ifnum#1=579015 %
\hatcurhtrCandxxxxxI
\else
\ifnum#1=579036 %
\hatcurhtrCandxxxxxJ
\else
\ifnum#1=579037 %
\hatcurhtrCandxxxxxK
\else
\ifnum#1=579039 %
\hatcurhtrCandxxxxxL
\else
\ifnum#1=579040 %
\hatcurhtrCandxxxxxM
\else
\ifnum#1=579041 %
\hatcurhtrCandxxxxxN
\else
\ifnum#1=579043 %
\hatcurhtrCandxxxxxO
\else
\ifnum#1=579044 %
\hatcurhtrCandxxxxxP
\else
\ifnum#1=579048 %
\hatcurhtrCandxxxxxQ
\else
\ifnum#1=579050 %
\hatcurhtrCandxxxxxR
\else
\ifnum#1=624002 %
\hatcurhtrCandxxxxxS
\else
\ifnum#1=624003 %
\hatcurhtrCandxxxxxT
\else
??????\fi
\fi
\fi
\fi
\fi
\fi
\fi
\fi
\fi
\fi
\fi
\fi
\fi
\fi
\fi
\fi
\fi
\fi
\fi
\fi
\fi
}
\newcommand{\hatcurLCbsqCand}[1]{\ifnum#1=578002 %
\hatcurLCbsqCandxxxxxA
\else
\ifnum#1=578003 %
\hatcurLCbsqCandxxxxxB
\else
\ifnum#1=578004 %
\hatcurLCbsqCandxxxxxC
\else
\ifnum#1=579001 %
\hatcurLCbsqCandxxxxxD
\else
\ifnum#1=579007 %
\hatcurLCbsqCandxxxxxE
\else
\ifnum#1=579008 %
\hatcurLCbsqCandxxxxxF
\else
\ifnum#1=579009 %
\hatcurLCbsqCandxxxxxG
\else
\ifnum#1=579010 %
\hatcurLCbsqCandxxxxxU
\else
\ifnum#1=579014 %
\hatcurLCbsqCandxxxxxH
\else
\ifnum#1=579015 %
\hatcurLCbsqCandxxxxxI
\else
\ifnum#1=579036 %
\hatcurLCbsqCandxxxxxJ
\else
\ifnum#1=579037 %
\hatcurLCbsqCandxxxxxK
\else
\ifnum#1=579039 %
\hatcurLCbsqCandxxxxxL
\else
\ifnum#1=579040 %
\hatcurLCbsqCandxxxxxM
\else
\ifnum#1=579041 %
\hatcurLCbsqCandxxxxxN
\else
\ifnum#1=579043 %
\hatcurLCbsqCandxxxxxO
\else
\ifnum#1=579044 %
\hatcurLCbsqCandxxxxxP
\else
\ifnum#1=579048 %
\hatcurLCbsqCandxxxxxQ
\else
\ifnum#1=579050 %
\hatcurLCbsqCandxxxxxR
\else
\ifnum#1=624002 %
\hatcurLCbsqCandxxxxxS
\else
\ifnum#1=624003 %
\hatcurLCbsqCandxxxxxT
\else
??????\fi
\fi
\fi
\fi
\fi
\fi
\fi
\fi
\fi
\fi
\fi
\fi
\fi
\fi
\fi
\fi
\fi
\fi
\fi
\fi
\fi
}
\newcommand{\hatcurLCdurCand}[1]{\ifnum#1=578002 %
\hatcurLCdurCandxxxxxA
\else
\ifnum#1=578003 %
\hatcurLCdurCandxxxxxB
\else
\ifnum#1=578004 %
\hatcurLCdurCandxxxxxC
\else
\ifnum#1=579001 %
\hatcurLCdurCandxxxxxD
\else
\ifnum#1=579007 %
\hatcurLCdurCandxxxxxE
\else
\ifnum#1=579008 %
\hatcurLCdurCandxxxxxF
\else
\ifnum#1=579009 %
\hatcurLCdurCandxxxxxG
\else
\ifnum#1=579010 %
\hatcurLCdurCandxxxxxU
\else
\ifnum#1=579014 %
\hatcurLCdurCandxxxxxH
\else
\ifnum#1=579015 %
\hatcurLCdurCandxxxxxI
\else
\ifnum#1=579036 %
\hatcurLCdurCandxxxxxJ
\else
\ifnum#1=579037 %
\hatcurLCdurCandxxxxxK
\else
\ifnum#1=579039 %
\hatcurLCdurCandxxxxxL
\else
\ifnum#1=579040 %
\hatcurLCdurCandxxxxxM
\else
\ifnum#1=579041 %
\hatcurLCdurCandxxxxxN
\else
\ifnum#1=579043 %
\hatcurLCdurCandxxxxxO
\else
\ifnum#1=579044 %
\hatcurLCdurCandxxxxxP
\else
\ifnum#1=579048 %
\hatcurLCdurCandxxxxxQ
\else
\ifnum#1=579050 %
\hatcurLCdurCandxxxxxR
\else
\ifnum#1=624002 %
\hatcurLCdurCandxxxxxS
\else
\ifnum#1=624003 %
\hatcurLCdurCandxxxxxT
\else
??????\fi
\fi
\fi
\fi
\fi
\fi
\fi
\fi
\fi
\fi
\fi
\fi
\fi
\fi
\fi
\fi
\fi
\fi
\fi
\fi
\fi
}
\newcommand{\hatcurLCPCand}[1]{\ifnum#1=578002 %
\hatcurLCPCandxxxxxA
\else
\ifnum#1=578003 %
\hatcurLCPCandxxxxxB
\else
\ifnum#1=578004 %
\hatcurLCPCandxxxxxC
\else
\ifnum#1=579001 %
\hatcurLCPCandxxxxxD
\else
\ifnum#1=579007 %
\hatcurLCPCandxxxxxE
\else
\ifnum#1=579008 %
\hatcurLCPCandxxxxxF
\else
\ifnum#1=579009 %
\hatcurLCPCandxxxxxG
\else
\ifnum#1=579010 %
\hatcurLCPCandxxxxxU
\else
\ifnum#1=579014 %
\hatcurLCPCandxxxxxH
\else
\ifnum#1=579015 %
\hatcurLCPCandxxxxxI
\else
\ifnum#1=579036 %
\hatcurLCPCandxxxxxJ
\else
\ifnum#1=579037 %
\hatcurLCPCandxxxxxK
\else
\ifnum#1=579039 %
\hatcurLCPCandxxxxxL
\else
\ifnum#1=579040 %
\hatcurLCPCandxxxxxM
\else
\ifnum#1=579041 %
\hatcurLCPCandxxxxxN
\else
\ifnum#1=579043 %
\hatcurLCPCandxxxxxO
\else
\ifnum#1=579044 %
\hatcurLCPCandxxxxxP
\else
\ifnum#1=579048 %
\hatcurLCPCandxxxxxQ
\else
\ifnum#1=579050 %
\hatcurLCPCandxxxxxR
\else
\ifnum#1=624002 %
\hatcurLCPCandxxxxxS
\else
\ifnum#1=624003 %
\hatcurLCPCandxxxxxT
\else
??????\fi
\fi
\fi
\fi
\fi
\fi
\fi
\fi
\fi
\fi
\fi
\fi
\fi
\fi
\fi
\fi
\fi
\fi
\fi
\fi
\fi
}
\newcommand{\hatcurLCrprstarCand}[1]{\ifnum#1=578002 %
\hatcurLCrprstarCandxxxxxA
\else
\ifnum#1=578003 %
\hatcurLCrprstarCandxxxxxB
\else
\ifnum#1=578004 %
\hatcurLCrprstarCandxxxxxC
\else
\ifnum#1=579001 %
\hatcurLCrprstarCandxxxxxD
\else
\ifnum#1=579007 %
\hatcurLCrprstarCandxxxxxE
\else
\ifnum#1=579008 %
\hatcurLCrprstarCandxxxxxF
\else
\ifnum#1=579009 %
\hatcurLCrprstarCandxxxxxG
\else
\ifnum#1=579010 %
\hatcurLCrprstarCandxxxxxU
\else
\ifnum#1=579014 %
\hatcurLCrprstarCandxxxxxH
\else
\ifnum#1=579015 %
\hatcurLCrprstarCandxxxxxI
\else
\ifnum#1=579036 %
\hatcurLCrprstarCandxxxxxJ
\else
\ifnum#1=579037 %
\hatcurLCrprstarCandxxxxxK
\else
\ifnum#1=579039 %
\hatcurLCrprstarCandxxxxxL
\else
\ifnum#1=579040 %
\hatcurLCrprstarCandxxxxxM
\else
\ifnum#1=579041 %
\hatcurLCrprstarCandxxxxxN
\else
\ifnum#1=579043 %
\hatcurLCrprstarCandxxxxxO
\else
\ifnum#1=579044 %
\hatcurLCrprstarCandxxxxxP
\else
\ifnum#1=579048 %
\hatcurLCrprstarCandxxxxxQ
\else
\ifnum#1=579050 %
\hatcurLCrprstarCandxxxxxR
\else
\ifnum#1=624002 %
\hatcurLCrprstarCandxxxxxS
\else
\ifnum#1=624003 %
\hatcurLCrprstarCandxxxxxT
\else
??????\fi
\fi
\fi
\fi
\fi
\fi
\fi
\fi
\fi
\fi
\fi
\fi
\fi
\fi
\fi
\fi
\fi
\fi
\fi
\fi
\fi
}
\newcommand{\hatcurLCTCand}[1]{\ifnum#1=578002 %
\hatcurLCTCandxxxxxA
\else
\ifnum#1=578003 %
\hatcurLCTCandxxxxxB
\else
\ifnum#1=578004 %
\hatcurLCTCandxxxxxC
\else
\ifnum#1=579001 %
\hatcurLCTCandxxxxxD
\else
\ifnum#1=579007 %
\hatcurLCTCandxxxxxE
\else
\ifnum#1=579008 %
\hatcurLCTCandxxxxxF
\else
\ifnum#1=579009 %
\hatcurLCTCandxxxxxG
\else
\ifnum#1=579010 %
\hatcurLCTCandxxxxxU
\else
\ifnum#1=579014 %
\hatcurLCTCandxxxxxH
\else
\ifnum#1=579015 %
\hatcurLCTCandxxxxxI
\else
\ifnum#1=579036 %
\hatcurLCTCandxxxxxJ
\else
\ifnum#1=579037 %
\hatcurLCTCandxxxxxK
\else
\ifnum#1=579039 %
\hatcurLCTCandxxxxxL
\else
\ifnum#1=579040 %
\hatcurLCTCandxxxxxM
\else
\ifnum#1=579041 %
\hatcurLCTCandxxxxxN
\else
\ifnum#1=579043 %
\hatcurLCTCandxxxxxO
\else
\ifnum#1=579044 %
\hatcurLCTCandxxxxxP
\else
\ifnum#1=579048 %
\hatcurLCTCandxxxxxQ
\else
\ifnum#1=579050 %
\hatcurLCTCandxxxxxR
\else
\ifnum#1=624002 %
\hatcurLCTCandxxxxxS
\else
\ifnum#1=624003 %
\hatcurLCTCandxxxxxT
\else
??????\fi
\fi
\fi
\fi
\fi
\fi
\fi
\fi
\fi
\fi
\fi
\fi
\fi
\fi
\fi
\fi
\fi
\fi
\fi
\fi
\fi
}
\newcommand{\hatcurSPECarcesdaterangelowresCand}[1]{\ifnum#1=578002 %
\hatcurSPECarcesdaterangelowresCandxxxxxA
\else
\ifnum#1=578003 %
\hatcurSPECarcesdaterangelowresCandxxxxxB
\else
\ifnum#1=578004 %
\hatcurSPECarcesdaterangelowresCandxxxxxC
\else
\ifnum#1=579001 %
\hatcurSPECarcesdaterangelowresCandxxxxxD
\else
\ifnum#1=579007 %
\hatcurSPECarcesdaterangelowresCandxxxxxE
\else
\ifnum#1=579008 %
\hatcurSPECarcesdaterangelowresCandxxxxxF
\else
\ifnum#1=579009 %
\hatcurSPECarcesdaterangelowresCandxxxxxG
\else
\ifnum#1=579010 %
\hatcurSPECarcesdaterangelowresCandxxxxxU
\else
\ifnum#1=579014 %
\hatcurSPECarcesdaterangelowresCandxxxxxH
\else
\ifnum#1=579015 %
\hatcurSPECarcesdaterangelowresCandxxxxxI
\else
\ifnum#1=579036 %
\hatcurSPECarcesdaterangelowresCandxxxxxJ
\else
\ifnum#1=579037 %
\hatcurSPECarcesdaterangelowresCandxxxxxK
\else
\ifnum#1=579039 %
\hatcurSPECarcesdaterangelowresCandxxxxxL
\else
\ifnum#1=579040 %
\hatcurSPECarcesdaterangelowresCandxxxxxM
\else
\ifnum#1=579041 %
\hatcurSPECarcesdaterangelowresCandxxxxxN
\else
\ifnum#1=579043 %
\hatcurSPECarcesdaterangelowresCandxxxxxO
\else
\ifnum#1=579044 %
\hatcurSPECarcesdaterangelowresCandxxxxxP
\else
\ifnum#1=579048 %
\hatcurSPECarcesdaterangelowresCandxxxxxQ
\else
\ifnum#1=579050 %
\hatcurSPECarcesdaterangelowresCandxxxxxR
\else
\ifnum#1=624002 %
\hatcurSPECarcesdaterangelowresCandxxxxxS
\else
\ifnum#1=624003 %
\hatcurSPECarcesdaterangelowresCandxxxxxT
\else
??????\fi
\fi
\fi
\fi
\fi
\fi
\fi
\fi
\fi
\fi
\fi
\fi
\fi
\fi
\fi
\fi
\fi
\fi
\fi
\fi
\fi
}
\newcommand{\hatcurSPECarcesloggCand}[1]{\ifnum#1=578002 %
\hatcurSPECarcesloggCandxxxxxA
\else
\ifnum#1=578003 %
\hatcurSPECarcesloggCandxxxxxB
\else
\ifnum#1=578004 %
\hatcurSPECarcesloggCandxxxxxC
\else
\ifnum#1=579001 %
\hatcurSPECarcesloggCandxxxxxD
\else
\ifnum#1=579007 %
\hatcurSPECarcesloggCandxxxxxE
\else
\ifnum#1=579008 %
\hatcurSPECarcesloggCandxxxxxF
\else
\ifnum#1=579009 %
\hatcurSPECarcesloggCandxxxxxG
\else
\ifnum#1=579010 %
\hatcurSPECarcesloggCandxxxxxU
\else
\ifnum#1=579014 %
\hatcurSPECarcesloggCandxxxxxH
\else
\ifnum#1=579015 %
\hatcurSPECarcesloggCandxxxxxI
\else
\ifnum#1=579036 %
\hatcurSPECarcesloggCandxxxxxJ
\else
\ifnum#1=579037 %
\hatcurSPECarcesloggCandxxxxxK
\else
\ifnum#1=579039 %
\hatcurSPECarcesloggCandxxxxxL
\else
\ifnum#1=579040 %
\hatcurSPECarcesloggCandxxxxxM
\else
\ifnum#1=579041 %
\hatcurSPECarcesloggCandxxxxxN
\else
\ifnum#1=579043 %
\hatcurSPECarcesloggCandxxxxxO
\else
\ifnum#1=579044 %
\hatcurSPECarcesloggCandxxxxxP
\else
\ifnum#1=579048 %
\hatcurSPECarcesloggCandxxxxxQ
\else
\ifnum#1=579050 %
\hatcurSPECarcesloggCandxxxxxR
\else
\ifnum#1=624002 %
\hatcurSPECarcesloggCandxxxxxS
\else
\ifnum#1=624003 %
\hatcurSPECarcesloggCandxxxxxT
\else
??????\fi
\fi
\fi
\fi
\fi
\fi
\fi
\fi
\fi
\fi
\fi
\fi
\fi
\fi
\fi
\fi
\fi
\fi
\fi
\fi
\fi
}
\newcommand{\hatcurSPECarcesNCand}[1]{\ifnum#1=578002 %
\hatcurSPECarcesNCandxxxxxA
\else
\ifnum#1=578003 %
\hatcurSPECarcesNCandxxxxxB
\else
\ifnum#1=578004 %
\hatcurSPECarcesNCandxxxxxC
\else
\ifnum#1=579001 %
\hatcurSPECarcesNCandxxxxxD
\else
\ifnum#1=579007 %
\hatcurSPECarcesNCandxxxxxE
\else
\ifnum#1=579008 %
\hatcurSPECarcesNCandxxxxxF
\else
\ifnum#1=579009 %
\hatcurSPECarcesNCandxxxxxG
\else
\ifnum#1=579010 %
\hatcurSPECarcesNCandxxxxxU
\else
\ifnum#1=579014 %
\hatcurSPECarcesNCandxxxxxH
\else
\ifnum#1=579015 %
\hatcurSPECarcesNCandxxxxxI
\else
\ifnum#1=579036 %
\hatcurSPECarcesNCandxxxxxJ
\else
\ifnum#1=579037 %
\hatcurSPECarcesNCandxxxxxK
\else
\ifnum#1=579039 %
\hatcurSPECarcesNCandxxxxxL
\else
\ifnum#1=579040 %
\hatcurSPECarcesNCandxxxxxM
\else
\ifnum#1=579041 %
\hatcurSPECarcesNCandxxxxxN
\else
\ifnum#1=579043 %
\hatcurSPECarcesNCandxxxxxO
\else
\ifnum#1=579044 %
\hatcurSPECarcesNCandxxxxxP
\else
\ifnum#1=579048 %
\hatcurSPECarcesNCandxxxxxQ
\else
\ifnum#1=579050 %
\hatcurSPECarcesNCandxxxxxR
\else
\ifnum#1=624002 %
\hatcurSPECarcesNCandxxxxxS
\else
\ifnum#1=624003 %
\hatcurSPECarcesNCandxxxxxT
\else
??????\fi
\fi
\fi
\fi
\fi
\fi
\fi
\fi
\fi
\fi
\fi
\fi
\fi
\fi
\fi
\fi
\fi
\fi
\fi
\fi
\fi
}
\newcommand{\hatcurSPECarcesRVGCand}[1]{\ifnum#1=578002 %
\hatcurSPECarcesRVGCandxxxxxA
\else
\ifnum#1=578003 %
\hatcurSPECarcesRVGCandxxxxxB
\else
\ifnum#1=578004 %
\hatcurSPECarcesRVGCandxxxxxC
\else
\ifnum#1=579001 %
\hatcurSPECarcesRVGCandxxxxxD
\else
\ifnum#1=579007 %
\hatcurSPECarcesRVGCandxxxxxE
\else
\ifnum#1=579008 %
\hatcurSPECarcesRVGCandxxxxxF
\else
\ifnum#1=579009 %
\hatcurSPECarcesRVGCandxxxxxG
\else
\ifnum#1=579010 %
\hatcurSPECarcesRVGCandxxxxxU
\else
\ifnum#1=579014 %
\hatcurSPECarcesRVGCandxxxxxH
\else
\ifnum#1=579015 %
\hatcurSPECarcesRVGCandxxxxxI
\else
\ifnum#1=579036 %
\hatcurSPECarcesRVGCandxxxxxJ
\else
\ifnum#1=579037 %
\hatcurSPECarcesRVGCandxxxxxK
\else
\ifnum#1=579039 %
\hatcurSPECarcesRVGCandxxxxxL
\else
\ifnum#1=579040 %
\hatcurSPECarcesRVGCandxxxxxM
\else
\ifnum#1=579041 %
\hatcurSPECarcesRVGCandxxxxxN
\else
\ifnum#1=579043 %
\hatcurSPECarcesRVGCandxxxxxO
\else
\ifnum#1=579044 %
\hatcurSPECarcesRVGCandxxxxxP
\else
\ifnum#1=579048 %
\hatcurSPECarcesRVGCandxxxxxQ
\else
\ifnum#1=579050 %
\hatcurSPECarcesRVGCandxxxxxR
\else
\ifnum#1=624002 %
\hatcurSPECarcesRVGCandxxxxxS
\else
\ifnum#1=624003 %
\hatcurSPECarcesRVGCandxxxxxT
\else
??????\fi
\fi
\fi
\fi
\fi
\fi
\fi
\fi
\fi
\fi
\fi
\fi
\fi
\fi
\fi
\fi
\fi
\fi
\fi
\fi
\fi
}
\newcommand{\hatcurSPECarcesRVKCand}[1]{\ifnum#1=578002 %
\hatcurSPECarcesRVKCandxxxxxA
\else
\ifnum#1=578003 %
\hatcurSPECarcesRVKCandxxxxxB
\else
\ifnum#1=578004 %
\hatcurSPECarcesRVKCandxxxxxC
\else
\ifnum#1=579001 %
\hatcurSPECarcesRVKCandxxxxxD
\else
\ifnum#1=579007 %
\hatcurSPECarcesRVKCandxxxxxE
\else
\ifnum#1=579008 %
\hatcurSPECarcesRVKCandxxxxxF
\else
\ifnum#1=579009 %
\hatcurSPECarcesRVKCandxxxxxG
\else
\ifnum#1=579010 %
\hatcurSPECarcesRVKCandxxxxxU
\else
\ifnum#1=579014 %
\hatcurSPECarcesRVKCandxxxxxH
\else
\ifnum#1=579015 %
\hatcurSPECarcesRVKCandxxxxxI
\else
\ifnum#1=579036 %
\hatcurSPECarcesRVKCandxxxxxJ
\else
\ifnum#1=579037 %
\hatcurSPECarcesRVKCandxxxxxK
\else
\ifnum#1=579039 %
\hatcurSPECarcesRVKCandxxxxxL
\else
\ifnum#1=579040 %
\hatcurSPECarcesRVKCandxxxxxM
\else
\ifnum#1=579041 %
\hatcurSPECarcesRVKCandxxxxxN
\else
\ifnum#1=579043 %
\hatcurSPECarcesRVKCandxxxxxO
\else
\ifnum#1=579044 %
\hatcurSPECarcesRVKCandxxxxxP
\else
\ifnum#1=579048 %
\hatcurSPECarcesRVKCandxxxxxQ
\else
\ifnum#1=579050 %
\hatcurSPECarcesRVKCandxxxxxR
\else
\ifnum#1=624002 %
\hatcurSPECarcesRVKCandxxxxxS
\else
\ifnum#1=624003 %
\hatcurSPECarcesRVKCandxxxxxT
\else
??????\fi
\fi
\fi
\fi
\fi
\fi
\fi
\fi
\fi
\fi
\fi
\fi
\fi
\fi
\fi
\fi
\fi
\fi
\fi
\fi
\fi
}
\newcommand{\hatcurSPECarcessnrangeCand}[1]{\ifnum#1=578002 %
\hatcurSPECarcessnrangeCandxxxxxA
\else
\ifnum#1=578003 %
\hatcurSPECarcessnrangeCandxxxxxB
\else
\ifnum#1=578004 %
\hatcurSPECarcessnrangeCandxxxxxC
\else
\ifnum#1=579001 %
\hatcurSPECarcessnrangeCandxxxxxD
\else
\ifnum#1=579007 %
\hatcurSPECarcessnrangeCandxxxxxE
\else
\ifnum#1=579008 %
\hatcurSPECarcessnrangeCandxxxxxF
\else
\ifnum#1=579009 %
\hatcurSPECarcessnrangeCandxxxxxG
\else
\ifnum#1=579010 %
\hatcurSPECarcessnrangeCandxxxxxU
\else
\ifnum#1=579014 %
\hatcurSPECarcessnrangeCandxxxxxH
\else
\ifnum#1=579015 %
\hatcurSPECarcessnrangeCandxxxxxI
\else
\ifnum#1=579036 %
\hatcurSPECarcessnrangeCandxxxxxJ
\else
\ifnum#1=579037 %
\hatcurSPECarcessnrangeCandxxxxxK
\else
\ifnum#1=579039 %
\hatcurSPECarcessnrangeCandxxxxxL
\else
\ifnum#1=579040 %
\hatcurSPECarcessnrangeCandxxxxxM
\else
\ifnum#1=579041 %
\hatcurSPECarcessnrangeCandxxxxxN
\else
\ifnum#1=579043 %
\hatcurSPECarcessnrangeCandxxxxxO
\else
\ifnum#1=579044 %
\hatcurSPECarcessnrangeCandxxxxxP
\else
\ifnum#1=579048 %
\hatcurSPECarcessnrangeCandxxxxxQ
\else
\ifnum#1=579050 %
\hatcurSPECarcessnrangeCandxxxxxR
\else
\ifnum#1=624002 %
\hatcurSPECarcessnrangeCandxxxxxS
\else
\ifnum#1=624003 %
\hatcurSPECarcessnrangeCandxxxxxT
\else
??????\fi
\fi
\fi
\fi
\fi
\fi
\fi
\fi
\fi
\fi
\fi
\fi
\fi
\fi
\fi
\fi
\fi
\fi
\fi
\fi
\fi
}
\newcommand{\hatcurSPECarcesTeffCand}[1]{\ifnum#1=578002 %
\hatcurSPECarcesTeffCandxxxxxA
\else
\ifnum#1=578003 %
\hatcurSPECarcesTeffCandxxxxxB
\else
\ifnum#1=578004 %
\hatcurSPECarcesTeffCandxxxxxC
\else
\ifnum#1=579001 %
\hatcurSPECarcesTeffCandxxxxxD
\else
\ifnum#1=579007 %
\hatcurSPECarcesTeffCandxxxxxE
\else
\ifnum#1=579008 %
\hatcurSPECarcesTeffCandxxxxxF
\else
\ifnum#1=579009 %
\hatcurSPECarcesTeffCandxxxxxG
\else
\ifnum#1=579010 %
\hatcurSPECarcesTeffCandxxxxxU
\else
\ifnum#1=579014 %
\hatcurSPECarcesTeffCandxxxxxH
\else
\ifnum#1=579015 %
\hatcurSPECarcesTeffCandxxxxxI
\else
\ifnum#1=579036 %
\hatcurSPECarcesTeffCandxxxxxJ
\else
\ifnum#1=579037 %
\hatcurSPECarcesTeffCandxxxxxK
\else
\ifnum#1=579039 %
\hatcurSPECarcesTeffCandxxxxxL
\else
\ifnum#1=579040 %
\hatcurSPECarcesTeffCandxxxxxM
\else
\ifnum#1=579041 %
\hatcurSPECarcesTeffCandxxxxxN
\else
\ifnum#1=579043 %
\hatcurSPECarcesTeffCandxxxxxO
\else
\ifnum#1=579044 %
\hatcurSPECarcesTeffCandxxxxxP
\else
\ifnum#1=579048 %
\hatcurSPECarcesTeffCandxxxxxQ
\else
\ifnum#1=579050 %
\hatcurSPECarcesTeffCandxxxxxR
\else
\ifnum#1=624002 %
\hatcurSPECarcesTeffCandxxxxxS
\else
\ifnum#1=624003 %
\hatcurSPECarcesTeffCandxxxxxT
\else
??????\fi
\fi
\fi
\fi
\fi
\fi
\fi
\fi
\fi
\fi
\fi
\fi
\fi
\fi
\fi
\fi
\fi
\fi
\fi
\fi
\fi
}
\newcommand{\hatcurSPECarcesvsiniCand}[1]{\ifnum#1=578002 %
\hatcurSPECarcesvsiniCandxxxxxA
\else
\ifnum#1=578003 %
\hatcurSPECarcesvsiniCandxxxxxB
\else
\ifnum#1=578004 %
\hatcurSPECarcesvsiniCandxxxxxC
\else
\ifnum#1=579001 %
\hatcurSPECarcesvsiniCandxxxxxD
\else
\ifnum#1=579007 %
\hatcurSPECarcesvsiniCandxxxxxE
\else
\ifnum#1=579008 %
\hatcurSPECarcesvsiniCandxxxxxF
\else
\ifnum#1=579009 %
\hatcurSPECarcesvsiniCandxxxxxG
\else
\ifnum#1=579010 %
\hatcurSPECarcesvsiniCandxxxxxU
\else
\ifnum#1=579014 %
\hatcurSPECarcesvsiniCandxxxxxH
\else
\ifnum#1=579015 %
\hatcurSPECarcesvsiniCandxxxxxI
\else
\ifnum#1=579036 %
\hatcurSPECarcesvsiniCandxxxxxJ
\else
\ifnum#1=579037 %
\hatcurSPECarcesvsiniCandxxxxxK
\else
\ifnum#1=579039 %
\hatcurSPECarcesvsiniCandxxxxxL
\else
\ifnum#1=579040 %
\hatcurSPECarcesvsiniCandxxxxxM
\else
\ifnum#1=579041 %
\hatcurSPECarcesvsiniCandxxxxxN
\else
\ifnum#1=579043 %
\hatcurSPECarcesvsiniCandxxxxxO
\else
\ifnum#1=579044 %
\hatcurSPECarcesvsiniCandxxxxxP
\else
\ifnum#1=579048 %
\hatcurSPECarcesvsiniCandxxxxxQ
\else
\ifnum#1=579050 %
\hatcurSPECarcesvsiniCandxxxxxR
\else
\ifnum#1=624002 %
\hatcurSPECarcesvsiniCandxxxxxS
\else
\ifnum#1=624003 %
\hatcurSPECarcesvsiniCandxxxxxT
\else
??????\fi
\fi
\fi
\fi
\fi
\fi
\fi
\fi
\fi
\fi
\fi
\fi
\fi
\fi
\fi
\fi
\fi
\fi
\fi
\fi
\fi
}
\newcommand{\hatcurSPECarceszfehCand}[1]{\ifnum#1=578002 %
\hatcurSPECarceszfehCandxxxxxA
\else
\ifnum#1=578003 %
\hatcurSPECarceszfehCandxxxxxB
\else
\ifnum#1=578004 %
\hatcurSPECarceszfehCandxxxxxC
\else
\ifnum#1=579001 %
\hatcurSPECarceszfehCandxxxxxD
\else
\ifnum#1=579007 %
\hatcurSPECarceszfehCandxxxxxE
\else
\ifnum#1=579008 %
\hatcurSPECarceszfehCandxxxxxF
\else
\ifnum#1=579009 %
\hatcurSPECarceszfehCandxxxxxG
\else
\ifnum#1=579010 %
\hatcurSPECarceszfehCandxxxxxU
\else
\ifnum#1=579014 %
\hatcurSPECarceszfehCandxxxxxH
\else
\ifnum#1=579015 %
\hatcurSPECarceszfehCandxxxxxI
\else
\ifnum#1=579036 %
\hatcurSPECarceszfehCandxxxxxJ
\else
\ifnum#1=579037 %
\hatcurSPECarceszfehCandxxxxxK
\else
\ifnum#1=579039 %
\hatcurSPECarceszfehCandxxxxxL
\else
\ifnum#1=579040 %
\hatcurSPECarceszfehCandxxxxxM
\else
\ifnum#1=579041 %
\hatcurSPECarceszfehCandxxxxxN
\else
\ifnum#1=579043 %
\hatcurSPECarceszfehCandxxxxxO
\else
\ifnum#1=579044 %
\hatcurSPECarceszfehCandxxxxxP
\else
\ifnum#1=579048 %
\hatcurSPECarceszfehCandxxxxxQ
\else
\ifnum#1=579050 %
\hatcurSPECarceszfehCandxxxxxR
\else
\ifnum#1=624002 %
\hatcurSPECarceszfehCandxxxxxS
\else
\ifnum#1=624003 %
\hatcurSPECarceszfehCandxxxxxT
\else
??????\fi
\fi
\fi
\fi
\fi
\fi
\fi
\fi
\fi
\fi
\fi
\fi
\fi
\fi
\fi
\fi
\fi
\fi
\fi
\fi
\fi
}
\newcommand{\hatcurSPECcoraliedaterangelowresCand}[1]{\ifnum#1=578002 %
\hatcurSPECcoraliedaterangelowresCandxxxxxA
\else
\ifnum#1=578003 %
\hatcurSPECcoraliedaterangelowresCandxxxxxB
\else
\ifnum#1=578004 %
\hatcurSPECcoraliedaterangelowresCandxxxxxC
\else
\ifnum#1=579001 %
\hatcurSPECcoraliedaterangelowresCandxxxxxD
\else
\ifnum#1=579007 %
\hatcurSPECcoraliedaterangelowresCandxxxxxE
\else
\ifnum#1=579008 %
\hatcurSPECcoraliedaterangelowresCandxxxxxF
\else
\ifnum#1=579009 %
\hatcurSPECcoraliedaterangelowresCandxxxxxG
\else
\ifnum#1=579010 %
\hatcurSPECcoraliedaterangelowresCandxxxxxU
\else
\ifnum#1=579014 %
\hatcurSPECcoraliedaterangelowresCandxxxxxH
\else
\ifnum#1=579015 %
\hatcurSPECcoraliedaterangelowresCandxxxxxI
\else
\ifnum#1=579036 %
\hatcurSPECcoraliedaterangelowresCandxxxxxJ
\else
\ifnum#1=579037 %
\hatcurSPECcoraliedaterangelowresCandxxxxxK
\else
\ifnum#1=579039 %
\hatcurSPECcoraliedaterangelowresCandxxxxxL
\else
\ifnum#1=579040 %
\hatcurSPECcoraliedaterangelowresCandxxxxxM
\else
\ifnum#1=579041 %
\hatcurSPECcoraliedaterangelowresCandxxxxxN
\else
\ifnum#1=579043 %
\hatcurSPECcoraliedaterangelowresCandxxxxxO
\else
\ifnum#1=579044 %
\hatcurSPECcoraliedaterangelowresCandxxxxxP
\else
\ifnum#1=579048 %
\hatcurSPECcoraliedaterangelowresCandxxxxxQ
\else
\ifnum#1=579050 %
\hatcurSPECcoraliedaterangelowresCandxxxxxR
\else
\ifnum#1=624002 %
\hatcurSPECcoraliedaterangelowresCandxxxxxS
\else
\ifnum#1=624003 %
\hatcurSPECcoraliedaterangelowresCandxxxxxT
\else
??????\fi
\fi
\fi
\fi
\fi
\fi
\fi
\fi
\fi
\fi
\fi
\fi
\fi
\fi
\fi
\fi
\fi
\fi
\fi
\fi
\fi
}
\newcommand{\hatcurSPECcoralieloggCand}[1]{\ifnum#1=578002 %
\hatcurSPECcoralieloggCandxxxxxA
\else
\ifnum#1=578003 %
\hatcurSPECcoralieloggCandxxxxxB
\else
\ifnum#1=578004 %
\hatcurSPECcoralieloggCandxxxxxC
\else
\ifnum#1=579001 %
\hatcurSPECcoralieloggCandxxxxxD
\else
\ifnum#1=579007 %
\hatcurSPECcoralieloggCandxxxxxE
\else
\ifnum#1=579008 %
\hatcurSPECcoralieloggCandxxxxxF
\else
\ifnum#1=579009 %
\hatcurSPECcoralieloggCandxxxxxG
\else
\ifnum#1=579010 %
\hatcurSPECcoralieloggCandxxxxxU
\else
\ifnum#1=579014 %
\hatcurSPECcoralieloggCandxxxxxH
\else
\ifnum#1=579015 %
\hatcurSPECcoralieloggCandxxxxxI
\else
\ifnum#1=579036 %
\hatcurSPECcoralieloggCandxxxxxJ
\else
\ifnum#1=579037 %
\hatcurSPECcoralieloggCandxxxxxK
\else
\ifnum#1=579039 %
\hatcurSPECcoralieloggCandxxxxxL
\else
\ifnum#1=579040 %
\hatcurSPECcoralieloggCandxxxxxM
\else
\ifnum#1=579041 %
\hatcurSPECcoralieloggCandxxxxxN
\else
\ifnum#1=579043 %
\hatcurSPECcoralieloggCandxxxxxO
\else
\ifnum#1=579044 %
\hatcurSPECcoralieloggCandxxxxxP
\else
\ifnum#1=579048 %
\hatcurSPECcoralieloggCandxxxxxQ
\else
\ifnum#1=579050 %
\hatcurSPECcoralieloggCandxxxxxR
\else
\ifnum#1=624002 %
\hatcurSPECcoralieloggCandxxxxxS
\else
\ifnum#1=624003 %
\hatcurSPECcoralieloggCandxxxxxT
\else
??????\fi
\fi
\fi
\fi
\fi
\fi
\fi
\fi
\fi
\fi
\fi
\fi
\fi
\fi
\fi
\fi
\fi
\fi
\fi
\fi
\fi
}
\newcommand{\hatcurSPECcoralieNCand}[1]{\ifnum#1=578002 %
\hatcurSPECcoralieNCandxxxxxA
\else
\ifnum#1=578003 %
\hatcurSPECcoralieNCandxxxxxB
\else
\ifnum#1=578004 %
\hatcurSPECcoralieNCandxxxxxC
\else
\ifnum#1=579001 %
\hatcurSPECcoralieNCandxxxxxD
\else
\ifnum#1=579007 %
\hatcurSPECcoralieNCandxxxxxE
\else
\ifnum#1=579008 %
\hatcurSPECcoralieNCandxxxxxF
\else
\ifnum#1=579009 %
\hatcurSPECcoralieNCandxxxxxG
\else
\ifnum#1=579010 %
\hatcurSPECcoralieNCandxxxxxU
\else
\ifnum#1=579014 %
\hatcurSPECcoralieNCandxxxxxH
\else
\ifnum#1=579015 %
\hatcurSPECcoralieNCandxxxxxI
\else
\ifnum#1=579036 %
\hatcurSPECcoralieNCandxxxxxJ
\else
\ifnum#1=579037 %
\hatcurSPECcoralieNCandxxxxxK
\else
\ifnum#1=579039 %
\hatcurSPECcoralieNCandxxxxxL
\else
\ifnum#1=579040 %
\hatcurSPECcoralieNCandxxxxxM
\else
\ifnum#1=579041 %
\hatcurSPECcoralieNCandxxxxxN
\else
\ifnum#1=579043 %
\hatcurSPECcoralieNCandxxxxxO
\else
\ifnum#1=579044 %
\hatcurSPECcoralieNCandxxxxxP
\else
\ifnum#1=579048 %
\hatcurSPECcoralieNCandxxxxxQ
\else
\ifnum#1=579050 %
\hatcurSPECcoralieNCandxxxxxR
\else
\ifnum#1=624002 %
\hatcurSPECcoralieNCandxxxxxS
\else
\ifnum#1=624003 %
\hatcurSPECcoralieNCandxxxxxT
\else
??????\fi
\fi
\fi
\fi
\fi
\fi
\fi
\fi
\fi
\fi
\fi
\fi
\fi
\fi
\fi
\fi
\fi
\fi
\fi
\fi
\fi
}
\newcommand{\hatcurSPECcoralieRVGCand}[1]{\ifnum#1=578002 %
\hatcurSPECcoralieRVGCandxxxxxA
\else
\ifnum#1=578003 %
\hatcurSPECcoralieRVGCandxxxxxB
\else
\ifnum#1=578004 %
\hatcurSPECcoralieRVGCandxxxxxC
\else
\ifnum#1=579001 %
\hatcurSPECcoralieRVGCandxxxxxD
\else
\ifnum#1=579007 %
\hatcurSPECcoralieRVGCandxxxxxE
\else
\ifnum#1=579008 %
\hatcurSPECcoralieRVGCandxxxxxF
\else
\ifnum#1=579009 %
\hatcurSPECcoralieRVGCandxxxxxG
\else
\ifnum#1=579010 %
\hatcurSPECcoralieRVGCandxxxxxU
\else
\ifnum#1=579014 %
\hatcurSPECcoralieRVGCandxxxxxH
\else
\ifnum#1=579015 %
\hatcurSPECcoralieRVGCandxxxxxI
\else
\ifnum#1=579036 %
\hatcurSPECcoralieRVGCandxxxxxJ
\else
\ifnum#1=579037 %
\hatcurSPECcoralieRVGCandxxxxxK
\else
\ifnum#1=579039 %
\hatcurSPECcoralieRVGCandxxxxxL
\else
\ifnum#1=579040 %
\hatcurSPECcoralieRVGCandxxxxxM
\else
\ifnum#1=579041 %
\hatcurSPECcoralieRVGCandxxxxxN
\else
\ifnum#1=579043 %
\hatcurSPECcoralieRVGCandxxxxxO
\else
\ifnum#1=579044 %
\hatcurSPECcoralieRVGCandxxxxxP
\else
\ifnum#1=579048 %
\hatcurSPECcoralieRVGCandxxxxxQ
\else
\ifnum#1=579050 %
\hatcurSPECcoralieRVGCandxxxxxR
\else
\ifnum#1=624002 %
\hatcurSPECcoralieRVGCandxxxxxS
\else
\ifnum#1=624003 %
\hatcurSPECcoralieRVGCandxxxxxT
\else
??????\fi
\fi
\fi
\fi
\fi
\fi
\fi
\fi
\fi
\fi
\fi
\fi
\fi
\fi
\fi
\fi
\fi
\fi
\fi
\fi
\fi
}
\newcommand{\hatcurSPECcoralieRVKCand}[1]{\ifnum#1=578002 %
\hatcurSPECcoralieRVKCandxxxxxA
\else
\ifnum#1=578003 %
\hatcurSPECcoralieRVKCandxxxxxB
\else
\ifnum#1=578004 %
\hatcurSPECcoralieRVKCandxxxxxC
\else
\ifnum#1=579001 %
\hatcurSPECcoralieRVKCandxxxxxD
\else
\ifnum#1=579007 %
\hatcurSPECcoralieRVKCandxxxxxE
\else
\ifnum#1=579008 %
\hatcurSPECcoralieRVKCandxxxxxF
\else
\ifnum#1=579009 %
\hatcurSPECcoralieRVKCandxxxxxG
\else
\ifnum#1=579010 %
\hatcurSPECcoralieRVKCandxxxxxU
\else
\ifnum#1=579014 %
\hatcurSPECcoralieRVKCandxxxxxH
\else
\ifnum#1=579015 %
\hatcurSPECcoralieRVKCandxxxxxI
\else
\ifnum#1=579036 %
\hatcurSPECcoralieRVKCandxxxxxJ
\else
\ifnum#1=579037 %
\hatcurSPECcoralieRVKCandxxxxxK
\else
\ifnum#1=579039 %
\hatcurSPECcoralieRVKCandxxxxxL
\else
\ifnum#1=579040 %
\hatcurSPECcoralieRVKCandxxxxxM
\else
\ifnum#1=579041 %
\hatcurSPECcoralieRVKCandxxxxxN
\else
\ifnum#1=579043 %
\hatcurSPECcoralieRVKCandxxxxxO
\else
\ifnum#1=579044 %
\hatcurSPECcoralieRVKCandxxxxxP
\else
\ifnum#1=579048 %
\hatcurSPECcoralieRVKCandxxxxxQ
\else
\ifnum#1=579050 %
\hatcurSPECcoralieRVKCandxxxxxR
\else
\ifnum#1=624002 %
\hatcurSPECcoralieRVKCandxxxxxS
\else
\ifnum#1=624003 %
\hatcurSPECcoralieRVKCandxxxxxT
\else
??????\fi
\fi
\fi
\fi
\fi
\fi
\fi
\fi
\fi
\fi
\fi
\fi
\fi
\fi
\fi
\fi
\fi
\fi
\fi
\fi
\fi
}
\newcommand{\hatcurSPECcoraliesnrangeCand}[1]{\ifnum#1=578002 %
\hatcurSPECcoraliesnrangeCandxxxxxA
\else
\ifnum#1=578003 %
\hatcurSPECcoraliesnrangeCandxxxxxB
\else
\ifnum#1=578004 %
\hatcurSPECcoraliesnrangeCandxxxxxC
\else
\ifnum#1=579001 %
\hatcurSPECcoraliesnrangeCandxxxxxD
\else
\ifnum#1=579007 %
\hatcurSPECcoraliesnrangeCandxxxxxE
\else
\ifnum#1=579008 %
\hatcurSPECcoraliesnrangeCandxxxxxF
\else
\ifnum#1=579009 %
\hatcurSPECcoraliesnrangeCandxxxxxG
\else
\ifnum#1=579010 %
\hatcurSPECcoraliesnrangeCandxxxxxU
\else
\ifnum#1=579014 %
\hatcurSPECcoraliesnrangeCandxxxxxH
\else
\ifnum#1=579015 %
\hatcurSPECcoraliesnrangeCandxxxxxI
\else
\ifnum#1=579036 %
\hatcurSPECcoraliesnrangeCandxxxxxJ
\else
\ifnum#1=579037 %
\hatcurSPECcoraliesnrangeCandxxxxxK
\else
\ifnum#1=579039 %
\hatcurSPECcoraliesnrangeCandxxxxxL
\else
\ifnum#1=579040 %
\hatcurSPECcoraliesnrangeCandxxxxxM
\else
\ifnum#1=579041 %
\hatcurSPECcoraliesnrangeCandxxxxxN
\else
\ifnum#1=579043 %
\hatcurSPECcoraliesnrangeCandxxxxxO
\else
\ifnum#1=579044 %
\hatcurSPECcoraliesnrangeCandxxxxxP
\else
\ifnum#1=579048 %
\hatcurSPECcoraliesnrangeCandxxxxxQ
\else
\ifnum#1=579050 %
\hatcurSPECcoraliesnrangeCandxxxxxR
\else
\ifnum#1=624002 %
\hatcurSPECcoraliesnrangeCandxxxxxS
\else
\ifnum#1=624003 %
\hatcurSPECcoraliesnrangeCandxxxxxT
\else
??????\fi
\fi
\fi
\fi
\fi
\fi
\fi
\fi
\fi
\fi
\fi
\fi
\fi
\fi
\fi
\fi
\fi
\fi
\fi
\fi
\fi
}
\newcommand{\hatcurSPECcoralieTeffCand}[1]{\ifnum#1=578002 %
\hatcurSPECcoralieTeffCandxxxxxA
\else
\ifnum#1=578003 %
\hatcurSPECcoralieTeffCandxxxxxB
\else
\ifnum#1=578004 %
\hatcurSPECcoralieTeffCandxxxxxC
\else
\ifnum#1=579001 %
\hatcurSPECcoralieTeffCandxxxxxD
\else
\ifnum#1=579007 %
\hatcurSPECcoralieTeffCandxxxxxE
\else
\ifnum#1=579008 %
\hatcurSPECcoralieTeffCandxxxxxF
\else
\ifnum#1=579009 %
\hatcurSPECcoralieTeffCandxxxxxG
\else
\ifnum#1=579010 %
\hatcurSPECcoralieTeffCandxxxxxU
\else
\ifnum#1=579014 %
\hatcurSPECcoralieTeffCandxxxxxH
\else
\ifnum#1=579015 %
\hatcurSPECcoralieTeffCandxxxxxI
\else
\ifnum#1=579036 %
\hatcurSPECcoralieTeffCandxxxxxJ
\else
\ifnum#1=579037 %
\hatcurSPECcoralieTeffCandxxxxxK
\else
\ifnum#1=579039 %
\hatcurSPECcoralieTeffCandxxxxxL
\else
\ifnum#1=579040 %
\hatcurSPECcoralieTeffCandxxxxxM
\else
\ifnum#1=579041 %
\hatcurSPECcoralieTeffCandxxxxxN
\else
\ifnum#1=579043 %
\hatcurSPECcoralieTeffCandxxxxxO
\else
\ifnum#1=579044 %
\hatcurSPECcoralieTeffCandxxxxxP
\else
\ifnum#1=579048 %
\hatcurSPECcoralieTeffCandxxxxxQ
\else
\ifnum#1=579050 %
\hatcurSPECcoralieTeffCandxxxxxR
\else
\ifnum#1=624002 %
\hatcurSPECcoralieTeffCandxxxxxS
\else
\ifnum#1=624003 %
\hatcurSPECcoralieTeffCandxxxxxT
\else
??????\fi
\fi
\fi
\fi
\fi
\fi
\fi
\fi
\fi
\fi
\fi
\fi
\fi
\fi
\fi
\fi
\fi
\fi
\fi
\fi
\fi
}
\newcommand{\hatcurSPECcoralievsiniCand}[1]{\ifnum#1=578002 %
\hatcurSPECcoralievsiniCandxxxxxA
\else
\ifnum#1=578003 %
\hatcurSPECcoralievsiniCandxxxxxB
\else
\ifnum#1=578004 %
\hatcurSPECcoralievsiniCandxxxxxC
\else
\ifnum#1=579001 %
\hatcurSPECcoralievsiniCandxxxxxD
\else
\ifnum#1=579007 %
\hatcurSPECcoralievsiniCandxxxxxE
\else
\ifnum#1=579008 %
\hatcurSPECcoralievsiniCandxxxxxF
\else
\ifnum#1=579009 %
\hatcurSPECcoralievsiniCandxxxxxG
\else
\ifnum#1=579010 %
\hatcurSPECcoralievsiniCandxxxxxU
\else
\ifnum#1=579014 %
\hatcurSPECcoralievsiniCandxxxxxH
\else
\ifnum#1=579015 %
\hatcurSPECcoralievsiniCandxxxxxI
\else
\ifnum#1=579036 %
\hatcurSPECcoralievsiniCandxxxxxJ
\else
\ifnum#1=579037 %
\hatcurSPECcoralievsiniCandxxxxxK
\else
\ifnum#1=579039 %
\hatcurSPECcoralievsiniCandxxxxxL
\else
\ifnum#1=579040 %
\hatcurSPECcoralievsiniCandxxxxxM
\else
\ifnum#1=579041 %
\hatcurSPECcoralievsiniCandxxxxxN
\else
\ifnum#1=579043 %
\hatcurSPECcoralievsiniCandxxxxxO
\else
\ifnum#1=579044 %
\hatcurSPECcoralievsiniCandxxxxxP
\else
\ifnum#1=579048 %
\hatcurSPECcoralievsiniCandxxxxxQ
\else
\ifnum#1=579050 %
\hatcurSPECcoralievsiniCandxxxxxR
\else
\ifnum#1=624002 %
\hatcurSPECcoralievsiniCandxxxxxS
\else
\ifnum#1=624003 %
\hatcurSPECcoralievsiniCandxxxxxT
\else
??????\fi
\fi
\fi
\fi
\fi
\fi
\fi
\fi
\fi
\fi
\fi
\fi
\fi
\fi
\fi
\fi
\fi
\fi
\fi
\fi
\fi
}
\newcommand{\hatcurSPECcoraliezfehCand}[1]{\ifnum#1=578002 %
\hatcurSPECcoraliezfehCandxxxxxA
\else
\ifnum#1=578003 %
\hatcurSPECcoraliezfehCandxxxxxB
\else
\ifnum#1=578004 %
\hatcurSPECcoraliezfehCandxxxxxC
\else
\ifnum#1=579001 %
\hatcurSPECcoraliezfehCandxxxxxD
\else
\ifnum#1=579007 %
\hatcurSPECcoraliezfehCandxxxxxE
\else
\ifnum#1=579008 %
\hatcurSPECcoraliezfehCandxxxxxF
\else
\ifnum#1=579009 %
\hatcurSPECcoraliezfehCandxxxxxG
\else
\ifnum#1=579010 %
\hatcurSPECcoraliezfehCandxxxxxU
\else
\ifnum#1=579014 %
\hatcurSPECcoraliezfehCandxxxxxH
\else
\ifnum#1=579015 %
\hatcurSPECcoraliezfehCandxxxxxI
\else
\ifnum#1=579036 %
\hatcurSPECcoraliezfehCandxxxxxJ
\else
\ifnum#1=579037 %
\hatcurSPECcoraliezfehCandxxxxxK
\else
\ifnum#1=579039 %
\hatcurSPECcoraliezfehCandxxxxxL
\else
\ifnum#1=579040 %
\hatcurSPECcoraliezfehCandxxxxxM
\else
\ifnum#1=579041 %
\hatcurSPECcoraliezfehCandxxxxxN
\else
\ifnum#1=579043 %
\hatcurSPECcoraliezfehCandxxxxxO
\else
\ifnum#1=579044 %
\hatcurSPECcoraliezfehCandxxxxxP
\else
\ifnum#1=579048 %
\hatcurSPECcoraliezfehCandxxxxxQ
\else
\ifnum#1=579050 %
\hatcurSPECcoraliezfehCandxxxxxR
\else
\ifnum#1=624002 %
\hatcurSPECcoraliezfehCandxxxxxS
\else
\ifnum#1=624003 %
\hatcurSPECcoraliezfehCandxxxxxT
\else
??????\fi
\fi
\fi
\fi
\fi
\fi
\fi
\fi
\fi
\fi
\fi
\fi
\fi
\fi
\fi
\fi
\fi
\fi
\fi
\fi
\fi
}
\newcommand{\hatcurSPECdupontdaterangelowresCand}[1]{\ifnum#1=578002 %
\hatcurSPECdupontdaterangelowresCandxxxxxA
\else
\ifnum#1=578003 %
\hatcurSPECdupontdaterangelowresCandxxxxxB
\else
\ifnum#1=578004 %
\hatcurSPECdupontdaterangelowresCandxxxxxC
\else
\ifnum#1=579001 %
\hatcurSPECdupontdaterangelowresCandxxxxxD
\else
\ifnum#1=579007 %
\hatcurSPECdupontdaterangelowresCandxxxxxE
\else
\ifnum#1=579008 %
\hatcurSPECdupontdaterangelowresCandxxxxxF
\else
\ifnum#1=579009 %
\hatcurSPECdupontdaterangelowresCandxxxxxG
\else
\ifnum#1=579010 %
\hatcurSPECdupontdaterangelowresCandxxxxxU
\else
\ifnum#1=579014 %
\hatcurSPECdupontdaterangelowresCandxxxxxH
\else
\ifnum#1=579015 %
\hatcurSPECdupontdaterangelowresCandxxxxxI
\else
\ifnum#1=579036 %
\hatcurSPECdupontdaterangelowresCandxxxxxJ
\else
\ifnum#1=579037 %
\hatcurSPECdupontdaterangelowresCandxxxxxK
\else
\ifnum#1=579039 %
\hatcurSPECdupontdaterangelowresCandxxxxxL
\else
\ifnum#1=579040 %
\hatcurSPECdupontdaterangelowresCandxxxxxM
\else
\ifnum#1=579041 %
\hatcurSPECdupontdaterangelowresCandxxxxxN
\else
\ifnum#1=579043 %
\hatcurSPECdupontdaterangelowresCandxxxxxO
\else
\ifnum#1=579044 %
\hatcurSPECdupontdaterangelowresCandxxxxxP
\else
\ifnum#1=579048 %
\hatcurSPECdupontdaterangelowresCandxxxxxQ
\else
\ifnum#1=579050 %
\hatcurSPECdupontdaterangelowresCandxxxxxR
\else
\ifnum#1=624002 %
\hatcurSPECdupontdaterangelowresCandxxxxxS
\else
\ifnum#1=624003 %
\hatcurSPECdupontdaterangelowresCandxxxxxT
\else
??????\fi
\fi
\fi
\fi
\fi
\fi
\fi
\fi
\fi
\fi
\fi
\fi
\fi
\fi
\fi
\fi
\fi
\fi
\fi
\fi
\fi
}
\newcommand{\hatcurSPECdupontloggCand}[1]{\ifnum#1=578002 %
\hatcurSPECdupontloggCandxxxxxA
\else
\ifnum#1=578003 %
\hatcurSPECdupontloggCandxxxxxB
\else
\ifnum#1=578004 %
\hatcurSPECdupontloggCandxxxxxC
\else
\ifnum#1=579001 %
\hatcurSPECdupontloggCandxxxxxD
\else
\ifnum#1=579007 %
\hatcurSPECdupontloggCandxxxxxE
\else
\ifnum#1=579008 %
\hatcurSPECdupontloggCandxxxxxF
\else
\ifnum#1=579009 %
\hatcurSPECdupontloggCandxxxxxG
\else
\ifnum#1=579010 %
\hatcurSPECdupontloggCandxxxxxU
\else
\ifnum#1=579014 %
\hatcurSPECdupontloggCandxxxxxH
\else
\ifnum#1=579015 %
\hatcurSPECdupontloggCandxxxxxI
\else
\ifnum#1=579036 %
\hatcurSPECdupontloggCandxxxxxJ
\else
\ifnum#1=579037 %
\hatcurSPECdupontloggCandxxxxxK
\else
\ifnum#1=579039 %
\hatcurSPECdupontloggCandxxxxxL
\else
\ifnum#1=579040 %
\hatcurSPECdupontloggCandxxxxxM
\else
\ifnum#1=579041 %
\hatcurSPECdupontloggCandxxxxxN
\else
\ifnum#1=579043 %
\hatcurSPECdupontloggCandxxxxxO
\else
\ifnum#1=579044 %
\hatcurSPECdupontloggCandxxxxxP
\else
\ifnum#1=579048 %
\hatcurSPECdupontloggCandxxxxxQ
\else
\ifnum#1=579050 %
\hatcurSPECdupontloggCandxxxxxR
\else
\ifnum#1=624002 %
\hatcurSPECdupontloggCandxxxxxS
\else
\ifnum#1=624003 %
\hatcurSPECdupontloggCandxxxxxT
\else
??????\fi
\fi
\fi
\fi
\fi
\fi
\fi
\fi
\fi
\fi
\fi
\fi
\fi
\fi
\fi
\fi
\fi
\fi
\fi
\fi
\fi
}
\newcommand{\hatcurSPECdupontNCand}[1]{\ifnum#1=578002 %
\hatcurSPECdupontNCandxxxxxA
\else
\ifnum#1=578003 %
\hatcurSPECdupontNCandxxxxxB
\else
\ifnum#1=578004 %
\hatcurSPECdupontNCandxxxxxC
\else
\ifnum#1=579001 %
\hatcurSPECdupontNCandxxxxxD
\else
\ifnum#1=579007 %
\hatcurSPECdupontNCandxxxxxE
\else
\ifnum#1=579008 %
\hatcurSPECdupontNCandxxxxxF
\else
\ifnum#1=579009 %
\hatcurSPECdupontNCandxxxxxG
\else
\ifnum#1=579010 %
\hatcurSPECdupontNCandxxxxxU
\else
\ifnum#1=579014 %
\hatcurSPECdupontNCandxxxxxH
\else
\ifnum#1=579015 %
\hatcurSPECdupontNCandxxxxxI
\else
\ifnum#1=579036 %
\hatcurSPECdupontNCandxxxxxJ
\else
\ifnum#1=579037 %
\hatcurSPECdupontNCandxxxxxK
\else
\ifnum#1=579039 %
\hatcurSPECdupontNCandxxxxxL
\else
\ifnum#1=579040 %
\hatcurSPECdupontNCandxxxxxM
\else
\ifnum#1=579041 %
\hatcurSPECdupontNCandxxxxxN
\else
\ifnum#1=579043 %
\hatcurSPECdupontNCandxxxxxO
\else
\ifnum#1=579044 %
\hatcurSPECdupontNCandxxxxxP
\else
\ifnum#1=579048 %
\hatcurSPECdupontNCandxxxxxQ
\else
\ifnum#1=579050 %
\hatcurSPECdupontNCandxxxxxR
\else
\ifnum#1=624002 %
\hatcurSPECdupontNCandxxxxxS
\else
\ifnum#1=624003 %
\hatcurSPECdupontNCandxxxxxT
\else
??????\fi
\fi
\fi
\fi
\fi
\fi
\fi
\fi
\fi
\fi
\fi
\fi
\fi
\fi
\fi
\fi
\fi
\fi
\fi
\fi
\fi
}
\newcommand{\hatcurSPECdupontRVGCand}[1]{\ifnum#1=578002 %
\hatcurSPECdupontRVGCandxxxxxA
\else
\ifnum#1=578003 %
\hatcurSPECdupontRVGCandxxxxxB
\else
\ifnum#1=578004 %
\hatcurSPECdupontRVGCandxxxxxC
\else
\ifnum#1=579001 %
\hatcurSPECdupontRVGCandxxxxxD
\else
\ifnum#1=579007 %
\hatcurSPECdupontRVGCandxxxxxE
\else
\ifnum#1=579008 %
\hatcurSPECdupontRVGCandxxxxxF
\else
\ifnum#1=579009 %
\hatcurSPECdupontRVGCandxxxxxG
\else
\ifnum#1=579010 %
\hatcurSPECdupontRVGCandxxxxxU
\else
\ifnum#1=579014 %
\hatcurSPECdupontRVGCandxxxxxH
\else
\ifnum#1=579015 %
\hatcurSPECdupontRVGCandxxxxxI
\else
\ifnum#1=579036 %
\hatcurSPECdupontRVGCandxxxxxJ
\else
\ifnum#1=579037 %
\hatcurSPECdupontRVGCandxxxxxK
\else
\ifnum#1=579039 %
\hatcurSPECdupontRVGCandxxxxxL
\else
\ifnum#1=579040 %
\hatcurSPECdupontRVGCandxxxxxM
\else
\ifnum#1=579041 %
\hatcurSPECdupontRVGCandxxxxxN
\else
\ifnum#1=579043 %
\hatcurSPECdupontRVGCandxxxxxO
\else
\ifnum#1=579044 %
\hatcurSPECdupontRVGCandxxxxxP
\else
\ifnum#1=579048 %
\hatcurSPECdupontRVGCandxxxxxQ
\else
\ifnum#1=579050 %
\hatcurSPECdupontRVGCandxxxxxR
\else
\ifnum#1=624002 %
\hatcurSPECdupontRVGCandxxxxxS
\else
\ifnum#1=624003 %
\hatcurSPECdupontRVGCandxxxxxT
\else
??????\fi
\fi
\fi
\fi
\fi
\fi
\fi
\fi
\fi
\fi
\fi
\fi
\fi
\fi
\fi
\fi
\fi
\fi
\fi
\fi
\fi
}
\newcommand{\hatcurSPECdupontRVKCand}[1]{\ifnum#1=578002 %
\hatcurSPECdupontRVKCandxxxxxA
\else
\ifnum#1=578003 %
\hatcurSPECdupontRVKCandxxxxxB
\else
\ifnum#1=578004 %
\hatcurSPECdupontRVKCandxxxxxC
\else
\ifnum#1=579001 %
\hatcurSPECdupontRVKCandxxxxxD
\else
\ifnum#1=579007 %
\hatcurSPECdupontRVKCandxxxxxE
\else
\ifnum#1=579008 %
\hatcurSPECdupontRVKCandxxxxxF
\else
\ifnum#1=579009 %
\hatcurSPECdupontRVKCandxxxxxG
\else
\ifnum#1=579010 %
\hatcurSPECdupontRVKCandxxxxxU
\else
\ifnum#1=579014 %
\hatcurSPECdupontRVKCandxxxxxH
\else
\ifnum#1=579015 %
\hatcurSPECdupontRVKCandxxxxxI
\else
\ifnum#1=579036 %
\hatcurSPECdupontRVKCandxxxxxJ
\else
\ifnum#1=579037 %
\hatcurSPECdupontRVKCandxxxxxK
\else
\ifnum#1=579039 %
\hatcurSPECdupontRVKCandxxxxxL
\else
\ifnum#1=579040 %
\hatcurSPECdupontRVKCandxxxxxM
\else
\ifnum#1=579041 %
\hatcurSPECdupontRVKCandxxxxxN
\else
\ifnum#1=579043 %
\hatcurSPECdupontRVKCandxxxxxO
\else
\ifnum#1=579044 %
\hatcurSPECdupontRVKCandxxxxxP
\else
\ifnum#1=579048 %
\hatcurSPECdupontRVKCandxxxxxQ
\else
\ifnum#1=579050 %
\hatcurSPECdupontRVKCandxxxxxR
\else
\ifnum#1=624002 %
\hatcurSPECdupontRVKCandxxxxxS
\else
\ifnum#1=624003 %
\hatcurSPECdupontRVKCandxxxxxT
\else
??????\fi
\fi
\fi
\fi
\fi
\fi
\fi
\fi
\fi
\fi
\fi
\fi
\fi
\fi
\fi
\fi
\fi
\fi
\fi
\fi
\fi
}
\newcommand{\hatcurSPECdupontsnrangeCand}[1]{\ifnum#1=578002 %
\hatcurSPECdupontsnrangeCandxxxxxA
\else
\ifnum#1=578003 %
\hatcurSPECdupontsnrangeCandxxxxxB
\else
\ifnum#1=578004 %
\hatcurSPECdupontsnrangeCandxxxxxC
\else
\ifnum#1=579001 %
\hatcurSPECdupontsnrangeCandxxxxxD
\else
\ifnum#1=579007 %
\hatcurSPECdupontsnrangeCandxxxxxE
\else
\ifnum#1=579008 %
\hatcurSPECdupontsnrangeCandxxxxxF
\else
\ifnum#1=579009 %
\hatcurSPECdupontsnrangeCandxxxxxG
\else
\ifnum#1=579010 %
\hatcurSPECdupontsnrangeCandxxxxxU
\else
\ifnum#1=579014 %
\hatcurSPECdupontsnrangeCandxxxxxH
\else
\ifnum#1=579015 %
\hatcurSPECdupontsnrangeCandxxxxxI
\else
\ifnum#1=579036 %
\hatcurSPECdupontsnrangeCandxxxxxJ
\else
\ifnum#1=579037 %
\hatcurSPECdupontsnrangeCandxxxxxK
\else
\ifnum#1=579039 %
\hatcurSPECdupontsnrangeCandxxxxxL
\else
\ifnum#1=579040 %
\hatcurSPECdupontsnrangeCandxxxxxM
\else
\ifnum#1=579041 %
\hatcurSPECdupontsnrangeCandxxxxxN
\else
\ifnum#1=579043 %
\hatcurSPECdupontsnrangeCandxxxxxO
\else
\ifnum#1=579044 %
\hatcurSPECdupontsnrangeCandxxxxxP
\else
\ifnum#1=579048 %
\hatcurSPECdupontsnrangeCandxxxxxQ
\else
\ifnum#1=579050 %
\hatcurSPECdupontsnrangeCandxxxxxR
\else
\ifnum#1=624002 %
\hatcurSPECdupontsnrangeCandxxxxxS
\else
\ifnum#1=624003 %
\hatcurSPECdupontsnrangeCandxxxxxT
\else
??????\fi
\fi
\fi
\fi
\fi
\fi
\fi
\fi
\fi
\fi
\fi
\fi
\fi
\fi
\fi
\fi
\fi
\fi
\fi
\fi
\fi
}
\newcommand{\hatcurSPECdupontTeffCand}[1]{\ifnum#1=578002 %
\hatcurSPECdupontTeffCandxxxxxA
\else
\ifnum#1=578003 %
\hatcurSPECdupontTeffCandxxxxxB
\else
\ifnum#1=578004 %
\hatcurSPECdupontTeffCandxxxxxC
\else
\ifnum#1=579001 %
\hatcurSPECdupontTeffCandxxxxxD
\else
\ifnum#1=579007 %
\hatcurSPECdupontTeffCandxxxxxE
\else
\ifnum#1=579008 %
\hatcurSPECdupontTeffCandxxxxxF
\else
\ifnum#1=579009 %
\hatcurSPECdupontTeffCandxxxxxG
\else
\ifnum#1=579010 %
\hatcurSPECdupontTeffCandxxxxxU
\else
\ifnum#1=579014 %
\hatcurSPECdupontTeffCandxxxxxH
\else
\ifnum#1=579015 %
\hatcurSPECdupontTeffCandxxxxxI
\else
\ifnum#1=579036 %
\hatcurSPECdupontTeffCandxxxxxJ
\else
\ifnum#1=579037 %
\hatcurSPECdupontTeffCandxxxxxK
\else
\ifnum#1=579039 %
\hatcurSPECdupontTeffCandxxxxxL
\else
\ifnum#1=579040 %
\hatcurSPECdupontTeffCandxxxxxM
\else
\ifnum#1=579041 %
\hatcurSPECdupontTeffCandxxxxxN
\else
\ifnum#1=579043 %
\hatcurSPECdupontTeffCandxxxxxO
\else
\ifnum#1=579044 %
\hatcurSPECdupontTeffCandxxxxxP
\else
\ifnum#1=579048 %
\hatcurSPECdupontTeffCandxxxxxQ
\else
\ifnum#1=579050 %
\hatcurSPECdupontTeffCandxxxxxR
\else
\ifnum#1=624002 %
\hatcurSPECdupontTeffCandxxxxxS
\else
\ifnum#1=624003 %
\hatcurSPECdupontTeffCandxxxxxT
\else
??????\fi
\fi
\fi
\fi
\fi
\fi
\fi
\fi
\fi
\fi
\fi
\fi
\fi
\fi
\fi
\fi
\fi
\fi
\fi
\fi
\fi
}
\newcommand{\hatcurSPECdupontvsiniCand}[1]{\ifnum#1=578002 %
\hatcurSPECdupontvsiniCandxxxxxA
\else
\ifnum#1=578003 %
\hatcurSPECdupontvsiniCandxxxxxB
\else
\ifnum#1=578004 %
\hatcurSPECdupontvsiniCandxxxxxC
\else
\ifnum#1=579001 %
\hatcurSPECdupontvsiniCandxxxxxD
\else
\ifnum#1=579007 %
\hatcurSPECdupontvsiniCandxxxxxE
\else
\ifnum#1=579008 %
\hatcurSPECdupontvsiniCandxxxxxF
\else
\ifnum#1=579009 %
\hatcurSPECdupontvsiniCandxxxxxG
\else
\ifnum#1=579010 %
\hatcurSPECdupontvsiniCandxxxxxU
\else
\ifnum#1=579014 %
\hatcurSPECdupontvsiniCandxxxxxH
\else
\ifnum#1=579015 %
\hatcurSPECdupontvsiniCandxxxxxI
\else
\ifnum#1=579036 %
\hatcurSPECdupontvsiniCandxxxxxJ
\else
\ifnum#1=579037 %
\hatcurSPECdupontvsiniCandxxxxxK
\else
\ifnum#1=579039 %
\hatcurSPECdupontvsiniCandxxxxxL
\else
\ifnum#1=579040 %
\hatcurSPECdupontvsiniCandxxxxxM
\else
\ifnum#1=579041 %
\hatcurSPECdupontvsiniCandxxxxxN
\else
\ifnum#1=579043 %
\hatcurSPECdupontvsiniCandxxxxxO
\else
\ifnum#1=579044 %
\hatcurSPECdupontvsiniCandxxxxxP
\else
\ifnum#1=579048 %
\hatcurSPECdupontvsiniCandxxxxxQ
\else
\ifnum#1=579050 %
\hatcurSPECdupontvsiniCandxxxxxR
\else
\ifnum#1=624002 %
\hatcurSPECdupontvsiniCandxxxxxS
\else
\ifnum#1=624003 %
\hatcurSPECdupontvsiniCandxxxxxT
\else
??????\fi
\fi
\fi
\fi
\fi
\fi
\fi
\fi
\fi
\fi
\fi
\fi
\fi
\fi
\fi
\fi
\fi
\fi
\fi
\fi
\fi
}
\newcommand{\hatcurSPECdupontzfehCand}[1]{\ifnum#1=578002 %
\hatcurSPECdupontzfehCandxxxxxA
\else
\ifnum#1=578003 %
\hatcurSPECdupontzfehCandxxxxxB
\else
\ifnum#1=578004 %
\hatcurSPECdupontzfehCandxxxxxC
\else
\ifnum#1=579001 %
\hatcurSPECdupontzfehCandxxxxxD
\else
\ifnum#1=579007 %
\hatcurSPECdupontzfehCandxxxxxE
\else
\ifnum#1=579008 %
\hatcurSPECdupontzfehCandxxxxxF
\else
\ifnum#1=579009 %
\hatcurSPECdupontzfehCandxxxxxG
\else
\ifnum#1=579010 %
\hatcurSPECdupontzfehCandxxxxxU
\else
\ifnum#1=579014 %
\hatcurSPECdupontzfehCandxxxxxH
\else
\ifnum#1=579015 %
\hatcurSPECdupontzfehCandxxxxxI
\else
\ifnum#1=579036 %
\hatcurSPECdupontzfehCandxxxxxJ
\else
\ifnum#1=579037 %
\hatcurSPECdupontzfehCandxxxxxK
\else
\ifnum#1=579039 %
\hatcurSPECdupontzfehCandxxxxxL
\else
\ifnum#1=579040 %
\hatcurSPECdupontzfehCandxxxxxM
\else
\ifnum#1=579041 %
\hatcurSPECdupontzfehCandxxxxxN
\else
\ifnum#1=579043 %
\hatcurSPECdupontzfehCandxxxxxO
\else
\ifnum#1=579044 %
\hatcurSPECdupontzfehCandxxxxxP
\else
\ifnum#1=579048 %
\hatcurSPECdupontzfehCandxxxxxQ
\else
\ifnum#1=579050 %
\hatcurSPECdupontzfehCandxxxxxR
\else
\ifnum#1=624002 %
\hatcurSPECdupontzfehCandxxxxxS
\else
\ifnum#1=624003 %
\hatcurSPECdupontzfehCandxxxxxT
\else
??????\fi
\fi
\fi
\fi
\fi
\fi
\fi
\fi
\fi
\fi
\fi
\fi
\fi
\fi
\fi
\fi
\fi
\fi
\fi
\fi
\fi
}
\newcommand{\hatcurSPECferosdaterangelowresCand}[1]{\ifnum#1=578002 %
\hatcurSPECferosdaterangelowresCandxxxxxA
\else
\ifnum#1=578003 %
\hatcurSPECferosdaterangelowresCandxxxxxB
\else
\ifnum#1=578004 %
\hatcurSPECferosdaterangelowresCandxxxxxC
\else
\ifnum#1=579001 %
\hatcurSPECferosdaterangelowresCandxxxxxD
\else
\ifnum#1=579007 %
\hatcurSPECferosdaterangelowresCandxxxxxE
\else
\ifnum#1=579008 %
\hatcurSPECferosdaterangelowresCandxxxxxF
\else
\ifnum#1=579009 %
\hatcurSPECferosdaterangelowresCandxxxxxG
\else
\ifnum#1=579010 %
\hatcurSPECferosdaterangelowresCandxxxxxU
\else
\ifnum#1=579014 %
\hatcurSPECferosdaterangelowresCandxxxxxH
\else
\ifnum#1=579015 %
\hatcurSPECferosdaterangelowresCandxxxxxI
\else
\ifnum#1=579036 %
\hatcurSPECferosdaterangelowresCandxxxxxJ
\else
\ifnum#1=579037 %
\hatcurSPECferosdaterangelowresCandxxxxxK
\else
\ifnum#1=579039 %
\hatcurSPECferosdaterangelowresCandxxxxxL
\else
\ifnum#1=579040 %
\hatcurSPECferosdaterangelowresCandxxxxxM
\else
\ifnum#1=579041 %
\hatcurSPECferosdaterangelowresCandxxxxxN
\else
\ifnum#1=579043 %
\hatcurSPECferosdaterangelowresCandxxxxxO
\else
\ifnum#1=579044 %
\hatcurSPECferosdaterangelowresCandxxxxxP
\else
\ifnum#1=579048 %
\hatcurSPECferosdaterangelowresCandxxxxxQ
\else
\ifnum#1=579050 %
\hatcurSPECferosdaterangelowresCandxxxxxR
\else
\ifnum#1=624002 %
\hatcurSPECferosdaterangelowresCandxxxxxS
\else
\ifnum#1=624003 %
\hatcurSPECferosdaterangelowresCandxxxxxT
\else
??????\fi
\fi
\fi
\fi
\fi
\fi
\fi
\fi
\fi
\fi
\fi
\fi
\fi
\fi
\fi
\fi
\fi
\fi
\fi
\fi
\fi
}
\newcommand{\hatcurSPECferosloggCand}[1]{\ifnum#1=578002 %
\hatcurSPECferosloggCandxxxxxA
\else
\ifnum#1=578003 %
\hatcurSPECferosloggCandxxxxxB
\else
\ifnum#1=578004 %
\hatcurSPECferosloggCandxxxxxC
\else
\ifnum#1=579001 %
\hatcurSPECferosloggCandxxxxxD
\else
\ifnum#1=579007 %
\hatcurSPECferosloggCandxxxxxE
\else
\ifnum#1=579008 %
\hatcurSPECferosloggCandxxxxxF
\else
\ifnum#1=579009 %
\hatcurSPECferosloggCandxxxxxG
\else
\ifnum#1=579010 %
\hatcurSPECferosloggCandxxxxxU
\else
\ifnum#1=579014 %
\hatcurSPECferosloggCandxxxxxH
\else
\ifnum#1=579015 %
\hatcurSPECferosloggCandxxxxxI
\else
\ifnum#1=579036 %
\hatcurSPECferosloggCandxxxxxJ
\else
\ifnum#1=579037 %
\hatcurSPECferosloggCandxxxxxK
\else
\ifnum#1=579039 %
\hatcurSPECferosloggCandxxxxxL
\else
\ifnum#1=579040 %
\hatcurSPECferosloggCandxxxxxM
\else
\ifnum#1=579041 %
\hatcurSPECferosloggCandxxxxxN
\else
\ifnum#1=579043 %
\hatcurSPECferosloggCandxxxxxO
\else
\ifnum#1=579044 %
\hatcurSPECferosloggCandxxxxxP
\else
\ifnum#1=579048 %
\hatcurSPECferosloggCandxxxxxQ
\else
\ifnum#1=579050 %
\hatcurSPECferosloggCandxxxxxR
\else
\ifnum#1=624002 %
\hatcurSPECferosloggCandxxxxxS
\else
\ifnum#1=624003 %
\hatcurSPECferosloggCandxxxxxT
\else
??????\fi
\fi
\fi
\fi
\fi
\fi
\fi
\fi
\fi
\fi
\fi
\fi
\fi
\fi
\fi
\fi
\fi
\fi
\fi
\fi
\fi
}
\newcommand{\hatcurSPECferosNCand}[1]{\ifnum#1=578002 %
\hatcurSPECferosNCandxxxxxA
\else
\ifnum#1=578003 %
\hatcurSPECferosNCandxxxxxB
\else
\ifnum#1=578004 %
\hatcurSPECferosNCandxxxxxC
\else
\ifnum#1=579001 %
\hatcurSPECferosNCandxxxxxD
\else
\ifnum#1=579007 %
\hatcurSPECferosNCandxxxxxE
\else
\ifnum#1=579008 %
\hatcurSPECferosNCandxxxxxF
\else
\ifnum#1=579009 %
\hatcurSPECferosNCandxxxxxG
\else
\ifnum#1=579010 %
\hatcurSPECferosNCandxxxxxU
\else
\ifnum#1=579014 %
\hatcurSPECferosNCandxxxxxH
\else
\ifnum#1=579015 %
\hatcurSPECferosNCandxxxxxI
\else
\ifnum#1=579036 %
\hatcurSPECferosNCandxxxxxJ
\else
\ifnum#1=579037 %
\hatcurSPECferosNCandxxxxxK
\else
\ifnum#1=579039 %
\hatcurSPECferosNCandxxxxxL
\else
\ifnum#1=579040 %
\hatcurSPECferosNCandxxxxxM
\else
\ifnum#1=579041 %
\hatcurSPECferosNCandxxxxxN
\else
\ifnum#1=579043 %
\hatcurSPECferosNCandxxxxxO
\else
\ifnum#1=579044 %
\hatcurSPECferosNCandxxxxxP
\else
\ifnum#1=579048 %
\hatcurSPECferosNCandxxxxxQ
\else
\ifnum#1=579050 %
\hatcurSPECferosNCandxxxxxR
\else
\ifnum#1=624002 %
\hatcurSPECferosNCandxxxxxS
\else
\ifnum#1=624003 %
\hatcurSPECferosNCandxxxxxT
\else
??????\fi
\fi
\fi
\fi
\fi
\fi
\fi
\fi
\fi
\fi
\fi
\fi
\fi
\fi
\fi
\fi
\fi
\fi
\fi
\fi
\fi
}
\newcommand{\hatcurSPECferosRVGCand}[1]{\ifnum#1=578002 %
\hatcurSPECferosRVGCandxxxxxA
\else
\ifnum#1=578003 %
\hatcurSPECferosRVGCandxxxxxB
\else
\ifnum#1=578004 %
\hatcurSPECferosRVGCandxxxxxC
\else
\ifnum#1=579001 %
\hatcurSPECferosRVGCandxxxxxD
\else
\ifnum#1=579007 %
\hatcurSPECferosRVGCandxxxxxE
\else
\ifnum#1=579008 %
\hatcurSPECferosRVGCandxxxxxF
\else
\ifnum#1=579009 %
\hatcurSPECferosRVGCandxxxxxG
\else
\ifnum#1=579010 %
\hatcurSPECferosRVGCandxxxxxU
\else
\ifnum#1=579014 %
\hatcurSPECferosRVGCandxxxxxH
\else
\ifnum#1=579015 %
\hatcurSPECferosRVGCandxxxxxI
\else
\ifnum#1=579036 %
\hatcurSPECferosRVGCandxxxxxJ
\else
\ifnum#1=579037 %
\hatcurSPECferosRVGCandxxxxxK
\else
\ifnum#1=579039 %
\hatcurSPECferosRVGCandxxxxxL
\else
\ifnum#1=579040 %
\hatcurSPECferosRVGCandxxxxxM
\else
\ifnum#1=579041 %
\hatcurSPECferosRVGCandxxxxxN
\else
\ifnum#1=579043 %
\hatcurSPECferosRVGCandxxxxxO
\else
\ifnum#1=579044 %
\hatcurSPECferosRVGCandxxxxxP
\else
\ifnum#1=579048 %
\hatcurSPECferosRVGCandxxxxxQ
\else
\ifnum#1=579050 %
\hatcurSPECferosRVGCandxxxxxR
\else
\ifnum#1=624002 %
\hatcurSPECferosRVGCandxxxxxS
\else
\ifnum#1=624003 %
\hatcurSPECferosRVGCandxxxxxT
\else
??????\fi
\fi
\fi
\fi
\fi
\fi
\fi
\fi
\fi
\fi
\fi
\fi
\fi
\fi
\fi
\fi
\fi
\fi
\fi
\fi
\fi
}
\newcommand{\hatcurSPECferosRVKCand}[1]{\ifnum#1=578002 %
\hatcurSPECferosRVKCandxxxxxA
\else
\ifnum#1=578003 %
\hatcurSPECferosRVKCandxxxxxB
\else
\ifnum#1=578004 %
\hatcurSPECferosRVKCandxxxxxC
\else
\ifnum#1=579001 %
\hatcurSPECferosRVKCandxxxxxD
\else
\ifnum#1=579007 %
\hatcurSPECferosRVKCandxxxxxE
\else
\ifnum#1=579008 %
\hatcurSPECferosRVKCandxxxxxF
\else
\ifnum#1=579009 %
\hatcurSPECferosRVKCandxxxxxG
\else
\ifnum#1=579010 %
\hatcurSPECferosRVKCandxxxxxU
\else
\ifnum#1=579014 %
\hatcurSPECferosRVKCandxxxxxH
\else
\ifnum#1=579015 %
\hatcurSPECferosRVKCandxxxxxI
\else
\ifnum#1=579036 %
\hatcurSPECferosRVKCandxxxxxJ
\else
\ifnum#1=579037 %
\hatcurSPECferosRVKCandxxxxxK
\else
\ifnum#1=579039 %
\hatcurSPECferosRVKCandxxxxxL
\else
\ifnum#1=579040 %
\hatcurSPECferosRVKCandxxxxxM
\else
\ifnum#1=579041 %
\hatcurSPECferosRVKCandxxxxxN
\else
\ifnum#1=579043 %
\hatcurSPECferosRVKCandxxxxxO
\else
\ifnum#1=579044 %
\hatcurSPECferosRVKCandxxxxxP
\else
\ifnum#1=579048 %
\hatcurSPECferosRVKCandxxxxxQ
\else
\ifnum#1=579050 %
\hatcurSPECferosRVKCandxxxxxR
\else
\ifnum#1=624002 %
\hatcurSPECferosRVKCandxxxxxS
\else
\ifnum#1=624003 %
\hatcurSPECferosRVKCandxxxxxT
\else
??????\fi
\fi
\fi
\fi
\fi
\fi
\fi
\fi
\fi
\fi
\fi
\fi
\fi
\fi
\fi
\fi
\fi
\fi
\fi
\fi
\fi
}
\newcommand{\hatcurSPECferossnrangeCand}[1]{\ifnum#1=578002 %
\hatcurSPECferossnrangeCandxxxxxA
\else
\ifnum#1=578003 %
\hatcurSPECferossnrangeCandxxxxxB
\else
\ifnum#1=578004 %
\hatcurSPECferossnrangeCandxxxxxC
\else
\ifnum#1=579001 %
\hatcurSPECferossnrangeCandxxxxxD
\else
\ifnum#1=579007 %
\hatcurSPECferossnrangeCandxxxxxE
\else
\ifnum#1=579008 %
\hatcurSPECferossnrangeCandxxxxxF
\else
\ifnum#1=579009 %
\hatcurSPECferossnrangeCandxxxxxG
\else
\ifnum#1=579010 %
\hatcurSPECferossnrangeCandxxxxxU
\else
\ifnum#1=579014 %
\hatcurSPECferossnrangeCandxxxxxH
\else
\ifnum#1=579015 %
\hatcurSPECferossnrangeCandxxxxxI
\else
\ifnum#1=579036 %
\hatcurSPECferossnrangeCandxxxxxJ
\else
\ifnum#1=579037 %
\hatcurSPECferossnrangeCandxxxxxK
\else
\ifnum#1=579039 %
\hatcurSPECferossnrangeCandxxxxxL
\else
\ifnum#1=579040 %
\hatcurSPECferossnrangeCandxxxxxM
\else
\ifnum#1=579041 %
\hatcurSPECferossnrangeCandxxxxxN
\else
\ifnum#1=579043 %
\hatcurSPECferossnrangeCandxxxxxO
\else
\ifnum#1=579044 %
\hatcurSPECferossnrangeCandxxxxxP
\else
\ifnum#1=579048 %
\hatcurSPECferossnrangeCandxxxxxQ
\else
\ifnum#1=579050 %
\hatcurSPECferossnrangeCandxxxxxR
\else
\ifnum#1=624002 %
\hatcurSPECferossnrangeCandxxxxxS
\else
\ifnum#1=624003 %
\hatcurSPECferossnrangeCandxxxxxT
\else
??????\fi
\fi
\fi
\fi
\fi
\fi
\fi
\fi
\fi
\fi
\fi
\fi
\fi
\fi
\fi
\fi
\fi
\fi
\fi
\fi
\fi
}
\newcommand{\hatcurSPECferosTeffCand}[1]{\ifnum#1=578002 %
\hatcurSPECferosTeffCandxxxxxA
\else
\ifnum#1=578003 %
\hatcurSPECferosTeffCandxxxxxB
\else
\ifnum#1=578004 %
\hatcurSPECferosTeffCandxxxxxC
\else
\ifnum#1=579001 %
\hatcurSPECferosTeffCandxxxxxD
\else
\ifnum#1=579007 %
\hatcurSPECferosTeffCandxxxxxE
\else
\ifnum#1=579008 %
\hatcurSPECferosTeffCandxxxxxF
\else
\ifnum#1=579009 %
\hatcurSPECferosTeffCandxxxxxG
\else
\ifnum#1=579010 %
\hatcurSPECferosTeffCandxxxxxU
\else
\ifnum#1=579014 %
\hatcurSPECferosTeffCandxxxxxH
\else
\ifnum#1=579015 %
\hatcurSPECferosTeffCandxxxxxI
\else
\ifnum#1=579036 %
\hatcurSPECferosTeffCandxxxxxJ
\else
\ifnum#1=579037 %
\hatcurSPECferosTeffCandxxxxxK
\else
\ifnum#1=579039 %
\hatcurSPECferosTeffCandxxxxxL
\else
\ifnum#1=579040 %
\hatcurSPECferosTeffCandxxxxxM
\else
\ifnum#1=579041 %
\hatcurSPECferosTeffCandxxxxxN
\else
\ifnum#1=579043 %
\hatcurSPECferosTeffCandxxxxxO
\else
\ifnum#1=579044 %
\hatcurSPECferosTeffCandxxxxxP
\else
\ifnum#1=579048 %
\hatcurSPECferosTeffCandxxxxxQ
\else
\ifnum#1=579050 %
\hatcurSPECferosTeffCandxxxxxR
\else
\ifnum#1=624002 %
\hatcurSPECferosTeffCandxxxxxS
\else
\ifnum#1=624003 %
\hatcurSPECferosTeffCandxxxxxT
\else
??????\fi
\fi
\fi
\fi
\fi
\fi
\fi
\fi
\fi
\fi
\fi
\fi
\fi
\fi
\fi
\fi
\fi
\fi
\fi
\fi
\fi
}
\newcommand{\hatcurSPECferosvsiniCand}[1]{\ifnum#1=578002 %
\hatcurSPECferosvsiniCandxxxxxA
\else
\ifnum#1=578003 %
\hatcurSPECferosvsiniCandxxxxxB
\else
\ifnum#1=578004 %
\hatcurSPECferosvsiniCandxxxxxC
\else
\ifnum#1=579001 %
\hatcurSPECferosvsiniCandxxxxxD
\else
\ifnum#1=579007 %
\hatcurSPECferosvsiniCandxxxxxE
\else
\ifnum#1=579008 %
\hatcurSPECferosvsiniCandxxxxxF
\else
\ifnum#1=579009 %
\hatcurSPECferosvsiniCandxxxxxG
\else
\ifnum#1=579010 %
\hatcurSPECferosvsiniCandxxxxxU
\else
\ifnum#1=579014 %
\hatcurSPECferosvsiniCandxxxxxH
\else
\ifnum#1=579015 %
\hatcurSPECferosvsiniCandxxxxxI
\else
\ifnum#1=579036 %
\hatcurSPECferosvsiniCandxxxxxJ
\else
\ifnum#1=579037 %
\hatcurSPECferosvsiniCandxxxxxK
\else
\ifnum#1=579039 %
\hatcurSPECferosvsiniCandxxxxxL
\else
\ifnum#1=579040 %
\hatcurSPECferosvsiniCandxxxxxM
\else
\ifnum#1=579041 %
\hatcurSPECferosvsiniCandxxxxxN
\else
\ifnum#1=579043 %
\hatcurSPECferosvsiniCandxxxxxO
\else
\ifnum#1=579044 %
\hatcurSPECferosvsiniCandxxxxxP
\else
\ifnum#1=579048 %
\hatcurSPECferosvsiniCandxxxxxQ
\else
\ifnum#1=579050 %
\hatcurSPECferosvsiniCandxxxxxR
\else
\ifnum#1=624002 %
\hatcurSPECferosvsiniCandxxxxxS
\else
\ifnum#1=624003 %
\hatcurSPECferosvsiniCandxxxxxT
\else
??????\fi
\fi
\fi
\fi
\fi
\fi
\fi
\fi
\fi
\fi
\fi
\fi
\fi
\fi
\fi
\fi
\fi
\fi
\fi
\fi
\fi
}
\newcommand{\hatcurSPECferoszfehCand}[1]{\ifnum#1=578002 %
\hatcurSPECferoszfehCandxxxxxA
\else
\ifnum#1=578003 %
\hatcurSPECferoszfehCandxxxxxB
\else
\ifnum#1=578004 %
\hatcurSPECferoszfehCandxxxxxC
\else
\ifnum#1=579001 %
\hatcurSPECferoszfehCandxxxxxD
\else
\ifnum#1=579007 %
\hatcurSPECferoszfehCandxxxxxE
\else
\ifnum#1=579008 %
\hatcurSPECferoszfehCandxxxxxF
\else
\ifnum#1=579009 %
\hatcurSPECferoszfehCandxxxxxG
\else
\ifnum#1=579010 %
\hatcurSPECferoszfehCandxxxxxU
\else
\ifnum#1=579014 %
\hatcurSPECferoszfehCandxxxxxH
\else
\ifnum#1=579015 %
\hatcurSPECferoszfehCandxxxxxI
\else
\ifnum#1=579036 %
\hatcurSPECferoszfehCandxxxxxJ
\else
\ifnum#1=579037 %
\hatcurSPECferoszfehCandxxxxxK
\else
\ifnum#1=579039 %
\hatcurSPECferoszfehCandxxxxxL
\else
\ifnum#1=579040 %
\hatcurSPECferoszfehCandxxxxxM
\else
\ifnum#1=579041 %
\hatcurSPECferoszfehCandxxxxxN
\else
\ifnum#1=579043 %
\hatcurSPECferoszfehCandxxxxxO
\else
\ifnum#1=579044 %
\hatcurSPECferoszfehCandxxxxxP
\else
\ifnum#1=579048 %
\hatcurSPECferoszfehCandxxxxxQ
\else
\ifnum#1=579050 %
\hatcurSPECferoszfehCandxxxxxR
\else
\ifnum#1=624002 %
\hatcurSPECferoszfehCandxxxxxS
\else
\ifnum#1=624003 %
\hatcurSPECferoszfehCandxxxxxT
\else
??????\fi
\fi
\fi
\fi
\fi
\fi
\fi
\fi
\fi
\fi
\fi
\fi
\fi
\fi
\fi
\fi
\fi
\fi
\fi
\fi
\fi
}
\newcommand{\hatcurSPECfiesdaterangelowresCand}[1]{\ifnum#1=578002 %
\hatcurSPECfiesdaterangelowresCandxxxxxA
\else
\ifnum#1=578003 %
\hatcurSPECfiesdaterangelowresCandxxxxxB
\else
\ifnum#1=578004 %
\hatcurSPECfiesdaterangelowresCandxxxxxC
\else
\ifnum#1=579001 %
\hatcurSPECfiesdaterangelowresCandxxxxxD
\else
\ifnum#1=579007 %
\hatcurSPECfiesdaterangelowresCandxxxxxE
\else
\ifnum#1=579008 %
\hatcurSPECfiesdaterangelowresCandxxxxxF
\else
\ifnum#1=579009 %
\hatcurSPECfiesdaterangelowresCandxxxxxG
\else
\ifnum#1=579010 %
\hatcurSPECfiesdaterangelowresCandxxxxxU
\else
\ifnum#1=579014 %
\hatcurSPECfiesdaterangelowresCandxxxxxH
\else
\ifnum#1=579015 %
\hatcurSPECfiesdaterangelowresCandxxxxxI
\else
\ifnum#1=579036 %
\hatcurSPECfiesdaterangelowresCandxxxxxJ
\else
\ifnum#1=579037 %
\hatcurSPECfiesdaterangelowresCandxxxxxK
\else
\ifnum#1=579039 %
\hatcurSPECfiesdaterangelowresCandxxxxxL
\else
\ifnum#1=579040 %
\hatcurSPECfiesdaterangelowresCandxxxxxM
\else
\ifnum#1=579041 %
\hatcurSPECfiesdaterangelowresCandxxxxxN
\else
\ifnum#1=579043 %
\hatcurSPECfiesdaterangelowresCandxxxxxO
\else
\ifnum#1=579044 %
\hatcurSPECfiesdaterangelowresCandxxxxxP
\else
\ifnum#1=579048 %
\hatcurSPECfiesdaterangelowresCandxxxxxQ
\else
\ifnum#1=579050 %
\hatcurSPECfiesdaterangelowresCandxxxxxR
\else
\ifnum#1=624002 %
\hatcurSPECfiesdaterangelowresCandxxxxxS
\else
\ifnum#1=624003 %
\hatcurSPECfiesdaterangelowresCandxxxxxT
\else
??????\fi
\fi
\fi
\fi
\fi
\fi
\fi
\fi
\fi
\fi
\fi
\fi
\fi
\fi
\fi
\fi
\fi
\fi
\fi
\fi
\fi
}
\newcommand{\hatcurSPECfiesloggCand}[1]{\ifnum#1=578002 %
\hatcurSPECfiesloggCandxxxxxA
\else
\ifnum#1=578003 %
\hatcurSPECfiesloggCandxxxxxB
\else
\ifnum#1=578004 %
\hatcurSPECfiesloggCandxxxxxC
\else
\ifnum#1=579001 %
\hatcurSPECfiesloggCandxxxxxD
\else
\ifnum#1=579007 %
\hatcurSPECfiesloggCandxxxxxE
\else
\ifnum#1=579008 %
\hatcurSPECfiesloggCandxxxxxF
\else
\ifnum#1=579009 %
\hatcurSPECfiesloggCandxxxxxG
\else
\ifnum#1=579010 %
\hatcurSPECfiesloggCandxxxxxU
\else
\ifnum#1=579014 %
\hatcurSPECfiesloggCandxxxxxH
\else
\ifnum#1=579015 %
\hatcurSPECfiesloggCandxxxxxI
\else
\ifnum#1=579036 %
\hatcurSPECfiesloggCandxxxxxJ
\else
\ifnum#1=579037 %
\hatcurSPECfiesloggCandxxxxxK
\else
\ifnum#1=579039 %
\hatcurSPECfiesloggCandxxxxxL
\else
\ifnum#1=579040 %
\hatcurSPECfiesloggCandxxxxxM
\else
\ifnum#1=579041 %
\hatcurSPECfiesloggCandxxxxxN
\else
\ifnum#1=579043 %
\hatcurSPECfiesloggCandxxxxxO
\else
\ifnum#1=579044 %
\hatcurSPECfiesloggCandxxxxxP
\else
\ifnum#1=579048 %
\hatcurSPECfiesloggCandxxxxxQ
\else
\ifnum#1=579050 %
\hatcurSPECfiesloggCandxxxxxR
\else
\ifnum#1=624002 %
\hatcurSPECfiesloggCandxxxxxS
\else
\ifnum#1=624003 %
\hatcurSPECfiesloggCandxxxxxT
\else
??????\fi
\fi
\fi
\fi
\fi
\fi
\fi
\fi
\fi
\fi
\fi
\fi
\fi
\fi
\fi
\fi
\fi
\fi
\fi
\fi
\fi
}
\newcommand{\hatcurSPECfiesNCand}[1]{\ifnum#1=578002 %
\hatcurSPECfiesNCandxxxxxA
\else
\ifnum#1=578003 %
\hatcurSPECfiesNCandxxxxxB
\else
\ifnum#1=578004 %
\hatcurSPECfiesNCandxxxxxC
\else
\ifnum#1=579001 %
\hatcurSPECfiesNCandxxxxxD
\else
\ifnum#1=579007 %
\hatcurSPECfiesNCandxxxxxE
\else
\ifnum#1=579008 %
\hatcurSPECfiesNCandxxxxxF
\else
\ifnum#1=579009 %
\hatcurSPECfiesNCandxxxxxG
\else
\ifnum#1=579010 %
\hatcurSPECfiesNCandxxxxxU
\else
\ifnum#1=579014 %
\hatcurSPECfiesNCandxxxxxH
\else
\ifnum#1=579015 %
\hatcurSPECfiesNCandxxxxxI
\else
\ifnum#1=579036 %
\hatcurSPECfiesNCandxxxxxJ
\else
\ifnum#1=579037 %
\hatcurSPECfiesNCandxxxxxK
\else
\ifnum#1=579039 %
\hatcurSPECfiesNCandxxxxxL
\else
\ifnum#1=579040 %
\hatcurSPECfiesNCandxxxxxM
\else
\ifnum#1=579041 %
\hatcurSPECfiesNCandxxxxxN
\else
\ifnum#1=579043 %
\hatcurSPECfiesNCandxxxxxO
\else
\ifnum#1=579044 %
\hatcurSPECfiesNCandxxxxxP
\else
\ifnum#1=579048 %
\hatcurSPECfiesNCandxxxxxQ
\else
\ifnum#1=579050 %
\hatcurSPECfiesNCandxxxxxR
\else
\ifnum#1=624002 %
\hatcurSPECfiesNCandxxxxxS
\else
\ifnum#1=624003 %
\hatcurSPECfiesNCandxxxxxT
\else
??????\fi
\fi
\fi
\fi
\fi
\fi
\fi
\fi
\fi
\fi
\fi
\fi
\fi
\fi
\fi
\fi
\fi
\fi
\fi
\fi
\fi
}
\newcommand{\hatcurSPECfiesRVGCand}[1]{\ifnum#1=578002 %
\hatcurSPECfiesRVGCandxxxxxA
\else
\ifnum#1=578003 %
\hatcurSPECfiesRVGCandxxxxxB
\else
\ifnum#1=578004 %
\hatcurSPECfiesRVGCandxxxxxC
\else
\ifnum#1=579001 %
\hatcurSPECfiesRVGCandxxxxxD
\else
\ifnum#1=579007 %
\hatcurSPECfiesRVGCandxxxxxE
\else
\ifnum#1=579008 %
\hatcurSPECfiesRVGCandxxxxxF
\else
\ifnum#1=579009 %
\hatcurSPECfiesRVGCandxxxxxG
\else
\ifnum#1=579010 %
\hatcurSPECfiesRVGCandxxxxxU
\else
\ifnum#1=579014 %
\hatcurSPECfiesRVGCandxxxxxH
\else
\ifnum#1=579015 %
\hatcurSPECfiesRVGCandxxxxxI
\else
\ifnum#1=579036 %
\hatcurSPECfiesRVGCandxxxxxJ
\else
\ifnum#1=579037 %
\hatcurSPECfiesRVGCandxxxxxK
\else
\ifnum#1=579039 %
\hatcurSPECfiesRVGCandxxxxxL
\else
\ifnum#1=579040 %
\hatcurSPECfiesRVGCandxxxxxM
\else
\ifnum#1=579041 %
\hatcurSPECfiesRVGCandxxxxxN
\else
\ifnum#1=579043 %
\hatcurSPECfiesRVGCandxxxxxO
\else
\ifnum#1=579044 %
\hatcurSPECfiesRVGCandxxxxxP
\else
\ifnum#1=579048 %
\hatcurSPECfiesRVGCandxxxxxQ
\else
\ifnum#1=579050 %
\hatcurSPECfiesRVGCandxxxxxR
\else
\ifnum#1=624002 %
\hatcurSPECfiesRVGCandxxxxxS
\else
\ifnum#1=624003 %
\hatcurSPECfiesRVGCandxxxxxT
\else
??????\fi
\fi
\fi
\fi
\fi
\fi
\fi
\fi
\fi
\fi
\fi
\fi
\fi
\fi
\fi
\fi
\fi
\fi
\fi
\fi
\fi
}
\newcommand{\hatcurSPECfiesRVKCand}[1]{\ifnum#1=578002 %
\hatcurSPECfiesRVKCandxxxxxA
\else
\ifnum#1=578003 %
\hatcurSPECfiesRVKCandxxxxxB
\else
\ifnum#1=578004 %
\hatcurSPECfiesRVKCandxxxxxC
\else
\ifnum#1=579001 %
\hatcurSPECfiesRVKCandxxxxxD
\else
\ifnum#1=579007 %
\hatcurSPECfiesRVKCandxxxxxE
\else
\ifnum#1=579008 %
\hatcurSPECfiesRVKCandxxxxxF
\else
\ifnum#1=579009 %
\hatcurSPECfiesRVKCandxxxxxG
\else
\ifnum#1=579010 %
\hatcurSPECfiesRVKCandxxxxxU
\else
\ifnum#1=579014 %
\hatcurSPECfiesRVKCandxxxxxH
\else
\ifnum#1=579015 %
\hatcurSPECfiesRVKCandxxxxxI
\else
\ifnum#1=579036 %
\hatcurSPECfiesRVKCandxxxxxJ
\else
\ifnum#1=579037 %
\hatcurSPECfiesRVKCandxxxxxK
\else
\ifnum#1=579039 %
\hatcurSPECfiesRVKCandxxxxxL
\else
\ifnum#1=579040 %
\hatcurSPECfiesRVKCandxxxxxM
\else
\ifnum#1=579041 %
\hatcurSPECfiesRVKCandxxxxxN
\else
\ifnum#1=579043 %
\hatcurSPECfiesRVKCandxxxxxO
\else
\ifnum#1=579044 %
\hatcurSPECfiesRVKCandxxxxxP
\else
\ifnum#1=579048 %
\hatcurSPECfiesRVKCandxxxxxQ
\else
\ifnum#1=579050 %
\hatcurSPECfiesRVKCandxxxxxR
\else
\ifnum#1=624002 %
\hatcurSPECfiesRVKCandxxxxxS
\else
\ifnum#1=624003 %
\hatcurSPECfiesRVKCandxxxxxT
\else
??????\fi
\fi
\fi
\fi
\fi
\fi
\fi
\fi
\fi
\fi
\fi
\fi
\fi
\fi
\fi
\fi
\fi
\fi
\fi
\fi
\fi
}
\newcommand{\hatcurSPECfiessnrangeCand}[1]{\ifnum#1=578002 %
\hatcurSPECfiessnrangeCandxxxxxA
\else
\ifnum#1=578003 %
\hatcurSPECfiessnrangeCandxxxxxB
\else
\ifnum#1=578004 %
\hatcurSPECfiessnrangeCandxxxxxC
\else
\ifnum#1=579001 %
\hatcurSPECfiessnrangeCandxxxxxD
\else
\ifnum#1=579007 %
\hatcurSPECfiessnrangeCandxxxxxE
\else
\ifnum#1=579008 %
\hatcurSPECfiessnrangeCandxxxxxF
\else
\ifnum#1=579009 %
\hatcurSPECfiessnrangeCandxxxxxG
\else
\ifnum#1=579010 %
\hatcurSPECfiessnrangeCandxxxxxU
\else
\ifnum#1=579014 %
\hatcurSPECfiessnrangeCandxxxxxH
\else
\ifnum#1=579015 %
\hatcurSPECfiessnrangeCandxxxxxI
\else
\ifnum#1=579036 %
\hatcurSPECfiessnrangeCandxxxxxJ
\else
\ifnum#1=579037 %
\hatcurSPECfiessnrangeCandxxxxxK
\else
\ifnum#1=579039 %
\hatcurSPECfiessnrangeCandxxxxxL
\else
\ifnum#1=579040 %
\hatcurSPECfiessnrangeCandxxxxxM
\else
\ifnum#1=579041 %
\hatcurSPECfiessnrangeCandxxxxxN
\else
\ifnum#1=579043 %
\hatcurSPECfiessnrangeCandxxxxxO
\else
\ifnum#1=579044 %
\hatcurSPECfiessnrangeCandxxxxxP
\else
\ifnum#1=579048 %
\hatcurSPECfiessnrangeCandxxxxxQ
\else
\ifnum#1=579050 %
\hatcurSPECfiessnrangeCandxxxxxR
\else
\ifnum#1=624002 %
\hatcurSPECfiessnrangeCandxxxxxS
\else
\ifnum#1=624003 %
\hatcurSPECfiessnrangeCandxxxxxT
\else
??????\fi
\fi
\fi
\fi
\fi
\fi
\fi
\fi
\fi
\fi
\fi
\fi
\fi
\fi
\fi
\fi
\fi
\fi
\fi
\fi
\fi
}
\newcommand{\hatcurSPECfiesTeffCand}[1]{\ifnum#1=578002 %
\hatcurSPECfiesTeffCandxxxxxA
\else
\ifnum#1=578003 %
\hatcurSPECfiesTeffCandxxxxxB
\else
\ifnum#1=578004 %
\hatcurSPECfiesTeffCandxxxxxC
\else
\ifnum#1=579001 %
\hatcurSPECfiesTeffCandxxxxxD
\else
\ifnum#1=579007 %
\hatcurSPECfiesTeffCandxxxxxE
\else
\ifnum#1=579008 %
\hatcurSPECfiesTeffCandxxxxxF
\else
\ifnum#1=579009 %
\hatcurSPECfiesTeffCandxxxxxG
\else
\ifnum#1=579010 %
\hatcurSPECfiesTeffCandxxxxxU
\else
\ifnum#1=579014 %
\hatcurSPECfiesTeffCandxxxxxH
\else
\ifnum#1=579015 %
\hatcurSPECfiesTeffCandxxxxxI
\else
\ifnum#1=579036 %
\hatcurSPECfiesTeffCandxxxxxJ
\else
\ifnum#1=579037 %
\hatcurSPECfiesTeffCandxxxxxK
\else
\ifnum#1=579039 %
\hatcurSPECfiesTeffCandxxxxxL
\else
\ifnum#1=579040 %
\hatcurSPECfiesTeffCandxxxxxM
\else
\ifnum#1=579041 %
\hatcurSPECfiesTeffCandxxxxxN
\else
\ifnum#1=579043 %
\hatcurSPECfiesTeffCandxxxxxO
\else
\ifnum#1=579044 %
\hatcurSPECfiesTeffCandxxxxxP
\else
\ifnum#1=579048 %
\hatcurSPECfiesTeffCandxxxxxQ
\else
\ifnum#1=579050 %
\hatcurSPECfiesTeffCandxxxxxR
\else
\ifnum#1=624002 %
\hatcurSPECfiesTeffCandxxxxxS
\else
\ifnum#1=624003 %
\hatcurSPECfiesTeffCandxxxxxT
\else
??????\fi
\fi
\fi
\fi
\fi
\fi
\fi
\fi
\fi
\fi
\fi
\fi
\fi
\fi
\fi
\fi
\fi
\fi
\fi
\fi
\fi
}
\newcommand{\hatcurSPECfiesvsiniCand}[1]{\ifnum#1=578002 %
\hatcurSPECfiesvsiniCandxxxxxA
\else
\ifnum#1=578003 %
\hatcurSPECfiesvsiniCandxxxxxB
\else
\ifnum#1=578004 %
\hatcurSPECfiesvsiniCandxxxxxC
\else
\ifnum#1=579001 %
\hatcurSPECfiesvsiniCandxxxxxD
\else
\ifnum#1=579007 %
\hatcurSPECfiesvsiniCandxxxxxE
\else
\ifnum#1=579008 %
\hatcurSPECfiesvsiniCandxxxxxF
\else
\ifnum#1=579009 %
\hatcurSPECfiesvsiniCandxxxxxG
\else
\ifnum#1=579010 %
\hatcurSPECfiesvsiniCandxxxxxU
\else
\ifnum#1=579014 %
\hatcurSPECfiesvsiniCandxxxxxH
\else
\ifnum#1=579015 %
\hatcurSPECfiesvsiniCandxxxxxI
\else
\ifnum#1=579036 %
\hatcurSPECfiesvsiniCandxxxxxJ
\else
\ifnum#1=579037 %
\hatcurSPECfiesvsiniCandxxxxxK
\else
\ifnum#1=579039 %
\hatcurSPECfiesvsiniCandxxxxxL
\else
\ifnum#1=579040 %
\hatcurSPECfiesvsiniCandxxxxxM
\else
\ifnum#1=579041 %
\hatcurSPECfiesvsiniCandxxxxxN
\else
\ifnum#1=579043 %
\hatcurSPECfiesvsiniCandxxxxxO
\else
\ifnum#1=579044 %
\hatcurSPECfiesvsiniCandxxxxxP
\else
\ifnum#1=579048 %
\hatcurSPECfiesvsiniCandxxxxxQ
\else
\ifnum#1=579050 %
\hatcurSPECfiesvsiniCandxxxxxR
\else
\ifnum#1=624002 %
\hatcurSPECfiesvsiniCandxxxxxS
\else
\ifnum#1=624003 %
\hatcurSPECfiesvsiniCandxxxxxT
\else
??????\fi
\fi
\fi
\fi
\fi
\fi
\fi
\fi
\fi
\fi
\fi
\fi
\fi
\fi
\fi
\fi
\fi
\fi
\fi
\fi
\fi
}
\newcommand{\hatcurSPECfieszfehCand}[1]{\ifnum#1=578002 %
\hatcurSPECfieszfehCandxxxxxA
\else
\ifnum#1=578003 %
\hatcurSPECfieszfehCandxxxxxB
\else
\ifnum#1=578004 %
\hatcurSPECfieszfehCandxxxxxC
\else
\ifnum#1=579001 %
\hatcurSPECfieszfehCandxxxxxD
\else
\ifnum#1=579007 %
\hatcurSPECfieszfehCandxxxxxE
\else
\ifnum#1=579008 %
\hatcurSPECfieszfehCandxxxxxF
\else
\ifnum#1=579009 %
\hatcurSPECfieszfehCandxxxxxG
\else
\ifnum#1=579010 %
\hatcurSPECfieszfehCandxxxxxU
\else
\ifnum#1=579014 %
\hatcurSPECfieszfehCandxxxxxH
\else
\ifnum#1=579015 %
\hatcurSPECfieszfehCandxxxxxI
\else
\ifnum#1=579036 %
\hatcurSPECfieszfehCandxxxxxJ
\else
\ifnum#1=579037 %
\hatcurSPECfieszfehCandxxxxxK
\else
\ifnum#1=579039 %
\hatcurSPECfieszfehCandxxxxxL
\else
\ifnum#1=579040 %
\hatcurSPECfieszfehCandxxxxxM
\else
\ifnum#1=579041 %
\hatcurSPECfieszfehCandxxxxxN
\else
\ifnum#1=579043 %
\hatcurSPECfieszfehCandxxxxxO
\else
\ifnum#1=579044 %
\hatcurSPECfieszfehCandxxxxxP
\else
\ifnum#1=579048 %
\hatcurSPECfieszfehCandxxxxxQ
\else
\ifnum#1=579050 %
\hatcurSPECfieszfehCandxxxxxR
\else
\ifnum#1=624002 %
\hatcurSPECfieszfehCandxxxxxS
\else
\ifnum#1=624003 %
\hatcurSPECfieszfehCandxxxxxT
\else
??????\fi
\fi
\fi
\fi
\fi
\fi
\fi
\fi
\fi
\fi
\fi
\fi
\fi
\fi
\fi
\fi
\fi
\fi
\fi
\fi
\fi
}
\newcommand{\hatcurSPECharpsdaterangelowresCand}[1]{\ifnum#1=578002 %
\hatcurSPECharpsdaterangelowresCandxxxxxA
\else
\ifnum#1=578003 %
\hatcurSPECharpsdaterangelowresCandxxxxxB
\else
\ifnum#1=578004 %
\hatcurSPECharpsdaterangelowresCandxxxxxC
\else
\ifnum#1=579001 %
\hatcurSPECharpsdaterangelowresCandxxxxxD
\else
\ifnum#1=579007 %
\hatcurSPECharpsdaterangelowresCandxxxxxE
\else
\ifnum#1=579008 %
\hatcurSPECharpsdaterangelowresCandxxxxxF
\else
\ifnum#1=579009 %
\hatcurSPECharpsdaterangelowresCandxxxxxG
\else
\ifnum#1=579010 %
\hatcurSPECharpsdaterangelowresCandxxxxxU
\else
\ifnum#1=579014 %
\hatcurSPECharpsdaterangelowresCandxxxxxH
\else
\ifnum#1=579015 %
\hatcurSPECharpsdaterangelowresCandxxxxxI
\else
\ifnum#1=579036 %
\hatcurSPECharpsdaterangelowresCandxxxxxJ
\else
\ifnum#1=579037 %
\hatcurSPECharpsdaterangelowresCandxxxxxK
\else
\ifnum#1=579039 %
\hatcurSPECharpsdaterangelowresCandxxxxxL
\else
\ifnum#1=579040 %
\hatcurSPECharpsdaterangelowresCandxxxxxM
\else
\ifnum#1=579041 %
\hatcurSPECharpsdaterangelowresCandxxxxxN
\else
\ifnum#1=579043 %
\hatcurSPECharpsdaterangelowresCandxxxxxO
\else
\ifnum#1=579044 %
\hatcurSPECharpsdaterangelowresCandxxxxxP
\else
\ifnum#1=579048 %
\hatcurSPECharpsdaterangelowresCandxxxxxQ
\else
\ifnum#1=579050 %
\hatcurSPECharpsdaterangelowresCandxxxxxR
\else
\ifnum#1=624002 %
\hatcurSPECharpsdaterangelowresCandxxxxxS
\else
\ifnum#1=624003 %
\hatcurSPECharpsdaterangelowresCandxxxxxT
\else
??????\fi
\fi
\fi
\fi
\fi
\fi
\fi
\fi
\fi
\fi
\fi
\fi
\fi
\fi
\fi
\fi
\fi
\fi
\fi
\fi
\fi
}
\newcommand{\hatcurSPECharpsloggCand}[1]{\ifnum#1=578002 %
\hatcurSPECharpsloggCandxxxxxA
\else
\ifnum#1=578003 %
\hatcurSPECharpsloggCandxxxxxB
\else
\ifnum#1=578004 %
\hatcurSPECharpsloggCandxxxxxC
\else
\ifnum#1=579001 %
\hatcurSPECharpsloggCandxxxxxD
\else
\ifnum#1=579007 %
\hatcurSPECharpsloggCandxxxxxE
\else
\ifnum#1=579008 %
\hatcurSPECharpsloggCandxxxxxF
\else
\ifnum#1=579009 %
\hatcurSPECharpsloggCandxxxxxG
\else
\ifnum#1=579010 %
\hatcurSPECharpsloggCandxxxxxU
\else
\ifnum#1=579014 %
\hatcurSPECharpsloggCandxxxxxH
\else
\ifnum#1=579015 %
\hatcurSPECharpsloggCandxxxxxI
\else
\ifnum#1=579036 %
\hatcurSPECharpsloggCandxxxxxJ
\else
\ifnum#1=579037 %
\hatcurSPECharpsloggCandxxxxxK
\else
\ifnum#1=579039 %
\hatcurSPECharpsloggCandxxxxxL
\else
\ifnum#1=579040 %
\hatcurSPECharpsloggCandxxxxxM
\else
\ifnum#1=579041 %
\hatcurSPECharpsloggCandxxxxxN
\else
\ifnum#1=579043 %
\hatcurSPECharpsloggCandxxxxxO
\else
\ifnum#1=579044 %
\hatcurSPECharpsloggCandxxxxxP
\else
\ifnum#1=579048 %
\hatcurSPECharpsloggCandxxxxxQ
\else
\ifnum#1=579050 %
\hatcurSPECharpsloggCandxxxxxR
\else
\ifnum#1=624002 %
\hatcurSPECharpsloggCandxxxxxS
\else
\ifnum#1=624003 %
\hatcurSPECharpsloggCandxxxxxT
\else
??????\fi
\fi
\fi
\fi
\fi
\fi
\fi
\fi
\fi
\fi
\fi
\fi
\fi
\fi
\fi
\fi
\fi
\fi
\fi
\fi
\fi
}
\newcommand{\hatcurSPECharpsNCand}[1]{\ifnum#1=578002 %
\hatcurSPECharpsNCandxxxxxA
\else
\ifnum#1=578003 %
\hatcurSPECharpsNCandxxxxxB
\else
\ifnum#1=578004 %
\hatcurSPECharpsNCandxxxxxC
\else
\ifnum#1=579001 %
\hatcurSPECharpsNCandxxxxxD
\else
\ifnum#1=579007 %
\hatcurSPECharpsNCandxxxxxE
\else
\ifnum#1=579008 %
\hatcurSPECharpsNCandxxxxxF
\else
\ifnum#1=579009 %
\hatcurSPECharpsNCandxxxxxG
\else
\ifnum#1=579010 %
\hatcurSPECharpsNCandxxxxxU
\else
\ifnum#1=579014 %
\hatcurSPECharpsNCandxxxxxH
\else
\ifnum#1=579015 %
\hatcurSPECharpsNCandxxxxxI
\else
\ifnum#1=579036 %
\hatcurSPECharpsNCandxxxxxJ
\else
\ifnum#1=579037 %
\hatcurSPECharpsNCandxxxxxK
\else
\ifnum#1=579039 %
\hatcurSPECharpsNCandxxxxxL
\else
\ifnum#1=579040 %
\hatcurSPECharpsNCandxxxxxM
\else
\ifnum#1=579041 %
\hatcurSPECharpsNCandxxxxxN
\else
\ifnum#1=579043 %
\hatcurSPECharpsNCandxxxxxO
\else
\ifnum#1=579044 %
\hatcurSPECharpsNCandxxxxxP
\else
\ifnum#1=579048 %
\hatcurSPECharpsNCandxxxxxQ
\else
\ifnum#1=579050 %
\hatcurSPECharpsNCandxxxxxR
\else
\ifnum#1=624002 %
\hatcurSPECharpsNCandxxxxxS
\else
\ifnum#1=624003 %
\hatcurSPECharpsNCandxxxxxT
\else
??????\fi
\fi
\fi
\fi
\fi
\fi
\fi
\fi
\fi
\fi
\fi
\fi
\fi
\fi
\fi
\fi
\fi
\fi
\fi
\fi
\fi
}
\newcommand{\hatcurSPECharpsRVGCand}[1]{\ifnum#1=578002 %
\hatcurSPECharpsRVGCandxxxxxA
\else
\ifnum#1=578003 %
\hatcurSPECharpsRVGCandxxxxxB
\else
\ifnum#1=578004 %
\hatcurSPECharpsRVGCandxxxxxC
\else
\ifnum#1=579001 %
\hatcurSPECharpsRVGCandxxxxxD
\else
\ifnum#1=579007 %
\hatcurSPECharpsRVGCandxxxxxE
\else
\ifnum#1=579008 %
\hatcurSPECharpsRVGCandxxxxxF
\else
\ifnum#1=579009 %
\hatcurSPECharpsRVGCandxxxxxG
\else
\ifnum#1=579010 %
\hatcurSPECharpsRVGCandxxxxxU
\else
\ifnum#1=579014 %
\hatcurSPECharpsRVGCandxxxxxH
\else
\ifnum#1=579015 %
\hatcurSPECharpsRVGCandxxxxxI
\else
\ifnum#1=579036 %
\hatcurSPECharpsRVGCandxxxxxJ
\else
\ifnum#1=579037 %
\hatcurSPECharpsRVGCandxxxxxK
\else
\ifnum#1=579039 %
\hatcurSPECharpsRVGCandxxxxxL
\else
\ifnum#1=579040 %
\hatcurSPECharpsRVGCandxxxxxM
\else
\ifnum#1=579041 %
\hatcurSPECharpsRVGCandxxxxxN
\else
\ifnum#1=579043 %
\hatcurSPECharpsRVGCandxxxxxO
\else
\ifnum#1=579044 %
\hatcurSPECharpsRVGCandxxxxxP
\else
\ifnum#1=579048 %
\hatcurSPECharpsRVGCandxxxxxQ
\else
\ifnum#1=579050 %
\hatcurSPECharpsRVGCandxxxxxR
\else
\ifnum#1=624002 %
\hatcurSPECharpsRVGCandxxxxxS
\else
\ifnum#1=624003 %
\hatcurSPECharpsRVGCandxxxxxT
\else
??????\fi
\fi
\fi
\fi
\fi
\fi
\fi
\fi
\fi
\fi
\fi
\fi
\fi
\fi
\fi
\fi
\fi
\fi
\fi
\fi
\fi
}
\newcommand{\hatcurSPECharpsRVKCand}[1]{\ifnum#1=578002 %
\hatcurSPECharpsRVKCandxxxxxA
\else
\ifnum#1=578003 %
\hatcurSPECharpsRVKCandxxxxxB
\else
\ifnum#1=578004 %
\hatcurSPECharpsRVKCandxxxxxC
\else
\ifnum#1=579001 %
\hatcurSPECharpsRVKCandxxxxxD
\else
\ifnum#1=579007 %
\hatcurSPECharpsRVKCandxxxxxE
\else
\ifnum#1=579008 %
\hatcurSPECharpsRVKCandxxxxxF
\else
\ifnum#1=579009 %
\hatcurSPECharpsRVKCandxxxxxG
\else
\ifnum#1=579010 %
\hatcurSPECharpsRVKCandxxxxxU
\else
\ifnum#1=579014 %
\hatcurSPECharpsRVKCandxxxxxH
\else
\ifnum#1=579015 %
\hatcurSPECharpsRVKCandxxxxxI
\else
\ifnum#1=579036 %
\hatcurSPECharpsRVKCandxxxxxJ
\else
\ifnum#1=579037 %
\hatcurSPECharpsRVKCandxxxxxK
\else
\ifnum#1=579039 %
\hatcurSPECharpsRVKCandxxxxxL
\else
\ifnum#1=579040 %
\hatcurSPECharpsRVKCandxxxxxM
\else
\ifnum#1=579041 %
\hatcurSPECharpsRVKCandxxxxxN
\else
\ifnum#1=579043 %
\hatcurSPECharpsRVKCandxxxxxO
\else
\ifnum#1=579044 %
\hatcurSPECharpsRVKCandxxxxxP
\else
\ifnum#1=579048 %
\hatcurSPECharpsRVKCandxxxxxQ
\else
\ifnum#1=579050 %
\hatcurSPECharpsRVKCandxxxxxR
\else
\ifnum#1=624002 %
\hatcurSPECharpsRVKCandxxxxxS
\else
\ifnum#1=624003 %
\hatcurSPECharpsRVKCandxxxxxT
\else
??????\fi
\fi
\fi
\fi
\fi
\fi
\fi
\fi
\fi
\fi
\fi
\fi
\fi
\fi
\fi
\fi
\fi
\fi
\fi
\fi
\fi
}
\newcommand{\hatcurSPECharpssnrangeCand}[1]{\ifnum#1=578002 %
\hatcurSPECharpssnrangeCandxxxxxA
\else
\ifnum#1=578003 %
\hatcurSPECharpssnrangeCandxxxxxB
\else
\ifnum#1=578004 %
\hatcurSPECharpssnrangeCandxxxxxC
\else
\ifnum#1=579001 %
\hatcurSPECharpssnrangeCandxxxxxD
\else
\ifnum#1=579007 %
\hatcurSPECharpssnrangeCandxxxxxE
\else
\ifnum#1=579008 %
\hatcurSPECharpssnrangeCandxxxxxF
\else
\ifnum#1=579009 %
\hatcurSPECharpssnrangeCandxxxxxG
\else
\ifnum#1=579010 %
\hatcurSPECharpssnrangeCandxxxxxU
\else
\ifnum#1=579014 %
\hatcurSPECharpssnrangeCandxxxxxH
\else
\ifnum#1=579015 %
\hatcurSPECharpssnrangeCandxxxxxI
\else
\ifnum#1=579036 %
\hatcurSPECharpssnrangeCandxxxxxJ
\else
\ifnum#1=579037 %
\hatcurSPECharpssnrangeCandxxxxxK
\else
\ifnum#1=579039 %
\hatcurSPECharpssnrangeCandxxxxxL
\else
\ifnum#1=579040 %
\hatcurSPECharpssnrangeCandxxxxxM
\else
\ifnum#1=579041 %
\hatcurSPECharpssnrangeCandxxxxxN
\else
\ifnum#1=579043 %
\hatcurSPECharpssnrangeCandxxxxxO
\else
\ifnum#1=579044 %
\hatcurSPECharpssnrangeCandxxxxxP
\else
\ifnum#1=579048 %
\hatcurSPECharpssnrangeCandxxxxxQ
\else
\ifnum#1=579050 %
\hatcurSPECharpssnrangeCandxxxxxR
\else
\ifnum#1=624002 %
\hatcurSPECharpssnrangeCandxxxxxS
\else
\ifnum#1=624003 %
\hatcurSPECharpssnrangeCandxxxxxT
\else
??????\fi
\fi
\fi
\fi
\fi
\fi
\fi
\fi
\fi
\fi
\fi
\fi
\fi
\fi
\fi
\fi
\fi
\fi
\fi
\fi
\fi
}
\newcommand{\hatcurSPECharpsTeffCand}[1]{\ifnum#1=578002 %
\hatcurSPECharpsTeffCandxxxxxA
\else
\ifnum#1=578003 %
\hatcurSPECharpsTeffCandxxxxxB
\else
\ifnum#1=578004 %
\hatcurSPECharpsTeffCandxxxxxC
\else
\ifnum#1=579001 %
\hatcurSPECharpsTeffCandxxxxxD
\else
\ifnum#1=579007 %
\hatcurSPECharpsTeffCandxxxxxE
\else
\ifnum#1=579008 %
\hatcurSPECharpsTeffCandxxxxxF
\else
\ifnum#1=579009 %
\hatcurSPECharpsTeffCandxxxxxG
\else
\ifnum#1=579010 %
\hatcurSPECharpsTeffCandxxxxxU
\else
\ifnum#1=579014 %
\hatcurSPECharpsTeffCandxxxxxH
\else
\ifnum#1=579015 %
\hatcurSPECharpsTeffCandxxxxxI
\else
\ifnum#1=579036 %
\hatcurSPECharpsTeffCandxxxxxJ
\else
\ifnum#1=579037 %
\hatcurSPECharpsTeffCandxxxxxK
\else
\ifnum#1=579039 %
\hatcurSPECharpsTeffCandxxxxxL
\else
\ifnum#1=579040 %
\hatcurSPECharpsTeffCandxxxxxM
\else
\ifnum#1=579041 %
\hatcurSPECharpsTeffCandxxxxxN
\else
\ifnum#1=579043 %
\hatcurSPECharpsTeffCandxxxxxO
\else
\ifnum#1=579044 %
\hatcurSPECharpsTeffCandxxxxxP
\else
\ifnum#1=579048 %
\hatcurSPECharpsTeffCandxxxxxQ
\else
\ifnum#1=579050 %
\hatcurSPECharpsTeffCandxxxxxR
\else
\ifnum#1=624002 %
\hatcurSPECharpsTeffCandxxxxxS
\else
\ifnum#1=624003 %
\hatcurSPECharpsTeffCandxxxxxT
\else
??????\fi
\fi
\fi
\fi
\fi
\fi
\fi
\fi
\fi
\fi
\fi
\fi
\fi
\fi
\fi
\fi
\fi
\fi
\fi
\fi
\fi
}
\newcommand{\hatcurSPECharpsvsiniCand}[1]{\ifnum#1=578002 %
\hatcurSPECharpsvsiniCandxxxxxA
\else
\ifnum#1=578003 %
\hatcurSPECharpsvsiniCandxxxxxB
\else
\ifnum#1=578004 %
\hatcurSPECharpsvsiniCandxxxxxC
\else
\ifnum#1=579001 %
\hatcurSPECharpsvsiniCandxxxxxD
\else
\ifnum#1=579007 %
\hatcurSPECharpsvsiniCandxxxxxE
\else
\ifnum#1=579008 %
\hatcurSPECharpsvsiniCandxxxxxF
\else
\ifnum#1=579009 %
\hatcurSPECharpsvsiniCandxxxxxG
\else
\ifnum#1=579010 %
\hatcurSPECharpsvsiniCandxxxxxU
\else
\ifnum#1=579014 %
\hatcurSPECharpsvsiniCandxxxxxH
\else
\ifnum#1=579015 %
\hatcurSPECharpsvsiniCandxxxxxI
\else
\ifnum#1=579036 %
\hatcurSPECharpsvsiniCandxxxxxJ
\else
\ifnum#1=579037 %
\hatcurSPECharpsvsiniCandxxxxxK
\else
\ifnum#1=579039 %
\hatcurSPECharpsvsiniCandxxxxxL
\else
\ifnum#1=579040 %
\hatcurSPECharpsvsiniCandxxxxxM
\else
\ifnum#1=579041 %
\hatcurSPECharpsvsiniCandxxxxxN
\else
\ifnum#1=579043 %
\hatcurSPECharpsvsiniCandxxxxxO
\else
\ifnum#1=579044 %
\hatcurSPECharpsvsiniCandxxxxxP
\else
\ifnum#1=579048 %
\hatcurSPECharpsvsiniCandxxxxxQ
\else
\ifnum#1=579050 %
\hatcurSPECharpsvsiniCandxxxxxR
\else
\ifnum#1=624002 %
\hatcurSPECharpsvsiniCandxxxxxS
\else
\ifnum#1=624003 %
\hatcurSPECharpsvsiniCandxxxxxT
\else
??????\fi
\fi
\fi
\fi
\fi
\fi
\fi
\fi
\fi
\fi
\fi
\fi
\fi
\fi
\fi
\fi
\fi
\fi
\fi
\fi
\fi
}
\newcommand{\hatcurSPECharpszfehCand}[1]{\ifnum#1=578002 %
\hatcurSPECharpszfehCandxxxxxA
\else
\ifnum#1=578003 %
\hatcurSPECharpszfehCandxxxxxB
\else
\ifnum#1=578004 %
\hatcurSPECharpszfehCandxxxxxC
\else
\ifnum#1=579001 %
\hatcurSPECharpszfehCandxxxxxD
\else
\ifnum#1=579007 %
\hatcurSPECharpszfehCandxxxxxE
\else
\ifnum#1=579008 %
\hatcurSPECharpszfehCandxxxxxF
\else
\ifnum#1=579009 %
\hatcurSPECharpszfehCandxxxxxG
\else
\ifnum#1=579010 %
\hatcurSPECharpszfehCandxxxxxU
\else
\ifnum#1=579014 %
\hatcurSPECharpszfehCandxxxxxH
\else
\ifnum#1=579015 %
\hatcurSPECharpszfehCandxxxxxI
\else
\ifnum#1=579036 %
\hatcurSPECharpszfehCandxxxxxJ
\else
\ifnum#1=579037 %
\hatcurSPECharpszfehCandxxxxxK
\else
\ifnum#1=579039 %
\hatcurSPECharpszfehCandxxxxxL
\else
\ifnum#1=579040 %
\hatcurSPECharpszfehCandxxxxxM
\else
\ifnum#1=579041 %
\hatcurSPECharpszfehCandxxxxxN
\else
\ifnum#1=579043 %
\hatcurSPECharpszfehCandxxxxxO
\else
\ifnum#1=579044 %
\hatcurSPECharpszfehCandxxxxxP
\else
\ifnum#1=579048 %
\hatcurSPECharpszfehCandxxxxxQ
\else
\ifnum#1=579050 %
\hatcurSPECharpszfehCandxxxxxR
\else
\ifnum#1=624002 %
\hatcurSPECharpszfehCandxxxxxS
\else
\ifnum#1=624003 %
\hatcurSPECharpszfehCandxxxxxT
\else
??????\fi
\fi
\fi
\fi
\fi
\fi
\fi
\fi
\fi
\fi
\fi
\fi
\fi
\fi
\fi
\fi
\fi
\fi
\fi
\fi
\fi
}
\newcommand{\hatcurSPECpfsdaterangelowresCand}[1]{\ifnum#1=578002 %
\hatcurSPECpfsdaterangelowresCandxxxxxA
\else
\ifnum#1=578003 %
\hatcurSPECpfsdaterangelowresCandxxxxxB
\else
\ifnum#1=578004 %
\hatcurSPECpfsdaterangelowresCandxxxxxC
\else
\ifnum#1=579001 %
\hatcurSPECpfsdaterangelowresCandxxxxxD
\else
\ifnum#1=579007 %
\hatcurSPECpfsdaterangelowresCandxxxxxE
\else
\ifnum#1=579008 %
\hatcurSPECpfsdaterangelowresCandxxxxxF
\else
\ifnum#1=579009 %
\hatcurSPECpfsdaterangelowresCandxxxxxG
\else
\ifnum#1=579010 %
\hatcurSPECpfsdaterangelowresCandxxxxxU
\else
\ifnum#1=579014 %
\hatcurSPECpfsdaterangelowresCandxxxxxH
\else
\ifnum#1=579015 %
\hatcurSPECpfsdaterangelowresCandxxxxxI
\else
\ifnum#1=579036 %
\hatcurSPECpfsdaterangelowresCandxxxxxJ
\else
\ifnum#1=579037 %
\hatcurSPECpfsdaterangelowresCandxxxxxK
\else
\ifnum#1=579039 %
\hatcurSPECpfsdaterangelowresCandxxxxxL
\else
\ifnum#1=579040 %
\hatcurSPECpfsdaterangelowresCandxxxxxM
\else
\ifnum#1=579041 %
\hatcurSPECpfsdaterangelowresCandxxxxxN
\else
\ifnum#1=579043 %
\hatcurSPECpfsdaterangelowresCandxxxxxO
\else
\ifnum#1=579044 %
\hatcurSPECpfsdaterangelowresCandxxxxxP
\else
\ifnum#1=579048 %
\hatcurSPECpfsdaterangelowresCandxxxxxQ
\else
\ifnum#1=579050 %
\hatcurSPECpfsdaterangelowresCandxxxxxR
\else
\ifnum#1=624002 %
\hatcurSPECpfsdaterangelowresCandxxxxxS
\else
\ifnum#1=624003 %
\hatcurSPECpfsdaterangelowresCandxxxxxT
\else
??????\fi
\fi
\fi
\fi
\fi
\fi
\fi
\fi
\fi
\fi
\fi
\fi
\fi
\fi
\fi
\fi
\fi
\fi
\fi
\fi
\fi
}
\newcommand{\hatcurSPECpfsloggCand}[1]{\ifnum#1=578002 %
\hatcurSPECpfsloggCandxxxxxA
\else
\ifnum#1=578003 %
\hatcurSPECpfsloggCandxxxxxB
\else
\ifnum#1=578004 %
\hatcurSPECpfsloggCandxxxxxC
\else
\ifnum#1=579001 %
\hatcurSPECpfsloggCandxxxxxD
\else
\ifnum#1=579007 %
\hatcurSPECpfsloggCandxxxxxE
\else
\ifnum#1=579008 %
\hatcurSPECpfsloggCandxxxxxF
\else
\ifnum#1=579009 %
\hatcurSPECpfsloggCandxxxxxG
\else
\ifnum#1=579010 %
\hatcurSPECpfsloggCandxxxxxU
\else
\ifnum#1=579014 %
\hatcurSPECpfsloggCandxxxxxH
\else
\ifnum#1=579015 %
\hatcurSPECpfsloggCandxxxxxI
\else
\ifnum#1=579036 %
\hatcurSPECpfsloggCandxxxxxJ
\else
\ifnum#1=579037 %
\hatcurSPECpfsloggCandxxxxxK
\else
\ifnum#1=579039 %
\hatcurSPECpfsloggCandxxxxxL
\else
\ifnum#1=579040 %
\hatcurSPECpfsloggCandxxxxxM
\else
\ifnum#1=579041 %
\hatcurSPECpfsloggCandxxxxxN
\else
\ifnum#1=579043 %
\hatcurSPECpfsloggCandxxxxxO
\else
\ifnum#1=579044 %
\hatcurSPECpfsloggCandxxxxxP
\else
\ifnum#1=579048 %
\hatcurSPECpfsloggCandxxxxxQ
\else
\ifnum#1=579050 %
\hatcurSPECpfsloggCandxxxxxR
\else
\ifnum#1=624002 %
\hatcurSPECpfsloggCandxxxxxS
\else
\ifnum#1=624003 %
\hatcurSPECpfsloggCandxxxxxT
\else
??????\fi
\fi
\fi
\fi
\fi
\fi
\fi
\fi
\fi
\fi
\fi
\fi
\fi
\fi
\fi
\fi
\fi
\fi
\fi
\fi
\fi
}
\newcommand{\hatcurSPECpfsNCand}[1]{\ifnum#1=578002 %
\hatcurSPECpfsNCandxxxxxA
\else
\ifnum#1=578003 %
\hatcurSPECpfsNCandxxxxxB
\else
\ifnum#1=578004 %
\hatcurSPECpfsNCandxxxxxC
\else
\ifnum#1=579001 %
\hatcurSPECpfsNCandxxxxxD
\else
\ifnum#1=579007 %
\hatcurSPECpfsNCandxxxxxE
\else
\ifnum#1=579008 %
\hatcurSPECpfsNCandxxxxxF
\else
\ifnum#1=579009 %
\hatcurSPECpfsNCandxxxxxG
\else
\ifnum#1=579010 %
\hatcurSPECpfsNCandxxxxxU
\else
\ifnum#1=579014 %
\hatcurSPECpfsNCandxxxxxH
\else
\ifnum#1=579015 %
\hatcurSPECpfsNCandxxxxxI
\else
\ifnum#1=579036 %
\hatcurSPECpfsNCandxxxxxJ
\else
\ifnum#1=579037 %
\hatcurSPECpfsNCandxxxxxK
\else
\ifnum#1=579039 %
\hatcurSPECpfsNCandxxxxxL
\else
\ifnum#1=579040 %
\hatcurSPECpfsNCandxxxxxM
\else
\ifnum#1=579041 %
\hatcurSPECpfsNCandxxxxxN
\else
\ifnum#1=579043 %
\hatcurSPECpfsNCandxxxxxO
\else
\ifnum#1=579044 %
\hatcurSPECpfsNCandxxxxxP
\else
\ifnum#1=579048 %
\hatcurSPECpfsNCandxxxxxQ
\else
\ifnum#1=579050 %
\hatcurSPECpfsNCandxxxxxR
\else
\ifnum#1=624002 %
\hatcurSPECpfsNCandxxxxxS
\else
\ifnum#1=624003 %
\hatcurSPECpfsNCandxxxxxT
\else
??????\fi
\fi
\fi
\fi
\fi
\fi
\fi
\fi
\fi
\fi
\fi
\fi
\fi
\fi
\fi
\fi
\fi
\fi
\fi
\fi
\fi
}
\newcommand{\hatcurSPECpfsRVGCand}[1]{\ifnum#1=578002 %
\hatcurSPECpfsRVGCandxxxxxA
\else
\ifnum#1=578003 %
\hatcurSPECpfsRVGCandxxxxxB
\else
\ifnum#1=578004 %
\hatcurSPECpfsRVGCandxxxxxC
\else
\ifnum#1=579001 %
\hatcurSPECpfsRVGCandxxxxxD
\else
\ifnum#1=579007 %
\hatcurSPECpfsRVGCandxxxxxE
\else
\ifnum#1=579008 %
\hatcurSPECpfsRVGCandxxxxxF
\else
\ifnum#1=579009 %
\hatcurSPECpfsRVGCandxxxxxG
\else
\ifnum#1=579010 %
\hatcurSPECpfsRVGCandxxxxxU
\else
\ifnum#1=579014 %
\hatcurSPECpfsRVGCandxxxxxH
\else
\ifnum#1=579015 %
\hatcurSPECpfsRVGCandxxxxxI
\else
\ifnum#1=579036 %
\hatcurSPECpfsRVGCandxxxxxJ
\else
\ifnum#1=579037 %
\hatcurSPECpfsRVGCandxxxxxK
\else
\ifnum#1=579039 %
\hatcurSPECpfsRVGCandxxxxxL
\else
\ifnum#1=579040 %
\hatcurSPECpfsRVGCandxxxxxM
\else
\ifnum#1=579041 %
\hatcurSPECpfsRVGCandxxxxxN
\else
\ifnum#1=579043 %
\hatcurSPECpfsRVGCandxxxxxO
\else
\ifnum#1=579044 %
\hatcurSPECpfsRVGCandxxxxxP
\else
\ifnum#1=579048 %
\hatcurSPECpfsRVGCandxxxxxQ
\else
\ifnum#1=579050 %
\hatcurSPECpfsRVGCandxxxxxR
\else
\ifnum#1=624002 %
\hatcurSPECpfsRVGCandxxxxxS
\else
\ifnum#1=624003 %
\hatcurSPECpfsRVGCandxxxxxT
\else
??????\fi
\fi
\fi
\fi
\fi
\fi
\fi
\fi
\fi
\fi
\fi
\fi
\fi
\fi
\fi
\fi
\fi
\fi
\fi
\fi
\fi
}
\newcommand{\hatcurSPECpfsRVKCand}[1]{\ifnum#1=578002 %
\hatcurSPECpfsRVKCandxxxxxA
\else
\ifnum#1=578003 %
\hatcurSPECpfsRVKCandxxxxxB
\else
\ifnum#1=578004 %
\hatcurSPECpfsRVKCandxxxxxC
\else
\ifnum#1=579001 %
\hatcurSPECpfsRVKCandxxxxxD
\else
\ifnum#1=579007 %
\hatcurSPECpfsRVKCandxxxxxE
\else
\ifnum#1=579008 %
\hatcurSPECpfsRVKCandxxxxxF
\else
\ifnum#1=579009 %
\hatcurSPECpfsRVKCandxxxxxG
\else
\ifnum#1=579010 %
\hatcurSPECpfsRVKCandxxxxxU
\else
\ifnum#1=579014 %
\hatcurSPECpfsRVKCandxxxxxH
\else
\ifnum#1=579015 %
\hatcurSPECpfsRVKCandxxxxxI
\else
\ifnum#1=579036 %
\hatcurSPECpfsRVKCandxxxxxJ
\else
\ifnum#1=579037 %
\hatcurSPECpfsRVKCandxxxxxK
\else
\ifnum#1=579039 %
\hatcurSPECpfsRVKCandxxxxxL
\else
\ifnum#1=579040 %
\hatcurSPECpfsRVKCandxxxxxM
\else
\ifnum#1=579041 %
\hatcurSPECpfsRVKCandxxxxxN
\else
\ifnum#1=579043 %
\hatcurSPECpfsRVKCandxxxxxO
\else
\ifnum#1=579044 %
\hatcurSPECpfsRVKCandxxxxxP
\else
\ifnum#1=579048 %
\hatcurSPECpfsRVKCandxxxxxQ
\else
\ifnum#1=579050 %
\hatcurSPECpfsRVKCandxxxxxR
\else
\ifnum#1=624002 %
\hatcurSPECpfsRVKCandxxxxxS
\else
\ifnum#1=624003 %
\hatcurSPECpfsRVKCandxxxxxT
\else
??????\fi
\fi
\fi
\fi
\fi
\fi
\fi
\fi
\fi
\fi
\fi
\fi
\fi
\fi
\fi
\fi
\fi
\fi
\fi
\fi
\fi
}
\newcommand{\hatcurSPECpfssnrangeCand}[1]{\ifnum#1=578002 %
\hatcurSPECpfssnrangeCandxxxxxA
\else
\ifnum#1=578003 %
\hatcurSPECpfssnrangeCandxxxxxB
\else
\ifnum#1=578004 %
\hatcurSPECpfssnrangeCandxxxxxC
\else
\ifnum#1=579001 %
\hatcurSPECpfssnrangeCandxxxxxD
\else
\ifnum#1=579007 %
\hatcurSPECpfssnrangeCandxxxxxE
\else
\ifnum#1=579008 %
\hatcurSPECpfssnrangeCandxxxxxF
\else
\ifnum#1=579009 %
\hatcurSPECpfssnrangeCandxxxxxG
\else
\ifnum#1=579010 %
\hatcurSPECpfssnrangeCandxxxxxU
\else
\ifnum#1=579014 %
\hatcurSPECpfssnrangeCandxxxxxH
\else
\ifnum#1=579015 %
\hatcurSPECpfssnrangeCandxxxxxI
\else
\ifnum#1=579036 %
\hatcurSPECpfssnrangeCandxxxxxJ
\else
\ifnum#1=579037 %
\hatcurSPECpfssnrangeCandxxxxxK
\else
\ifnum#1=579039 %
\hatcurSPECpfssnrangeCandxxxxxL
\else
\ifnum#1=579040 %
\hatcurSPECpfssnrangeCandxxxxxM
\else
\ifnum#1=579041 %
\hatcurSPECpfssnrangeCandxxxxxN
\else
\ifnum#1=579043 %
\hatcurSPECpfssnrangeCandxxxxxO
\else
\ifnum#1=579044 %
\hatcurSPECpfssnrangeCandxxxxxP
\else
\ifnum#1=579048 %
\hatcurSPECpfssnrangeCandxxxxxQ
\else
\ifnum#1=579050 %
\hatcurSPECpfssnrangeCandxxxxxR
\else
\ifnum#1=624002 %
\hatcurSPECpfssnrangeCandxxxxxS
\else
\ifnum#1=624003 %
\hatcurSPECpfssnrangeCandxxxxxT
\else
??????\fi
\fi
\fi
\fi
\fi
\fi
\fi
\fi
\fi
\fi
\fi
\fi
\fi
\fi
\fi
\fi
\fi
\fi
\fi
\fi
\fi
}
\newcommand{\hatcurSPECpfsTeffCand}[1]{\ifnum#1=578002 %
\hatcurSPECpfsTeffCandxxxxxA
\else
\ifnum#1=578003 %
\hatcurSPECpfsTeffCandxxxxxB
\else
\ifnum#1=578004 %
\hatcurSPECpfsTeffCandxxxxxC
\else
\ifnum#1=579001 %
\hatcurSPECpfsTeffCandxxxxxD
\else
\ifnum#1=579007 %
\hatcurSPECpfsTeffCandxxxxxE
\else
\ifnum#1=579008 %
\hatcurSPECpfsTeffCandxxxxxF
\else
\ifnum#1=579009 %
\hatcurSPECpfsTeffCandxxxxxG
\else
\ifnum#1=579010 %
\hatcurSPECpfsTeffCandxxxxxU
\else
\ifnum#1=579014 %
\hatcurSPECpfsTeffCandxxxxxH
\else
\ifnum#1=579015 %
\hatcurSPECpfsTeffCandxxxxxI
\else
\ifnum#1=579036 %
\hatcurSPECpfsTeffCandxxxxxJ
\else
\ifnum#1=579037 %
\hatcurSPECpfsTeffCandxxxxxK
\else
\ifnum#1=579039 %
\hatcurSPECpfsTeffCandxxxxxL
\else
\ifnum#1=579040 %
\hatcurSPECpfsTeffCandxxxxxM
\else
\ifnum#1=579041 %
\hatcurSPECpfsTeffCandxxxxxN
\else
\ifnum#1=579043 %
\hatcurSPECpfsTeffCandxxxxxO
\else
\ifnum#1=579044 %
\hatcurSPECpfsTeffCandxxxxxP
\else
\ifnum#1=579048 %
\hatcurSPECpfsTeffCandxxxxxQ
\else
\ifnum#1=579050 %
\hatcurSPECpfsTeffCandxxxxxR
\else
\ifnum#1=624002 %
\hatcurSPECpfsTeffCandxxxxxS
\else
\ifnum#1=624003 %
\hatcurSPECpfsTeffCandxxxxxT
\else
??????\fi
\fi
\fi
\fi
\fi
\fi
\fi
\fi
\fi
\fi
\fi
\fi
\fi
\fi
\fi
\fi
\fi
\fi
\fi
\fi
\fi
}
\newcommand{\hatcurSPECpfsvsiniCand}[1]{\ifnum#1=578002 %
\hatcurSPECpfsvsiniCandxxxxxA
\else
\ifnum#1=578003 %
\hatcurSPECpfsvsiniCandxxxxxB
\else
\ifnum#1=578004 %
\hatcurSPECpfsvsiniCandxxxxxC
\else
\ifnum#1=579001 %
\hatcurSPECpfsvsiniCandxxxxxD
\else
\ifnum#1=579007 %
\hatcurSPECpfsvsiniCandxxxxxE
\else
\ifnum#1=579008 %
\hatcurSPECpfsvsiniCandxxxxxF
\else
\ifnum#1=579009 %
\hatcurSPECpfsvsiniCandxxxxxG
\else
\ifnum#1=579010 %
\hatcurSPECpfsvsiniCandxxxxxU
\else
\ifnum#1=579014 %
\hatcurSPECpfsvsiniCandxxxxxH
\else
\ifnum#1=579015 %
\hatcurSPECpfsvsiniCandxxxxxI
\else
\ifnum#1=579036 %
\hatcurSPECpfsvsiniCandxxxxxJ
\else
\ifnum#1=579037 %
\hatcurSPECpfsvsiniCandxxxxxK
\else
\ifnum#1=579039 %
\hatcurSPECpfsvsiniCandxxxxxL
\else
\ifnum#1=579040 %
\hatcurSPECpfsvsiniCandxxxxxM
\else
\ifnum#1=579041 %
\hatcurSPECpfsvsiniCandxxxxxN
\else
\ifnum#1=579043 %
\hatcurSPECpfsvsiniCandxxxxxO
\else
\ifnum#1=579044 %
\hatcurSPECpfsvsiniCandxxxxxP
\else
\ifnum#1=579048 %
\hatcurSPECpfsvsiniCandxxxxxQ
\else
\ifnum#1=579050 %
\hatcurSPECpfsvsiniCandxxxxxR
\else
\ifnum#1=624002 %
\hatcurSPECpfsvsiniCandxxxxxS
\else
\ifnum#1=624003 %
\hatcurSPECpfsvsiniCandxxxxxT
\else
??????\fi
\fi
\fi
\fi
\fi
\fi
\fi
\fi
\fi
\fi
\fi
\fi
\fi
\fi
\fi
\fi
\fi
\fi
\fi
\fi
\fi
}
\newcommand{\hatcurSPECpfszfehCand}[1]{\ifnum#1=578002 %
\hatcurSPECpfszfehCandxxxxxA
\else
\ifnum#1=578003 %
\hatcurSPECpfszfehCandxxxxxB
\else
\ifnum#1=578004 %
\hatcurSPECpfszfehCandxxxxxC
\else
\ifnum#1=579001 %
\hatcurSPECpfszfehCandxxxxxD
\else
\ifnum#1=579007 %
\hatcurSPECpfszfehCandxxxxxE
\else
\ifnum#1=579008 %
\hatcurSPECpfszfehCandxxxxxF
\else
\ifnum#1=579009 %
\hatcurSPECpfszfehCandxxxxxG
\else
\ifnum#1=579010 %
\hatcurSPECpfszfehCandxxxxxU
\else
\ifnum#1=579014 %
\hatcurSPECpfszfehCandxxxxxH
\else
\ifnum#1=579015 %
\hatcurSPECpfszfehCandxxxxxI
\else
\ifnum#1=579036 %
\hatcurSPECpfszfehCandxxxxxJ
\else
\ifnum#1=579037 %
\hatcurSPECpfszfehCandxxxxxK
\else
\ifnum#1=579039 %
\hatcurSPECpfszfehCandxxxxxL
\else
\ifnum#1=579040 %
\hatcurSPECpfszfehCandxxxxxM
\else
\ifnum#1=579041 %
\hatcurSPECpfszfehCandxxxxxN
\else
\ifnum#1=579043 %
\hatcurSPECpfszfehCandxxxxxO
\else
\ifnum#1=579044 %
\hatcurSPECpfszfehCandxxxxxP
\else
\ifnum#1=579048 %
\hatcurSPECpfszfehCandxxxxxQ
\else
\ifnum#1=579050 %
\hatcurSPECpfszfehCandxxxxxR
\else
\ifnum#1=624002 %
\hatcurSPECpfszfehCandxxxxxS
\else
\ifnum#1=624003 %
\hatcurSPECpfszfehCandxxxxxT
\else
??????\fi
\fi
\fi
\fi
\fi
\fi
\fi
\fi
\fi
\fi
\fi
\fi
\fi
\fi
\fi
\fi
\fi
\fi
\fi
\fi
\fi
}
\newcommand{\hatcurSPECWiFeSdaterangelowresCand}[1]{\ifnum#1=578002 %
\hatcurSPECWiFeSdaterangelowresCandxxxxxA
\else
\ifnum#1=578003 %
\hatcurSPECWiFeSdaterangelowresCandxxxxxB
\else
\ifnum#1=578004 %
\hatcurSPECWiFeSdaterangelowresCandxxxxxC
\else
\ifnum#1=579001 %
\hatcurSPECWiFeSdaterangelowresCandxxxxxD
\else
\ifnum#1=579007 %
\hatcurSPECWiFeSdaterangelowresCandxxxxxE
\else
\ifnum#1=579008 %
\hatcurSPECWiFeSdaterangelowresCandxxxxxF
\else
\ifnum#1=579009 %
\hatcurSPECWiFeSdaterangelowresCandxxxxxG
\else
\ifnum#1=579010 %
\hatcurSPECWiFeSdaterangelowresCandxxxxxU
\else
\ifnum#1=579014 %
\hatcurSPECWiFeSdaterangelowresCandxxxxxH
\else
\ifnum#1=579015 %
\hatcurSPECWiFeSdaterangelowresCandxxxxxI
\else
\ifnum#1=579036 %
\hatcurSPECWiFeSdaterangelowresCandxxxxxJ
\else
\ifnum#1=579037 %
\hatcurSPECWiFeSdaterangelowresCandxxxxxK
\else
\ifnum#1=579039 %
\hatcurSPECWiFeSdaterangelowresCandxxxxxL
\else
\ifnum#1=579040 %
\hatcurSPECWiFeSdaterangelowresCandxxxxxM
\else
\ifnum#1=579041 %
\hatcurSPECWiFeSdaterangelowresCandxxxxxN
\else
\ifnum#1=579043 %
\hatcurSPECWiFeSdaterangelowresCandxxxxxO
\else
\ifnum#1=579044 %
\hatcurSPECWiFeSdaterangelowresCandxxxxxP
\else
\ifnum#1=579048 %
\hatcurSPECWiFeSdaterangelowresCandxxxxxQ
\else
\ifnum#1=579050 %
\hatcurSPECWiFeSdaterangelowresCandxxxxxR
\else
\ifnum#1=624002 %
\hatcurSPECWiFeSdaterangelowresCandxxxxxS
\else
\ifnum#1=624003 %
\hatcurSPECWiFeSdaterangelowresCandxxxxxT
\else
??????\fi
\fi
\fi
\fi
\fi
\fi
\fi
\fi
\fi
\fi
\fi
\fi
\fi
\fi
\fi
\fi
\fi
\fi
\fi
\fi
\fi
}
\newcommand{\hatcurSPECWiFeSdaterangemedresCand}[1]{\ifnum#1=578002 %
\hatcurSPECWiFeSdaterangemedresCandxxxxxA
\else
\ifnum#1=578003 %
\hatcurSPECWiFeSdaterangemedresCandxxxxxB
\else
\ifnum#1=578004 %
\hatcurSPECWiFeSdaterangemedresCandxxxxxC
\else
\ifnum#1=579001 %
\hatcurSPECWiFeSdaterangemedresCandxxxxxD
\else
\ifnum#1=579007 %
\hatcurSPECWiFeSdaterangemedresCandxxxxxE
\else
\ifnum#1=579008 %
\hatcurSPECWiFeSdaterangemedresCandxxxxxF
\else
\ifnum#1=579009 %
\hatcurSPECWiFeSdaterangemedresCandxxxxxG
\else
\ifnum#1=579010 %
\hatcurSPECWiFeSdaterangemedresCandxxxxxU
\else
\ifnum#1=579014 %
\hatcurSPECWiFeSdaterangemedresCandxxxxxH
\else
\ifnum#1=579015 %
\hatcurSPECWiFeSdaterangemedresCandxxxxxI
\else
\ifnum#1=579036 %
\hatcurSPECWiFeSdaterangemedresCandxxxxxJ
\else
\ifnum#1=579037 %
\hatcurSPECWiFeSdaterangemedresCandxxxxxK
\else
\ifnum#1=579039 %
\hatcurSPECWiFeSdaterangemedresCandxxxxxL
\else
\ifnum#1=579040 %
\hatcurSPECWiFeSdaterangemedresCandxxxxxM
\else
\ifnum#1=579041 %
\hatcurSPECWiFeSdaterangemedresCandxxxxxN
\else
\ifnum#1=579043 %
\hatcurSPECWiFeSdaterangemedresCandxxxxxO
\else
\ifnum#1=579044 %
\hatcurSPECWiFeSdaterangemedresCandxxxxxP
\else
\ifnum#1=579048 %
\hatcurSPECWiFeSdaterangemedresCandxxxxxQ
\else
\ifnum#1=579050 %
\hatcurSPECWiFeSdaterangemedresCandxxxxxR
\else
\ifnum#1=624002 %
\hatcurSPECWiFeSdaterangemedresCandxxxxxS
\else
\ifnum#1=624003 %
\hatcurSPECWiFeSdaterangemedresCandxxxxxT
\else
??????\fi
\fi
\fi
\fi
\fi
\fi
\fi
\fi
\fi
\fi
\fi
\fi
\fi
\fi
\fi
\fi
\fi
\fi
\fi
\fi
\fi
}
\newcommand{\hatcurSPECWiFeSloggCand}[1]{\ifnum#1=578002 %
\hatcurSPECWiFeSloggCandxxxxxA
\else
\ifnum#1=578003 %
\hatcurSPECWiFeSloggCandxxxxxB
\else
\ifnum#1=578004 %
\hatcurSPECWiFeSloggCandxxxxxC
\else
\ifnum#1=579001 %
\hatcurSPECWiFeSloggCandxxxxxD
\else
\ifnum#1=579007 %
\hatcurSPECWiFeSloggCandxxxxxE
\else
\ifnum#1=579008 %
\hatcurSPECWiFeSloggCandxxxxxF
\else
\ifnum#1=579009 %
\hatcurSPECWiFeSloggCandxxxxxG
\else
\ifnum#1=579010 %
\hatcurSPECWiFeSloggCandxxxxxU
\else
\ifnum#1=579014 %
\hatcurSPECWiFeSloggCandxxxxxH
\else
\ifnum#1=579015 %
\hatcurSPECWiFeSloggCandxxxxxI
\else
\ifnum#1=579036 %
\hatcurSPECWiFeSloggCandxxxxxJ
\else
\ifnum#1=579037 %
\hatcurSPECWiFeSloggCandxxxxxK
\else
\ifnum#1=579039 %
\hatcurSPECWiFeSloggCandxxxxxL
\else
\ifnum#1=579040 %
\hatcurSPECWiFeSloggCandxxxxxM
\else
\ifnum#1=579041 %
\hatcurSPECWiFeSloggCandxxxxxN
\else
\ifnum#1=579043 %
\hatcurSPECWiFeSloggCandxxxxxO
\else
\ifnum#1=579044 %
\hatcurSPECWiFeSloggCandxxxxxP
\else
\ifnum#1=579048 %
\hatcurSPECWiFeSloggCandxxxxxQ
\else
\ifnum#1=579050 %
\hatcurSPECWiFeSloggCandxxxxxR
\else
\ifnum#1=624002 %
\hatcurSPECWiFeSloggCandxxxxxS
\else
\ifnum#1=624003 %
\hatcurSPECWiFeSloggCandxxxxxT
\else
??????\fi
\fi
\fi
\fi
\fi
\fi
\fi
\fi
\fi
\fi
\fi
\fi
\fi
\fi
\fi
\fi
\fi
\fi
\fi
\fi
\fi
}
\newcommand{\hatcurSPECWiFeSNlowresCand}[1]{\ifnum#1=578002 %
\hatcurSPECWiFeSNlowresCandxxxxxA
\else
\ifnum#1=578003 %
\hatcurSPECWiFeSNlowresCandxxxxxB
\else
\ifnum#1=578004 %
\hatcurSPECWiFeSNlowresCandxxxxxC
\else
\ifnum#1=579001 %
\hatcurSPECWiFeSNlowresCandxxxxxD
\else
\ifnum#1=579007 %
\hatcurSPECWiFeSNlowresCandxxxxxE
\else
\ifnum#1=579008 %
\hatcurSPECWiFeSNlowresCandxxxxxF
\else
\ifnum#1=579009 %
\hatcurSPECWiFeSNlowresCandxxxxxG
\else
\ifnum#1=579010 %
\hatcurSPECWiFeSNlowresCandxxxxxU
\else
\ifnum#1=579014 %
\hatcurSPECWiFeSNlowresCandxxxxxH
\else
\ifnum#1=579015 %
\hatcurSPECWiFeSNlowresCandxxxxxI
\else
\ifnum#1=579036 %
\hatcurSPECWiFeSNlowresCandxxxxxJ
\else
\ifnum#1=579037 %
\hatcurSPECWiFeSNlowresCandxxxxxK
\else
\ifnum#1=579039 %
\hatcurSPECWiFeSNlowresCandxxxxxL
\else
\ifnum#1=579040 %
\hatcurSPECWiFeSNlowresCandxxxxxM
\else
\ifnum#1=579041 %
\hatcurSPECWiFeSNlowresCandxxxxxN
\else
\ifnum#1=579043 %
\hatcurSPECWiFeSNlowresCandxxxxxO
\else
\ifnum#1=579044 %
\hatcurSPECWiFeSNlowresCandxxxxxP
\else
\ifnum#1=579048 %
\hatcurSPECWiFeSNlowresCandxxxxxQ
\else
\ifnum#1=579050 %
\hatcurSPECWiFeSNlowresCandxxxxxR
\else
\ifnum#1=624002 %
\hatcurSPECWiFeSNlowresCandxxxxxS
\else
\ifnum#1=624003 %
\hatcurSPECWiFeSNlowresCandxxxxxT
\else
??????\fi
\fi
\fi
\fi
\fi
\fi
\fi
\fi
\fi
\fi
\fi
\fi
\fi
\fi
\fi
\fi
\fi
\fi
\fi
\fi
\fi
}
\newcommand{\hatcurSPECWiFeSNmedresCand}[1]{\ifnum#1=578002 %
\hatcurSPECWiFeSNmedresCandxxxxxA
\else
\ifnum#1=578003 %
\hatcurSPECWiFeSNmedresCandxxxxxB
\else
\ifnum#1=578004 %
\hatcurSPECWiFeSNmedresCandxxxxxC
\else
\ifnum#1=579001 %
\hatcurSPECWiFeSNmedresCandxxxxxD
\else
\ifnum#1=579007 %
\hatcurSPECWiFeSNmedresCandxxxxxE
\else
\ifnum#1=579008 %
\hatcurSPECWiFeSNmedresCandxxxxxF
\else
\ifnum#1=579009 %
\hatcurSPECWiFeSNmedresCandxxxxxG
\else
\ifnum#1=579010 %
\hatcurSPECWiFeSNmedresCandxxxxxU
\else
\ifnum#1=579014 %
\hatcurSPECWiFeSNmedresCandxxxxxH
\else
\ifnum#1=579015 %
\hatcurSPECWiFeSNmedresCandxxxxxI
\else
\ifnum#1=579036 %
\hatcurSPECWiFeSNmedresCandxxxxxJ
\else
\ifnum#1=579037 %
\hatcurSPECWiFeSNmedresCandxxxxxK
\else
\ifnum#1=579039 %
\hatcurSPECWiFeSNmedresCandxxxxxL
\else
\ifnum#1=579040 %
\hatcurSPECWiFeSNmedresCandxxxxxM
\else
\ifnum#1=579041 %
\hatcurSPECWiFeSNmedresCandxxxxxN
\else
\ifnum#1=579043 %
\hatcurSPECWiFeSNmedresCandxxxxxO
\else
\ifnum#1=579044 %
\hatcurSPECWiFeSNmedresCandxxxxxP
\else
\ifnum#1=579048 %
\hatcurSPECWiFeSNmedresCandxxxxxQ
\else
\ifnum#1=579050 %
\hatcurSPECWiFeSNmedresCandxxxxxR
\else
\ifnum#1=624002 %
\hatcurSPECWiFeSNmedresCandxxxxxS
\else
\ifnum#1=624003 %
\hatcurSPECWiFeSNmedresCandxxxxxT
\else
??????\fi
\fi
\fi
\fi
\fi
\fi
\fi
\fi
\fi
\fi
\fi
\fi
\fi
\fi
\fi
\fi
\fi
\fi
\fi
\fi
\fi
}
\newcommand{\hatcurSPECWiFeSRVGCand}[1]{\ifnum#1=578002 %
\hatcurSPECWiFeSRVGCandxxxxxA
\else
\ifnum#1=578003 %
\hatcurSPECWiFeSRVGCandxxxxxB
\else
\ifnum#1=578004 %
\hatcurSPECWiFeSRVGCandxxxxxC
\else
\ifnum#1=579001 %
\hatcurSPECWiFeSRVGCandxxxxxD
\else
\ifnum#1=579007 %
\hatcurSPECWiFeSRVGCandxxxxxE
\else
\ifnum#1=579008 %
\hatcurSPECWiFeSRVGCandxxxxxF
\else
\ifnum#1=579009 %
\hatcurSPECWiFeSRVGCandxxxxxG
\else
\ifnum#1=579010 %
\hatcurSPECWiFeSRVGCandxxxxxU
\else
\ifnum#1=579014 %
\hatcurSPECWiFeSRVGCandxxxxxH
\else
\ifnum#1=579015 %
\hatcurSPECWiFeSRVGCandxxxxxI
\else
\ifnum#1=579036 %
\hatcurSPECWiFeSRVGCandxxxxxJ
\else
\ifnum#1=579037 %
\hatcurSPECWiFeSRVGCandxxxxxK
\else
\ifnum#1=579039 %
\hatcurSPECWiFeSRVGCandxxxxxL
\else
\ifnum#1=579040 %
\hatcurSPECWiFeSRVGCandxxxxxM
\else
\ifnum#1=579041 %
\hatcurSPECWiFeSRVGCandxxxxxN
\else
\ifnum#1=579043 %
\hatcurSPECWiFeSRVGCandxxxxxO
\else
\ifnum#1=579044 %
\hatcurSPECWiFeSRVGCandxxxxxP
\else
\ifnum#1=579048 %
\hatcurSPECWiFeSRVGCandxxxxxQ
\else
\ifnum#1=579050 %
\hatcurSPECWiFeSRVGCandxxxxxR
\else
\ifnum#1=624002 %
\hatcurSPECWiFeSRVGCandxxxxxS
\else
\ifnum#1=624003 %
\hatcurSPECWiFeSRVGCandxxxxxT
\else
??????\fi
\fi
\fi
\fi
\fi
\fi
\fi
\fi
\fi
\fi
\fi
\fi
\fi
\fi
\fi
\fi
\fi
\fi
\fi
\fi
\fi
}
\newcommand{\hatcurSPECWiFeSRVKCand}[1]{\ifnum#1=578002 %
\hatcurSPECWiFeSRVKCandxxxxxA
\else
\ifnum#1=578003 %
\hatcurSPECWiFeSRVKCandxxxxxB
\else
\ifnum#1=578004 %
\hatcurSPECWiFeSRVKCandxxxxxC
\else
\ifnum#1=579001 %
\hatcurSPECWiFeSRVKCandxxxxxD
\else
\ifnum#1=579007 %
\hatcurSPECWiFeSRVKCandxxxxxE
\else
\ifnum#1=579008 %
\hatcurSPECWiFeSRVKCandxxxxxF
\else
\ifnum#1=579009 %
\hatcurSPECWiFeSRVKCandxxxxxG
\else
\ifnum#1=579010 %
\hatcurSPECWiFeSRVKCandxxxxxU
\else
\ifnum#1=579014 %
\hatcurSPECWiFeSRVKCandxxxxxH
\else
\ifnum#1=579015 %
\hatcurSPECWiFeSRVKCandxxxxxI
\else
\ifnum#1=579036 %
\hatcurSPECWiFeSRVKCandxxxxxJ
\else
\ifnum#1=579037 %
\hatcurSPECWiFeSRVKCandxxxxxK
\else
\ifnum#1=579039 %
\hatcurSPECWiFeSRVKCandxxxxxL
\else
\ifnum#1=579040 %
\hatcurSPECWiFeSRVKCandxxxxxM
\else
\ifnum#1=579041 %
\hatcurSPECWiFeSRVKCandxxxxxN
\else
\ifnum#1=579043 %
\hatcurSPECWiFeSRVKCandxxxxxO
\else
\ifnum#1=579044 %
\hatcurSPECWiFeSRVKCandxxxxxP
\else
\ifnum#1=579048 %
\hatcurSPECWiFeSRVKCandxxxxxQ
\else
\ifnum#1=579050 %
\hatcurSPECWiFeSRVKCandxxxxxR
\else
\ifnum#1=624002 %
\hatcurSPECWiFeSRVKCandxxxxxS
\else
\ifnum#1=624003 %
\hatcurSPECWiFeSRVKCandxxxxxT
\else
??????\fi
\fi
\fi
\fi
\fi
\fi
\fi
\fi
\fi
\fi
\fi
\fi
\fi
\fi
\fi
\fi
\fi
\fi
\fi
\fi
\fi
}
\newcommand{\hatcurSPECWiFeSsnrangelowresCand}[1]{\ifnum#1=578002 %
\hatcurSPECWiFeSsnrangelowresCandxxxxxA
\else
\ifnum#1=578003 %
\hatcurSPECWiFeSsnrangelowresCandxxxxxB
\else
\ifnum#1=578004 %
\hatcurSPECWiFeSsnrangelowresCandxxxxxC
\else
\ifnum#1=579001 %
\hatcurSPECWiFeSsnrangelowresCandxxxxxD
\else
\ifnum#1=579007 %
\hatcurSPECWiFeSsnrangelowresCandxxxxxE
\else
\ifnum#1=579008 %
\hatcurSPECWiFeSsnrangelowresCandxxxxxF
\else
\ifnum#1=579009 %
\hatcurSPECWiFeSsnrangelowresCandxxxxxG
\else
\ifnum#1=579010 %
\hatcurSPECWiFeSsnrangelowresCandxxxxxU
\else
\ifnum#1=579014 %
\hatcurSPECWiFeSsnrangelowresCandxxxxxH
\else
\ifnum#1=579015 %
\hatcurSPECWiFeSsnrangelowresCandxxxxxI
\else
\ifnum#1=579036 %
\hatcurSPECWiFeSsnrangelowresCandxxxxxJ
\else
\ifnum#1=579037 %
\hatcurSPECWiFeSsnrangelowresCandxxxxxK
\else
\ifnum#1=579039 %
\hatcurSPECWiFeSsnrangelowresCandxxxxxL
\else
\ifnum#1=579040 %
\hatcurSPECWiFeSsnrangelowresCandxxxxxM
\else
\ifnum#1=579041 %
\hatcurSPECWiFeSsnrangelowresCandxxxxxN
\else
\ifnum#1=579043 %
\hatcurSPECWiFeSsnrangelowresCandxxxxxO
\else
\ifnum#1=579044 %
\hatcurSPECWiFeSsnrangelowresCandxxxxxP
\else
\ifnum#1=579048 %
\hatcurSPECWiFeSsnrangelowresCandxxxxxQ
\else
\ifnum#1=579050 %
\hatcurSPECWiFeSsnrangelowresCandxxxxxR
\else
\ifnum#1=624002 %
\hatcurSPECWiFeSsnrangelowresCandxxxxxS
\else
\ifnum#1=624003 %
\hatcurSPECWiFeSsnrangelowresCandxxxxxT
\else
??????\fi
\fi
\fi
\fi
\fi
\fi
\fi
\fi
\fi
\fi
\fi
\fi
\fi
\fi
\fi
\fi
\fi
\fi
\fi
\fi
\fi
}
\newcommand{\hatcurSPECWiFeSsnrangemedresCand}[1]{\ifnum#1=578002 %
\hatcurSPECWiFeSsnrangemedresCandxxxxxA
\else
\ifnum#1=578003 %
\hatcurSPECWiFeSsnrangemedresCandxxxxxB
\else
\ifnum#1=578004 %
\hatcurSPECWiFeSsnrangemedresCandxxxxxC
\else
\ifnum#1=579001 %
\hatcurSPECWiFeSsnrangemedresCandxxxxxD
\else
\ifnum#1=579007 %
\hatcurSPECWiFeSsnrangemedresCandxxxxxE
\else
\ifnum#1=579008 %
\hatcurSPECWiFeSsnrangemedresCandxxxxxF
\else
\ifnum#1=579009 %
\hatcurSPECWiFeSsnrangemedresCandxxxxxG
\else
\ifnum#1=579010 %
\hatcurSPECWiFeSsnrangemedresCandxxxxxU
\else
\ifnum#1=579014 %
\hatcurSPECWiFeSsnrangemedresCandxxxxxH
\else
\ifnum#1=579015 %
\hatcurSPECWiFeSsnrangemedresCandxxxxxI
\else
\ifnum#1=579036 %
\hatcurSPECWiFeSsnrangemedresCandxxxxxJ
\else
\ifnum#1=579037 %
\hatcurSPECWiFeSsnrangemedresCandxxxxxK
\else
\ifnum#1=579039 %
\hatcurSPECWiFeSsnrangemedresCandxxxxxL
\else
\ifnum#1=579040 %
\hatcurSPECWiFeSsnrangemedresCandxxxxxM
\else
\ifnum#1=579041 %
\hatcurSPECWiFeSsnrangemedresCandxxxxxN
\else
\ifnum#1=579043 %
\hatcurSPECWiFeSsnrangemedresCandxxxxxO
\else
\ifnum#1=579044 %
\hatcurSPECWiFeSsnrangemedresCandxxxxxP
\else
\ifnum#1=579048 %
\hatcurSPECWiFeSsnrangemedresCandxxxxxQ
\else
\ifnum#1=579050 %
\hatcurSPECWiFeSsnrangemedresCandxxxxxR
\else
\ifnum#1=624002 %
\hatcurSPECWiFeSsnrangemedresCandxxxxxS
\else
\ifnum#1=624003 %
\hatcurSPECWiFeSsnrangemedresCandxxxxxT
\else
??????\fi
\fi
\fi
\fi
\fi
\fi
\fi
\fi
\fi
\fi
\fi
\fi
\fi
\fi
\fi
\fi
\fi
\fi
\fi
\fi
\fi
}
\newcommand{\hatcurSPECWiFeSTeffCand}[1]{\ifnum#1=578002 %
\hatcurSPECWiFeSTeffCandxxxxxA
\else
\ifnum#1=578003 %
\hatcurSPECWiFeSTeffCandxxxxxB
\else
\ifnum#1=578004 %
\hatcurSPECWiFeSTeffCandxxxxxC
\else
\ifnum#1=579001 %
\hatcurSPECWiFeSTeffCandxxxxxD
\else
\ifnum#1=579007 %
\hatcurSPECWiFeSTeffCandxxxxxE
\else
\ifnum#1=579008 %
\hatcurSPECWiFeSTeffCandxxxxxF
\else
\ifnum#1=579009 %
\hatcurSPECWiFeSTeffCandxxxxxG
\else
\ifnum#1=579010 %
\hatcurSPECWiFeSTeffCandxxxxxU
\else
\ifnum#1=579014 %
\hatcurSPECWiFeSTeffCandxxxxxH
\else
\ifnum#1=579015 %
\hatcurSPECWiFeSTeffCandxxxxxI
\else
\ifnum#1=579036 %
\hatcurSPECWiFeSTeffCandxxxxxJ
\else
\ifnum#1=579037 %
\hatcurSPECWiFeSTeffCandxxxxxK
\else
\ifnum#1=579039 %
\hatcurSPECWiFeSTeffCandxxxxxL
\else
\ifnum#1=579040 %
\hatcurSPECWiFeSTeffCandxxxxxM
\else
\ifnum#1=579041 %
\hatcurSPECWiFeSTeffCandxxxxxN
\else
\ifnum#1=579043 %
\hatcurSPECWiFeSTeffCandxxxxxO
\else
\ifnum#1=579044 %
\hatcurSPECWiFeSTeffCandxxxxxP
\else
\ifnum#1=579048 %
\hatcurSPECWiFeSTeffCandxxxxxQ
\else
\ifnum#1=579050 %
\hatcurSPECWiFeSTeffCandxxxxxR
\else
\ifnum#1=624002 %
\hatcurSPECWiFeSTeffCandxxxxxS
\else
\ifnum#1=624003 %
\hatcurSPECWiFeSTeffCandxxxxxT
\else
??????\fi
\fi
\fi
\fi
\fi
\fi
\fi
\fi
\fi
\fi
\fi
\fi
\fi
\fi
\fi
\fi
\fi
\fi
\fi
\fi
\fi
}
\newcommand{\hatcurSPECWiFeSzfehCand}[1]{\ifnum#1=578002 %
\hatcurSPECWiFeSzfehCandxxxxxA
\else
\ifnum#1=578003 %
\hatcurSPECWiFeSzfehCandxxxxxB
\else
\ifnum#1=578004 %
\hatcurSPECWiFeSzfehCandxxxxxC
\else
\ifnum#1=579001 %
\hatcurSPECWiFeSzfehCandxxxxxD
\else
\ifnum#1=579007 %
\hatcurSPECWiFeSzfehCandxxxxxE
\else
\ifnum#1=579008 %
\hatcurSPECWiFeSzfehCandxxxxxF
\else
\ifnum#1=579009 %
\hatcurSPECWiFeSzfehCandxxxxxG
\else
\ifnum#1=579010 %
\hatcurSPECWiFeSzfehCandxxxxxU
\else
\ifnum#1=579014 %
\hatcurSPECWiFeSzfehCandxxxxxH
\else
\ifnum#1=579015 %
\hatcurSPECWiFeSzfehCandxxxxxI
\else
\ifnum#1=579036 %
\hatcurSPECWiFeSzfehCandxxxxxJ
\else
\ifnum#1=579037 %
\hatcurSPECWiFeSzfehCandxxxxxK
\else
\ifnum#1=579039 %
\hatcurSPECWiFeSzfehCandxxxxxL
\else
\ifnum#1=579040 %
\hatcurSPECWiFeSzfehCandxxxxxM
\else
\ifnum#1=579041 %
\hatcurSPECWiFeSzfehCandxxxxxN
\else
\ifnum#1=579043 %
\hatcurSPECWiFeSzfehCandxxxxxO
\else
\ifnum#1=579044 %
\hatcurSPECWiFeSzfehCandxxxxxP
\else
\ifnum#1=579048 %
\hatcurSPECWiFeSzfehCandxxxxxQ
\else
\ifnum#1=579050 %
\hatcurSPECWiFeSzfehCandxxxxxR
\else
\ifnum#1=624002 %
\hatcurSPECWiFeSzfehCandxxxxxS
\else
\ifnum#1=624003 %
\hatcurSPECWiFeSzfehCandxxxxxT
\else
??????\fi
\fi
\fi
\fi
\fi
\fi
\fi
\fi
\fi
\fi
\fi
\fi
\fi
\fi
\fi
\fi
\fi
\fi
\fi
\fi
\fi
}

\newcommand{\GAIAa}{14.702}
\newcommand{\GAIAb}{15.288}
\newcommand{\GAIAc}{15.610}
\newcommand{\GAIAd}{18.182}

\newcommand{\hatcurCCra}{\ensuremath{19^{\mathrm h}25^{\mathrm m}54.84{\mathrm s}}}                     
\newcommand{\hatcurCCdec}{\ensuremath{-23{\arcdeg}12{\arcmin}10.0{\arcsec}}}                    
\newcommand{\hatcurCCmag}{14.386}                         
\newcommand{\hatcurCCtwomass}{2MASS~19255488-2312100}     
\newcommand{\hatcurCCtassmv}{\ensuremath{14.386\pm0.020}} 
\newcommand{\hatcurCCtassmB}{\ensuremath{15.060\pm0.030}} 
\newcommand{\hatcurCCtassmg}{\ensuremath{14.675\pm0.010}} 
\newcommand{\hatcurCCtassmr}{\ensuremath{14.231\pm0.010}} 
\newcommand{\hatcurCCtassmrshort}{\ensuremath{14.231}}      
\newcommand{\hatcurCCtassmi}{\ensuremath{14.146\pm0.050}} 
\newcommand{\hatcurCCtwomassJmag}{\ensuremath{13.181\pm0.026}} 
\newcommand{\hatcurCCtwomassHmag}{\ensuremath{12.855\pm0.025}} 
\newcommand{\hatcurCCtwomassKmag}{\ensuremath{12.809\pm0.026}} 
\newcommand{\hatcurLCrprstar}{\ensuremath{0.10942\pm0.00073}} 
\newcommand{\hatcurLCbsq}{\ensuremath{0.165_{-0.043}^{+0.038}}} 
\newcommand{\hatcurLCimp}{\ensuremath{0.406_{-0.057}^{+0.044}}} 
\newcommand{\hatcurLCzeta}{\ensuremath{15.799\pm0.057}}   
\newcommand{\hatcurLCdur}{\ensuremath{0.14301\pm0.00068}} 
\newcommand{\hatcurLCingdur}{\ensuremath{0.01661\pm0.00085}} 
\newcommand{\hatcurLCP}{\ensuremath{4.1752379\pm0.0000021}} 
\newcommand{\hatcurLCPprec}{\ensuremath{4.1752379}}       
\newcommand{\hatcurLCPshort}{\ensuremath{4.1752}}         
\newcommand{\hatcurLCT}{\ensuremath{2457289.411210\pm0.000069}} 
\newcommand{\hatcurSMEiteff}{\ensuremath{5970\pm160}}     
\newcommand{\hatcurSMEizfeh}{\ensuremath{0.13\pm0.10}}    
\newcommand{\hatcurSMEizfehshort}{\ensuremath{0.13}}      
\newcommand{\hatcurSMEilogg}{\ensuremath{4.35\pm0.26}}    
\newcommand{\hatcurSMEivsin}{\ensuremath{5.47\pm0.59}}    
\newcommand{\hatcurSMEivmac}{\ensuremath{0.0}}            
\newcommand{\hatcurSMEivmic}{\ensuremath{0.0}}            
\newcommand{\hatcurSMEiiteff}{\ensuremath{5970\pm160}}     
\newcommand{\hatcurSMEiizfeh}{\ensuremath{0.13\pm0.10}}    
\newcommand{\hatcurSMEiivsin}{\ensuremath{5.47\pm0.59}}    
\newcommand{\hatcurLBii}{\ensuremath{0.2416}}             
\newcommand{\hatcurLBiii}{\ensuremath{0.3477}}            
\newcommand{\hatcurLBir}{\ensuremath{0.3254}}             
\newcommand{\hatcurLBiir}{\ensuremath{0.3470}}            
\newcommand{\hatcurISOm}{\ensuremath{1.135\pm0.067}}      
\newcommand{\hatcurISOmlong}{\ensuremath{1.135\pm0.067}}  
\newcommand{\hatcurISOr}{\ensuremath{1.186\pm0.036}}      
\newcommand{\hatcurISOrlong}{\ensuremath{1.186\pm0.036}}  
\newcommand{\hatcurISOlogg}{\ensuremath{4.344\pm0.020}}   
\newcommand{\hatcurISOlum}{\ensuremath{1.60\pm0.24}}      
\newcommand{\hatcurISOmv}{\ensuremath{4.29\pm0.19}}       
\newcommand{\hatcurISOage}{\ensuremath{3.4_{-1.4}^{+1.9}}} 
\newcommand{\hatcurISOMK}{\ensuremath{2.873\pm0.096}}     
\newcommand{\hatcurRVK}{\ensuremath{324\pm45}}            
\newcommand{\hatcurRVjitter}{\ensuremath{105\pm26}}       
\newcommand{\hatcurPPi}{\ensuremath{87.57\pm0.36}}        
\newcommand{\hatcurPPlogg}{\ensuremath{3.636\pm0.067}}    
\newcommand{\hatcurPPar}{\ensuremath{9.59\pm0.21}}        
\newcommand{\hatcurPParel}{\ensuremath{0.0529\pm0.0011}}  
\newcommand{\hatcurPPrho}{\ensuremath{1.72\pm0.29}}       
\newcommand{\hatcurPPm}{\ensuremath{2.79\pm0.40}}         
\newcommand{\hatcurPPmlong}{\ensuremath{2.79\pm0.40}}     
\newcommand{\hatcurPPr}{\ensuremath{1.263\pm0.045}}       
\newcommand{\hatcurPPrlong}{\ensuremath{1.263\pm0.045}}   
\newcommand{\hatcurPPmrcorr}{\ensuremath{0.20}}           
\newcommand{\hatcurPPteff}{\ensuremath{1363\pm40}}        
\newcommand{\hatcurPPtheta}{\ensuremath{0.206\pm0.030}}   
\newcommand{\hatcurPPfluxavg}{\ensuremath{7.79\pm0.92}}   
\newcommand{\hatcurPPfluxavgdim}{\ensuremath{8}}          
\newcommand{\hatcurXAv}{\ensuremath{0.14\pm0.11}}         
\newcommand{\hatcurXdistred}{\ensuremath{977\pm42}}       
\newcommand{\hatcurCCpmra}{\ensuremath{-1.5\pm3.0}}       
\newcommand{\hatcurCCpmdec}{\ensuremath{-7.5\pm3.0}}      

\newcommand{\hatcur}{HATS-36}
\newcommand{\hatcurb}{HATS-36b}
\newcommand{\hatcurRVgammaabs}{\ensuremath{-24.392\pm31}}

\newcommand{\hs}{HATSouth}

\newcommand{\hatcurlumind}{\rhostar}
\newcommand{\hatcurjhkfilset}{ESO}

\newcommand{\hatcurSMEversion}{ii}                                       
\newcommand{\hatcurSMEteff}{\ifthenelse{\equal{\hatcurSMEversion}{i}}{\hatcurSMEiteff}{\hatcurSMEiiteff}}
\newcommand{\hatcurSMEzfeh}{\ifthenelse{\equal{\hatcurSMEversion}{i}}{\hatcurSMEizfeh}{\hatcurSMEiizfeh}}
\newcommand{\hatcurSMEzfehshort}{\ifthenelse{\equal{\hatcurSMEversion}{i}}{\hatcurSMEizfehshort}{\hatcurSMEiizfehshort}}
\newcommand{\hatcurSMElogg}{\ifthenelse{\equal{\hatcurSMEversion}{i}}{\hatcurSMEilogg}{\hatcurSMEiilogg}}
\newcommand{\hatcurSMEvsin}{\ifthenelse{\equal{\hatcurSMEversion}{i}}{\hatcurSMEivsin}{\hatcurSMEiivsin}}
\newcommand{\hatcurSMEvmac}{\ifthenelse{\equal{\hatcurSMEversion}{i}}{\hatcurSMEivmac}{\hatcurSMEiivmac}}
\newcommand{\hatcurSMEvmic}{\ifthenelse{\equal{\hatcurSMEversion}{i}}{\hatcurSMEivmic}{\hatcurSMEiivmic}}
\newcommand{\hatcurRVeccentwosiglimeccen}{<0.294}

\newcommand{\kepler}{$Kepler$}
\newcommand{\gaia}{\textit{Gaia}}
\newcommand{\kk}{K2}
\newcommand{\hskk}{HS-K2C7}
\newcommand{\sso}{Siding Spring Observatory (SSO)}
\newcommand{\lco}{Las Campanas Observatory (LCO)}
\newcommand{\hess}{High Energy Spectroscopic Survey (H.E.S.S.)}
\newcommand{\fieldra}{\ensuremath{19^{\mathrm h}30^{\mathrm m}00{\mathrm s}}}
\newcommand{\fielddec}{\ensuremath{-22{\arcdeg}30{\arcmin}00{\arcsec}}}
\newcommand{\kkfieldra}{\ensuremath{19^{\mathrm h}11^{\mathrm m}19{\mathrm s}}}
\newcommand{\kkfielddec}{\ensuremath{-23{\arcdeg}21{\arcmin}36{\arcsec}}}
\newcommand{\kkdates}{2015 October 4 to 2015 December 26}
\newcommand{\GOplanets}{GO7066}
\newcommand{\GOcands}{GO7067}
\newcommand{\hatsdates}{2010 March to 2011 August}
\newcommand{\totalims}{10137}
\newcommand{\cadence}{300}
\newcommand{\rprs}{\ensuremath{R_{p}/\rstar}}
\newcommand{\hatcurEPIC}{EPIC 215969174}

\newcommand{\kkrelease}{2016 April 20}
\newcommand{\everest}{\textsc{\lowercase{everest}}}
\newcommand{\hatcurCCkep}{\ensuremath{14.300\pm 0.030}} 
\newcommand{\hatcurCCgaia}{\ensuremath{14.148\pm 0.013}}

\newboolean{emulateapj}
\setboolean{emulateapj}{true}

\newboolean{rvtablelong}
\setboolean{rvtablelong}{true}

\newboolean{astroph}
\setboolean{astroph}{true}

\shortauthors{Bayliss et al.}
\shorttitle{
\hatcur\lowercase{b}
}
\ifthenelse{\boolean{emulateapj}}{
    \newcommand{\titledag}{$\dagger$}
}{
    \newcommand{\titledag}{\dagger}
}

\begin{document}
\submitted{2017 June 12: Submitted to AJ}
\title{\hatcur\lowercase{b} and 24 other transiting/eclipsing systems from the HATSouth - \kk\ Campaign 7 program \altaffilmark{\titledag}}

\author{
    D.~Bayliss\altaffilmark{1},
    J.~D.~Hartman\altaffilmark{2},
    G.~Zhou\altaffilmark{3},
    G.~\'A.~Bakos\altaffilmark{2,$\star$,$\star\star$},    
    A.~Vanderburg\altaffilmark{3},
    J.~Bento\altaffilmark{4},
    L.~Mancini\altaffilmark{5,6,7},
    S.~Ciceri\altaffilmark{6},
    R.~Brahm\altaffilmark{8,9},
    A.~Jord\'an\altaffilmark{9,8},
    N.~Espinoza\altaffilmark{9,8},
    M.~Rabus\altaffilmark{6,9},
    T.~G.~Tan\altaffilmark{10},
    K.~Penev\altaffilmark{2},
    W.~Bhatti\altaffilmark{2},
    M.~de~Val-Borro\altaffilmark{2},
    V.~Suc\altaffilmark{9},
    Z.~Csubry\altaffilmark{2},
    Th.~Henning\altaffilmark{6},
    P.~Sarkis\altaffilmark{6},
    J.~L\'az\'ar\altaffilmark{11},
    I.~Papp\altaffilmark{11},
    P.~S\'ari\altaffilmark{11}
}

\altaffiltext{1}{Observatoire Astronomique de l'Universit\'e de
  Geneve, 51 ch.  des Maillettes, 1290 Versoix, Switzerland; email:
  daniel.bayliss@unige.ch}
\altaffiltext{2}{Department of Astrophysical Sciences, Princeton
  University, NJ 08544, USA.}  \altaffiltext{3}{Harvard-Smithsonian
  Center for Astrophysics, 60 Garden St., Cambridge, MA 02138, USA.}
\altaffiltext{4}{Research School of Astronomy and Astrophysics,
  Australian National University, Canberra, ACT 2611, Australia.}
\altaffiltext{5}{Dipartimento di Fisica, Universit\`a di Roma Tor
  Vergata, Via della Ricerca Scientifica 1, 00133 -- Roma, Italy}
\altaffiltext{6}{Max Planck Institute for Astronomy, K\"{o}nigstuhl
  17, 69117 Heidelberg, Germany} \altaffiltext{7}{INAF -- Osservatorio
  Astrofisico di Torino, Via Osservatorio 20, 10025 -- Pino Torinese,
  Italy} \altaffiltext{8}{Millennium Institute of Astrophysics,
  Av. Vicu\~na Mackenna 4860, 7820436 Macul, Santiago, Chile}
\altaffiltext{9}{Instituto de Astrof\'isica, Facultad de F\'isica,
  Pontificia Universidad Cat\'olica de Chile, Av.\ Vicu\~na Mackenna
  4860, 7820436 Macul, Santiago, Chile} \altaffiltext{10}{Perth
  Exoplanet Survey Telescope, Perth, Australia.}
\altaffiltext{11}{Hungarian Astronomical Association, 1451 Budapest,
  Hungary} \altaffiltext{$\star$}{Alfred P.~Sloan Research Fellow}
\altaffiltext{$\star\star$}{Packard Fellow}
\altaffiltext{$\dagger$}{The HATSouth network is operated by a
  collaboration consisting of Princeton University (PU), the Max
  Planck Institute f\"ur Astronomie (MPIA), the Australian National
  University (ANU), and the Pontificia Universidad Cat\'olica de Chile
  (PUC).  The station at \lco\ of the Carnegie Institute is operated
  by PU in conjunction with PUC, the station at the \hess\ site is
  operated in conjunction with MPIA, and the station at \sso\ is
  operated jointly with ANU.  Based in part on observations made with
  the MPG~2.2\,m Telescope at the ESO Observatory in La Silla.  }


\begin{abstract}
\setcounter{footnote}{10}
We report on the result of a campaign to monitor 25 HATSouth
candidates using the \kepler{} space telescope during Campaign 7 of
the \kk\ mission.  We discover \hatcurb{} (\hatcurEPIC b), a
\hj\ with a mass of $\hatcurPPm$\,\mjup{} and a radius of
$\hatcurPPr$\,\rjup{} which transits a solar-type G0V star
(V=\hatcurCCmag) in a $\hatcurLCPshort$\,d period.  We also refine the
properties of three previously discovered HATSouth transiting planets
(\hatcurnineb, \hatcurelevenb, and \hatcurtwelveb) and search the
\kk\ data for TTVs and additional transiting planets in these systems.
In addition we also report on a further three systems that remain as
Jupiter-radius transiting exoplanet candidates.  These candidates do
not have determined masses, however pass all of our other vetting
observations.  Finally we report on the 18 candidates which we are now
able to classify as eclipsing binary or blended eclipsing binary
systems based on a combination of the HATSouth data, the \kk\ data,
and follow-up ground-based photometry and spectroscopy.  These range
in periods from 0.7\,days to 16.7\,days, and down to 1.5\,mmag in
eclipse depths.  Our results show the power of combining
ground-based imaging and spectroscopy with higher precision
space-based photometry, and serve as an illustration as to what will
be possible when combining ground-based observations with TESS data.
\setcounter{footnote}{0}
\end{abstract}



\section{Introduction}
\label{sec:introduction}
Transiting planet systems are the most valuable systems for exoplanet
studies due to the wide range of characterisation observations that
can been made for them, both during transit and also in
secondary eclipse.  However since the geometric probability of transit
is low, large numbers of stars must be monitored in order to discover such
systems.  This task began over a decade ago with lenses-based
wide-field ground-based surveys, most successfully HATNet
\citep{bakos:2004:hatnet}, WASP \citep{2006PASP..118.1407P}, and
KELT \citep{2007PASP..119..923P}.  These were followed by space-based
surveys CoRoT \citep{2008AA...482L..17B} and Kepler
\citep{2010Sci...327..977B}.  In this paper we combine data from two
ongoing transit surveys: HATSouth \citep[][hereafter HS]{bakos:2013:hatsouth} and
\kk\ \citep{2014PASP..126..398H}.  While the HATSouth survey has been
monitoring selected fields in the southern Hemisphere over the last
six years, the \kk\ survey has been monitoring selected ecliptic plane
fields since 2014.  \kk\ Campaign 7 (C7) monitored a field centered at
$\alpha=\kkfieldra$, $\delta=\kkfielddec$ between \kkdates.  By
chance, approximately 25\% of this field had already been monitored by the HATSouth
survey five years earlier between \hatsdates.  It was therefore
possible for us to propose for \kk\ targets based on the HATSouth
\lcs\ that showed transit-like signals, and this was done as part of
the \kk\ Guest Observer program (\GOcands; PI: Bakos).  In addition we
were also able to select as \kk\ targets the HATSouth transiting
planets that had already been confirmed as bona fide planets, namely
\hatcurnineb{} \citep{2015AJ....150...33B}, \hatcurelevenb{} and \hatcurtwelveb{}
\citep{2016AJ....152...88R}; this formed the \kk\ Guest Observer
program \GOplanets\ (PI:Bakos).  In total 25 stars were monitored by
\kk{} as part of these programs, which we collectively name ``\hskk''
targets.  Details for each of the \hskk\ candidates are set out in
Table~\ref{tab:candidates}.

\ifthenelse{\boolean{emulateapj}}{
        \begin{deluxetable*}{lrrrrr} }{
        \begin{deluxetable}{lrrrRr} 
    } \tablewidth{0pc} \tablecaption{\hskk\
  Candidates \label{tab:candidates}} \tablehead{ \colhead{EPIC ID} &
  \colhead{Mag} & \colhead{RA} & \colhead{Dec} & \colhead{Period} &
  \colhead{\rprs} \\ \colhead{(HS-ID)} & \colhead{\band{r}} &
  \colhead{} & \colhead{} & \colhead{(days)} & \colhead{} } \startdata
   \hatcurEPICnine\          (\hatcurnine)             &  \hatcurCCtassmrnine     & \hatcurCCranine     & \hatcurCCdecnine   & \hatcurLCPnine    &    \hatcurLCrprstarnine\\
   \hatcurEPICeleven\        (\hatcureleven)           &  \hatcurCCtassmreleven   & \hatcurCCraeleven   & \hatcurCCdeceleven & \hatcurLCPeleven  &    \hatcurLCrprstareleven\\
   \hatcurEPICtwelve\        (\hatcurtwelve)           &  \hatcurCCtassmrtwelve   & \hatcurCCratwelve   & \hatcurCCdectwelve & \hatcurLCPtwelve  &    \hatcurLCrprstartwelve\\
   \hatcurEPIC\               (\hatcur)                  &  \hatcurCCtassmrshort     & \hatcurCCra          & \hatcurCCdec        & \hatcurLCP         &    \hatcurLCrprstar\\
   \hatcurCCEPICCand{578002}\ (\hatcurhtrCand{578002})   & \hatcurCCmagrCand{578002} & \hatcurCCraCand{578002}   & \hatcurCCdecCand{578002} & \hatcurLCPCand{578002}  &    \hatcurLCrprstarCand{578002}\\
   \hatcurCCEPICCand{578003}\ (\hatcurhtrCand{578003})   & \hatcurCCmagrCand{578003} & \hatcurCCraCand{578003}   & \hatcurCCdecCand{578003} & \hatcurLCPCand{578003}  &    \hatcurLCrprstarCand{578003}\\
   \hatcurCCEPICCand{578004}\ (\hatcurhtrCand{578004})   & \hatcurCCmagrCand{578004} & \hatcurCCraCand{578004}   & \hatcurCCdecCand{578004} & \hatcurLCPCand{578004}  &    \hatcurLCrprstarCand{578004}\\
   \hatcurCCEPICCand{579001}\ (\hatcurhtrCand{579001})   & \hatcurCCmagrCand{579001} & \hatcurCCraCand{579001}   & \hatcurCCdecCand{579001} & \hatcurLCPCand{579001}  &    \hatcurLCrprstarCand{579001}\\
   \hatcurCCEPICCand{579007}\ (\hatcurhtrCand{579007})   & \hatcurCCmagrCand{579007} & \hatcurCCraCand{579007}   & \hatcurCCdecCand{579007} & \hatcurLCPCand{579007}  &    \hatcurLCrprstarCand{579007}\\
   \hatcurCCEPICCand{579008}\ (\hatcurhtrCand{579008})   & \hatcurCCmagrCand{579008} & \hatcurCCraCand{579008}   & \hatcurCCdecCand{579008} & \hatcurLCPCand{579008}  &    \hatcurLCrprstarCand{579008}\\
   \hatcurCCEPICCand{579009}\ (\hatcurhtrCand{579009})   & \hatcurCCmagrCand{579009} & \hatcurCCraCand{579009}   & \hatcurCCdecCand{579009} & \hatcurLCPCand{579009}  &    \hatcurLCrprstarCand{579009}\\
   \hatcurCCEPICCand{579010}\ (\hatcurhtrCand{579010})   & \hatcurCCmagrCand{579010} & \hatcurCCraCand{579010}   & \hatcurCCdecCand{579010} & \hatcurLCPCand{579010}  &    \hatcurLCrprstarCand{579010}\\
   \hatcurCCEPICCand{579014}\ (\hatcurhtrCand{579014})   & \hatcurCCmagrCand{579014} & \hatcurCCraCand{579014}   & \hatcurCCdecCand{579014} & \hatcurLCPCand{579014}  &    \hatcurLCrprstarCand{579014}\\
   \hatcurCCEPICCand{579015}\ (\hatcurhtrCand{579015})   & \hatcurCCmagrCand{579015} & \hatcurCCraCand{579015}   & \hatcurCCdecCand{579015} & \hatcurLCPCand{579015}  &    \hatcurLCrprstarCand{579015}\\
   \hatcurCCEPICCand{579036}\ (\hatcurhtrCand{579036})   & \hatcurCCmagrCand{579036} & \hatcurCCraCand{579036}   & \hatcurCCdecCand{579036} & \hatcurLCPCand{579036}  &    \hatcurLCrprstarCand{579036}\\
   \hatcurCCEPICCand{579037}\ (\hatcurhtrCand{579037})   & \hatcurCCmagrCand{579037} & \hatcurCCraCand{579037}   & \hatcurCCdecCand{579037} & \hatcurLCPCand{579037}  &    \hatcurLCrprstarCand{579037}\\
   \hatcurCCEPICCand{579039}\ (\hatcurhtrCand{579039})   & \hatcurCCmagrCand{579039} & \hatcurCCraCand{579039}   & \hatcurCCdecCand{579039} & \hatcurLCPCand{579039}  &    \hatcurLCrprstarCand{579039}\\
   \hatcurCCEPICCand{579040}\ (\hatcurhtrCand{579040})   & \hatcurCCmagrCand{579040} & \hatcurCCraCand{579040}   & \hatcurCCdecCand{579040} & \hatcurLCPCand{579040}  &    \hatcurLCrprstarCand{579040}\\
   \hatcurCCEPICCand{579041}\ (\hatcurhtrCand{579041})   & \hatcurCCmagrCand{579041} & \hatcurCCraCand{579041}   & \hatcurCCdecCand{579041} & \hatcurLCPCand{579041}  &    \hatcurLCrprstarCand{579041}\\
   \hatcurCCEPICCand{579043}\ (\hatcurhtrCand{579043})   & \hatcurCCmagrCand{579043} & \hatcurCCraCand{579043}   & \hatcurCCdecCand{579043} & \hatcurLCPCand{579043}  &    \hatcurLCrprstarCand{579043}\\
   \hatcurCCEPICCand{579044}\ (\hatcurhtrCand{579044})   & \hatcurCCmagrCand{579044} & \hatcurCCraCand{579044}   & \hatcurCCdecCand{579044} & \hatcurLCPCand{579044}  &    \hatcurLCrprstarCand{579044}\\
   \hatcurCCEPICCand{579048}\ (\hatcurhtrCand{579048})   & \hatcurCCmagrCand{579048} & \hatcurCCraCand{579048}   & \hatcurCCdecCand{579048} & \hatcurLCPCand{579048}  &    \hatcurLCrprstarCand{579048}\\
   \hatcurCCEPICCand{579050}\ (\hatcurhtrCand{579050})   & \hatcurCCmagrCand{579050} & \hatcurCCraCand{579050}   & \hatcurCCdecCand{579050} & \hatcurLCPCand{579050}  &    \hatcurLCrprstarCand{579050}\\
   \hatcurCCEPICCand{624002}\ (\hatcurhtrCand{624002})   & \hatcurCCmagrCand{624002} & \hatcurCCraCand{624002}   & \hatcurCCdecCand{624002} & \hatcurLCPCand{624002}  &    \hatcurLCrprstarCand{624002}\\
   \hatcurCCEPICCand{624003}\ (\hatcurhtrCand{624003})   & \hatcurCCmagrCand{624003} & \hatcurCCraCand{624003}   & \hatcurCCdecCand{624003} & \hatcurLCPCand{624003}  &    \hatcurLCrprstarCand{624003}\\
[-1.5ex] \enddata \tablecomments{
        Periods and \rprs\ from HATSouth discovery \lcs.
} \ifthenelse{\boolean{emulateapj}}{ \end{deluxetable*}
}{ \end{deluxetable} }

This paper is organised as follows.  In Section~\ref{sec:obs} we
describe the observations made for each \hskk\ target, including the
initial HATSouth discovery photometry, reconnaissance spectroscopy and
photometry, radial velocity measurements, and the \kk{} photometry.
In Section~\ref{sec:analysis} we analyse each \hskk\ candidate,
including the discovery of a new transiting exoplanet (\hatcurb),
refinement of the parameters of three known HATSouth transiting planets
(\hatcurnineb, \hatcurelevenb, \hatcurtwelveb), identification of
three Jupiter-radii candidates, and classification of 18
\hskk\ candidates as eclipsing binaries or blended eclipsing binaries.
Finally in Section~\ref{sec:discussion} we discuss the results and
implications of this joint ground/space-based photometry project.

\section{Observations}
\label{sec:obs}

\subsection{HATSouth Photometry}
\label{sec:hsphotometry}
The HATSouth global telescope network consists of six HS4 units, two
each at \sso, \lco, and the \hess\ site.  Each HS4 unit holds four
Takahashi astrographs ($f$/2.8, 18 cm apertures) which are each
coupled to an Apogee U16M 4K$\times$4K CCD camera.  Imaging is performed
using Sloan r-band filters and with exposure times of 240\,s.  A full
description of the HATSouth telescope network hardware and operations
can be found in \citet{bakos:2013:hatsouth}.

As part of the HATSouth survey (\url{http://hatsouth.org/}), we
monitored a 64 square degree field (centered at \fieldra,\fielddec)
between \hatsdates.  In total we obtained \totalims\ images, with a
cadence of approximately \cadence\,s.  In Table~\ref{tab:photobs} we
set out a summary of the HATSouth observations of this field, including
the number of images from each site, the observation dates, and the
cadence.

\ifthenelse{\boolean{emulateapj}}{
    \begin{deluxetable*}{cccccc}
}{
    \begin{deluxetable}{cccccc}
} \tablewidth{0pc} \tabletypesize{\scriptsize} \tablecaption{ Summary
  of \hskk\ photometric observations
    \label{tab:photobs}
} \tablehead{\colhead{Facility} &
  \colhead{Target(s)} &
  \colhead{Date} &
  \colhead{Number} &
  \colhead{Cadence} \tablenotemark{a} &
  \colhead{Filter}
  \\
  \colhead{} &
  \colhead{} &
  \colhead{Range} &
  \colhead{of Images} &
  \colhead{(s)} &
  \colhead{}
  }\startdata
~~~~HS-1 (LCO)  & all & 2010 Mar -- 2011 Aug & 4293 & 300  & $r$ \\
~~~~HS-3 (HESS) & all & 2010 Mar -- 2011 Aug & 2556 & 303  & $r$ \\
~~~~HS-5 (SSO)  & all & 2010 Sep--2011 Aug & 3287 & 303  & $r$ \\
~~~~K2 LC (Kepler) & all & 2015 Oct 4 -- 2015 Dec 2  & 3754 & 1800 & $Kep$ \\
~~~~K2 SC (Kepler) & HATS-9  & 2015 Oct 4 -- 2015 Dec 2  & 117401 & 60.5 & $Kep$ \\
~~~~K2 SC (Kepler) & HATS-11 & 2015 Oct 4 -- 2015 Dec 2  & 117601 & 60.5 & $Kep$ \\
~~~~PEST        & HATS-36 & 2013 July 1 & 137 & 130 & $R_C$ \\
~~~~GROND       & HATS-36 & 2014 July 24 & 75 & 139 & $g,r,i,z$ \\
~~~~GROND       & HATS-36 & 2014 July 28 & 80 & 139 & $g,r,i$ \\
~~~~ANU2.3m     & HATS579-037 (EPIC 215234145) & 2012 Sept 8 & 138 & 134 & $I$\\
~~~~PEST        & HATS579-037 (EPIC 215234145) & 2013 July 7 & 194 & 130 & $R_C$\\
~~~~PEST        & HATS579-037 (EPIC 215234145) & 2015 Aug 11 & 168 & 131 & $I$\\
~~~~SWOPE       & HATS579-039 (EPIC 215353525) & 2013 Aug 20 & 67 & 211 & $I$\\
[-1.5ex] \enddata \tablenotetext{a}{
  The mode time difference rounded to the nearest second between
  consecutive points in each light curve.
} \ifthenelse{\boolean{emulateapj}}{
    \end{deluxetable*}
}{
    \end{deluxetable}
}

HATSouth raw images are reduced to \lcs\ using an automated aperture
photometric pipeline detailed in \cite{penev:2013:hats1}.  The
\lcs\ are detrended using External Parameter Decorrelation
\citep[EPD;][]{bakos:2010:hat11} and the Trend Filtering Algorithm
\citep[TFA;][]{kovacs:2005:tfa}.  These light curves are then combined
for transit-like features using the box-fitting least squares
algorithm \cite{kovacs:2002:BLS}.  We found 25 candidates with
transit-like signals which were also were on-silicon for the
\kk\ Campaign 7 (see Section~\ref{sec:k2}).  We designate
these as ``\hskk{} candidates'', and summarise them in
Table~\ref{tab:candidates}.\@  Three of these have already been
published as confirmed transiting planets
\hatcurnineb{} \citep{2015AJ....150...33B},
\hatcurelevenb{} \citep{2016AJ....152...88R}, and
\hatcurtwelveb{} \citep{2016AJ....152...88R}.  We therefore do not
discuss these further in this Section.

\begin{figure}[tbp]
\epsscale{1.0}
\plotone{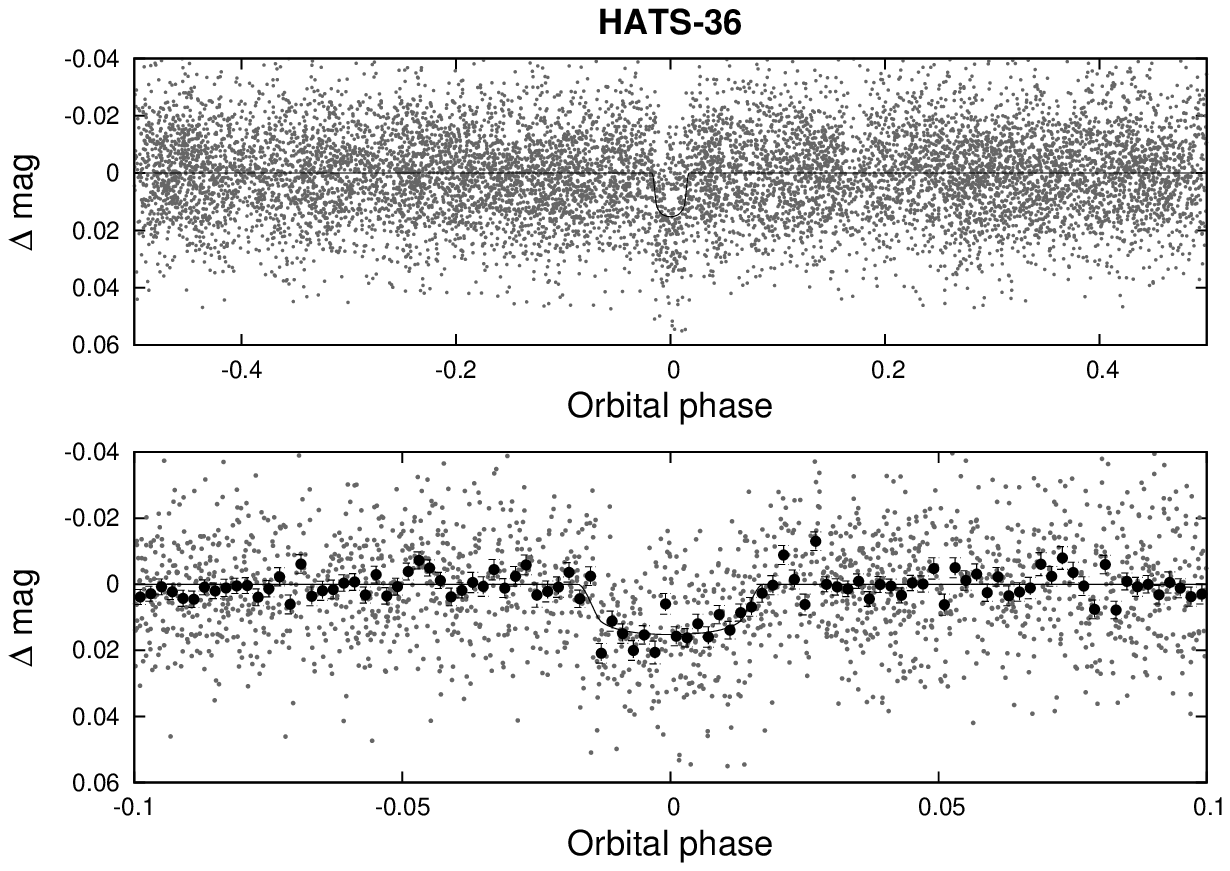}
\caption[]{
        Unbinned instrumental \band{r} \lc{} of \hatcur{} folded with
        at the period $P = \hatcurLCPprec$\,days.  The solid line shows the
        best-fit transit model (see \refsecl{analysis}).  Lower
        panel shows a zoom-in on the transit; the dark filled points here
        show the light curve binned in phase using a bin-size of 0.002.\\
\label{fig:hatsouth}}
\end{figure}

HATSouth light curves for all stars overlapping with the \kk\ Campaign
7, including the \hskk\ candidates, are publicly available at
\url{http://data.hatsurveys.org/}.  By way of example, we present the
HATSouth \lc\ for \hatcurb\ in Figure~\ref{fig:hatsouth}, which shows
the 18\,mmag transit-like dip when phase-folded at
$P=\hatcurLCPprec$\,days.  All HATSouth photometric data used in this
paper is set out in Table~\ref{tab:photometry}.  We note that as a
result of the EPD and TFA detrending, and also due to blending from
neighbors, the apparent transit depth in the HATSouth light curves is somewhat
shallower than that of the true depth in the Sloan~$r$ filter (the
apparent depth is typically 85\% that of the true depth).

\ifthenelse{\boolean{emulateapj}}{
        \begin{deluxetable*}{cccccccc} }{
        \begin{deluxetable}{ccccccc} 
    } \tablewidth{0pc} \tablecaption{Differential photometry of
  \hskk\ candidates \label{tab:photometry}} \tablehead{\colhead{\hskk} & \colhead{BJD} & \colhead{Mag} &
  \colhead{\ensuremath{\sigma_{\rm Mag}}} & \colhead{Mag} &
  \colhead{Filter} & \colhead{Instrument} \\ \colhead{ID}
  &\colhead{\hbox{~~~~(2\,400\,000$+$)~~~~}} & \colhead{} & \colhead{}
  & \colhead{(orig)} & \colhead{} & \colhead{} } \startdata
   HATS-36 & $ 55800.93945 $ & $   0.00446 $ & $   0.01045 $ & $ \cdots $ & $ r$ &         HS\\
   HATS-36 & $ 55788.41475 $ & $  -0.00216 $ & $   0.01768 $ & $ \cdots $ & $ r$ &         HS\\
   HATS-36 & $ 55755.01420 $ & $  -0.02741 $ & $   0.02600 $ & $ \cdots $ & $ r$ &         HS\\
   HATS-36 & $ 55679.86013 $ & $   0.00595 $ & $   0.01239 $ & $ \cdots $ & $ r$ &         HS\\
   HATS-36 & $ 55800.94295 $ & $  -0.00508 $ & $   0.00957 $ & $ \cdots $ & $ r$ &         HS\\
   HATS-36 & $ 55725.78922 $ & $  -0.00775 $ & $   0.01555 $ & $ \cdots $ & $ r$ &         HS\\
   HATS-36 & $ 55788.41822 $ & $  -0.00621 $ & $   0.01639 $ & $ \cdots $ & $ r$ &         HS\\
   HATS-36 & $ 55755.01796 $ & $  -0.02324 $ & $   0.03203 $ & $ \cdots $ & $ r$ &         HS\\
   HATS-36 & $ 55679.86505 $ & $  -0.00567 $ & $   0.01186 $ & $ \cdots $ & $ r$ &         HS\\
   HATS-36 & $ 55725.79268 $ & $  -0.02899 $ & $   0.01553 $ & $ \cdots $ & $ r$ &         HS\\
[-1.5ex] \enddata

\tablecomments{
        This table is available in a machine-readable form in the
        online journal.  A portion is shown here for guidance
        regarding its form and content. The data are also available on
        the HATSouth website at \url{http://www.hatsouth.org}.
} \ifthenelse{\boolean{emulateapj}}{ \end{deluxetable*}
}{ \end{deluxetable} }

\subsection{Reconnaissance Spectroscopy}
\label{sec:specfu}

As part of the usual HATSouth program to follow-up transiting planet
candidates, follow-up spectroscopy was obtained for the
\hskk\ candidates, primarily using WiFeS \citep{dopita:2007} on the
ANU 2.3\,m telescope at SSO.\@ Full details for this observing program
are given in \citet{bayliss:2013:hats3}.  In summary, we obtain a flux
calibrated, high S/N, $R = \lambda/\Delta \lambda = 3000$ spectrum for
each candidate in order to determine the spectral type and class of
the host star.  These spectra are compared with a grid of synthetic
templates from the MARCS model atmospheres \citep{2008AA...486..951G}
in order to estimate \teff.  The results for the \hskk\ candidates are
set out in Table~\ref{tab:summary}.  For scheduling reasons, we
obtained reconnaissance spectra for two candidates
(\hatcurhtrCand{579014} and \hatcurhtrCand{579036}) using FEROS
\citep{kaufer:1998} on the MPG 2.2\,m telescope at the ESO observatory in La Silla, Chile
(LSO).  Spectral parameters were derived from these spectra using the
CERES code \citep{brahm:2017:ceres} and these are also tabulated in
Table~ \ref{tab:summary}.  We note that the spectrum of
\hatcurhtrCand{579014}\ showed that the candidate was a spectroscopic
binary.  We were not able to obtain a spectrum for two candidates,
\hatcurhtrCand{579050} and \hatcurhtrCand{624003}, as they were too
optically faint (V=\hatcurCCmagVCand{579050} and
V=\hatcurCCmagVCand{624003} respectively).

We also use WiFeS on the ANU 2.3\,m telescope to obtain multi-epoch
medium resolution ($R = \lambda/\Delta \lambda = 7000$) spectra to
check for large amplitude (K$>$2\,\kms) radial velocity variations in
phase with the photometric signal.  This allows us to identify
candidates which are eclipsing binaries without the need of more
resource-intensive high-precision radial velocity monitoring.  For
those targets with multi-epoch spectra we list the measured
semi-amplitudes in Table~\ref{tab:summary}.  Again for the candidate
\hatcurhtrCand{579036}, FEROS was used instead of WiFeS to measure
the radial velocity semi-amplitude.

\ifthenelse{\boolean{emulateapj}}{
        \begin{deluxetable*}{lccccc} }{
        \begin{deluxetable}{lccccc} 
    } \tablewidth{0pc} \tablecaption{\hskk\ Candidate Classifications 
  \label{tab:summary}} \tablehead{ \colhead{EPIC ID} &
  \colhead{HS Recon.}&\colhead{HS Recon.}& \colhead{Sec. Eclipse}&
  \colhead{Class.}& \colhead{Comment} \\ \colhead{(HS-ID)} &
  \colhead{\teff\ (K)}&\colhead{K (\kms)}& \colhead{in \kk\ LC}&\colhead{} & \colhead{}}
\startdata
\hatcurEPICnine\ (\hatcurnine)                     & ...                             & ...    & NO                          &  TEP                      & \hatcurnineb\    \citep{2015AJ....150...33B} \\
\hatcurEPICeleven\ (\hatcureleven)                 & ...                             & ...    & NO                          &  TEP                      & \hatcurelevenb\ \citep{2016AJ....152...88R} \\
\hatcurEPICtwelve\ (\hatcurtwelve)                 & ...                             & ...    & NO                          &  TEP                      & \hatcurtwelveb\ \citep{2016AJ....152...88R} \\
\hatcurEPIC\ (\hatcur)                             & $6000\pm300$                    & $<2.0$ & NO                          &  TEP                      & \hatcurb\ (this work) \\
\hatcurCCEPICCand{578002} (\hatcurhtrCand{578002}) & \hatcurSPECWiFeSTeffCand{578002}& \hatcurSPECWiFeSRVKCand{578002}& YES &  \hatcurCLASSCand{578002} & ... \\
\hatcurCCEPICCand{578003} (\hatcurhtrCand{578003}) & \hatcurSPECWiFeSTeffCand{578003}& \hatcurSPECWiFeSRVKCand{578003}& YES &  \hatcurCLASSCand{578003} & Shallow \kk\ sec. eclipse \\
\hatcurCCEPICCand{578004} (\hatcurhtrCand{578004}) & \hatcurSPECWiFeSTeffCand{578004}& \hatcurSPECWiFeSRVKCand{578004}& YES &  \hatcurCLASSCand{578004} & Blended in wide \kk\ apertures\\
\hatcurCCEPICCand{579001} (\hatcurhtrCand{579001}) & \hatcurSPECWiFeSTeffCand{579001}& \hatcurSPECWiFeSRVKCand{579001}& YES &  \hatcurCLASSCand{579001} & ...\\
\hatcurCCEPICCand{579007} (\hatcurhtrCand{579007}) & \hatcurSPECWiFeSTeffCand{579007}& \hatcurSPECWiFeSRVKCand{579007}& YES &  \hatcurCLASSCand{579007} & Blended in wide \kk\ apertures\\
\hatcurCCEPICCand{579008} (\hatcurhtrCand{579008}) & \hatcurSPECWiFeSTeffCand{579008}& \hatcurSPECWiFeSRVKCand{579008}& YES &  \hatcurCLASSCand{579008} & Shallow \kk\ sec. eclipse \\
\hatcurCCEPICCand{579009} (\hatcurhtrCand{579009}) & \hatcurSPECWiFeSTeffCand{579009}& \hatcurSPECWiFeSRVKCand{579009}& NO  &  \hatcurCLASSCand{579009} & ... \\
\hatcurCCEPICCand{579010} (\hatcurhtrCand{579010}) & \hatcurSPECWiFeSTeffCand{579010}& \hatcurSPECWiFeSRVKCand{579010}& YES &  \hatcurCLASSCand{579010} & Blend with nearby EB (P=43\,d) \\
\hatcurCCEPICCand{579014} (\hatcurhtrCand{579014}) & \hatcurSPECWiFeSTeffCand{579014}& \hatcurSPECWiFeSRVKCand{579014}& YES &  \hatcurCLASSCand{579014} & Shallow \kk\ sec. eclipse and OOT \\
\hatcurCCEPICCand{579015} (\hatcurhtrCand{579015}) & \hatcurSPECWiFeSTeffCand{579015}& $13.497\pm0.011$                 & YES &  \hatcurCLASSCand{579015} & Shallow \kk\ sec. eclipse  \\
\hatcurCCEPICCand{579036} (\hatcurhtrCand{579036}) & \hatcurSPECWiFeSTeffCand{579036}& \hatcurSPECWiFeSRVKCand{579036}& YES &  \hatcurCLASSCand{579036} & Shallow \kk\ sec. eclipse  \\
\hatcurCCEPICCand{579037} (\hatcurhtrCand{579037}) & \hatcurSPECWiFeSTeffCand{579037}& \hatcurSPECWiFeSRVKCand{579037}& NO  &  \hatcurCLASSCand{579037} & Colour dependent depth, \kk\ OOT at P=$\times$2 \\
\hatcurCCEPICCand{579039} (\hatcurhtrCand{579039}) & \hatcurSPECWiFeSTeffCand{579039}& \hatcurSPECWiFeSRVKCand{579039}& YES &  \hatcurCLASSCand{579039} & K2 OOT  \\
\hatcurCCEPICCand{579040} (\hatcurhtrCand{579040}) & \hatcurSPECWiFeSTeffCand{579040}& \hatcurSPECWiFeSRVKCand{579040}& NO  &  \hatcurCLASSCand{579040} & P=3.9\,d candidate (depth=1.9\%) \\
\hatcurCCEPICCand{579041} (\hatcurhtrCand{579041}) & \hatcurSPECWiFeSTeffCand{579041}& \hatcurSPECWiFeSRVKCand{579041}& YES &  \hatcurCLASSCand{579041} & Shallow \kk\ sec. eclipse and OOT  \\
\hatcurCCEPICCand{579043} (\hatcurhtrCand{579043}) & \hatcurSPECWiFeSTeffCand{579043}& \hatcurSPECWiFeSRVKCand{579043}& YES &  \hatcurCLASSCand{579043} & \kk\ sec. eclipse and OOT at P=$\times$2 \\
\hatcurCCEPICCand{579044} (\hatcurhtrCand{579044}) & \hatcurSPECWiFeSTeffCand{579044}& \hatcurSPECWiFeSRVKCand{579044}& NO  &  \hatcurCLASSCand{579044} & P=1.3\,d candidate (depth=1.1\%) \\
\hatcurCCEPICCand{579048} (\hatcurhtrCand{579048}) & \hatcurSPECWiFeSTeffCand{579048}& \hatcurSPECWiFeSRVKCand{579048}& NO  &  \hatcurCLASSCand{579048} & P=10.1\,d candidate (depth=2.5\%) \\
\hatcurCCEPICCand{579050} (\hatcurhtrCand{579050}) & \hatcurSPECWiFeSTeffCand{579050}& \hatcurSPECWiFeSRVKCand{579050}& YES &  \hatcurCLASSCand{579050} & ... \\
\hatcurCCEPICCand{624002} (\hatcurhtrCand{624002}) & \hatcurSPECWiFeSTeffCand{624002}& \hatcurSPECWiFeSRVKCand{624002}& YES &  \hatcurCLASSCand{624002} & ...  \\
\hatcurCCEPICCand{624003} (\hatcurhtrCand{624003}) & \hatcurSPECWiFeSTeffCand{624003}& \hatcurSPECWiFeSRVKCand{624003}& YES &  \hatcurCLASSCand{624003} & Blended in wide \kk\ apertures \\
      [-1.5ex] \enddata
\ifthenelse{\boolean{emulateapj}}{ \end{deluxetable*} }{ \end{deluxetable} }

\subsection{Reconnaissance Photometry}
\label{sec:phfu}
In order to further rule out eclipsing binaries, and refine our
ephemerides, we obtained ground-based photometric follow-up for four
\hskk\ candidates.  In this section we detail all of these
observations.

\begin{figure*}[!ht]
\plotone{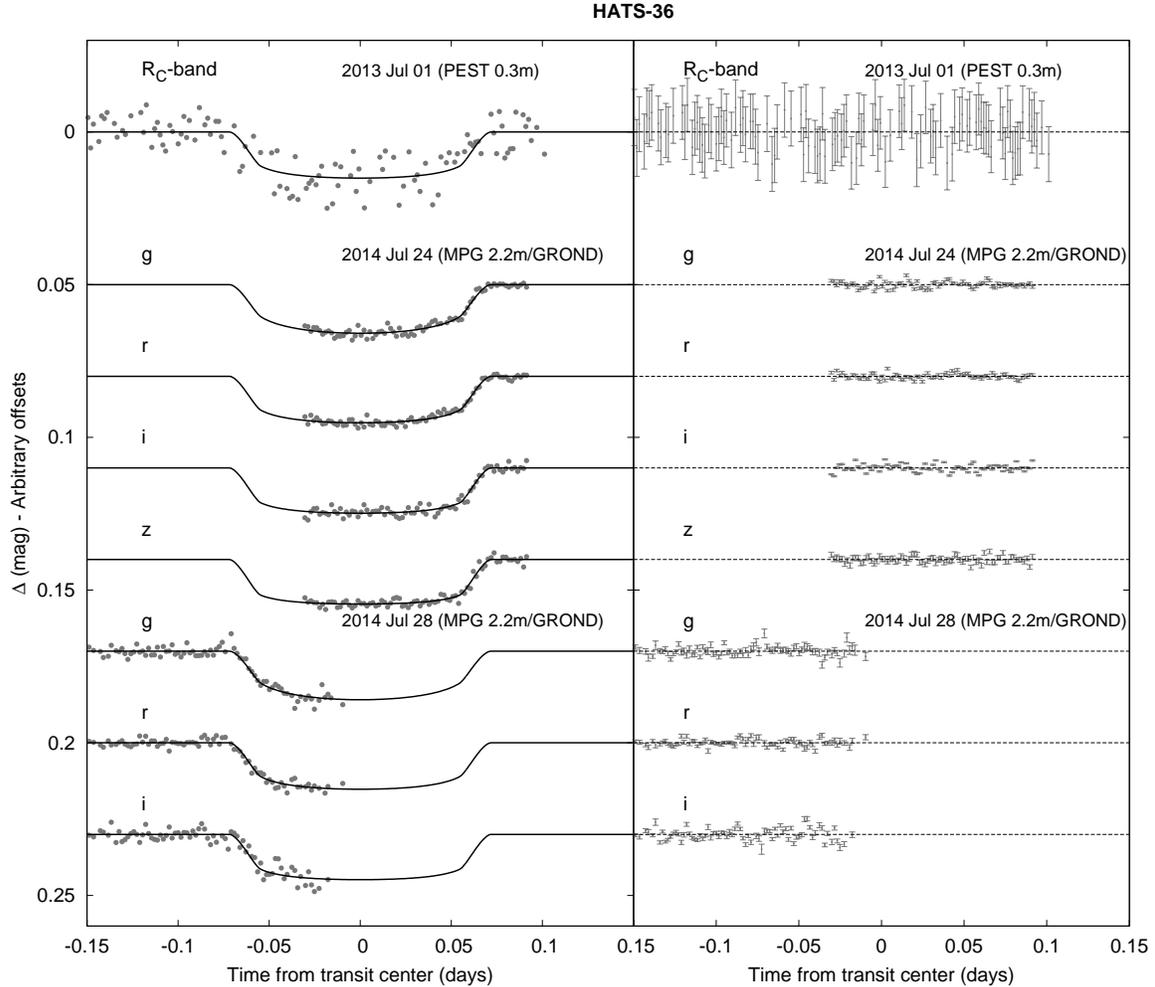}
\caption{
        Left: Unbinned ground-based follow-up transit \lcs{} of
        \hatcur{}.  The dates, filters and instruments used for each
        event are indicated.  The light curves have been detrended
        using the EPD process.  Curves after the first are shifted for
        clarity.  Our best fit is shown by the solid lines.  Right:
        Residuals from the fits in the same order as the curves at
        left.
\label{fig:lc}} \end{figure*}

\begin{figure*}[!ht]
\plotone{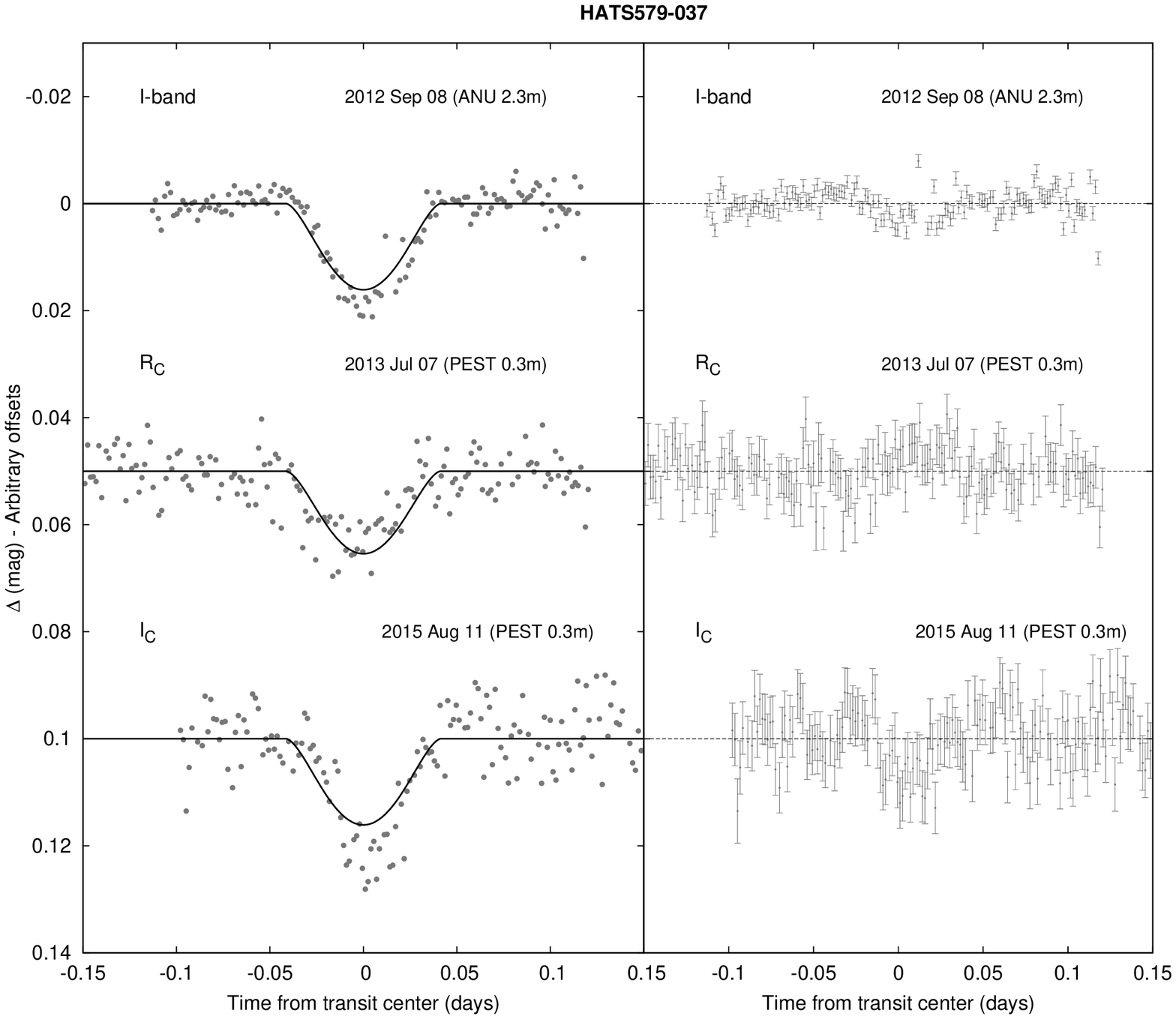}
\caption{
        Left: Unbinned ground-based follow-up transit \lcs{} of
        \hatcurhtrCand{579037}.  Plots as for Figure~\ref{fig:lc}.
        The over-plotted model is based on a global fit of all
        available data, and accounts for differences in limb darkening
        between the $R$- and $i$-bands, but assumes all other light
        curve parameters are the same. The poor fit of this model
        shows clear differences in the observed transit depths between
        the R-band and i-band observations, which allow us to classify
        this candidate as a eclipsing binary.\\
\label{fig:579037}} \end{figure*}

\begin{itemize}
\item \textit{\hatcur{} (\hatcurEPIC)}: we obtained initial
  photometric follow-up on the night of 2013 Jul 1 using the 0.3\,m
  Perth Exoplanet Survey Telescope (PEST) in Perth, Australia.  For a
  full description of the PEST facility see \cite{2014MNRAS.437.2831Z}
  and the PEST website (\url{http://pestobservatory.com}).  Imaging was
  carried out in the \band{R_{C}} with exposure times of 120\,s.  In
  total 137 exposures of \hatcur{} were taken.  Data was reduced via
  aperture photometry as described in \cite{2014MNRAS.437.2831Z}.  The
  resulting \lc\ is plotted in Figure~\ref{fig:lc} and the data is
  provided in Table~\ref{tab:photometry}.  A full transit is clearly
  detected with a depth and duration consistent with the HATSouth
  discovery data.  These data allowed us to refine the transit
  ephemeris.  The following year we observed two consecutive partial
  transits of \hatcurb{} with the multiband GROND imager
  \citep{greiner:2008} on the MPG 2.2\,m telescope at LSO in Chile.
  On the night of 2014 Jul 24 we observed a transit egress (75
  $\times$110\,s images in \band{g,r,i,z}), while on the night of 2014
  Jul 28 we observed an transit ingress (80$\times$ 110\,s images in
  \band{g,r,i}).  These data were reduced to \lcs\ via aperture
  photometry following the method set out in
  \cite{mohlerfischer:2013:hats2}.  The \lcs\ are plotted in
  Figure~\ref{fig:lc} and the data is set out in
  Table~\ref{tab:photometry}.  These high precision light curves
  confirmed the transit depth was colour independent and were
  consistent with a transiting planet.  Both the PEST and GROND
  \lcs\ for \hatcur{} are used in the global fitting described in
  Section~\ref{sec:analysis}.

\item \textit{\hatcurhtrCand{579037}(\hatcurCCEPICCand{579037})}: we
  obtained reconnaissance photometric follow-up from the ANU2.3\,m
  imaging camera at SSO on 2012 September 8 in \band{I} which showed a
  ``V''-shaped transit with a depth of 20\,mmag\ (compared with the
  10\,mmag\ transit observed in the HATSouth discovery data).  This
  colour dependent depth difference was confirmed with observations
  from the PEST 0.3\,m telescope which observed a 10\,mmag\ transit in
  \band{RC} on 2013 July 7, and a 20\,mmag\ transiting \band{I} on
  2015 August 11.  The photometric data for this candidate is set out
  in Table~\ref{tab:photometry}, while the follow-up light-curves are
  plotted in Figure~\ref{fig:579037}.  The color dependent depths are
  a strong indication that this candidate is an eclipsing
  binary.

\item \textit{\hatcurhtrCand{579007} (\hatcurCCEPICCand{579007})}: we
  obtained reconnaissance photometric follow-up from the ANU2.3\,m
  imaging camera at SSO on 2013 May 24, the imaging camera on the
  SWOPE 1\,m at LCO on 2013 August 21, and the PEST 0.3\,m PEST
  telescope on 2014 July 3 and 2015 August 27.  None of these
  observations showed a transit feature. It is likely that
  uncertainties in the HATSouth ephemeris for this candidate were
  responsible for us missing the transit event for this candidate
  during these photometric follow-up observations.

\item \textit{\hatcurhtrCand{579039} (\hatcurCCEPICCand{579039})}: we
  obtained reconnaissance photometric follow-up in \band{I} from the
  SWOPE 1\,m at LCO on 2013 August 20.  A deep (25\,mmag) V-shaped
  transit was observed, which was consistent with the \band{r}
  discovery data.  This observation confirmed the transit feature and
  refined the ephemeris for this candidate.  The data for this
  observation are set out in Table~\ref{tab:photometry}.
\end{itemize}

A summary of all photometric observations are set out in
Table~\ref{tab:photobs}.  All of the follow-up photometric data are set
out in Table~\ref{tab:photometry}.

\subsection{High Resolution Spectroscopy - Radial Velocities}
\label{sec:hispec}
\begin{figure} [ht]
\plotone{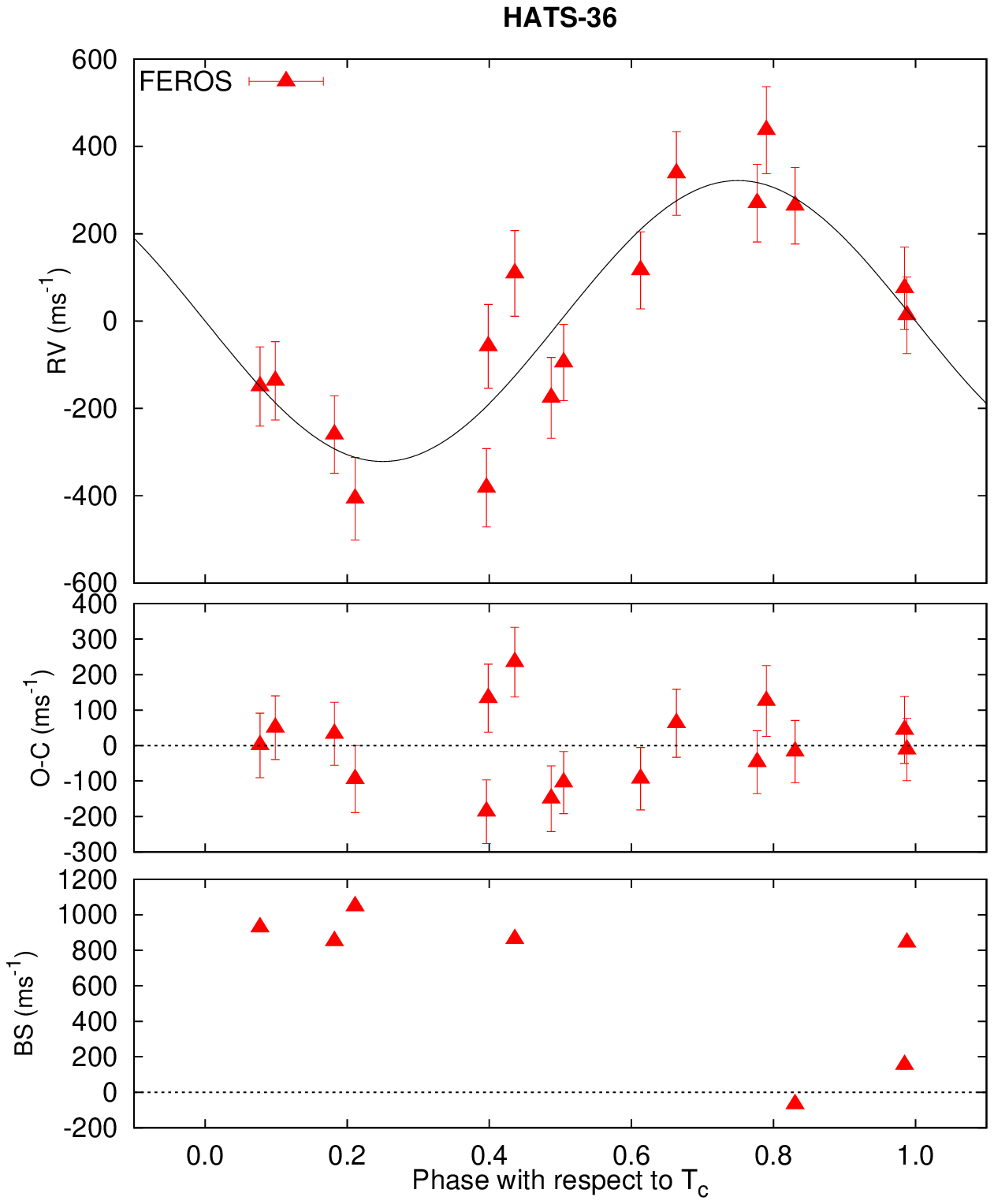}
\caption{
    {\em Top panel:} High-precision RV measurements for
    \hbox{\hatcur{}} from MPG\,2.2\,m/FEROS, together with our
    best-fit circular orbit model.  Zero phase corresponds to the time
    of mid-transit.  The center-of-mass velocity has been subtracted.
    {\em Second panel:} Velocity $O\!-\!C$ residuals from the best-fit
    model.  The error bars for each instrument include the jitter
    which is varied in the fit.  {\em Third panel:} Bisector spans
    (BS), with the mean value subtracted.  Note the different vertical
    scales of the panels.\\
\label{fig:rvbis}}
\end{figure}

We obtain radial velocity measurements for \hskk\ candidates that
remain after the reconnaissance spectroscopy and photometry set out in
Sections \ref{sec:specfu} \& \ref{sec:phfu} respectively.  An
additional magnitude constraint of V$<$14.5 is placed on candidates at
this stage, as the radial velocity monitoring of candidates fainter
than V=14.5 is beyond the reach of most telescopes/instruments.  Only
in exceptional cases such as an M-dwarf host \citep[e.g. HATS-6b;][]{2015AJ....149..166H}) do we attempt high precision radial
velocity measurements for such faint candidates.  By this criteria
just seven candidates remained: three of which have already been
published (\hatcurnineb, \hatcurelevenb, and \hatcurtwelveb) and four
which are presented below:

\begin{itemize}
\item \textit{\hatcur{} (\hatcurEPIC)}: We measured the radial
  velocity of \hatcur{} using FEROS on the MPG~2.2\,m at LSO between
  2013 July 16 and 2014 July 24. In total 16 measurements were made
  spread over the phase of the photometric period
  (\hatcurLCPshort\,day).  These data were reduced using the CERES
  FEROS echelle spectrograph pipeline described in
  \citet{brahm:2017:ceres}.  We find a radial velocity variation in
  phase with the photometric period and with an amplitude of
  K=\hatcurRVK \,\ms, indicating the transiting companion was of a
  planetary mass.  We present these radial velocity data in
  Table~\ref{tab:rvs} and plot the data along with a best fit circular
  orbit in Figure~\ref{fig:rvbis}.  These data are used in
  Section~\ref{sec:analysis} to model the global parameters of the
  system - primarily determining the mass of the transiting planet.

\item \textit{\hatcurhtrCand{579007} (\hatcurCCEPICCand{579007})}: We
  obtained multiple high resolution spectra of this target from
  several different instruments, however the data showed no radial
  velocity variation above 10\,\ms.  Coupled with the lack of a
  transit in the reconnaissance photometry (see
  Section~\ref{sec:phfu}), we put the monitoring of this candidate on
  hold until \kk\ data became available (see Section~\ref{sec:k2}).

\item \textit{\hatcurhtrCand{579015} (\hatcurCCEPICCand{579015})}: We
  obtained four FEROS observations for this candidate between 2016 May
  18 and 2017 May 31.  The observations show an in-phase radial
  velocity variation with $K=13.49715.68\pm0.011\,\kms$, indicating the
  companion is of stellar mass and the candidate is therefore an
  eclipsing binary.
\item \textit{\hatcurhtrCand{579037} (\hatcurCCEPICCand{579037})}: We
  obtained multiple high resolution spectra of this target from
  several different instruments, however the data showed no radial
  velocity variation above 10\,\ms.  Due to this lack of variation,
  along with the color-dependent transit depths (see Section
  \ref{sec:phfu}), we put this candidate on hold until \kk\ data
  became available (see Section~\ref{sec:k2}).
  
\end{itemize}

\ifthenelse{\boolean{emulateapj}}{
    \begin{deluxetable*}{lrrrrrr}
}{
    \begin{deluxetable}{lrrrrrr}
} \tablewidth{0pc} \tablecaption{
    Relative radial velocities and bisector span measurements of
    \hatcur{}.
    \label{tab:rvs}
} \tablehead{\colhead{BJD} & \colhead{RV\tablenotemark{a}} &
  \colhead{\ensuremath{\sigma_{\rm RV}}\tablenotemark{b}} &
  \colhead{BS} & \colhead{\ensuremath{\sigma_{\rm BS}}} &
  \colhead{Phase} & \colhead{Instrument}\\
    \colhead{\hbox{(2\,450\,000$+$)}} & \colhead{(\ms)} &
    \colhead{(\ms)} & \colhead{(\ms)} & \colhead{} & \colhead{} &
    \colhead{} } \startdata
 $ 6489.79939 $ & $  -175.91 $ & $    42.00 $ & \nodata      & \nodata      & $   0.487 $ & FEROS \\
 $ 6492.82238 $ & $  -406.91 $ & $    46.00 $ & $ 1047.0 $ & $   17.0 $ & $   0.211 $ & FEROS \\
 $ 6841.73036 $ & $   270.09 $ & $    33.00 $ & \nodata      & \nodata      & $   0.777 $ & FEROS \\
 $ 6842.59654 $ & $    75.09 $ & $    46.00 $ & $  153.0 $ & $   23.0 $ & $   0.985 $ & FEROS \\
 $ 6844.76808 $ & $   -94.91 $ & $    29.00 $ & \nodata      & \nodata      & $   0.505 $ & FEROS \\
 $ 6846.78538 $ & $    13.09 $ & $    30.00 $ & $  842.0 $ & $   15.0 $ & $   0.988 $ & FEROS \\
 $ 6847.59581 $ & $  -259.91 $ & $    33.00 $ & $  851.0 $ & $   16.0 $ & $   0.182 $ & FEROS \\
 $ 6852.67564 $ & $   -57.91 $ & $    49.00 $ & \nodata      & \nodata      & $   0.399 $ & FEROS \\
 $ 6852.83176 $ & $   109.09 $ & $    53.00 $ & $  862.0 $ & $   17.0 $ & $   0.436 $ & FEROS \\
 $ 6853.78104 $ & $   338.09 $ & $    49.00 $ & \nodata      & \nodata      & $   0.663 $ & FEROS \\
 $ 6855.59903 $ & $  -136.91 $ & $    35.00 $ & \nodata      & \nodata      & $   0.099 $ & FEROS \\
 $ 6856.84008 $ & $  -381.91 $ & $    35.00 $ & \nodata      & \nodata      & $   0.396 $ & FEROS \\
 $ 6857.74556 $ & $   116.09 $ & $    31.00 $ & \nodata      & \nodata      & $   0.613 $ & FEROS \\
 $ 6858.65437 $ & $   264.09 $ & $    30.00 $ & $  -68.0 $ & $   15.0 $ & $   0.831 $ & FEROS \\
 $ 6859.68462 $ & $  -149.91 $ & $    38.00 $ & $  929.0 $ & $   19.0 $ & $   0.077 $ & FEROS \\
 $ 6862.66053 $ & $   437.09 $ & $    56.00 $ & \nodata      & \nodata      & $   0.790 $ & FEROS \\
      [-1.5ex] \enddata
\tablenotetext{a}{
        The zero-point of these velocities is arbitrary. An overall
        offset $\gamma_{\rm rel}$ fitted separately to the FEROS
        velocities in \refsecl{analysis} has been subtracted.
} \tablenotetext{b}{
        Internal errors excluding the component of
        astrophysical/instrumental jitter considered in
        \refsecl{analysis}.
} \ifthenelse{\boolean{emulateapj}}{
    \end{deluxetable*}
}{
    \end{deluxetable}
}

\subsection{K2 Photometry}
\label{sec:k2}
The \kk\ mission \citep{2014PASP..126..398H} uses the \kepler\ space
telescope \citep{2010Sci...327..977B} to monitor selected fields in
the ecliptic plane for campaigns of approximately 80\,d each.  Between
2015 Oct 4 and 2015 Dec 2, \kk\ Campaign 7 monitored a field centered
on \kkfieldra, \kkfielddec.  Since this \kk\ field overlapped with the
previously monitored HATSouth field described in
\ref{sec:hsphotometry}, we were able to obtain \kk\ data for the 25
\hskk\ candidates listed in Table~\ref{tab:photobs} via the \kk\ Guest
Observer programs GO7066 and GO7067.  A summary of the \kk\ imaging is
set out in Table~\ref{tab:candidates}.

The \kk\ data was made available on \kkrelease.  We used three
different versions of the \kk\ light curves downloaded from the
Mikulski Archive for Space Telescopes (\url{http://archive.stsci.edu}).  We
used the \kk\ PDC light curves, the light curves produced by the
``self-flat-fielding" technique described in
\citet{2014PASP..126..948V}, and the \kk\ light curves from the
\everest\ open-source pipeline fully described in
\cite{2016AJ....152..100L, 2017arXiv170205488L}.  In addition we used the K2SC light curves
\citep{2016MNRAS.459.2408A} which were provided to us upon request
(B. Pope 2016, private communication).

All four of these data products are derived from the same raw pixel
data.  However the different apertures and detrending techniques
result in light curves with sometimes quite marked differences.  We
therefore analyzed all four in order to help categorize our
\hskk\ candidates.  Each \hskk\ candidate was detrended individually
to take into account the fact that each light curve potentially
contained a combiniation of \kk\ systematics, variability due to
stellar rotation, and elipsoidal variability.  We detrended the
light curves by first masking the transit/eclipse signal, and then
flattening the curve using a fifth order Savitzky–Golay filter
\citep{1964AnaCh..36.1627S}, with the window length selected to
detrend variations due to \kk\ systematics and variability due to
stellar rotation, but to avoid flattening any potential ellipsoidal
variation or secondary eclipse.  The resulting 25 light curves are
presented in the Appendix.  In most instances we utilized the Everest
light curves, however for some light curves the Everest algorithm
removed real astrophysical variability, so in those cases we used the
PDC light curves.  Features seen in these light curves, such as
secondary eclipses and out-of-transit variability, are noted in
Table~\ref{tab:summary}.

Due to the space-based environment, the large aperture (1\,m), and
near continuous 80 day coverage, all of the \kk\ light curves we used
are of very high precision - for \kk\ Campaign 7 the median 6.5-hr
combined differential photometric precision (CDPP) for a Kep=12 mag
dwarf star was 120\,ppm.  With this very high precision we are able to
see features not visible or ambiguous in the HATSouth discovery
light curves.  Most importantly, we can search for secondary eclipses,
out-of-transit ellipsoidal variation, and odd/even transit depth
differences.  These features, in an optical light curve and at
significant amplitudes, are characteristic of eclipsing binary systems
rather than transiting exoplanets.

The timing was such that the \kk\ observations and data followed
\textit{after} we had already completed the photometric and
spectroscopic follow-up of the \hskk\ candidates detailed in
Sections~\ref{sec:phfu}, \ref{sec:specfu}, \& \ref{sec:hispec}.  In
this respect some candidates had already been robustly identified as
either transiting exoplanets or eclipsing binaries before the
\kk\ data was analyzed.  However in other cases the \kk\ data were
critical to our classification of the candidate.  Here we detail the
findings for each \hskk\ candidate.

\begin{figure} [ht]
\plotone{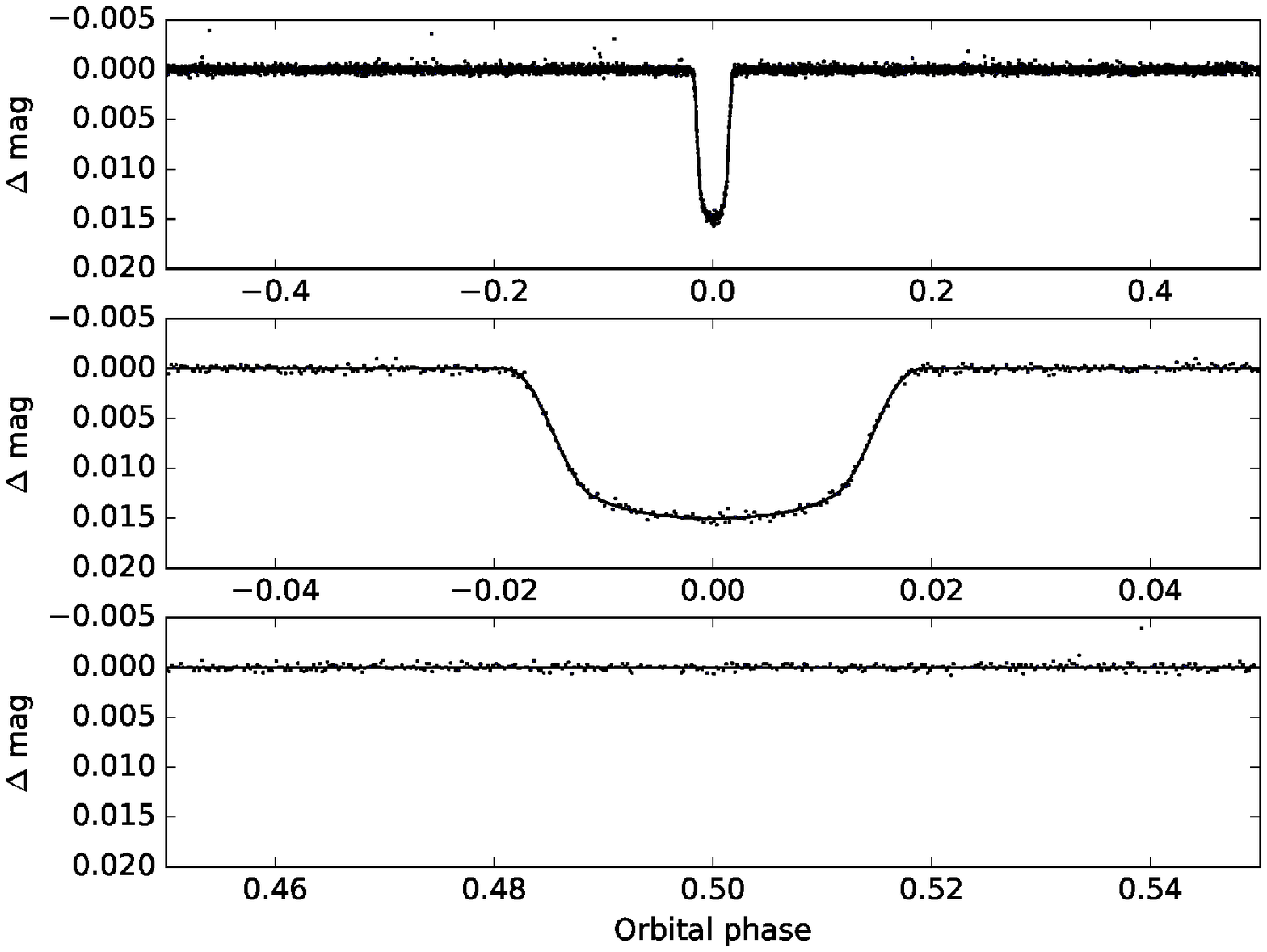}
\caption{
  \kk\ \band{Kep} \lc{} (\everest) of \hatcur{} folded with the period
  $P = \hatcurLCPprec$\,days resulting from the global fit described
  in \refsecl{analysis}.  The top panel shows the full phase-wrapped
  \lc{}.  The middle panel shows a zoom-in on the transit.  The lower
  panel shows a zoom around phase 0.5 with no detection of a secondary
  eclipse.  Points are individual \kk\ measurements.  Black solid line is 
  best-fit global model described in \refsecl{analysis}.
\label{fig:k2}}
\end{figure}

\begin{itemize}
  \item \textit{\hatcur{} (\hatcurEPIC)}: The phase folded
    \kk\ light curve for \hatcur\ is presented in Figure~\ref{fig:k2},
    and the data is tabulated in Table~\ref{tab:photometry}.  It shows
    a 15mmag U-shaped transit consistent with the discovery and
    follow-up photometry presented in Sections~\ref{sec:hsphotometry}
    \& \ref{sec:phfu} respectively.  There is no secondary eclipse,
    odd/even depth difference, or out-of-transit variation present to
    the limits of the photometry.  This \lc{} is used in our global
    analysis of this newly discovered transiting exoplanet system in
    Section~\ref{sec:analysis}.
    
  \item \textit{Known HATSouth Planets}: The \kk{} data for
    \hatcurnine, \hatcureleven, and \hatcurtwelve{} are consistent
    with the previous published exoplanet discoveries - i.e. there was
    no evidence of any secondary eclipses, odd/even depth differences,
    or out-of-transit variations.  We analyze these \lcs{} for
    additional planets, TTVs and phase modulations in
    Section~\ref{sec:analysis}.

  \item \textit{HATSouth Candidates}: \hatcurhtrCand{579040},
    \hatcurhtrCand{579044}, and \hatcurhtrCand{579048} all have
    \kk\ \lcs\ that are consistent with the HATSouth discovery data
    with no sign of secondary eclipses, odd/even depth differences, or
    out-of-transit variations.  The light curves are set out in the Appendix. 

  \item \textit{Eclipsing Binaries}: The 18 remaining
    \hskk\ candidates are eclipsing binaries.  There are 16 candidates
    that show secondary eclipses in the \kk\ light curves indicating
    they are eclipsing binaries.  The details for each candidate are
    set out in Table~\ref{tab:summary}, and the light curves are set
    out in the Appendix.  For \hatcurhtrCand{579009} we do not
    detect a secondary eclipse, however we detected a high amplitude
    (K=20\,\kms) in-phase radial velocity variation (see
    Section~\ref{sec:hispec}) indicating it is an eclipsing binary.
    Likewise \hatcurhtrCand{579037} does not have a detectable
    secondary eclipse, however it had color dependent transit depths
    (see Figure~\ref{fig:579037}) and the \kk\ light curve also shows
    out-of-transit variation when phase-folded at $\times$2 the
    discovery period.  Therefore we classify this candidate as a
    eclipsing binary.
\end{itemize}

\section{Analysis}
\label{sec:analysis}
In this section we analyze the newly discovered transiting exoplanet
\hatcurb, the three known \hs\ with \kk\ data (\hatcurnineb,
\hatcurelevenb, and \hatcurtwelveb), and the three \hskk\ targets that
remain as transiting exoplanet candidates.

\subsection{\hatcurb\ - A high-mass transiting hot Jupiter}
\label{sec:hats36b}
\begin{figure}[tbp]
\plotone{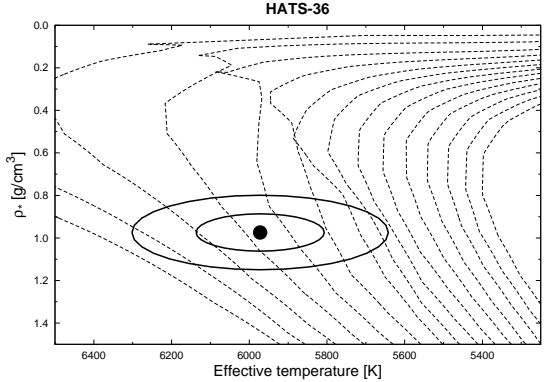}
\caption[]{
    Comparison between the measured values of \teffstar\ and
    \rhostar\ (from SPC applied to the HIRES spectra, and from our
    modelling of the light curves and RV data, respectively), and the
    Y$^{2}$ model isochrones from \citet{yi:2001}. The best-fit values
    (dark filled circle), and approximate 1$\sigma$ and 2$\sigma$
    confidence ellipsoids are shown. The values from our initial SPC
    iteration are shown with the open triangle. The Y$^{2}$ isochrones
    are shown for ages of 0.2\,Gyr, and 1.0 to 14.0\,Gyr in 1\,Gyr
    increments.
\label{fig:iso}}
\end{figure}

To derive the physical properties of \hatcur\ we obtain initial
stellar parameters from the high-resolution spectra of \hatcur{} from
FEROS, together with ZASPE \citep{2017MNRAS.tmp..146B}.  This provides a
first estimate of the temperature (\teffstar), surface gravity
(\loggstar), metallicity (\feh), and projected equatorial rotation
velocity (\vsini) of \hatcur.  The \teffstar\ and \feh\ values are
then used with the stellar density \rhostar{}, determined from the
combined light-curve and radial velocity analysis, to determine a
first estimate of the stellar physical parameters following the method
described in \citet{sozzetti:2007}.  We use the Yonsei-Yale isochrones
\citep[Y2;][]{yi:2001} to derive the stellar mass, radius and age that
best fit our estimated \teffstar, \feh\ and \rhostar{} values.  We
then determine a revised value of \loggstar\ and perform a second
iteration of ZASPE holding \loggstar\ fixed to this value while
fitting for \teffstar, \feh\ and \vsini.  We again compare this new
value of \rhostar\ to the Y2 isochrones to produce our final adopted
values for the physical stellar parameters.  Figure~\ref{fig:iso}
shows the final \loggstar\ value plotted against \teffstar, with
1-$\sigma$\ and 2-$\sigma$\ confidence ellipsoids and the appropriate
Y2 isochrones for various stellar ages.  The final parameters indicate
that \hatcur{} is a G0V dwarf star (\teffstar=\hatcurSMEiteff\,K,
\loggstar=\hatcurISOlogg) with a mass and radius of
\mstar=\hatcurISOr\,\msun\ and
\rstar=\hatcurISOm\,\rsun\ respectively.  The metallicity is slightly
above solar with \feh=\hatcurSMEiizfeh.  The rotational velocity
(\vsini=\hatcurSMEiivsin\,\kms) and age (\hatcurISOage\,Gyr) are not
atypical for a star of this type and class.  The full list of final
stellar parameters are set out in Table~\ref{tab:stellar}.


\ifthenelse{\boolean{emulateapj}}{
  \begin{deluxetable*}{lcr}
}{
  \begin{deluxetable}{lcr}
} \tablewidth{0pc} \tabletypesize{\scriptsize} \tablecaption{ Stellar
  Parameters for \hatcur{}
    \label{tab:stellar}
} \tablehead{ \multicolumn{1}{c}{~~~~~~~~Parameter~~~~~~~~} &
  \multicolumn{1}{c}{Value} &
  \multicolumn{1}{c}{Source\tablenotemark{a} } } \startdata
\noalign{\vskip -3pt}
\sidehead{Astrometric properties and cross-identifications}
~~~~2MASS-ID\dotfill & \hatcurCCtwomass{} & \\
~~~~R.A. (J2000)\dotfill & \hatcurCCra{} &
2MASS\\ ~~~~Dec. (J2000)\dotfill & \hatcurCCdec{} &
2MASS\\ ~~~~$\mu_{\rm R.A.}$ (\masy) & \hatcurCCpmra{} &
UCAC4\\ ~~~~$\mu_{\rm Dec.}$ (\masy) & \hatcurCCpmdec{} & UCAC4\\
\sidehead{Spectroscopic properties} ~~~~$\teffstar$ (K)\dotfill &
\hatcurSMEteff{} & ZASPE \\ ~~~~$\feh$\dotfill & \hatcurSMEzfeh{} &
ZASPE \\ ~~~~$\vsini$ (\kms)\dotfill & \hatcurSMEvsin{} & ZASPE
\\ ~~~~$\gamma_{\rm RV}$ (\kms)\dotfill & \hatcurRVgammaabs{} & FEROS
\\
\sidehead{Photometric properties} ~~~~$B$ (mag)\dotfill &
\hatcurCCtassmB{} & APASS \\ ~~~~$V$ (mag)\dotfill & \hatcurCCtassmv{}
& APASS \\ ~~~~$g$ (mag)\dotfill & \hatcurCCtassmg{} & APASS
\\ ~~~~$r$ (mag)\dotfill & \hatcurCCtassmr{} & APASS \\ ~~~~$i$
(mag)\dotfill & \hatcurCCtassmi{} & APASS \\ ~~~~$Kep$ (mag)\dotfill &
\hatcurCCkep{} & EPIC \\ ~~~~$g_{Gaia}$ (mag)\dotfill &
\hatcurCCgaia{} & \gaia{} DR1 \\ ~~~~$J$ (mag)\dotfill &
\hatcurCCtwomassJmag{} & 2MASS \\ ~~~~$H$ (mag)\dotfill &
\hatcurCCtwomassHmag{} & 2MASS \\ ~~~~$K_s$ (mag)\dotfill &
\hatcurCCtwomassKmag{} & 2MASS \\
\sidehead{Derived properties} ~~~~$\mstar$ ($\msun$)\dotfill &
\hatcurISOmlong{} & Y$^{2}$+\hatcurlumind{}+ZASPE\\ ~~~~$\rstar$
($\rsun$)\dotfill & \hatcurISOrlong{} &
Y$^{2}$+\hatcurlumind{}+ZASPE\\ ~~~~$\loggstar$ (cgs)\dotfill &
\hatcurISOlogg{} & Y$^{2}$+\hatcurlumind{}+ZASPE \\ ~~~~$\lstar$
($\lsun$)\dotfill & \hatcurISOlum{} & Y$^{2}$+\hatcurlumind{}+ZASPE
\\ ~~~~$M_V$ (mag)\dotfill & \hatcurISOmv{} &
Y$^{2}$+\hatcurlumind{}+ZASPE \\ ~~~~$M_K$ (mag,\hatcurjhkfilset{})&
\hatcurISOMK{} & Y$^{2}$+\hatcurlumind{}+ZASPE \\ ~~~~Age
(Gyr)\dotfill & \hatcurISOage{} & Y$^{2}$+\hatcurlumind{}+ZASPE
\\ ~~~~$A_{V}$ (mag)\dotfill & \hatcurXAv{} &
Y$^{2}$+\hatcurlumind{}+ZASPE\\ ~~~~Distance (pc)\dotfill &
\hatcurXdistred{} & Y$^{2}$+\hatcurlumind{}+ZASPE \enddata
\tablenotetext{a}{2MASS \citep{skrutskie:2006}; APASS
  \citep{henden:2009}; \gaia{} \citep{2016A&A...595A...4L}; ZASPE = Zonal
  Atmospherical Stellar Parameter Estimator routine for the analysis
  of high-resolution spectra \citep{2017MNRAS.tmp..146B}; Y$^{2}$
  isochrones \citep{yi:2001}} \ifthenelse{\boolean{emulateapj}}{
  \end{deluxetable*}
}{
  \end{deluxetable}
}

To exclude blended eclipsing binary scenarios for \hatcurb, we carried
out an analysis of all the data following the methodology set out in
\cite{hartman:2012:hat39hat41}.  We model the photometric data,
including the \kk\ observations, as an eclipsing binary system blended
with a third star.  The stars in the model are constrained using the
Padova isochrones \citep{girardi:2000}, and also must have a blended
spectrum consistent with the determined atmospheric parameters. We
also simulate composite cross-correlation functions (CCFs) and use
them to predict radial velocities and bisector spans for each blend
scenario.  All blend models tested can be rejected with $> 6\sigma$
confidence based on the photometry alone. Moreover, none of the blend
models tested would produce RV variations or non-variable bisectors
consistent with the observations.

To determine the properties of \hatcurb, we globally model the
photometric and spectroscopic data following \citet{pal:2008:hat7},
\citet{bakos:2010:hat11}, and \citet{hartman:2012:hat39hat41}.  We fit
\citet{mandel:2002} transit models to the light-curves, and a
Keplerian orbit is fit to the radial velocity measurements presented
in Section~\ref{sec:hispec}, allowing for RV jitter.  For the
ground-based light curves we fixed the quadratic limb-darkening
coefficients to tabulated values based on the stellar atmospheric
parameters, while for the \kk\ light curve we allowed the quadratic
limb darkening coefficients to vary in the fit. For the long-cadence
\kk\ observations we made use of the \everest\ light curves, with the
transit model numerically integrated over the exposure times. This
light curve showed large amplitude quasi-periodic variations, likely
due to a combination of low frequency systematic errors in the
\kk\ photometry, and the rotational modulation of stellar active
regions on the surface of \hatcur. We discuss the stellar activity
signature later in this section. In order to model these variations in
our analysis we made use of a Morlet-type wavelet basis and a low
order polynomial. This model has the form:
\begin{eqnarray}
  \sum_{i=0}^{N_{\rm poly}}c_{i}(t-T_{0})^i + \sum_{j=1}^{N_{\rm Morlet}}e^{-0.5((t-\tau_{j})/\sigma)^2} \nonumber \\
  \left( \sum_{k=1}^{N_{\rm harm}}(a_{j,k}\cos((t-T_{0})k\nu)+b_{j,k}\sin((t-T_{0})k\nu))\right)
\end{eqnarray}
where $t$ is the time of observation, $T_{0}$ is a fixed reference
epoch, $\tau_{j}$ are a fixed evenly spaced set of times for centering
the wavelets, $\sigma$ is a fixed wavelet width which we set equal to
$\tau_{j+1}-\tau_{j}$, $\nu$ is fixed to the dominant quasi-periodic
frequency, and $c_{i}$, $a_{j,k}$, and $b_{j,k}$ are linearly
optimized parameters in the model. For our analysis of \hatcurb\ we
adopted $N_{\rm poly} = 2$, $N_{\rm Morlet} = 7$ and $N_{\rm harm} =
4$.  The resulting \kk{} light curve is shown in
Figure~\ref{fig:k2}, while the data are part of the photometry set
presented in Table~\ref{tab:photometry}. The model is fit
simultaneously with the transit model as part of the global analysis.

We use a Differential Evolution Markov Chain Monte Carlo procedure to
determine the posterior distribution of the parameters.  We compare
the Bayesian evidence for an $e$=0 fixed eccentricity model to a
free-$e$\ model, and find that the fixed circular model is
preferred. We therefore adopt the circular orbit model.  The $95\%$
confidence upper limit on the eccentricity is $e < 0.294$.  The
planetary parameters resulting from this global fit are set out in
Table~\ref{tab:planetparam}.  \hatcurb\ is a high mass
(\mpl=\hatcurPPm\,\mjup), Jupiter-sized (\rpl=\hatcurPPr\,\rjup)
transiting planet with an orbital period of \hatcurLCP\,d.  It has a
bulk density of \hatcurPPrho\,\gcmc.

\begin{figure}[]
\plotone{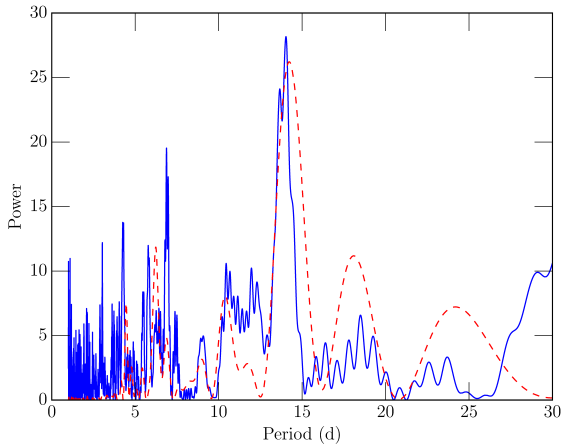}
\caption[]{
  The power spectrum from a Lomb-Scargle analysis of the HATSouth
  photometry (blue solid line) and the \kk\ photometry (red dashed
  line) for \hatcur{}.  The peak at $\sim 14$\,d is the likely rotational
  period of the star, and is prominent in both data sets.
\label{fig:rotation}}
\end{figure}

From our radial velocity monitoring we measure a high jitter
(\hatcurRVjitter\,\ms), hinting at strong stellar activity.  Jitter of
this amplitude is seen in other other hot Jupiter hosts, e.g. HATS-27b
with RV$_{jitter}$=72\,\ms \citep{2016AJ....152..108E}.  However in
this case we have a precise \kk\ light curve with which to examine the
underlying nature of this radial velocity jitter.  In order to do this
we take the \hatcur\ \kk\ light curve \citep{2014PASP..126..948V},
remove the transit events, and examine the longer time-scale
variability using a Lomb Scargle analysis.  From this analysis we
detect a strong modulation at 14.4~d with a peak-to-peak amplitude of
approximately $0.5\%$.  Such a rotational period is typical for a star
of the spectral type and class of \hatcur{}
\citep{2014ApJS..211...24M}.  We confirm this rotational signal is
also present in the HATSouth discovery light curve, again showing
modulation at P=14~d.  The Lomb-Scargle power spectrum is presented in
Figure~\ref{fig:rotation} for both the \kk\ data and the HATSouth data.  Assuming $\rstar=\hatcurISOrlong$, this
rotation period would result in an equatorial rotational of
$v_{e}=4.17\,\kms$, in contrast the spectroscopic derived
$\vsini=\hatcurSMEiivsin\,\kms$.  The fact that the derived rotational
period is $2-\sigma$ below the spectroscopic \vsini{} may be due to
non-equatorial spots and solar-like differential rotation.

\ifthenelse{\boolean{emulateapj}}{
  \begin{deluxetable*}{lc}
}{
  \begin{deluxetable}{lc}
} \tabletypesize{\scriptsize} \tablecaption{Parameters for the
  transiting planet \hatcurb{}.\label{tab:planetparam}} \tablehead{
  \multicolumn{1}{c}{~~~~~~~~Parameter~~~~~~~~} &
  \multicolumn{1}{c}{Value} } \startdata \noalign{\vskip -3pt}
\sidehead{\Lc{} parameters} ~~~$P$ (days) \dotfill & $\hatcurLCP{}$
\\ ~~~$T_c$ (${\rm BJD}$) \tablenotemark{} \dotfill & $\hatcurLCT{}$
\\ ~~~$T_{14}$ (days) \tablenotemark{} \dotfill & $\hatcurLCdur{}$
\\ ~~~$T_{12} = T_{34}$ (days) \tablenotemark{} \dotfill &
$\hatcurLCingdur{}$ \\ ~~~$\arstar$ \dotfill & $\hatcurPPar{}$
\\ ~~~$\zrstar$ \tablenotemark{} \dotfill & $\hatcurLCzeta{}$\phn
\\ ~~~$\rpl/\rstar$ \dotfill & $\hatcurLCrprstar{}$ \\ ~~~$b^2$
\dotfill & $\hatcurLCbsq{}$ \\ ~~~$b \equiv a \cos i/\rstar$ \dotfill
& $\hatcurLCimp{}$ \\ ~~~$i$ (deg) \dotfill & $\hatcurPPi{}$\phn \\

\sidehead{Limb-darkening coefficients \tablenotemark{}} ~~~$c_1,i$
(linear term) \dotfill & $\hatcurLBii{}$ \\ ~~~$c_2,i$ (quadratic
term) \dotfill & $\hatcurLBiii{}$ \\ ~~~$c_1,r$ \dotfill &
$\hatcurLBir{}$ \\ ~~~$c_2,r$ \dotfill & $\hatcurLBiir{}$
\\ ~~~$c_1,{\rm kep}$ \dotfill & $0.335 \pm 0.051$ \\ ~~~$c_2,{\rm
  kep}$ \dotfill & $0.291 \pm 0.090$ \\

\sidehead{RV parameters} ~~~$K$ (\ms) \dotfill &
$\hatcurRVK{}$\phn\phn \\ ~~~$e$ \tablenotemark{} \dotfill &
$\hatcurRVeccentwosiglimeccen{}$ \\ ~~~RV jitter (\ms)
\tablenotemark{} \dotfill & \hatcurRVjitter{} \\

\sidehead{Planetary parameters} ~~~$\mpl$ ($\mjup$) \dotfill &
$\hatcurPPmlong{}$ \\ ~~~$\rpl$ ($\rjup$) \dotfill &
$\hatcurPPrlong{}$ \\ ~~~$C(\mpl,\rpl)$ \tablenotemark{} \dotfill &
$\hatcurPPmrcorr{}$ \\ ~~~$\rhopl$ (\gcmc) \dotfill & $\hatcurPPrho{}$
\\ ~~~$\log g_p$ (cgs) \dotfill & $\hatcurPPlogg{}$ \\ ~~~$a$ (AU)
\dotfill & $\hatcurPParel{}$ \\ ~~~$T_{\rm eq}$ (K) \tablenotemark{}
\dotfill & $\hatcurPPteff{}$ \\ ~~~$\Theta$ \tablenotemark{} \dotfill
& $\hatcurPPtheta{}$ \\ ~~~$\langle F \rangle$
($10^{\hatcurPPfluxavgdim{}}$\ergscmsq) \tablenotemark{} \dotfill &
$\hatcurPPfluxavg{}$ \\ [-1.5ex] \enddata

\tablenotetext{}{For each parameter we give the median value and
  68.3\% (1$\sigma$) confidence intervals from the posterior
  distribution.  Detailed notes on parameters can be found in Table~6
  of \citet{2016AJ....152..108E}.}

\ifthenelse{\boolean{emulateapj}}{
  \end{deluxetable*}
}{
  \end{deluxetable}
}
%


\subsection{HATS-9b,11b,12b}
\label{sec:hats91112}
High precision \kk\ data allow us to revisit the three transiting
exoplanets already published by the HATSouth team, namely
\hatcurnineb{} \citep{2015AJ....150...33B} and \hatcurelevenb{} \&
\hatcurtwelveb{} \citep{2016AJ....152...88R}.  We augment our
originally published photometric and spectroscopic data with the new
\kk\ photometry and re-run the global modelling with the methodology
described in those papers and in Section~\ref{sec:hats36b}.  Here we
make use of the \citep{2016AJ....152..108E} light curves rather than
the \everest\ light curves, and we apply filter low frequency
variations due to stellar activity and instrumental errors prior to
fitting the transit model. For \hatcurnineb{} and \hatcurelevenb{}
short cadence data was available, and we made use of these
observations, rather than the long cadence, in our analysis.

The
results allow us to improve the precision for the planetary parameters
for these systems, and we list these in Table~\ref{tab:updateparam}.
Most of the revisions to the planetary parameters are relatively
minor.  The largest change is for the radius of \hatcurnineb, which is
revised upwards by almost 10$\%$.  This is primarily due to the
limited photometric follow-up that was available when the parameters
of \hatcurnineb{} were calculated in the original
analysis of \citet{2015AJ....150...33B}.

For the case of \hatcurelevenb{} our global modelling, assuming a
circular orbit, also finds a secondary eclipse with a planet-to-star
flux ratio of 0.0032$\pm$0.0012.

In addition to improving the parameters for these planets, we utilize
the \kk\ light curves to search for additional transiting planets.
The discoveries of WASP-47d and WASP-47e \citep{2015ApJ...812L..18B}
show that hot Jupiters can have nearby planetary neighbours.  After
removing the transit events from the hot Jupiters, we search the
\kk\ light curves of \hatcurnine, \hatcureleven, and \hatcurtwelve,
using the BLS search algorithm as discussed in Section
\ref{sec:hsphotometry}.  We do not find any significant evidence for
additional transiting planets in any of these systems.

Using the original light curves and the \kk\ data, we check for any
changes in the timing of the transits for these exoplanets.  We do
this by fitting the best fit transit model (using the parameters set
out in Table~\ref{tab:updateparam}) to each individual transit event,
and measure the difference between the best fit central transit time
and the expected transit time from a purely Keplerian orbit.  For all
three systems we find no evidence of any transit timing variations.

Finally we analyze the \kk\ light curves for evidence for of
additional variability. \hatcureleven{} and \hatcurtwelve{} only show
long term drift in the \kk\ data that are likely caused by systematics
from the spacecraft.  \hatcurnine{} shows an additional modulation
with a period of approximately 8.8~d and a peak-to-peak amplitude of
approximately $0.1\%$.  If this modulation is indeed due to stellar
rotation, it would imply an equatorial rotation of $v_{e}=8.6\,\kms$.
Given the spectroscopic \vsini{} of \hatcurnine{} is
$4.58\pm0.90\,\kms$, this would mean the spin axis of the star is
inclined with respect to our line of sight, and that the planet is on a
mis-aligned orbit.


\ifthenelse{\boolean{emulateapj}}{
  \begin{deluxetable*}{lcc}
}{
  \begin{deluxetable}{lcc}
} \tabletypesize{\scriptsize} \tablecaption{Updated planet parameters.
  \label{tab:updateparam}}
\tablehead{
    \multicolumn{1}{c}{~~~~~~~~Parameter~~~~~~~~} &
    \multicolumn{1}{c}{Previous Value} &
    \multicolumn{1}{c}{This work}
}
\startdata
\noalign{\vskip -3pt}
\sidehead{\textbf{\hatcurnineb{}}}
~~~$P$ (days)             \dotfill    & $\hatcurLCPnine{}$             &1.91531100 $\pm$ 0.00000094 \\
~~~$T_c$ (${\rm BJD}$)    \dotfill    & $\hatcurLCTnine{}$             &2457380.702470$\pm$ 0.000036  \\
~~~$T_{14}$ (days)         \dotfill    & $\hatcurLCdurnine{}$          & 0.14618 $\pm$ 0.00017 \\
~~~$\rpl/\rstar$          \dotfill    & $\hatcurLCrprstarnine{}$      & 0.08316 $\pm$ 0.00014  \\
~~~$i$ (deg)              \dotfill    & $\hatcurPPinine{}$            & 88.94 $\pm$ 0.44   \\
~~~$\mpl$ ($\mjup$)       \dotfill    & $\hatcurPPmlongnine{}$        & 0.816 $\pm$ 0.038   \\
~~~$\rpl$ ($\rjup$)       \dotfill    & $\hatcurPPrlongnine{}$        & 1.1724 $\pm$ 0.0098  \\
~~~$\rhopl$ (\gcmc)       \dotfill    & $\hatcurPPrhonine{}$          & 0.626 $_{-0.022}^{0.029}$  \\ [1.5ex]

\sidehead{\textbf{\hatcurelevenb{}}}
~~~$P$ (days)             \dotfill    & $\hatcurLCPeleven{}$             & 3.6191634 $\pm$ 0.0000031  \\
~~~$T_c$ (${\rm BJD}$)    \dotfill    & $\hatcurLCTeleven{}$             & 2457378.419100 $\pm$ 0.000069   \\
~~~$T_{14}$ (days)         \dotfill    & $\hatcurLCdureleven{}$           & 0.18202 $\pm$ 0.00025 \\
~~~$\rpl/\rstar$          \dotfill    & $\hatcurLCrprstareleven{}$       & 0.10721 $\pm$ 0.00028   \\
~~~$i$ (deg)              \dotfill    & $\hatcurPPieleven{}$             & 89.03 $_{-0.47}^{+0.36}$  \\
~~~$\mpl$ ($\mjup$)       \dotfill    & $\hatcurPPmlongeleven{}$         & 0.83 $\pm$ 0.10   \\
~~~$\rpl$ ($\rjup$)       \dotfill    & $\hatcurPPrlongeleven{}$         & 1.487 $\pm$ 0.031  \\
~~~$\rhopl$ (\gcmc)       \dotfill    & $\hatcurPPrhoeleven{}$           & 0.315 $\pm$ 0.037  \\ [1.5ex]

\sidehead{\textbf{\hatcurtwelveb{}}}
~~~$P$ (days)             \dotfill    & $\hatcurLCPtwelve{}$             & 3.1428347 $\pm$ 0.0000022  \\
~~~$T_c$ (${\rm BJD}$)    \dotfill    & $\hatcurLCTtwelve{}$             & 2457364.66541 $\pm$ 0.00012  \\
~~~$T_{14}$ (days)         \dotfill    & $\hatcurLCdurtwelve{}$           & 0.18849 $\pm$ 0.00064 \\
~~~$\rpl/\rstar$          \dotfill    & $\hatcurLCrprstartwelve{}$       & 0.06316 $\pm$ 0.00063  \\
~~~$i$ (deg)              \dotfill    & $\hatcurPPitwelve{}$             & 82.27 $\pm$ 0.64   \\
~~~$\mpl$ ($\mjup$)       \dotfill    & $\hatcurPPmlongtwelve{}$         & 2.390 $\pm$ 0.087  \\
~~~$\rpl$ ($\rjup$)       \dotfill    & $\hatcurPPrlongtwelve{}$         & 1.384 $\pm$ 0.059  \\
~~~$\rhopl$ (\gcmc)       \dotfill    & $\hatcurPPrhotwelve{}$           & 1.12 $\pm$ 0.14   \\ [-1.5ex]

\enddata
\ifthenelse{\boolean{emulateapj}}{
  \end{deluxetable*}
}{
  \end{deluxetable}
}
%


\subsection{Candidates}
\label{sec:k2candidates}
Three of the HATSouth candidates from the \hskk\ campaign remain
viable transiting exoplanet candidates, but have not been confirmed
via radial velocity measurements due to the faintness of the host
stars.  In this Section we discuss the details for each candidate.

\begin{itemize}
\item \textit{\hatcurhtrCand{579040} (\hatcurCCEPICCand{579040})}: The
  transiting candidate has a period of P=3.905\,d and shows a 17\,mmag
  ``U''-shaped transit.  The host star is a faint
  (V=\hatcurCCmagVCand{579040}) G9 dwarf
  (\teff=\hatcurSPECWiFeSTeffCand{579040}\,K).  Assuming a host star
  mass of 0.9\,\mstar, and with a radial velocity semi-amplitude of
  K\hatcurSPECWiFeSRVKCand{579040}\,\kms, we can constrain the
  companion mass to be less than 15\mjup\ for an unblended
  single-host/single companion system.  \gaia{} DR1
  \citep{2016A&A...595A...4L} shows a single source
  (G$_{Gaia}$=$\GAIAa$) having no neighbours within 15$\arcsec$\ down
  to the \gaia{} DR1 magnitude limit (G$_{Gaia}\sim20$) and separation
  limit ($\sim1\arcsec$). 
  
\item \textit{\hatcurhtrCand{579044} (\hatcurCCEPICCand{579044})}: The
  transiting candidate has a period of P=1.320\,d and shows a 10\,mmag
  ``U''-shaped transit.  The host star is faint
  (V=\hatcurCCmagVCand{579044}) G0 dwarf
  (\teff=\hatcurSPECWiFeSTeffCand{579044}\,K).  Assuming a host star
  with 1.1\,\mstar, and with a radial velocity semi-amplitude of
  K\hatcurSPECWiFeSRVKCand{579044}\,\kms, we can constrain the
  companion mass to be less than 11.5\mjup\ for an unblended
  single-host/single companion system.  \gaia{} DR1 shows the host is a
  single source (G$_{Gaia}$=$\GAIAb$), having no neighbours within
  15$\arcsec$.

\item \textit{\hatcurhtrCand{579048} (\hatcurCCEPICCand{579048})}: The
  transiting candidate has a relatively long period of P=10.148\,d and
  shows a 20\,mmag ``U''-shaped transit.  The host star is a
  faint (V=\hatcurCCmagVCand{579048}) G9 dwarf
  (\teff=\hatcurSPECWiFeSTeffCand{579048}\,K).  Assuming a host star
  mass of 0.9\,\mstar, with a radial velocity semi-amplitude of
  K\hatcurSPECWiFeSRVKCand{579048}\,\kms, we would determine a
  companion mass of 70\mjup\ - approximately at the lower limit for
  H-burning.  If confirmed, this would join the small population of
  known transiting brown dwarfs, and would follow the trend of having
  a longer orbital period than typical hot Jupiters
  \citep{2017AJ....153...15B}.  From the \gaia{} DR1 we see a primary
  source (G$_{Gaia}$=$\GAIAc$), with a neighbour at 9.5$\arcsec$
  (G$_{Gaia}$=$\GAIAd$).  However by analyzing multiple pixel-aperture
  sizes from the SFF \kk\ light curves \citep{2014PASP..126..948V}, we
  determine the detected transit signal originates from the primary
  candidate star rather than the fainter neighbor.
  
\end{itemize}


\section{Discussion}
\label{sec:discussion}
This is the first time we have vetted HATSouth candidates using the
high precision photometry afforded by the Kepler telescope, although
it has been done for HATNet under \kk\ program GO0116 (PI Bakos)
resulting in the discovery of HAT-P-56b \citep{2015AJ....150...85H}.
In the cases of both \hatcurb\ and HAT-P-56b the radial velocity
semi-amplitudes are high ($\sim300$\,\ms), but the stellar
jitter is also high ($\sim100$\,\ms) and thus the \kk\ data is especially
helpful in robustly confirming the nature of the systems.

\subsection{HATS-36b}
\label{sec:hats36}
\begin{figure}[tbp]
\plotone{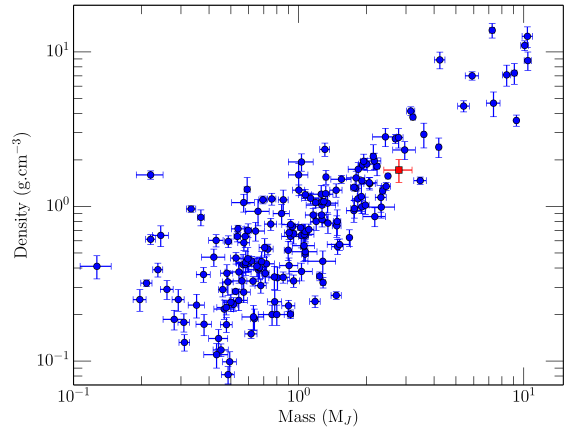}
\caption[]{
    Mass-density relationship of all well-characterised (density
    uncertainty $<$20\%) giant exoplanets.  Blue circles are data
    from NASA Exoplanet Archive (23 November 2016) and the red square
    is \hatcurb.
\label{fig:density}}
\end{figure}

\hatcurb\ is a hot Jupiter with a typical orbital period
(P=\hatcurLCP\,d).  The star is active, which we see manifest in both
the variability in the LC and the high jitter in the radial velocity
measurements.  Due to its high mass compared with the known
population of hot Jupiters, \hatcurb\ lies in a relatively sparsely
populated region of the mass-density relationship for gas giant
exoplanets (see Figure~\ref{fig:density}).  However its bulk density
fits well on the mass-density sequence of gas giants.

\subsection{Candidates}
\label{sec:candidates}
Secondary eclipses in the \kk\ allow us to robustly rule out 17 of
the 25 \hskk\ candidates.  However we have three candidates that remain
active.  All three candidates require future radial velocity
monitoring in order to determine if they are transiting
exoplanets/brown dwarfs or (blended) eclipsing binaries.  However such
a task is extremely difficult to due to faintness of the host stars.
\hatcurhtrCand{579048} (\hatcurCCEPICCand{579048}) is the most
promising as we have an indication of a velocity variation of
K=\hatcurSPECWiFeSRVKCand{579048}\,\kms.  However the host star is also
the faintest of the three candidates at
(V=\hatcurCCmagVCand{579048}).

\subsection{Outlook}
\label{sec:outlook}
This program shows the benefit of using high precision
\kk\ space-based photometry to vet candidates identified from
ground-based surveys.  This concept will naturally extend to the TESS
mission \citep{ricker:2014:TESS}.  The primary differences will
be twofold.  Firstly, TESS will monitor most stars for only 27
days, about one-third the duration of the K2 campaigns.  For many of
our \hskk\ candidates such duration would result in only one or two
transits to be observed.  In these cases the value of combining the
TESS data with ground-based monitoring is greatly enhanced.  Secondly,
the spatial resolution of TESS is just $21\arcsec.1~$pixel$^{-1}$,
meaning many blended systems that can be resolved in \kk, such as the
blended eclipsing binaries presented in this work, will not be readily
identifiable from TESS data alone.  In these cases ground based data
such as that from HATSouth ($3\arcsec.7~$pixel$^{-1}$) will be highly
beneficial in resolving the nature of the systems.


\acknowledgements 
\paragraph{Acknowledgements}
Development of the HATSouth project was funded by NSF MRI grant
NSF/AST-0723074, operations have been supported by NASA grants
NNX09AB29G, NNX12AH91H, and NNX17AB61G, and follow-up observations
receive partial support from grant NSF/AST-1108686.
This work has been carried out within the framework of the National
Centre for Competence in Research PlanetS supported by the Swiss
National Science Foundation. DB acknowledges the financial
support of the SNSF.
J.H.\ acknowledges support from NASA grant NNX14AE87G.
A.J.\ acknowledges support from FONDECYT project 1171208, BASAL CATA
PFB-06, and project IC120009 ``Millennium Institute of Astrophysics
(MAS)'' of the Millenium Science Initiative, Chilean Ministry of
Economy. N.E.\ is supported by CONICYT-PCHA/Doctorado
Nacional. R.B.\ and N.E.\ acknowledge support from project
IC120009 ``Millenium Institute of Astrophysics (MAS)'' of the
Millennium Science Initiative, Chilean Ministry of Economy.
V.S.\ acknowledges support form BASAL CATA PFB-06.  
A.V. is supported by the NSF Graduate Research Fellowship, Grant No. DGE 1144152.
This paper includes data collected by the K2 mission. Funding for the
K2 mission is provided by the NASA Science Mission directorate.
The K2 observations presented here were obtained through the GO program, 
with analysis supported by NASA grant NNX16AE68G.
This work is based on observations made with ESO Telescopes at the La
Silla Observatory.
This paper also uses observations obtained with facilities of the Las
Cumbres Observatory Global Telescope.
We acknowledge the use of the AAVSO Photometric All-Sky Survey (APASS),
funded by the Robert Martin Ayers Sciences Fund, and the SIMBAD
database, operated at CDS, Strasbourg, France.
Operations at the MPG~2.2\,m Telescope are jointly performed by the
Max Planck Gesellschaft and the European Southern Observatory.  The
imaging system GROND has been built by the high-energy group of MPE in
collaboration with the LSW Tautenburg and ESO\@.  We thank the MPG
2.2m telescope support team for their technical assistance during
observations."


\clearpage
\appendix
Here we present the \kk{} light curves for the 24 \hskk{} candidates
(\hatcur{} is excluded).  Light curves are from Everest (EV) or the
\kk\ PDC pipeline (PDC).  Classification is transiting exoplanet
(TEP), candidate (CAND), eclipsing binary (EB) or blended eclipsing
binary (BEB).\\
\plottwo{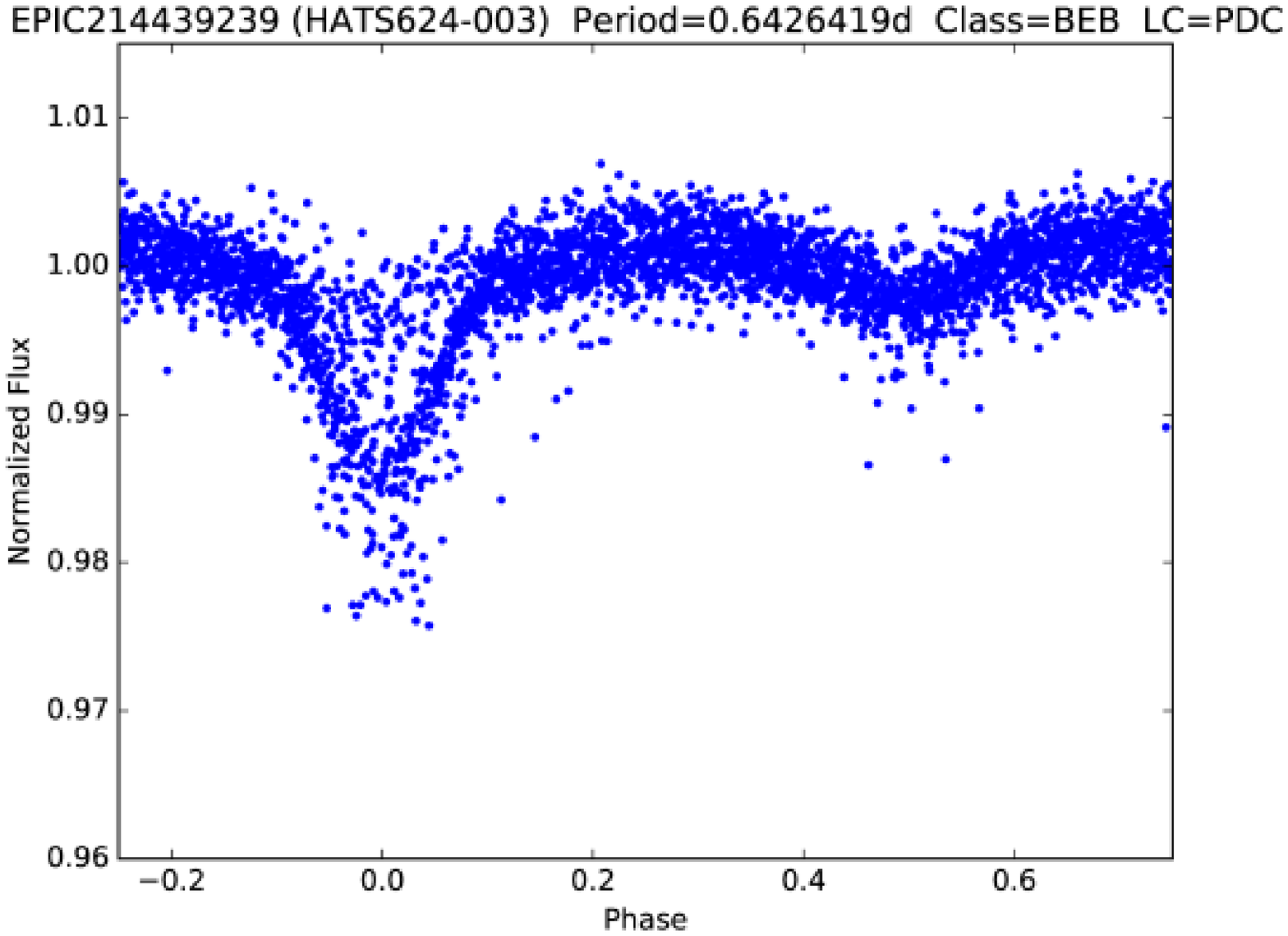}{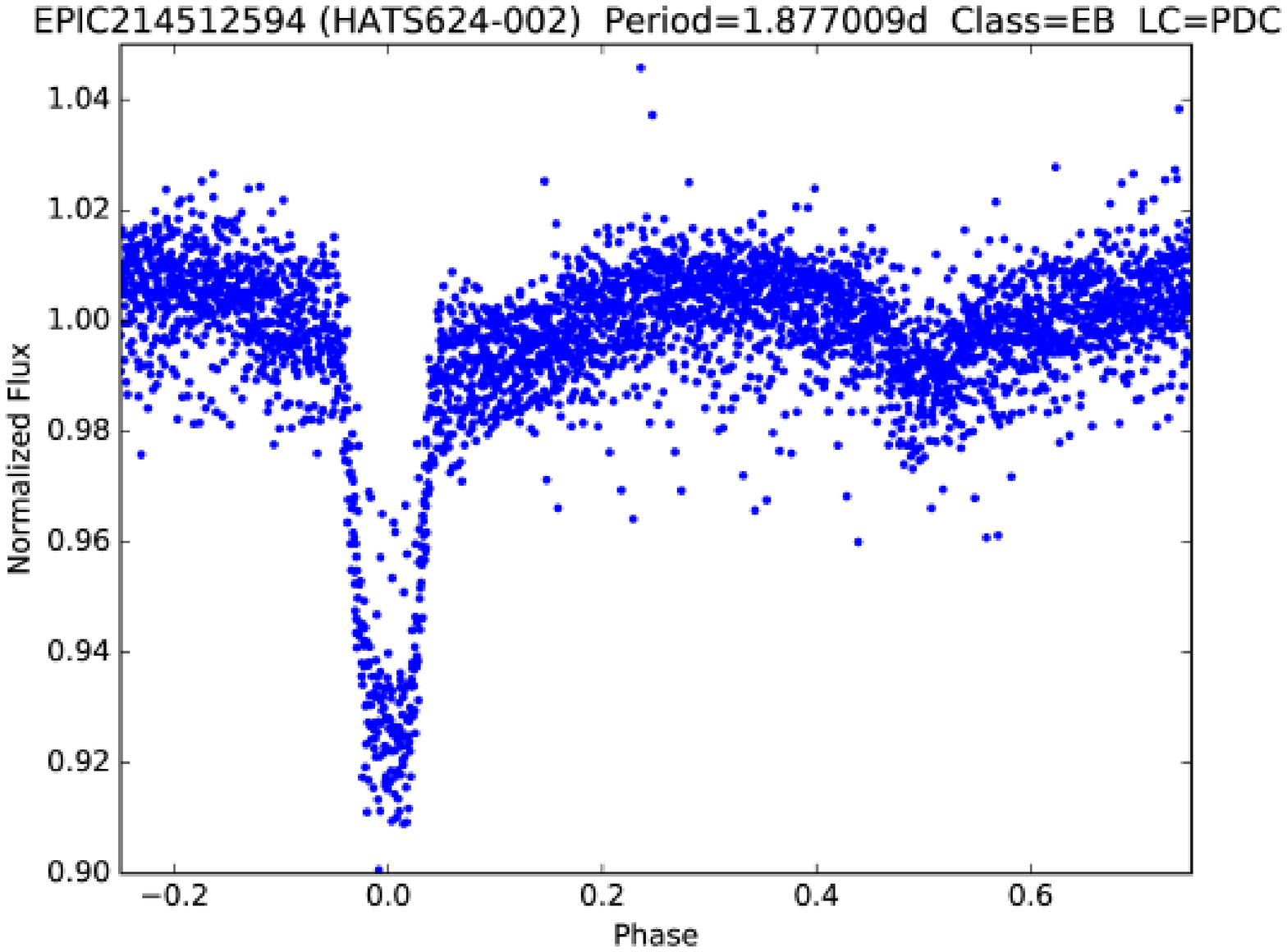}
\plottwo{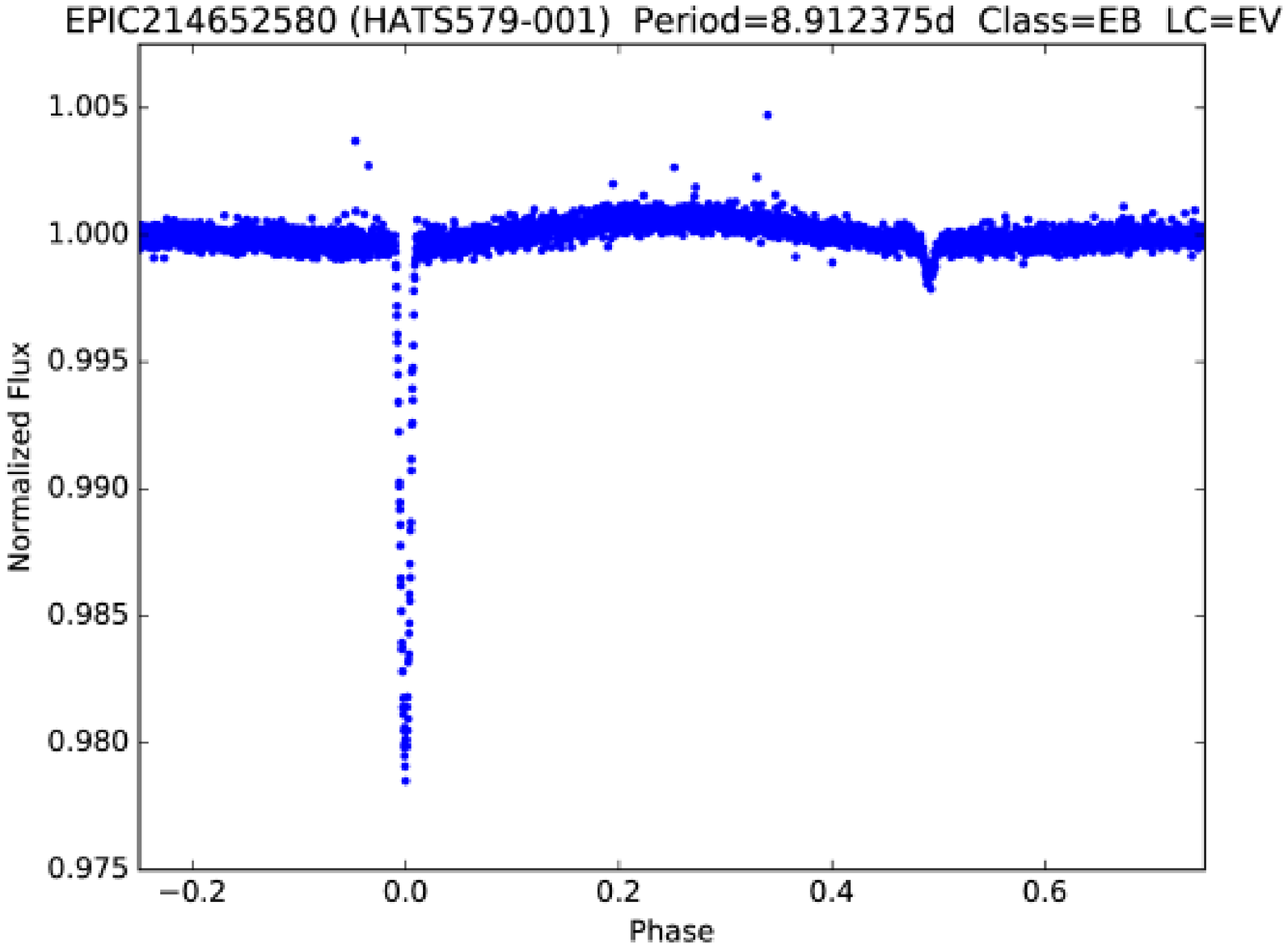}{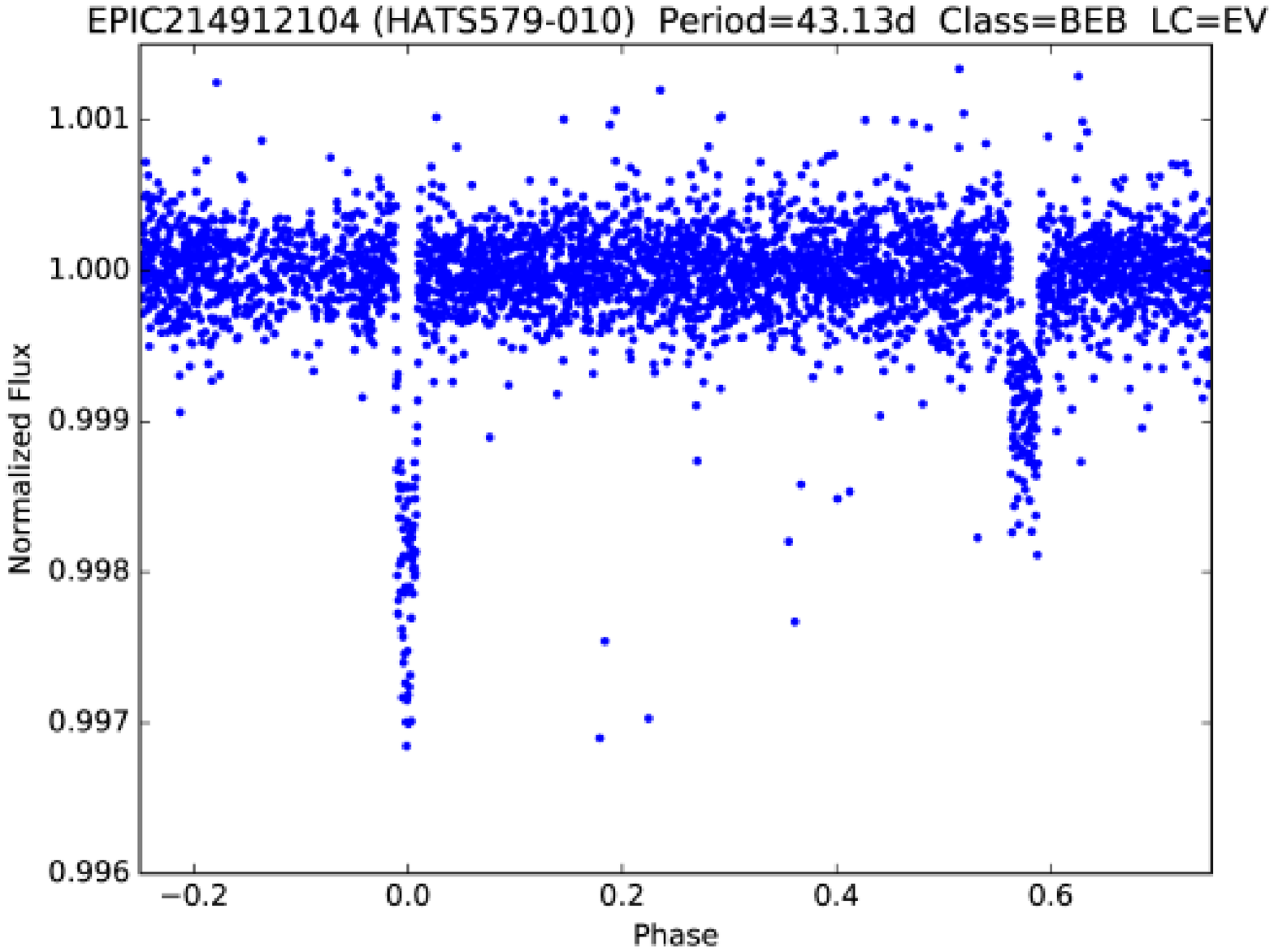}
\plottwo{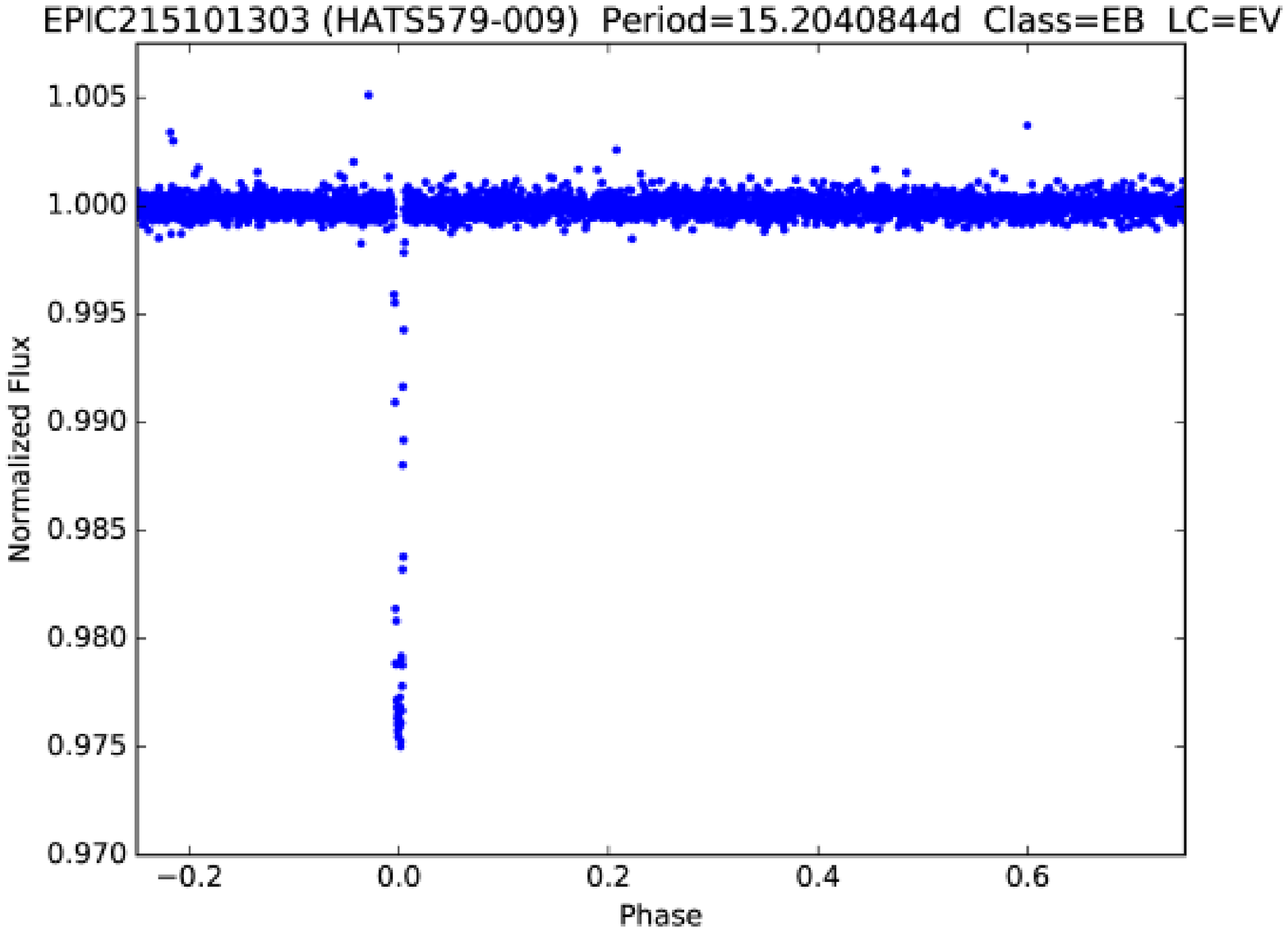}{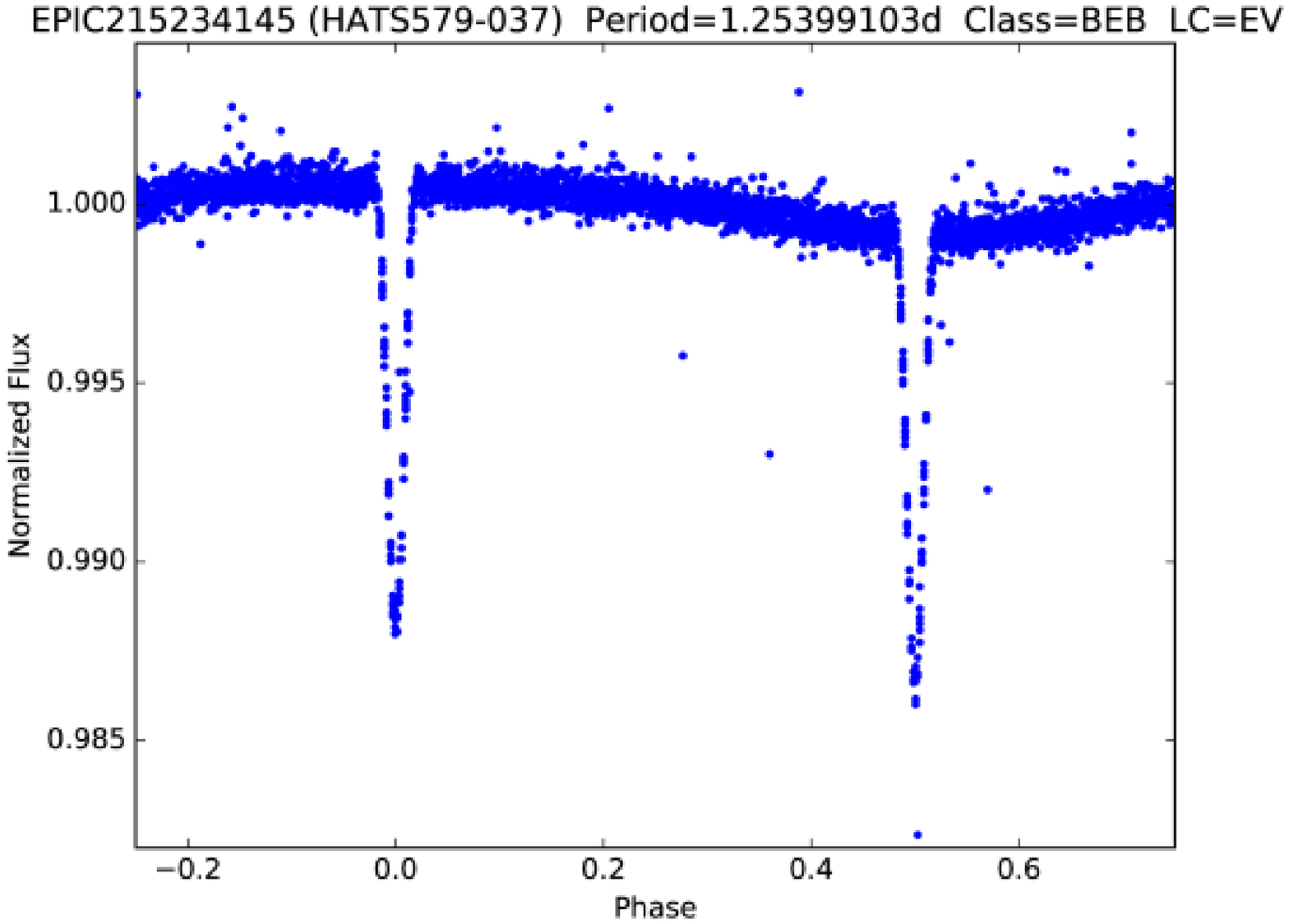}
\plottwo{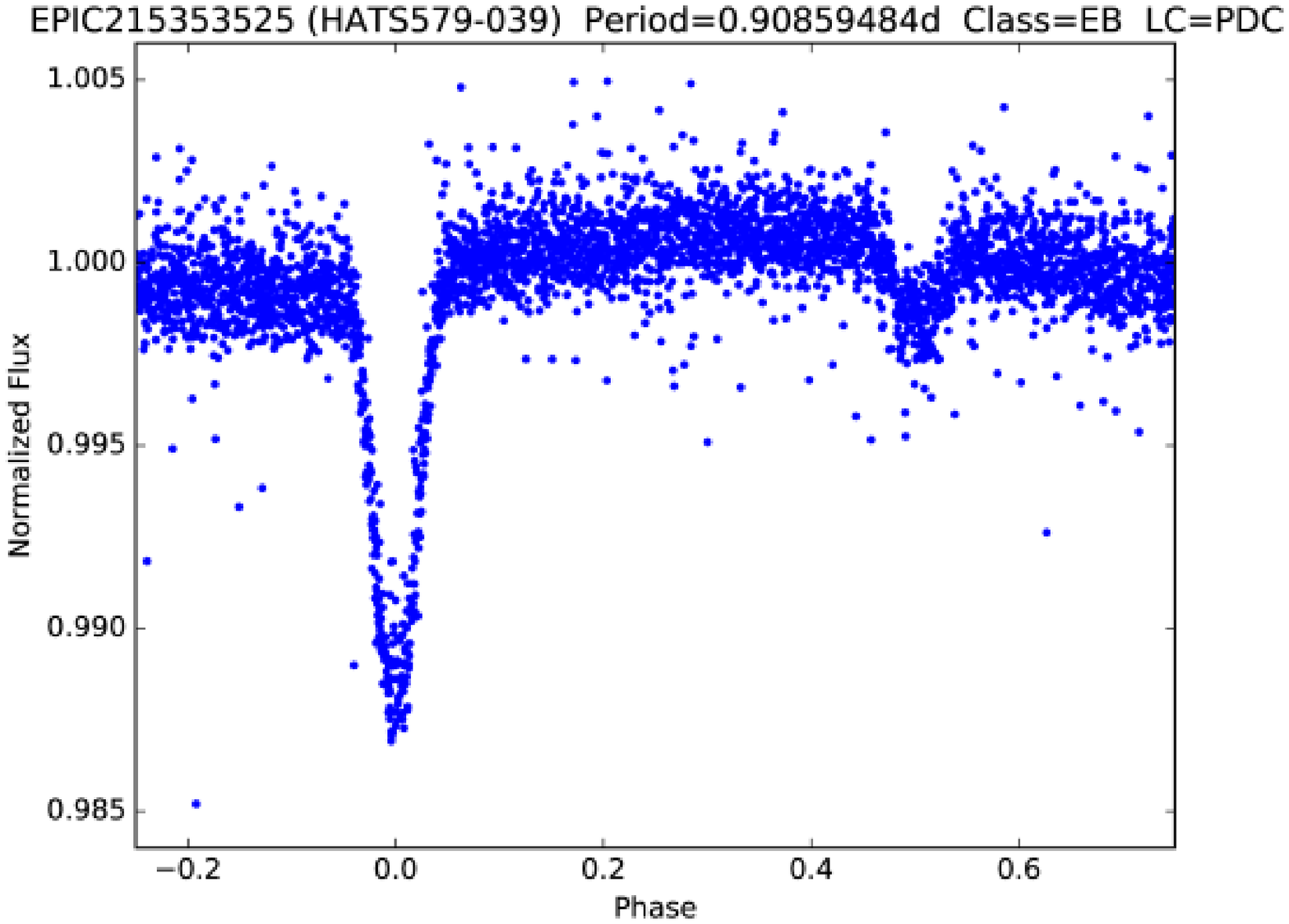}{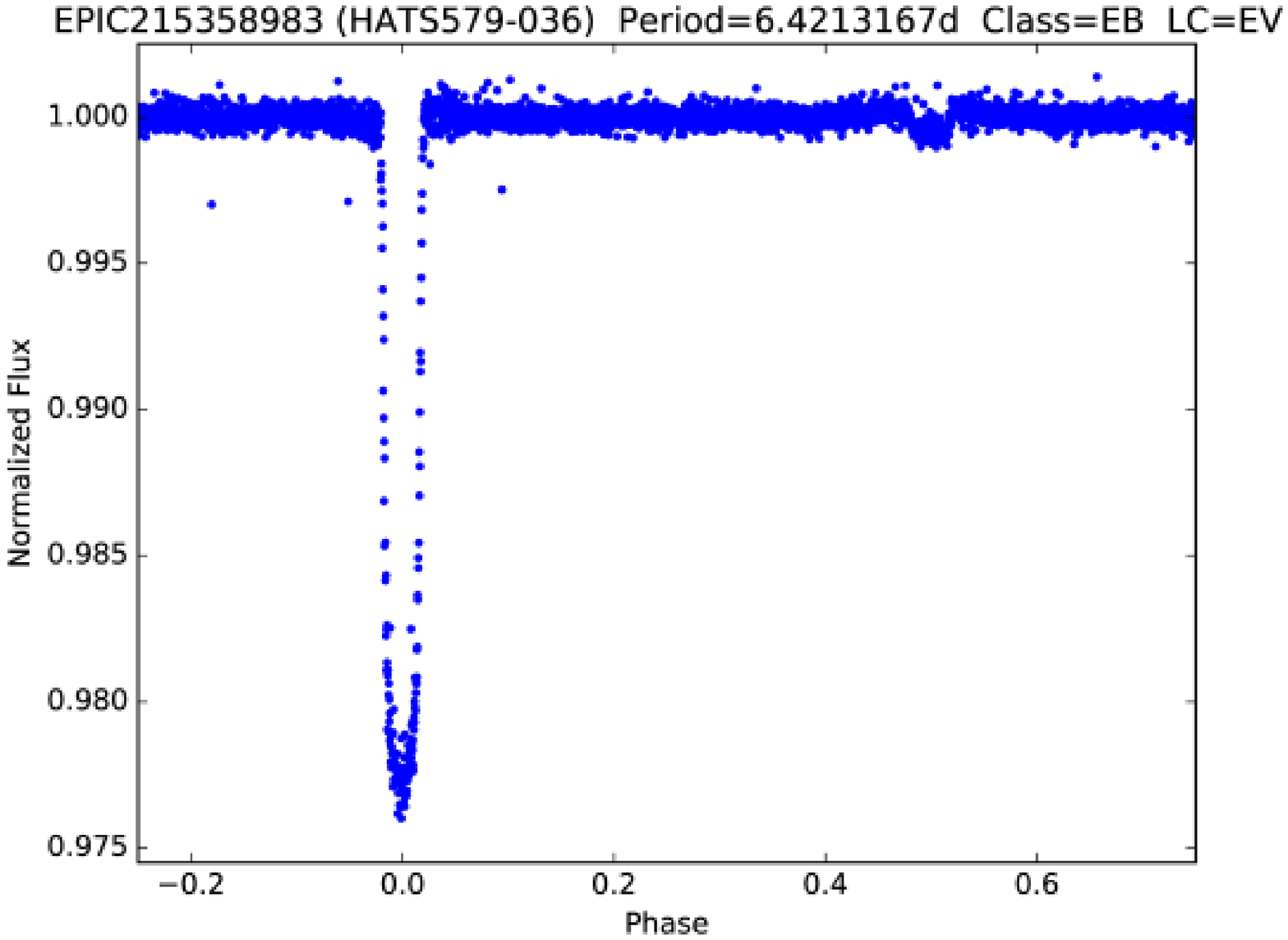}
\plottwo{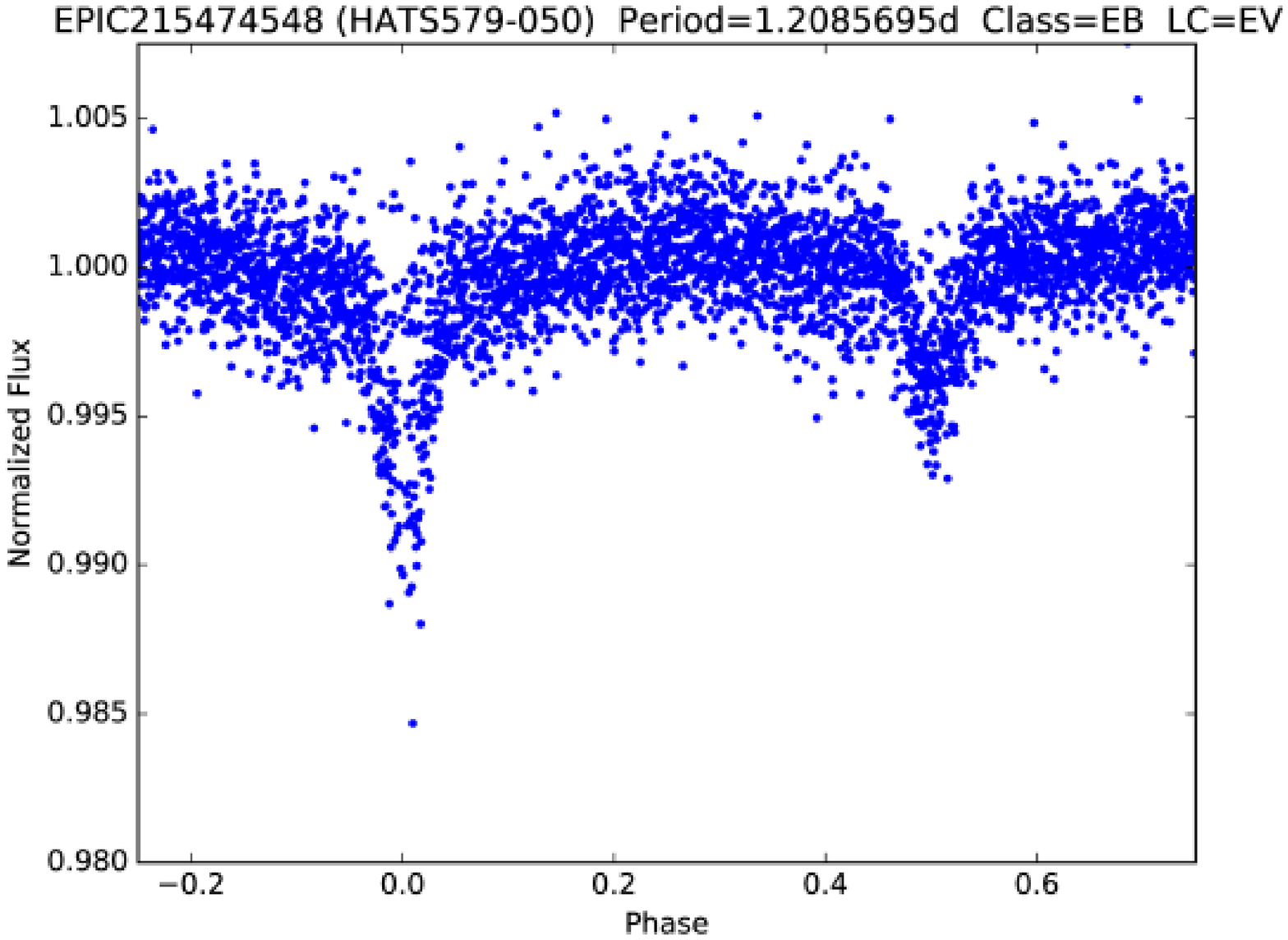}{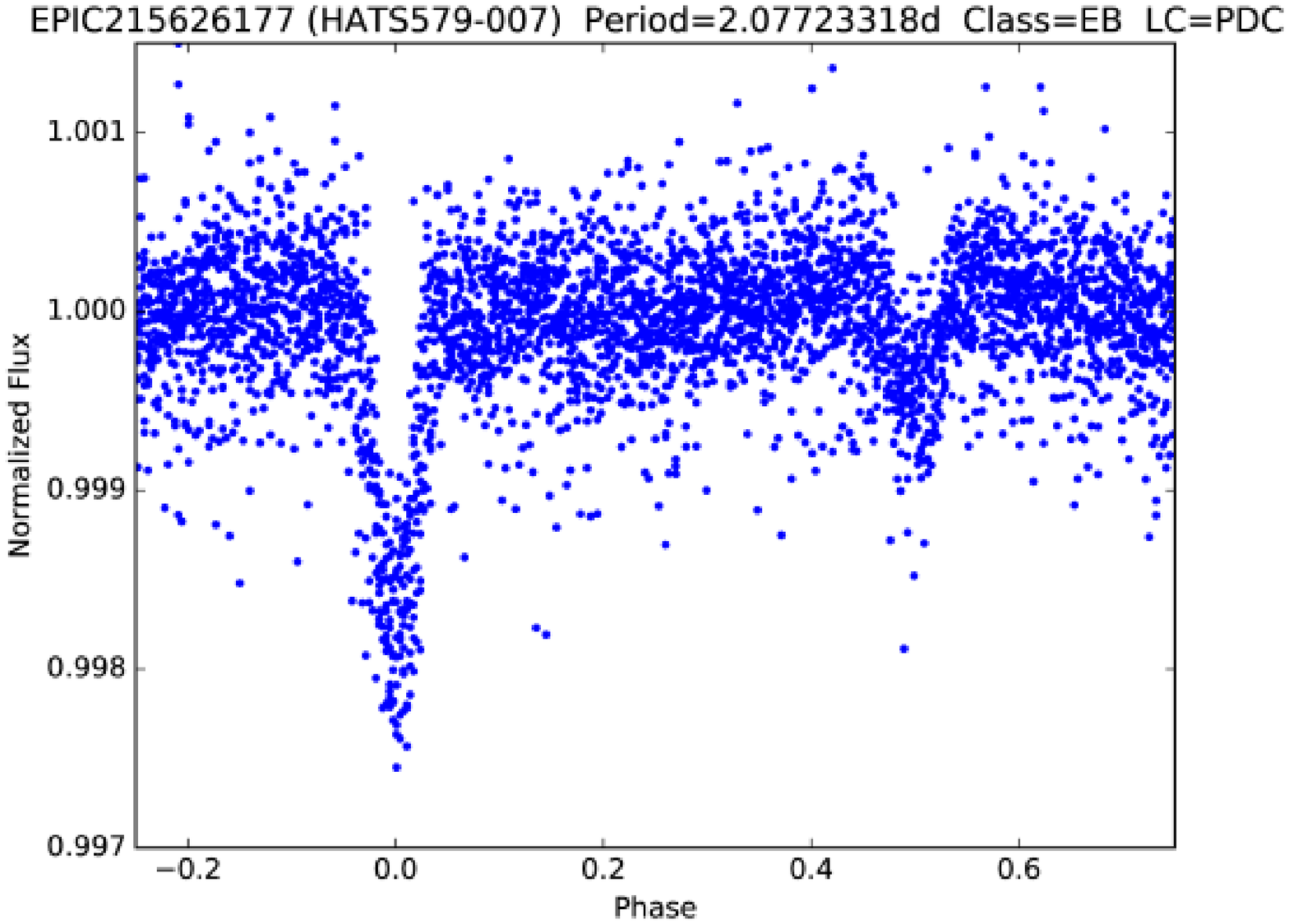}
\plottwo{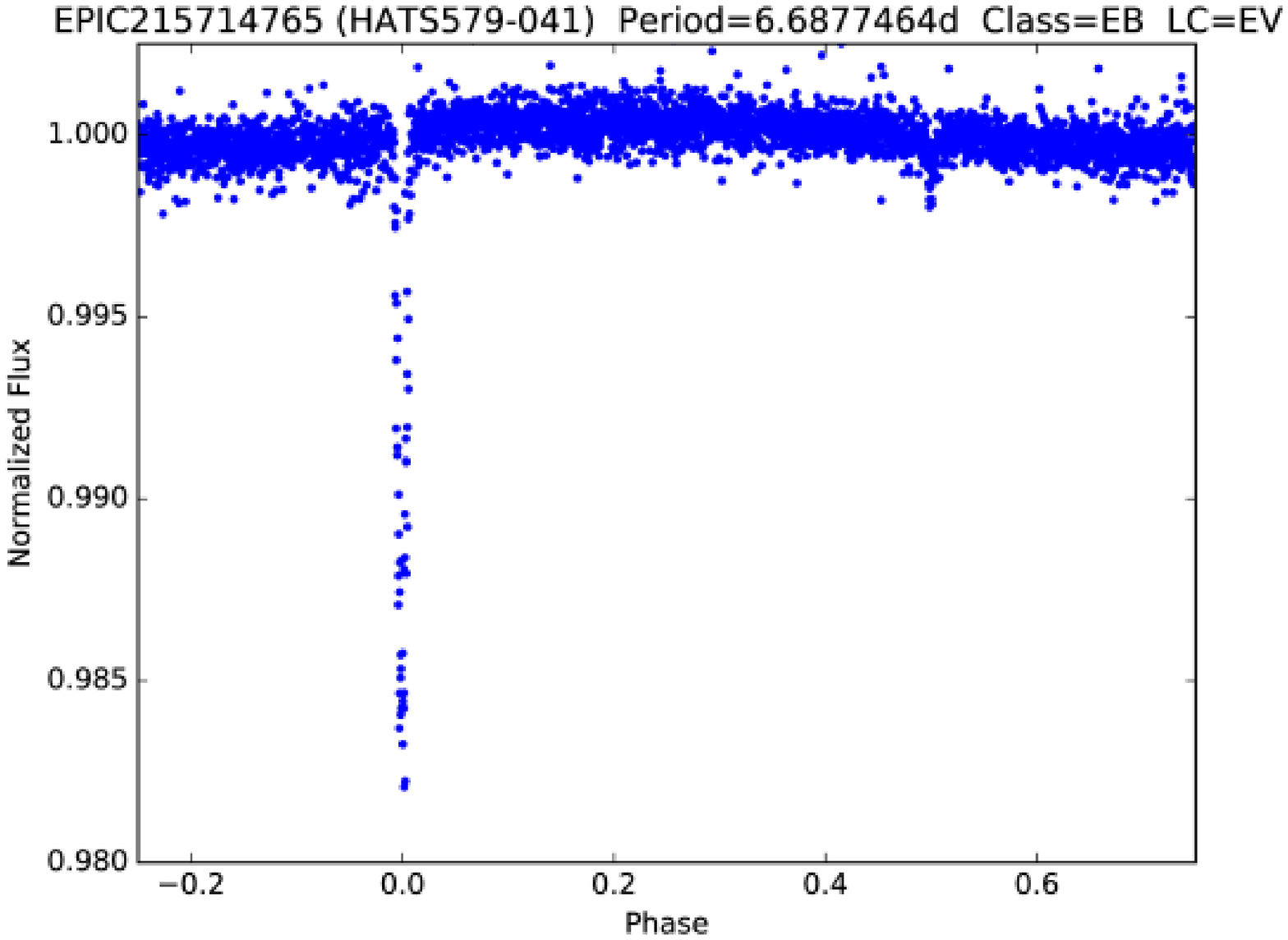}{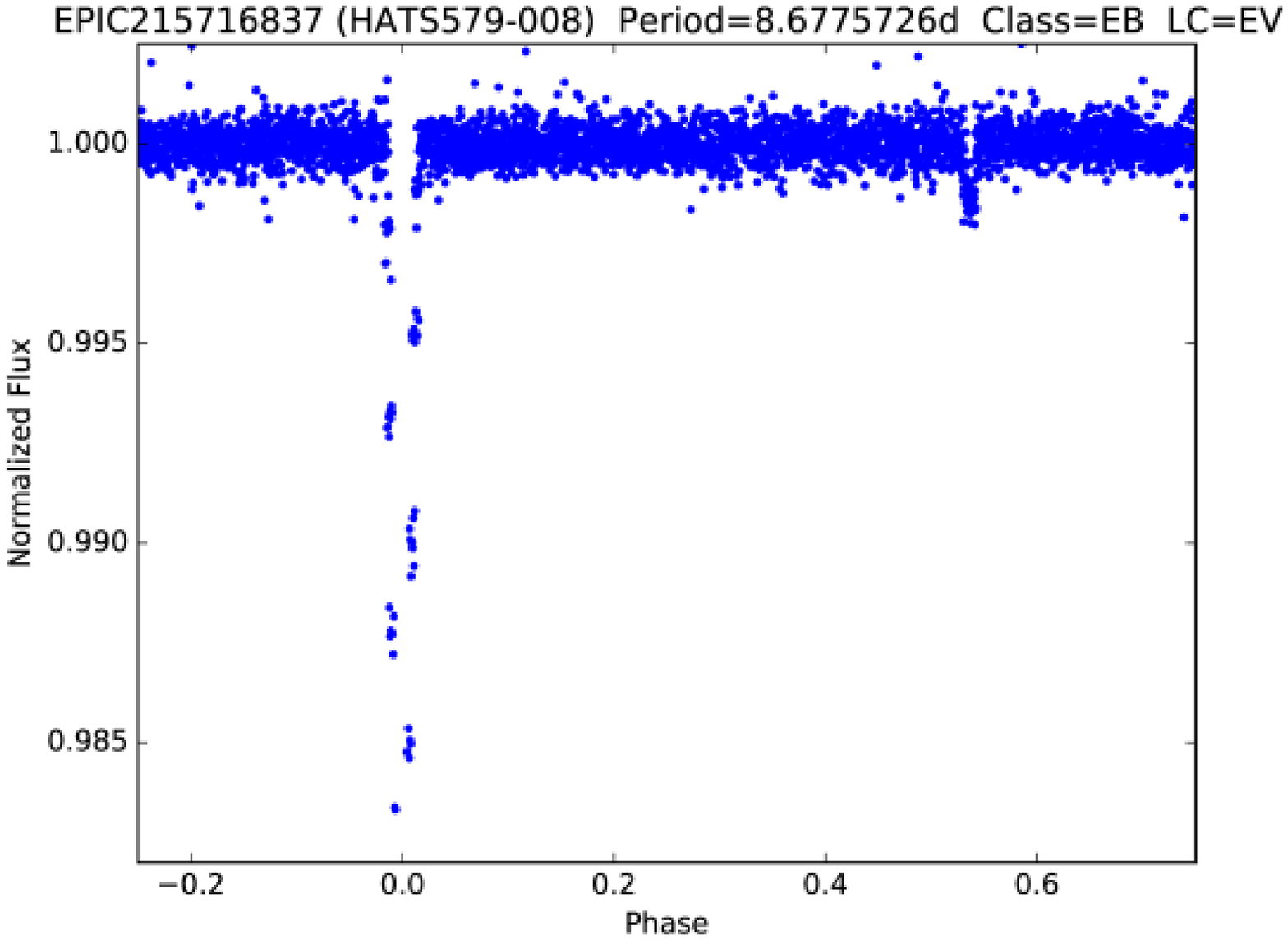}
\plottwo{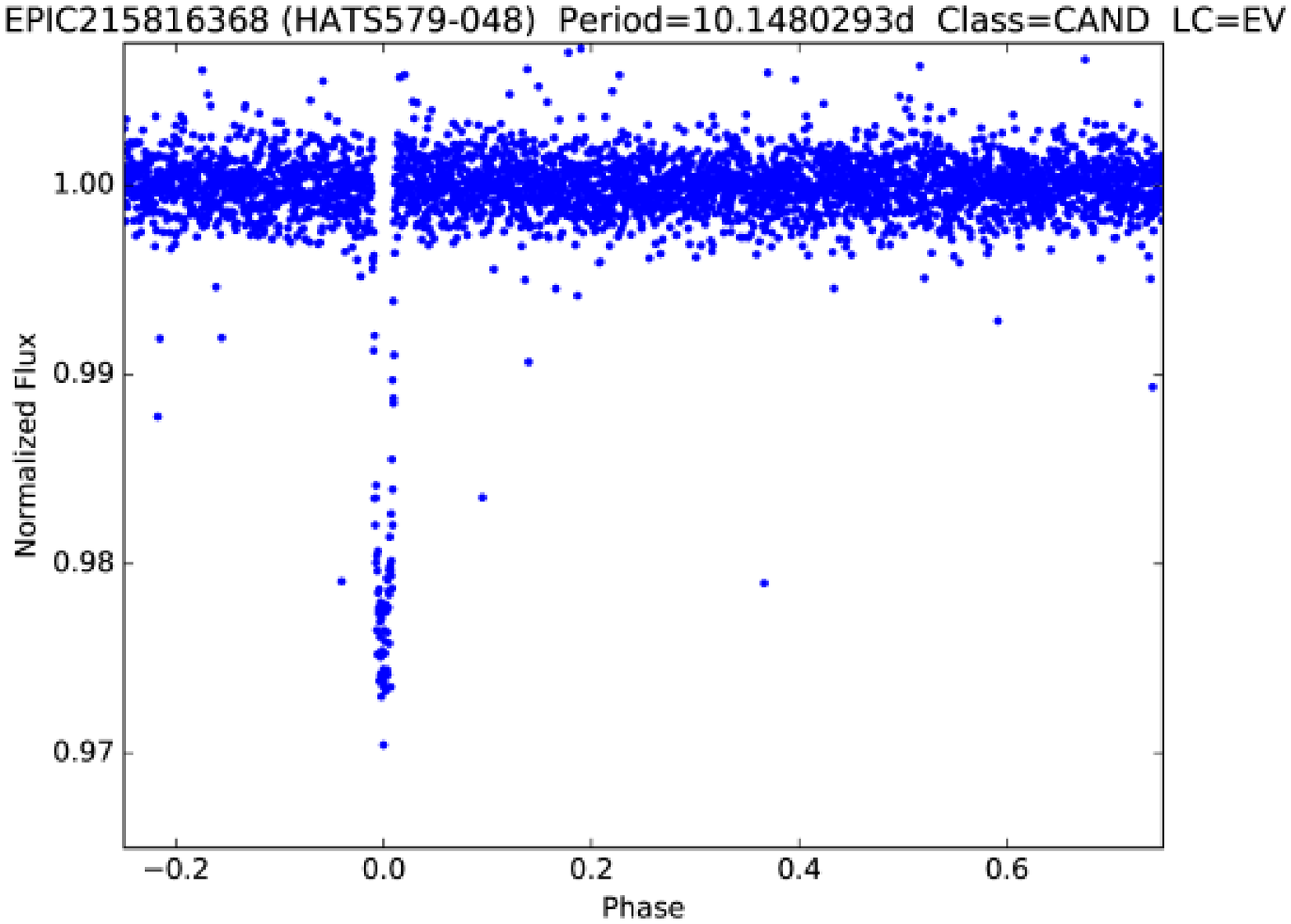}{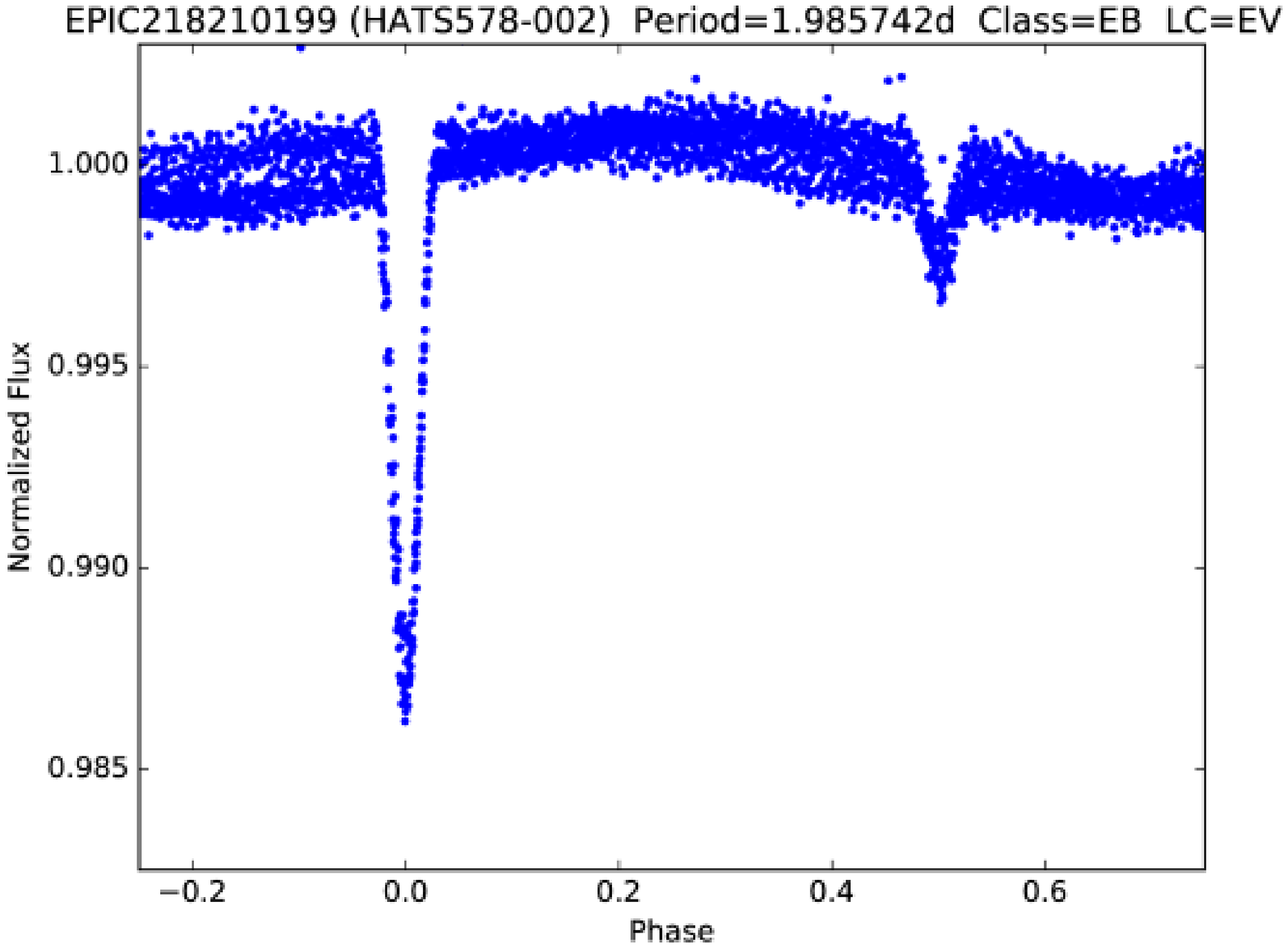}
\plottwo{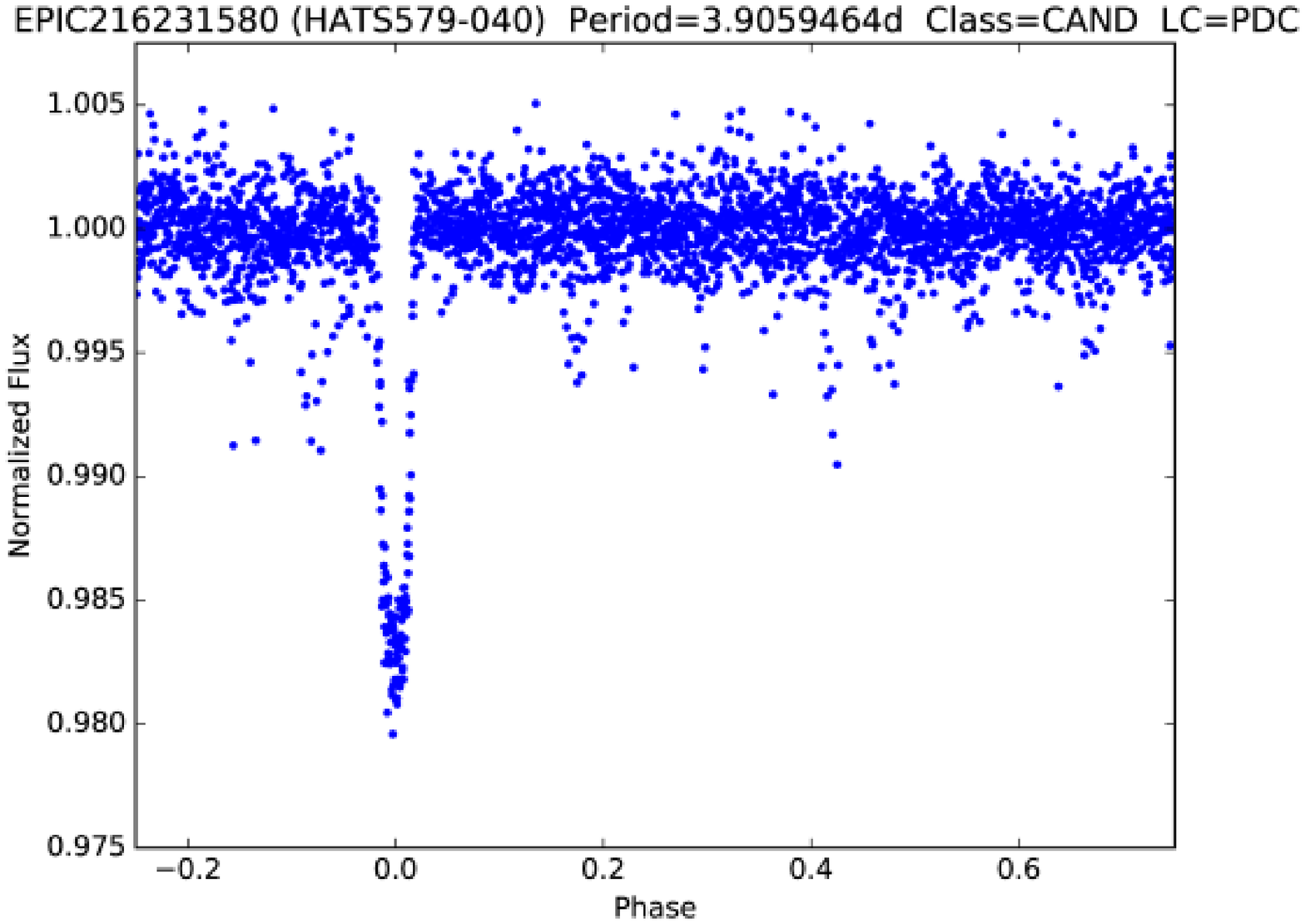}{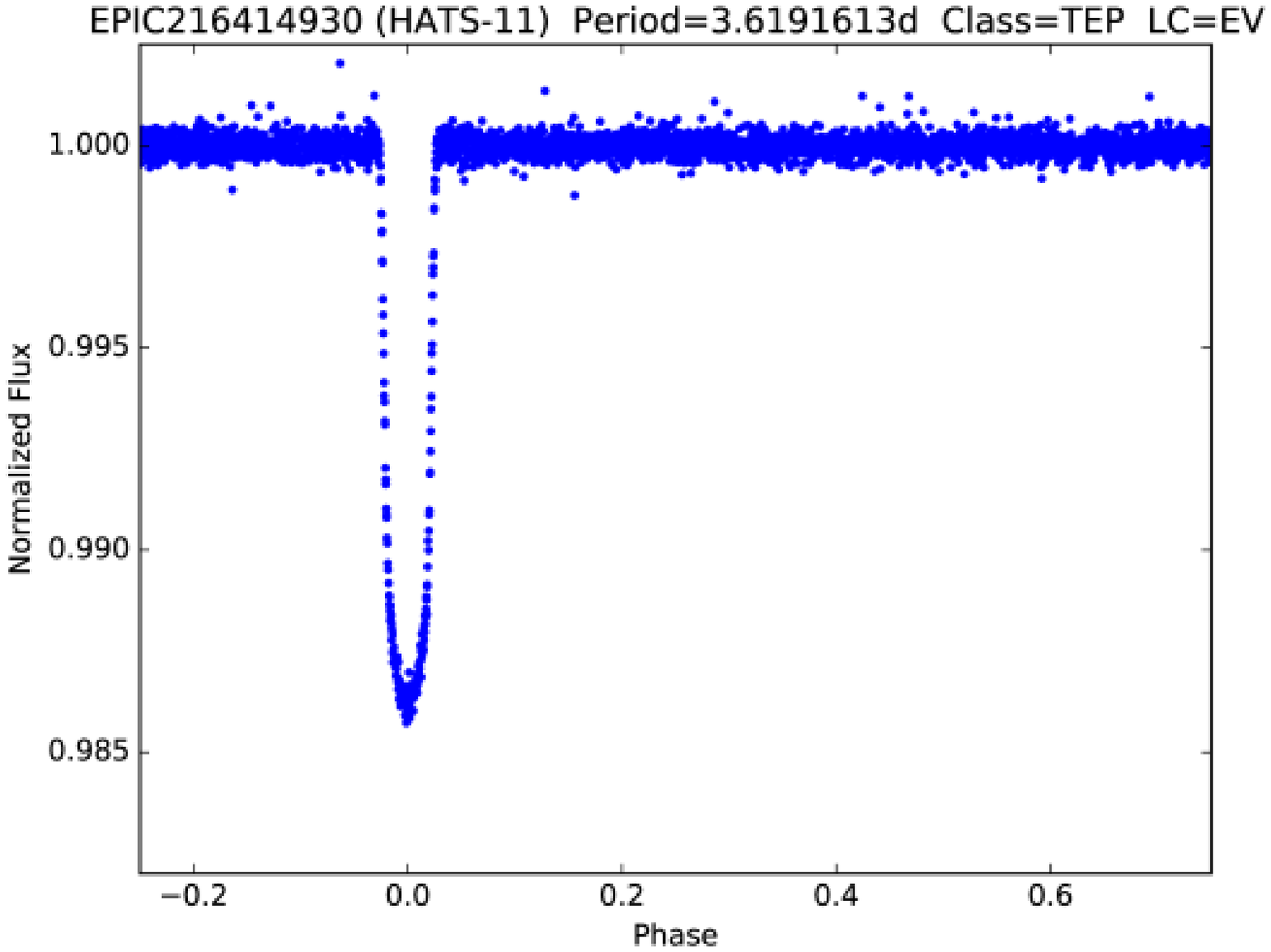}
\plottwo{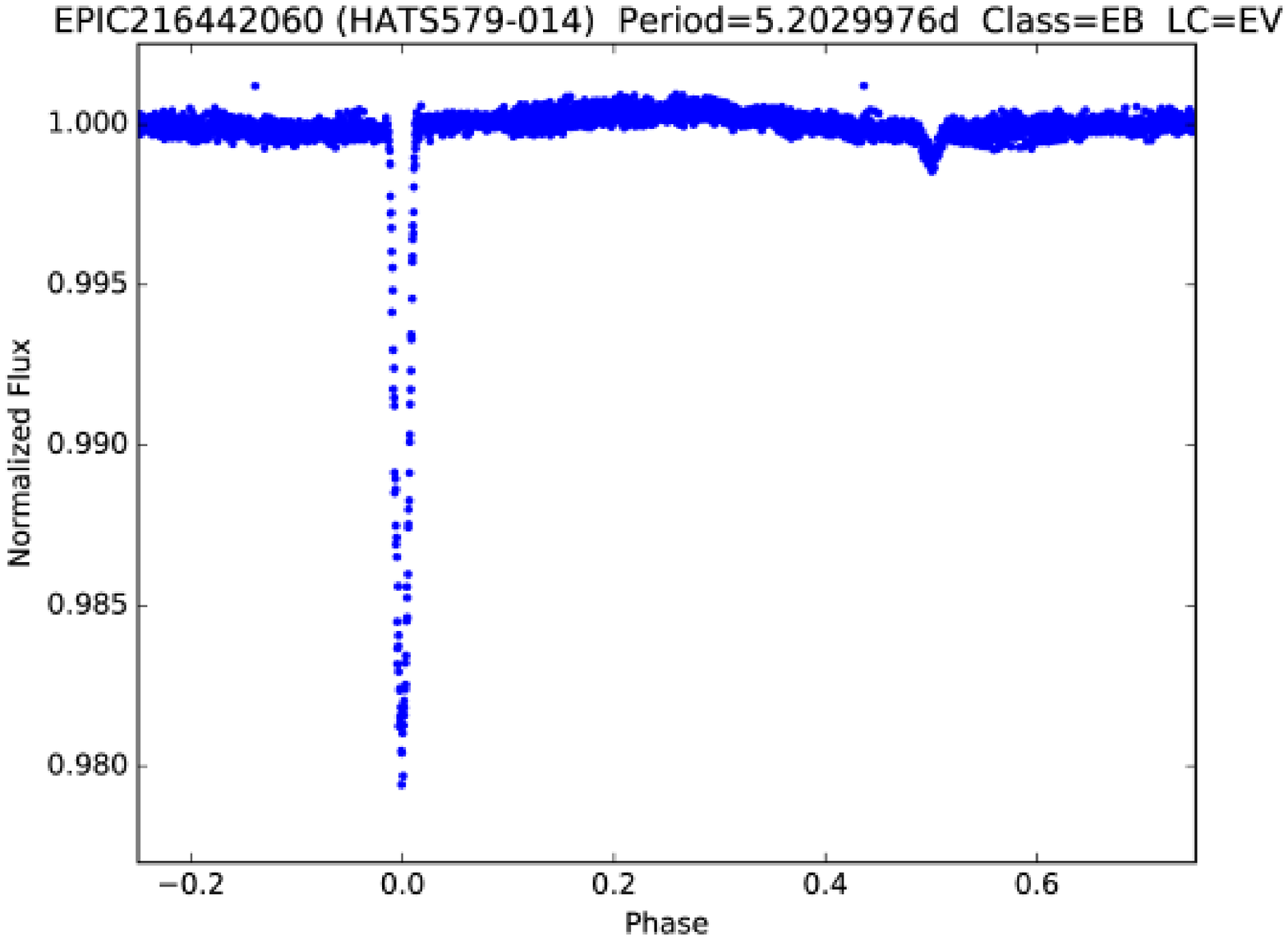}{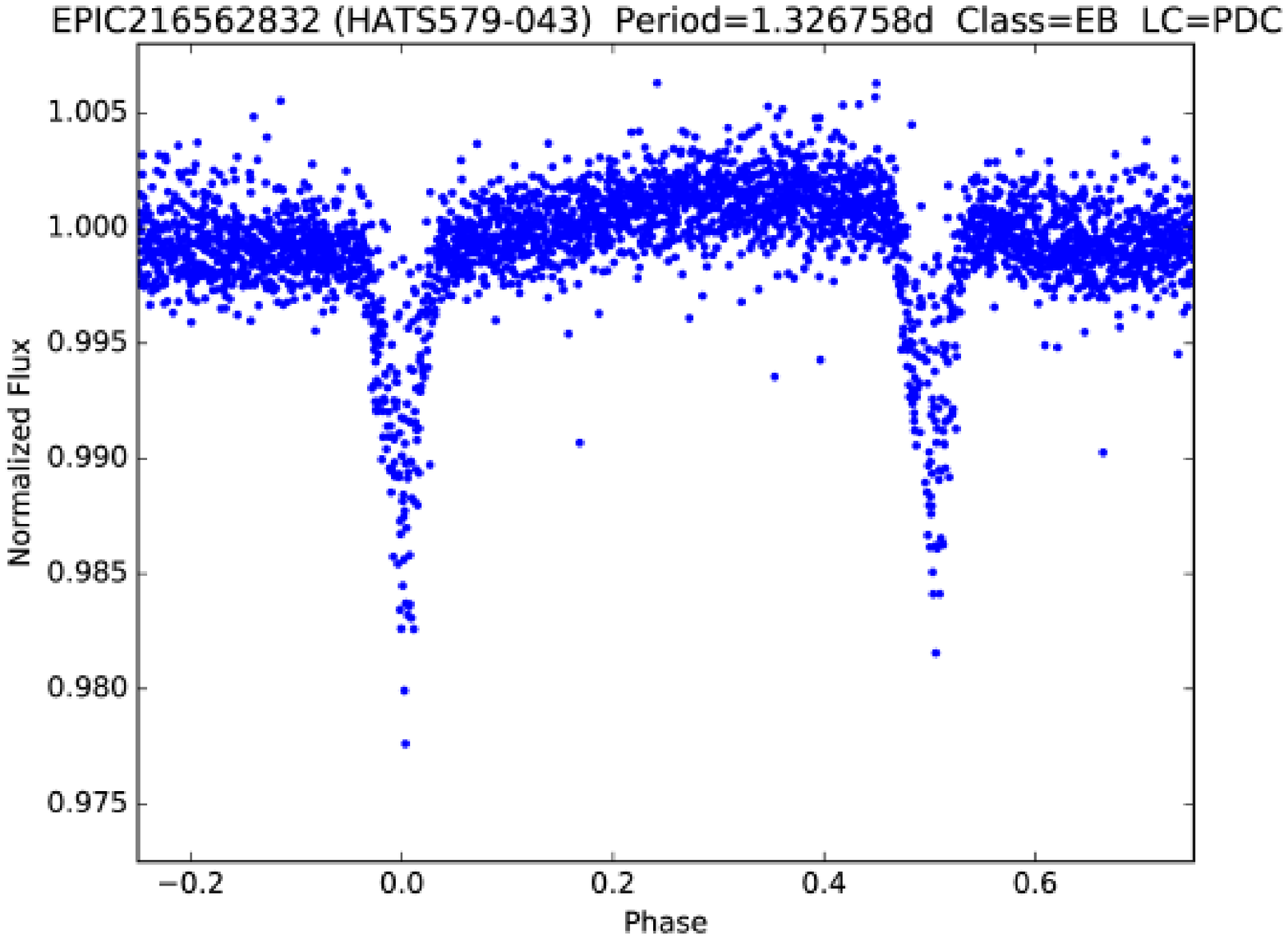}
\plottwo{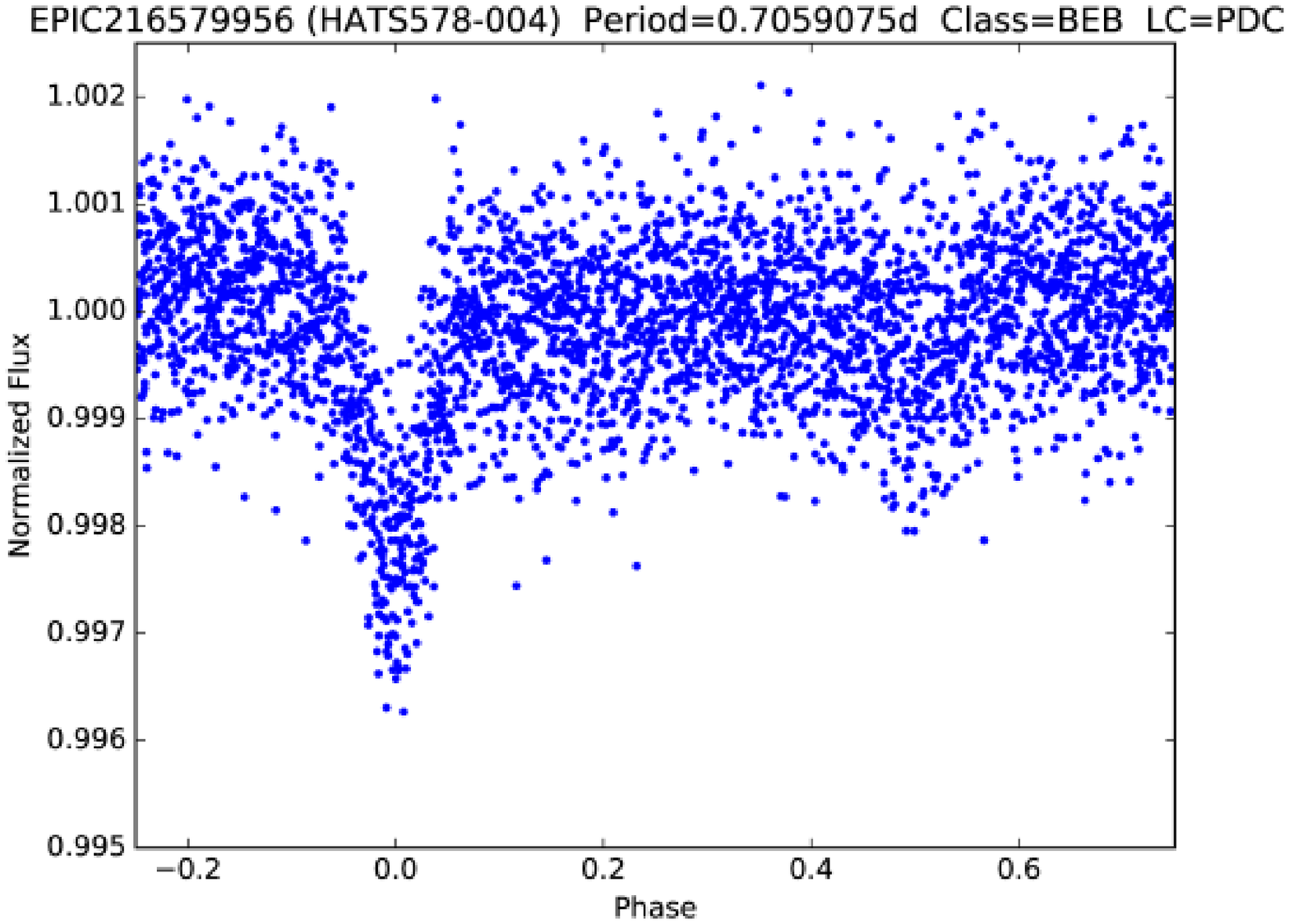}{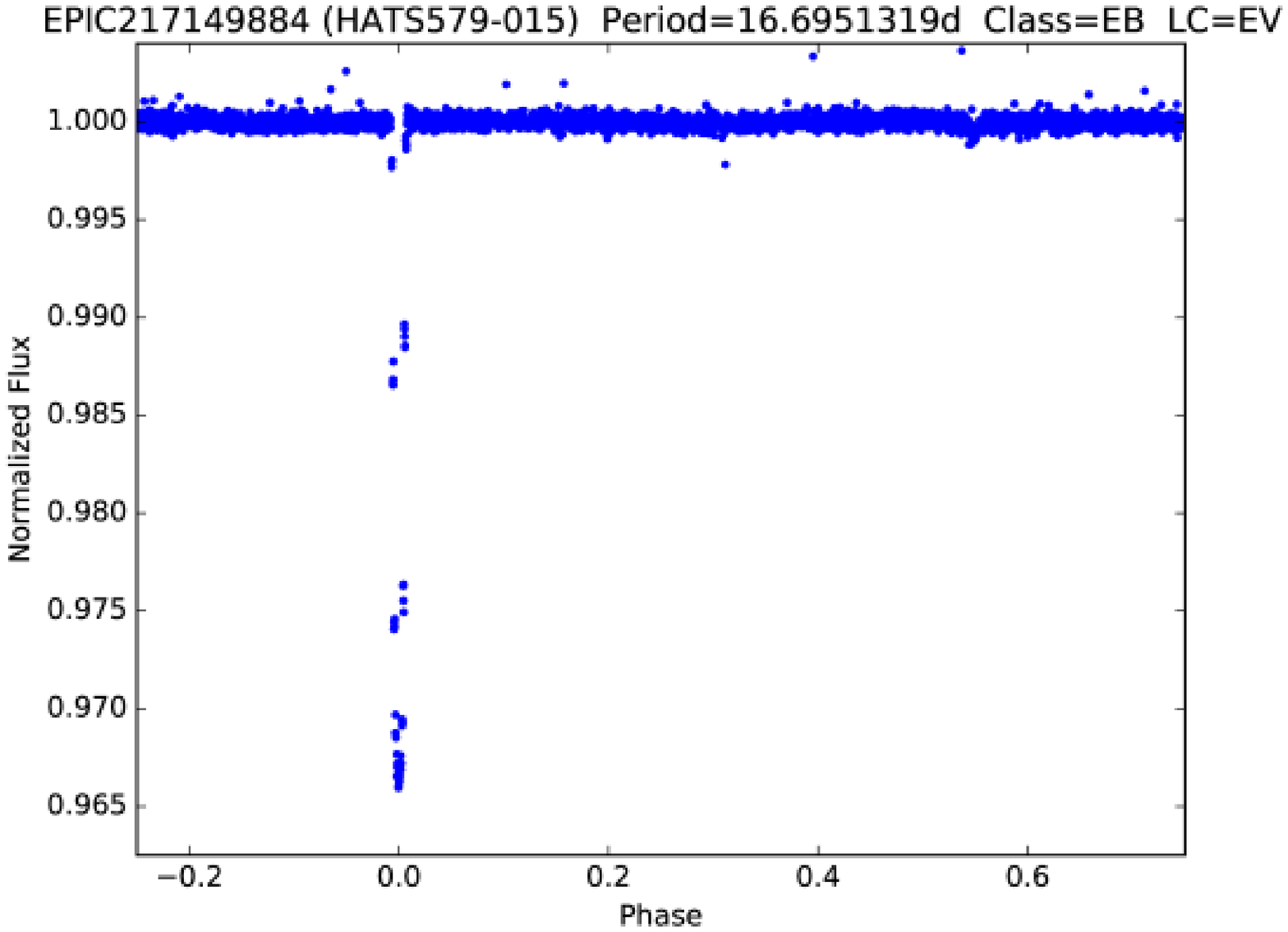}
\plottwo{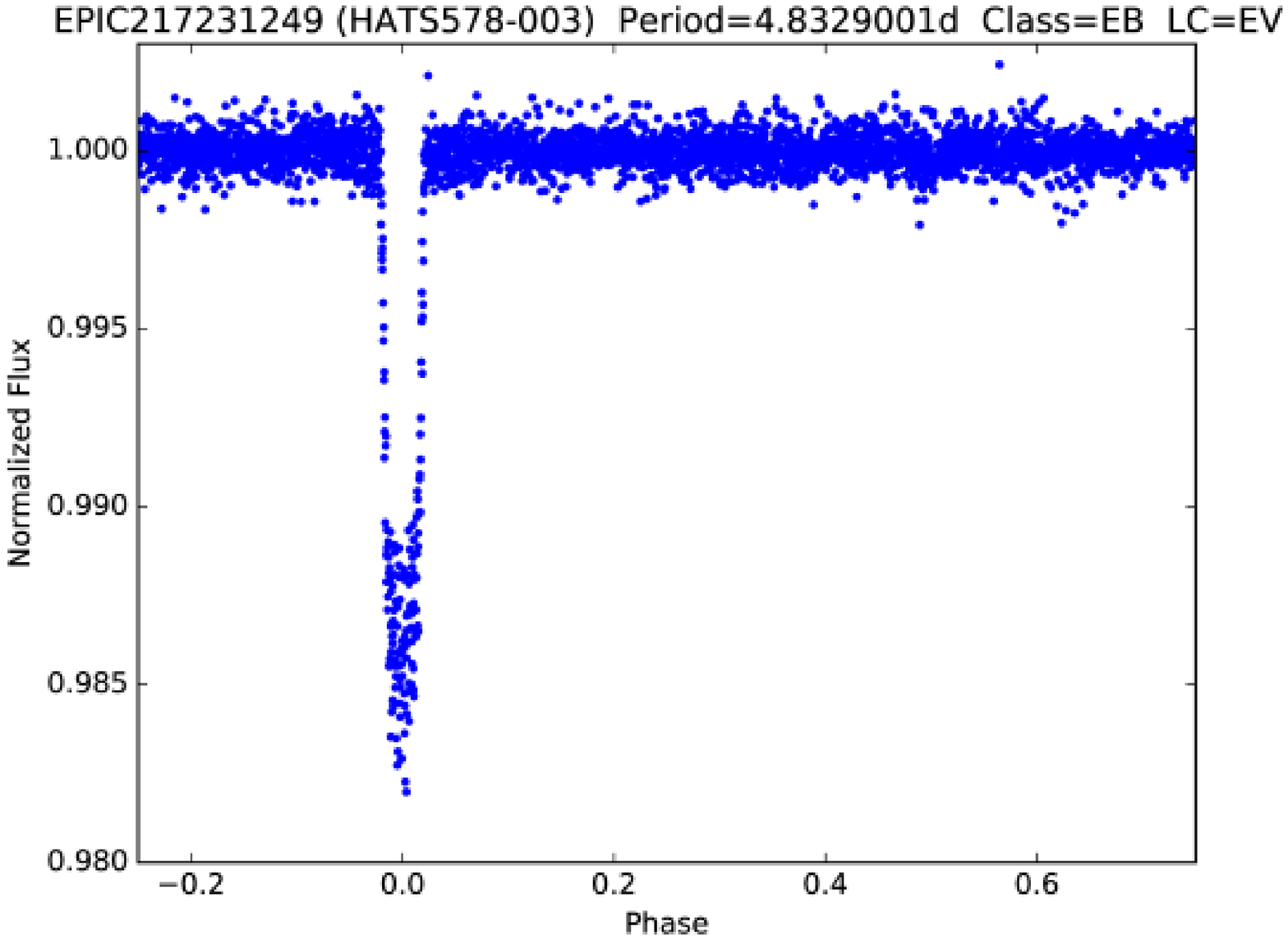}{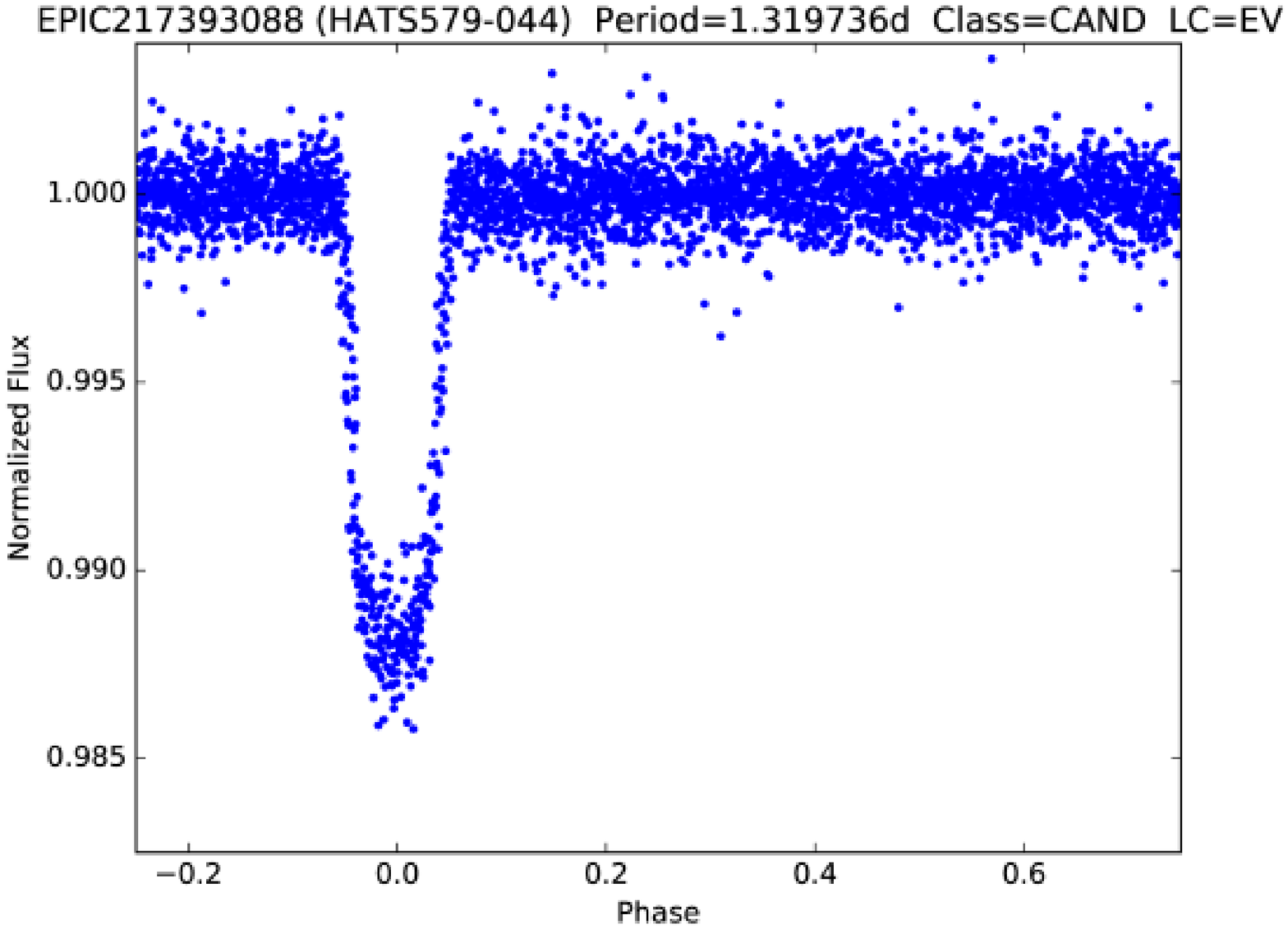}
\plottwo{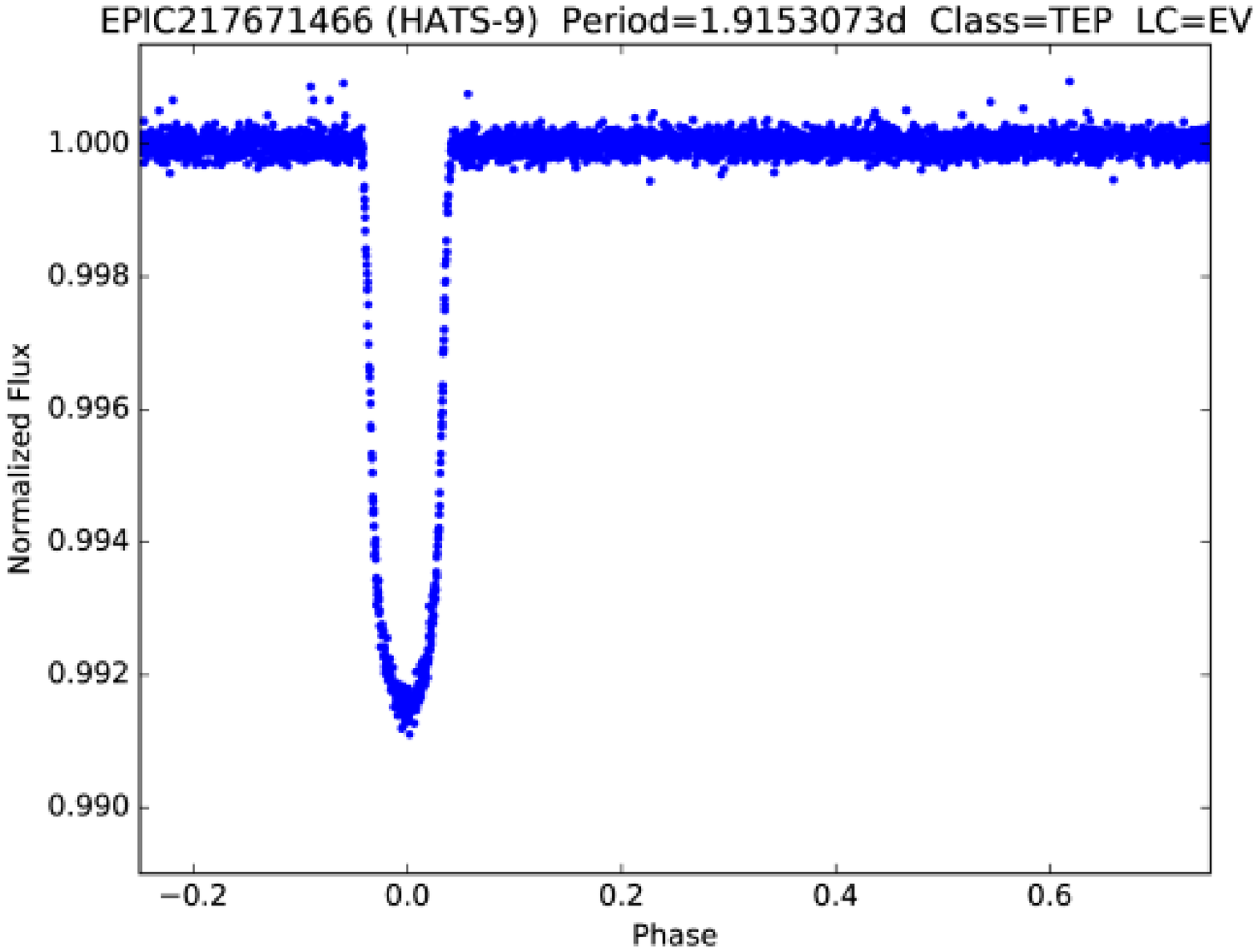}{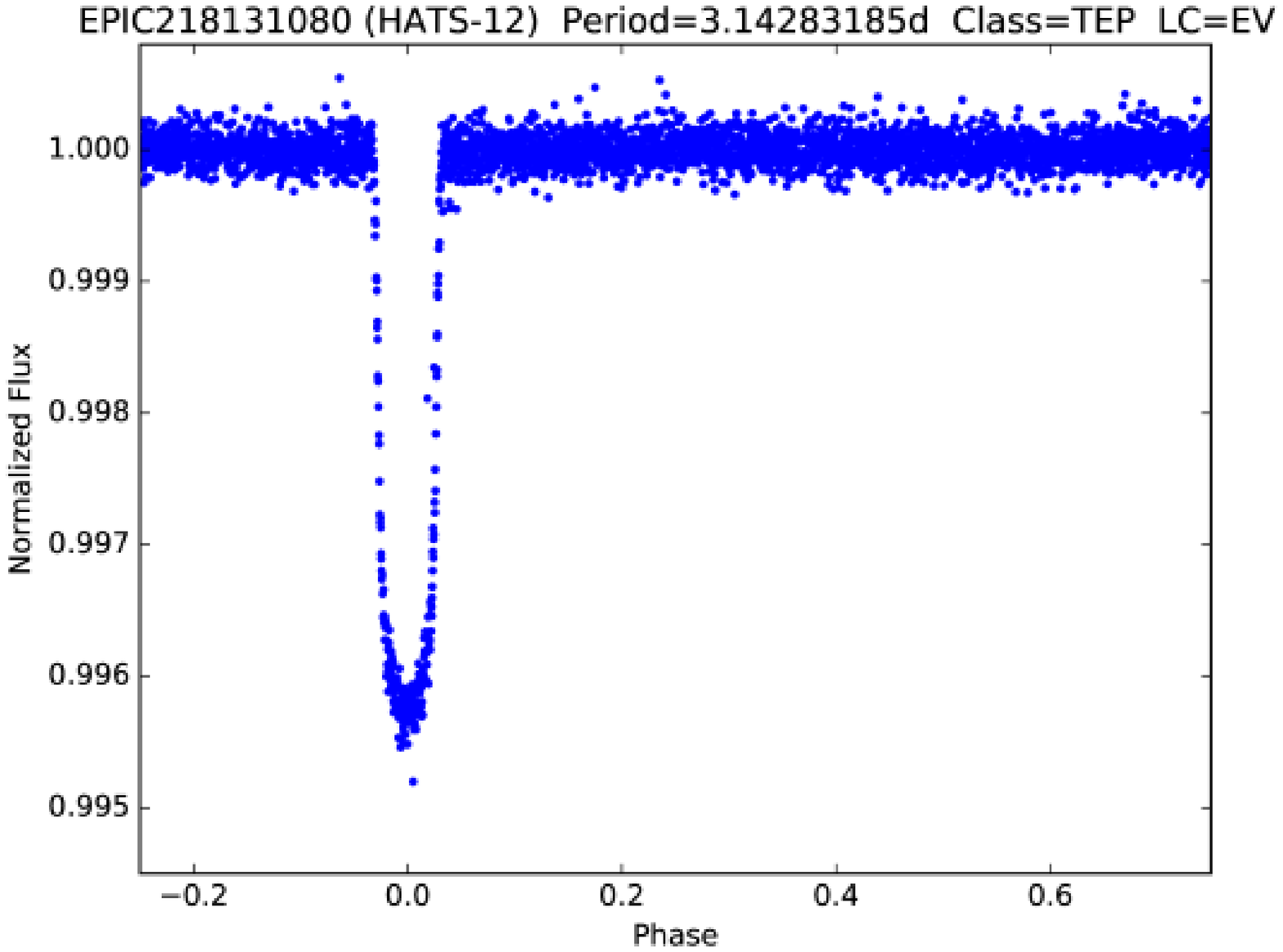}
\end{document}